\documentclass[10pt,twocolumn]{article}

\usepackage[a4paper,margin=1in]{geometry}
\usepackage[T1]{fontenc}
\usepackage{hyperref}
\usepackage{graphicx}
\usepackage{amsmath,amsfonts,amssymb,amsthm}
\usepackage[numbers,sort&compress]{natbib} 
\usepackage{booktabs}
\usepackage{multirow}
\usepackage{subcaption}
\usepackage{etoc}

\usepackage[table,x11names]{xcolor}
\definecolor{red}{HTML}{cc1100}

\usepackage{pifont}

\title{JetParticle-JEPA: An Efficient Self-Supervised Representation Learning method for Jet Tagging in High-Energy Physics}
\author{
Guillaume Letellier$^1$ and Antonin Vacheret$^2$ and Frederic Jurie$^1$\\
\small $^1$ GREYC, Normandy University, Unicaen, ENSICAEN, UMR CNRS 6072, F-14000 Caen, France\\
\small $^2$ LPC, Normandy University, Unicaen, ENSICAEN, IN2P3, UMR CNRS 6534, F-14000 Caen, France\\
{\tt\small{\{firstname.lastname\}@unicaen.fr},
vacheret@lpccaen.in2p3.fr}
}
\date{}

\begin{document}
\maketitle


\begin{abstract}
Jet tagging at the Large Hadron Collider increasingly relies on deep learning models trained on massive simulated datasets, leading to high computational costs and limited robustness to detector mismodeling. We introduce JetParticle-JEPA (JP-JEPA), a self-supervised Joint-Embedding Predictive Architecture that learns physically meaningful jet representations directly from continuous particle clouds without tokenization or reconstruction of raw inputs. Built on a Particle Transformer backbone, JP-JEPA predicts latent representations of masked particles while preserving fine-grained kinematic correlations. On the JetClass benchmark, JP-JEPA achieves performance comparable to fully supervised state-of-the-art methods on the full dataset, surpasses supervised baselines in low-label regimes, and significantly outperforms existing SSL approaches. On Top Quark and Quark-Gluon Tagging benchmarks, it remains on par with supervised methods. The learned representations also exhibit strong robustness to missing detector information and improved uncertainty behavior, highlighting JP-JEPA as a promising foundation-model framework for robust and data-efficient jet physics at the LHC.
\end{abstract}

\section{Introduction}
\label{sec:intro}

The quest to unravel the fundamental laws of nature at the Large Hadron Collider (LHC) depends increasingly on the ability to extract subtle signals from petabytes of particle collision data. Central to this endeavor is the analysis of jets—collimated sprays of hadrons produced by the fragmentation of high-energy quarks and gluons. Identifying the elementary particle that initiated a jet, a task known as jet tagging, is essential for separating rare physics processes, such as the production of the Higgs boson or top quarks, from the overwhelming background of quantum chromodynamic (QCD) processes. Deep learning has recently revolutionised this field, with architectures evolving from simple dense networks to sophisticated geometric models that process jets as point clouds or graphs~\cite{guest2018deep, shlomi2020graph, qu2022particle}. 

Despite these algorithmic advances, the deployment of deep learning in High-Energy Physics (HEP) faces a critical and growing bottleneck: an unsustainable reliance on supervised learning at scale. Recent work~\cite{bhimji2025omnilearned} demonstrates that reaching peak performance requires training state-of-the-art models on up to billions of labeled jet examples. These labels are inherently unobservable in real data and must be generated using sophisticated Monte Carlo simulations, consuming almost half of the vast compute power available for LHC analyses. Furthermore, this heavy reliance on ``virtual twins'' creates a difficult-to-quantify ``reality gap''. Synthetic events often differ from real-world data distributions, as they depend on theoretical approximations and the simulated fidelity of the instrument geometry. Consequently, models trained purely via supervised learning tend to overfit to simulation-specific artifacts, slowing down their deployment on real data and requiring complex calibrations or domain adaptation strategies~\cite{karagiorgi2022machine} to mitigate their intrinsic fragility.

To fully exploit the discovery potential of the LHC and its upgrade (High-Luminosity LHC), a paradigm shift is urgently needed. The HEP community requires learning frameworks capable of extracting robust, physically meaningful representations from limited simulated data and, ultimately, from vast streams of unlabeled experimental data. Inspired by the transformative success of Foundation Models in natural language processing and computer vision~\cite{radford2018improving, devlin2018bert, brown2020language}, Self-Supervised Learning (SSL) has emerged as a highly promising path toward universal jet representations. 

Early efforts to bring SSL to jet physics have largely adapted generative paradigms, such as Masked Autoencoders (MAE)~\cite{he2022masked} or autoregressive transformers~\cite{birk2024omnijet}. For instance, Masked Particle Modeling (MPM)~\cite{golling2024masked} trains a model to reconstruct missing particle attributes, while OmniJet-$\alpha$~\cite{birk2024omnijet} predicts the next constituent in a sequence. However, applying these methods to particle physics presents unique ontological challenges. Unlike discrete text tokens or rigid image pixels, particles are described by continuous kinematic variables ($p_T, \eta, \phi$). To bridge this gap, generative models have often relied on discretization techniques, such as Vector-Quantised Variational Autoencoders (VQ-VAE), to ``tokenize'' the continuous phase space~\cite{golling2024masked, leigh2025tokenization}. This forced quantization inevitably leads to a critical loss of fine-grained quantum mechanical information. Furthermore, generative objectives inherently compel the model to reconstruct every detail of the input. In the context of a particle detector, this forces the network to waste representational capacity on modeling stochastic noise or unpredictable detector artifacts rather than the underlying semantic and physical structure of the jet.

In computer vision, contrastive and predictive methods have successfully bypassed the pitfalls of pixel-level reconstruction \cite{chen2020simple,grill2020bootstrap,chen2021exploring}. Architectures relying on latent-space prediction, such as DINO~\cite{caron2021emerging,oquab2023dinov2,simeoni2025dinov3} and I-JEPA~\cite{assran2023self}, have shown that joint embedding is significantly more robust to noisy data than generative objectives~\cite{van2025joint}. Joint-Embedding Predictive Architectures (JEPAs) thus appear conceptually ideal for HEP. Yet, the few recent works pursuing this approach~\cite{katel2024learning,bardhan2025hep} rely heavily on compressing the input information prior to processing—either by restructuring particles into macroscopic subjets~\cite{katel2024learning} or by constructing localized spatial patches~\cite{bardhan2025hep}. While computationally convenient, this obscures the individual particle-to-particle interactions that are critical for identifying complex decay signatures~\cite{qu2022particle}.

To overcome these fundamental limitations, we introduce \textit{JetParticle-JEPA} (JP-JEPA), a self-supervised foundation model framework that learns directly from the raw set of constituent particles without any intermediate clustering, voxelization, or discrete tokenization. Our approach adapts the JEPA paradigm~\cite{assran2023self} to the irregular domain of particle clouds. Unlike generative methods, JP-JEPA trains a student encoder to predict the representations of masked particles in a high-level, abstract latent space produced by a momentum-updated teacher network. By completely avoiding input-space reconstruction, this predictive objective allows the model to actively ignore low-level instrumental noise, focusing instead on learning the global geometry and relativistic kinematics that define the physical properties of the jet.

\begin{figure*}[ht!]
    \centering
    \includegraphics[width=\linewidth]{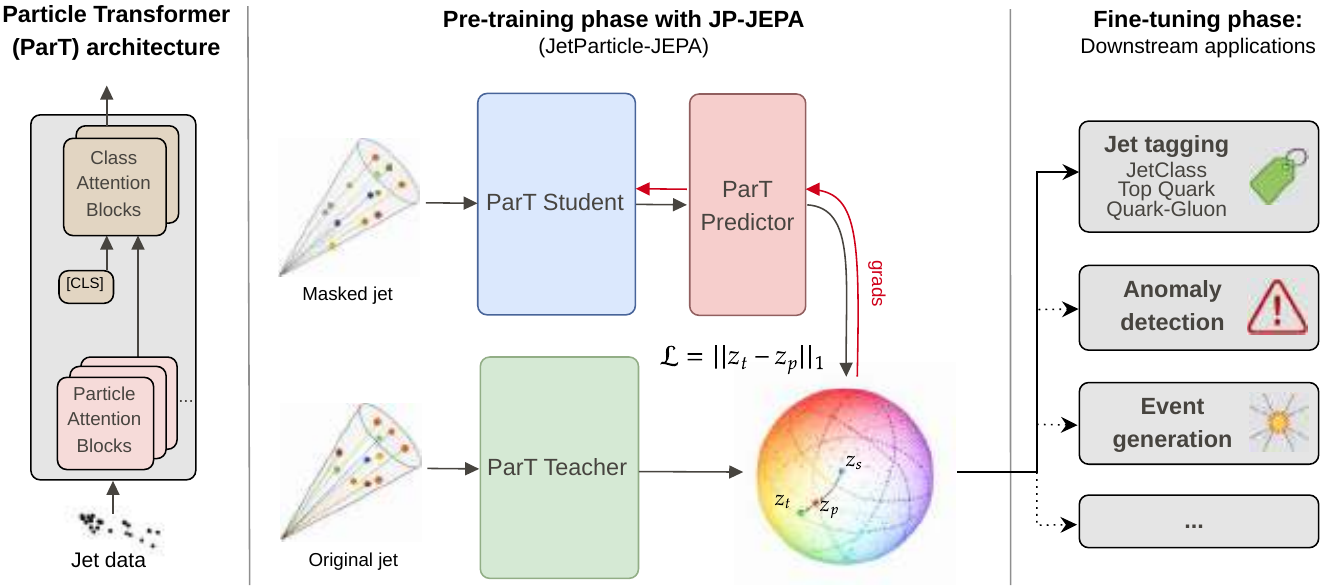}
    \caption{\textbf{Overview of the JetParticle-JEPA framework.} We pre-train our model by learning to predict the abstract latent representations of the masked target particles (grey) using only the visible context particles (colored) passed through a student network. By leveraging a Particle Transformer (ParT)~\cite{qu2022particle} backbone (left) and predicting in the latent space rather than reconstructing raw sensory inputs, the model avoids overfitting to detector noise. The resulting robust pre-trained encoder can be seamlessly fine-tuned for downstream tasks, such as jet tagging, anomaly detection, or event generation.}
    \label{fig:main}
\end{figure*}

We integrate this predictive framework with a Particle Transformer (ParT)~\cite{qu2022particle} backbone, exploiting its pairwise interaction mechanism to capture deep correlations between individual constituents (see Figure \ref{fig:main}). To promote physically consistent representations, we couple the latent-state prediction with a physics-informed multi-objective loss that includes Lorentz-vector regression and particle-identification objectives.

Our results demonstrate that preserving the particle-level continuum within a predictive SSL framework yields exceptional advantages. While JP-JEPA achieves state-of-the-art tagging performance on the standard JetClass benchmark—performing on par with fully supervised baselines trained on massive datasets—its true value lies beyond raw accuracy. We show that JP-JEPA drastically outperforms dedicated supervised models in low-data regimes and exhibits superior uncertainty calibration. More importantly, it demonstrates unprecedented robustness when confronted with missing particle information and simulated detector failures. By successfully decoupling model performance from idealized simulations, JP-JEPA establishes a highly scalable pathway toward foundation models capable of accelerating discovery at the LHC.

\section{Results}
\label{sec:results}

To validate whether predictive self-supervised learning circumvents the limitations of current supervised jet taggers, we evaluate JP-JEPA across four critical dimensions: the physical meaningfulness of its learned representations without labels, its data-efficiency in downstream tasks, its robustness to simulated detector failures, and its predictive calibration.

\subsection{Emergence of physical kinematics in the latent space}
\label{sec:results:jepa-physics}

Before assessing downstream classification, we investigate what the model intrinsically learns during the self-supervised pre-training phase. Since JP-JEPA is trained without any explicit class labels or physics constraints, a fundamental question is whether its latent representations naturally capture the underlying kinematics of the particle clouds. To answer this, we probe the geometry and dimensionality of the latent space using specific metrics and visualization tools.

\begin{figure*}[ht!]
    \centering
    \begin{subfigure}{0.31\textwidth}
        \centering
        \includegraphics[width=\textwidth]{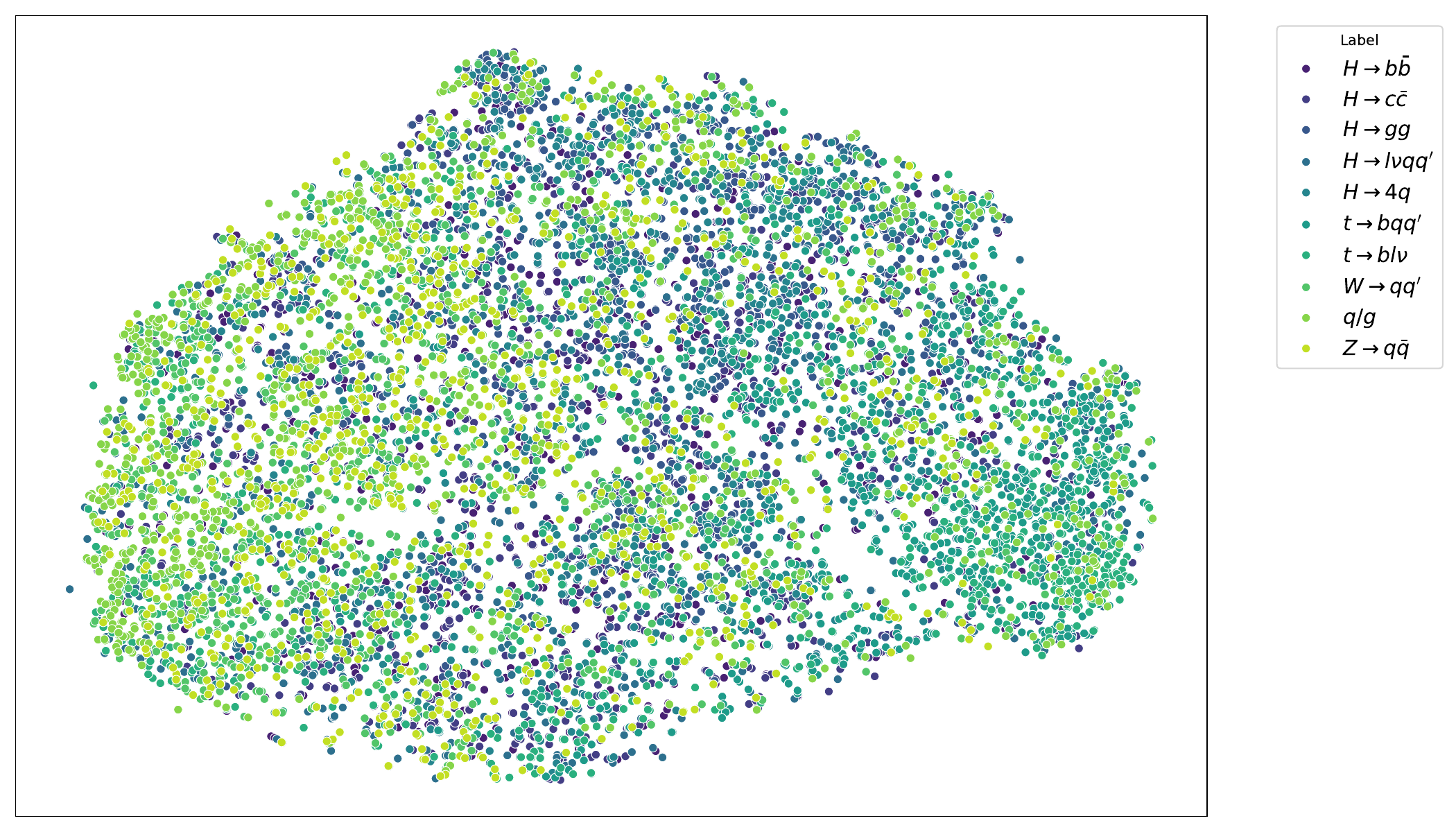}
        \caption{\textbf{Labels.}}
        \label{fig:embedding-viz-class}
    \end{subfigure}
    \hfill
    \begin{subfigure}{0.31\textwidth}
        \centering
        \includegraphics[width=\textwidth]{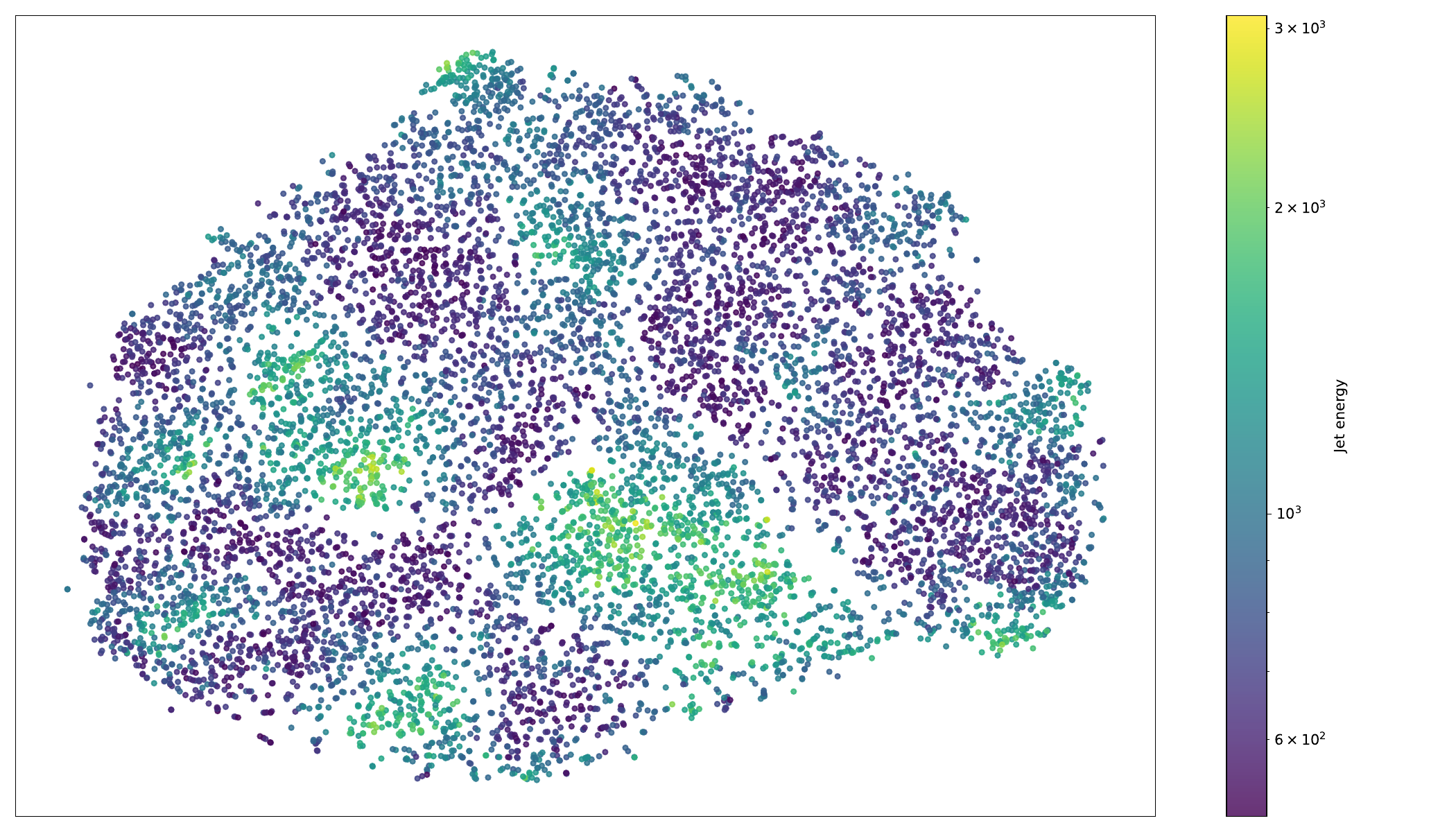}
        \caption{\textbf{Jet energy.}}
        \label{fig:embedding-viz-energy}
    \end{subfigure}
    \hfill
    \begin{subfigure}{0.31\textwidth}
        \centering
        \includegraphics[width=\textwidth]{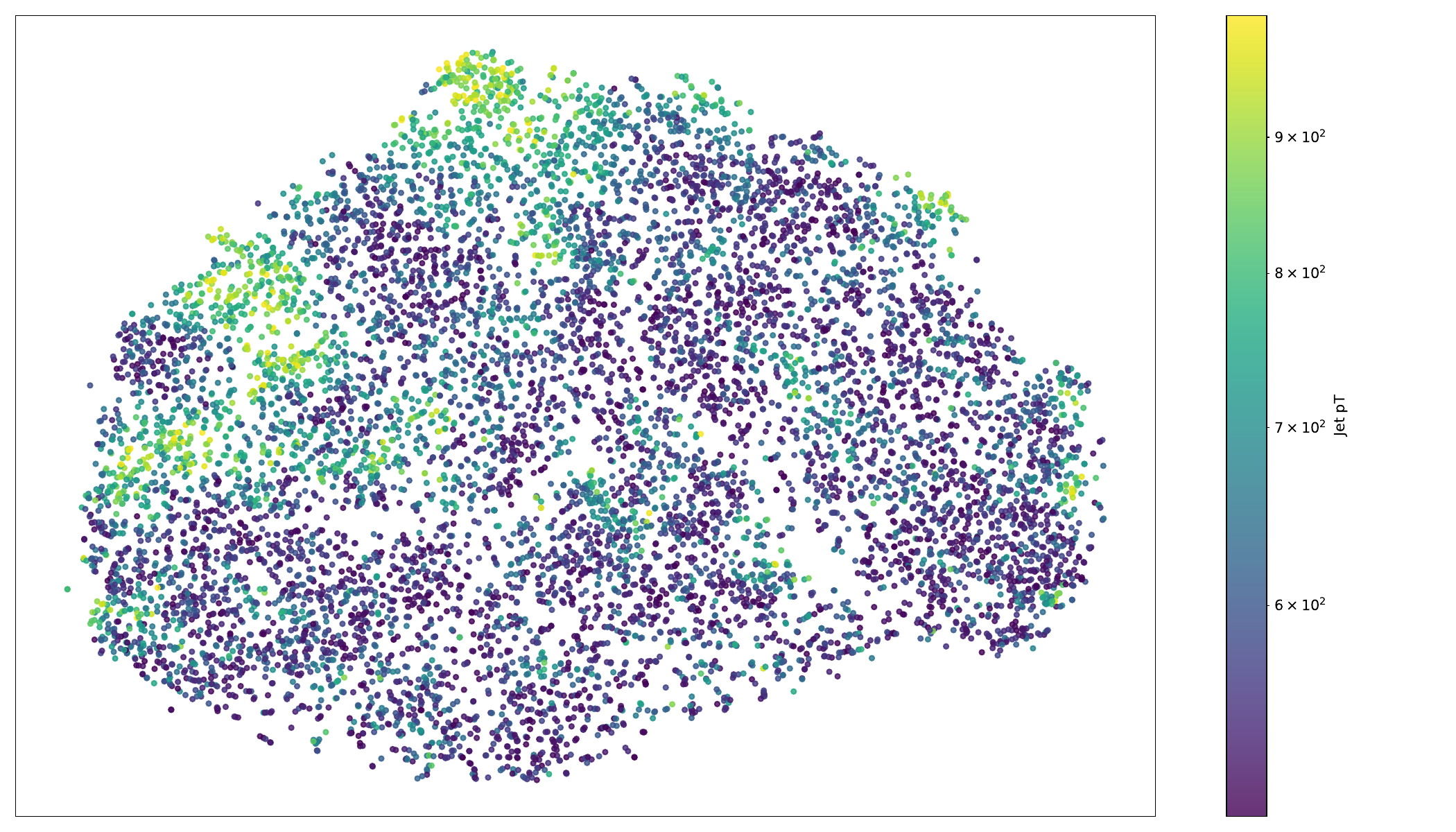}
        \caption{\textbf{Jet transverse momentum.}}
        \label{fig:embedding-viz-pt}
    \end{subfigure}
    
    \begin{subfigure}{0.49\linewidth}
        \centering
        \includegraphics[width=\linewidth]{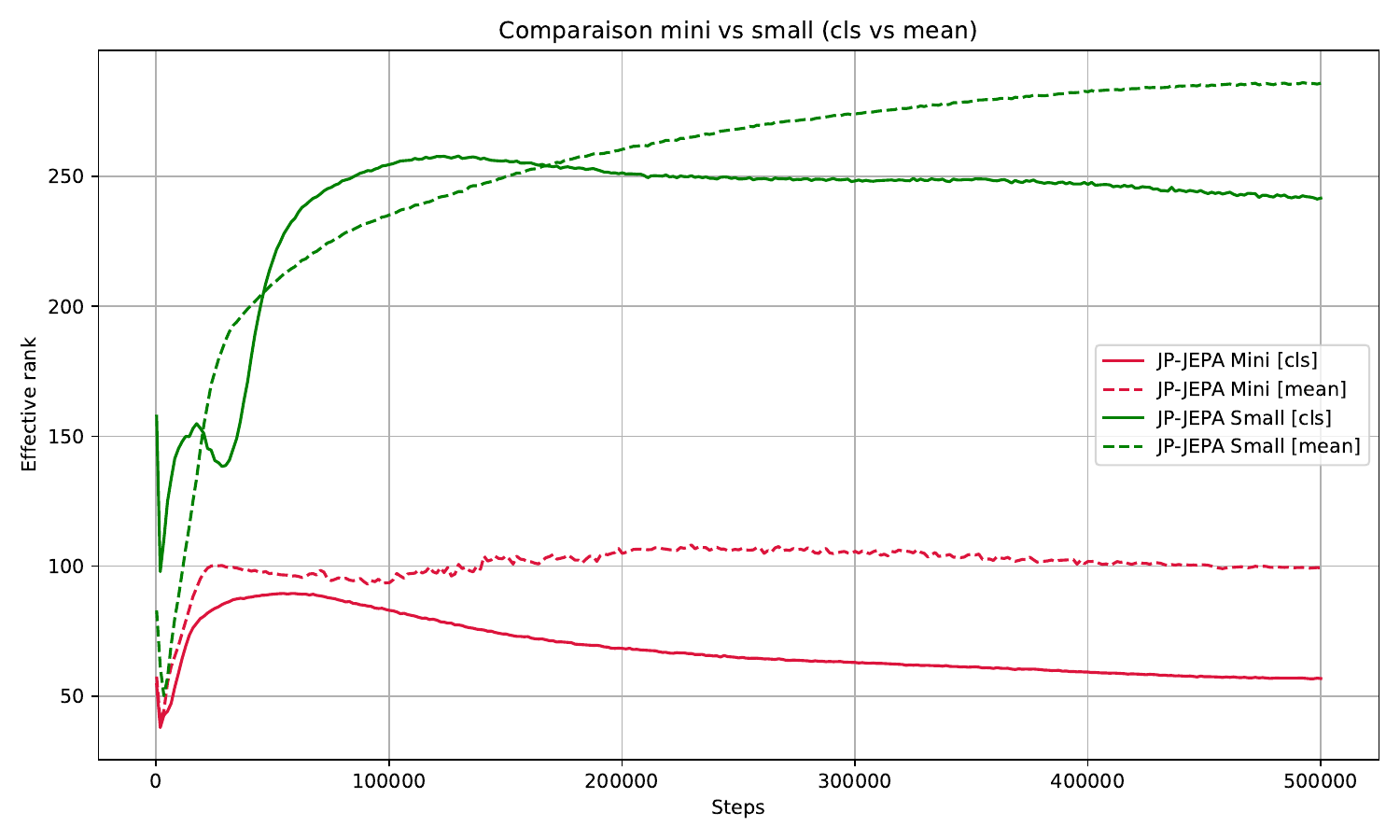}
        \caption{\textbf{Evolution of effective rank during pre-training.}}
        \label{fig:effective-rank}
    \end{subfigure}
    \hfill
    \begin{subfigure}{0.49\linewidth}
        \centering
        \includegraphics[width=\linewidth]{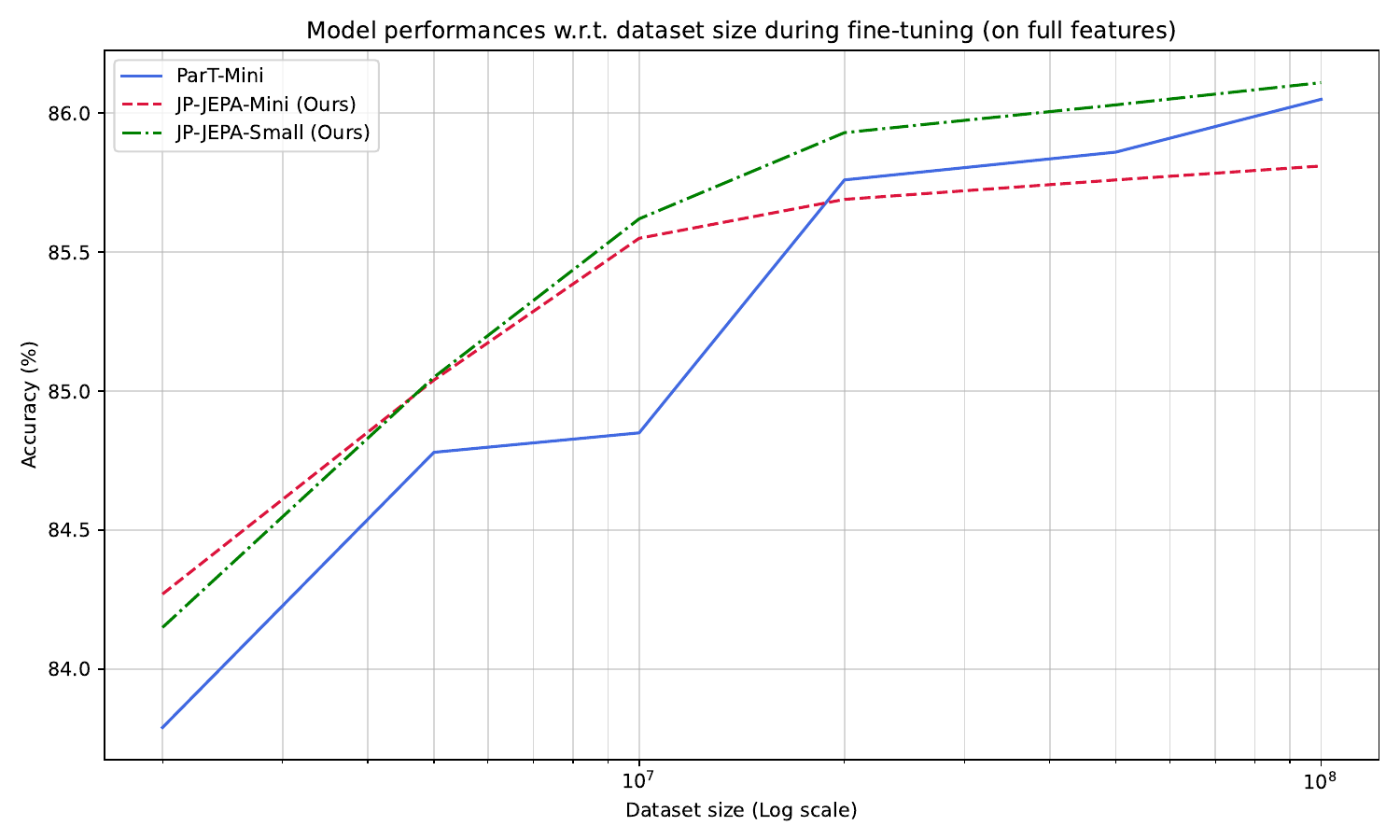}
        \caption{\textbf{Evolution of performances on JetClass w.r.t. fine-tuning dataset size.}}
        \label{fig:jetclass-vs-dataset-size}
    \end{subfigure}

    \begin{subfigure}{0.49\linewidth}
        \centering
        \includegraphics[width=\linewidth]{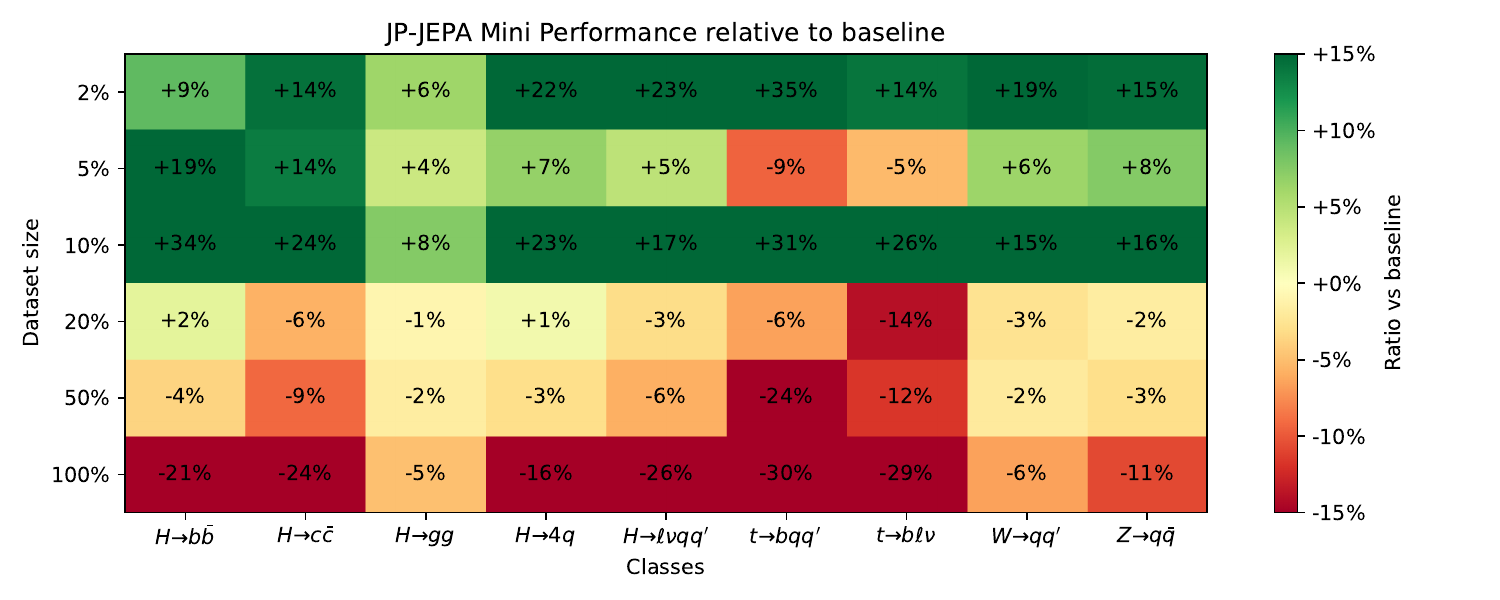}
        \caption{\textbf{Detailed rejection performances of JP-JEPA Mini relative to ParT on JetClass at optimum AUROC.}}
        \label{fig:rejection:mini}
    \end{subfigure}
    \hfill
    \begin{subfigure}{0.49\linewidth}
        \centering
        \includegraphics[width=\linewidth]{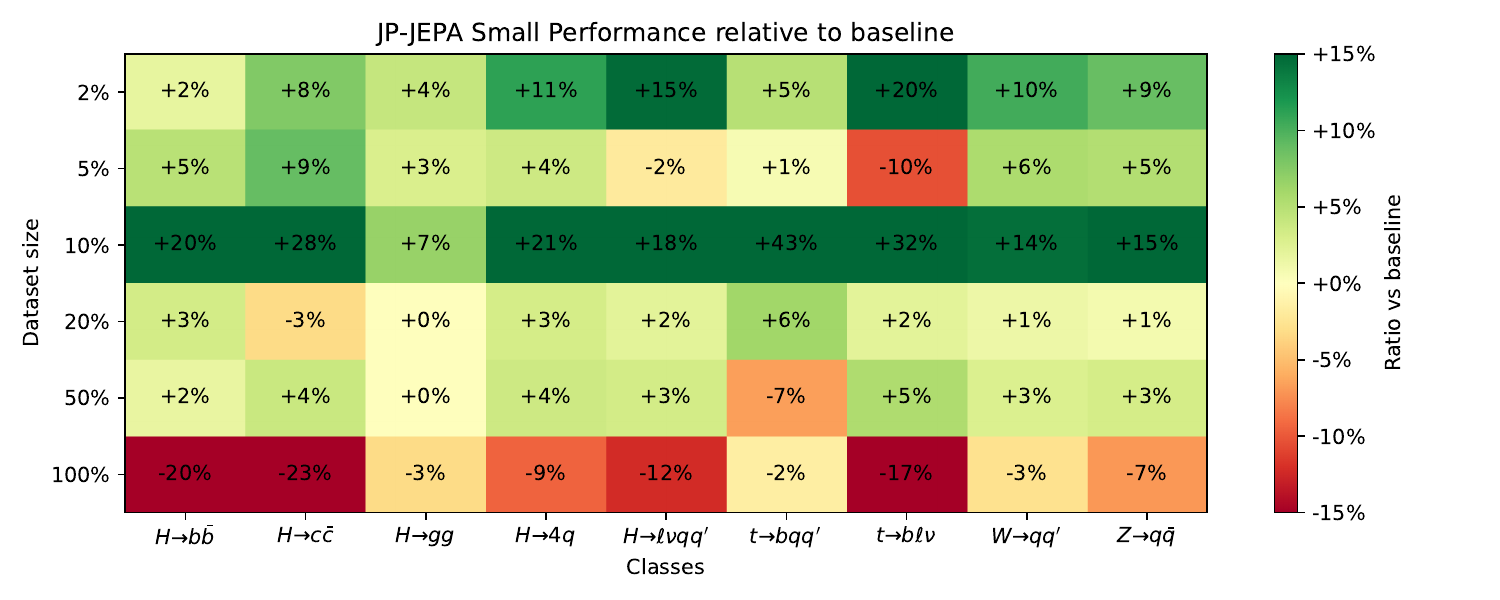}
        \caption{\textbf{Detailed rejection performances of JP-JEPA Small relative to ParT on JetClass at optimum AUROC.}}
        \label{fig:rejection:small}
    \end{subfigure}
    \caption{\textbf{Emergent structure and downstream performance of JP-JEPA representations.} (a) t-SNE projection of the JP-JEPA Small embeddings for 10{,}000 JetClass samples, uniformly drawn from the 10 classes and colored according to the ground-truth labels. (b) Same embedding projection colored by jet energy (GeV). (c) Same embedding projection colored by jet transverse momentum $p_T$ (GeV). (d) Evolution of the effective rank during pre-training. [cls] (solid lines) denotes the effective rank computed from the classification token at the output of the final ParT Clas Attention Blocks (CAB), while [mean] (dashed lines) corresponds to the effective rank computed from the mean particle-token representation at the output of the Particle Attention Blocks (PAB). (e) JetClass performance as a function of fine-tuning dataset size. JP-JEPA (dashed lines) is compared against fully supervised ParT training (solid lines). (f,g) Detailed rejection performance of the JP-JEPA Mini and Small models relative to ParT on JetClass at the optimal AUROC operating point.}
\end{figure*}

First, we chose metrics commonly used in SSL\footnote{Details regarding these metrics are available in the supplementary materials (Section \ref{sec:supp:ssl-metrics}).}, such as the effective rank~\cite{roy2007effective}. This metric provides an estimate of the effective number of dimensions used by a model, evaluating redundant information shared between them\footnote{The higher the value (bounded by the embedding dimensionality)}. ParT has two types of blocks (as shown on the left side of Figure \ref{fig:main}), PAB and CAB. We devise a dynamic metric for both to track particle-level and jet-level information. Figure \ref{fig:effective-rank} presents the evolution of these metrics (\texttt{[Mean]} for PAB and \texttt{[CLS]} token for CAB). These are presented for both the Mini and Small model pre-training on the entirety of JetClass.

The evolution of the effective rank variables seen during training are consistent with the increase in model capacity between both models with similar dynamics for the output of each type of ParT block. For the tokens produced by the Particle Attention Blocks (PAB), the effective rank continuously increases suggesting that the data reside on a very high-dimensional manifold with respect to the representation capacity of the model. A close look at the \texttt{[CLS]} token reveals two phases during training: the model initially used as many dimensions as needed to uniformly distribute the features over the latent space, likely encouraged by the KoLeo regularization~\cite{DBLP:conf/iclr/SablayrollesDSJ19}. It is then followed by a compression phase, progressively removing irrelevant dimensions in terms of mutual information~\cite{patel2024learning} showing a healthy pre-training phase with both local and global features being learned by the model. Scaling up the model from Mini to Small clearly shows benefits in representational capacity and is indicative of the better performance achieved in the downstream classification tasks. It is very likely that a larger model could improve further the performance in downstream tasks. We leave this study to future work investigating the scaling of models based on our framework.

Since the latent representation space remains abstract to humans after the pre-training phase, visualization techniques such as t-SNE~\cite{van2008visualizing} are used to project encoded features from a subset of the data to study the organization of the space. Figures \ref{fig:embedding-viz-class}, \ref{fig:embedding-viz-energy}, and \ref{fig:embedding-viz-pt} show the same examples encoded by JP-JEPA and then projected using t-SNE, colored according to three variables to which the model did not have access during pre-training. Figure \ref{fig:embedding-viz-class} colors the points according to their associated labels. We observe that large structures begin to emerge, but that they are not separable at all. This therefore confirms that the model does not improperly learn to classify during the pre-training phase. Nevertheless, this remains promising because the emergence of some structure suggests that using a larger quantity of data could improve class separability. 

Crucially, despite the lack of labels, the learned representations appear to organize predominantly according to fundamental physical properties rather than class-level semantic structure. Indeed, Figures \ref{fig:embedding-viz-energy} and \ref{fig:embedding-viz-pt} respectively show encoded events colored according to the jet energy and the jet transverse momentum. We observe regions containing a large number of points, which cannot result from a random organization. Thus, our model successfully learns how to naturally structure its output space according to physical variables related to the jet.

\subsection{Data-efficient learning for downstream jet tagging}
\label{sec:results:main-results}

Having established that JP-JEPA naturally encodes physical kinematics, we evaluate its primary purpose as a foundation model: data-efficiency in downstream tasks. To perform direct comparisons, we build upon the Particle Transformer (ParT)~\cite{qu2022particle} architecture, pre-train our JP-JEPA model on JetClass~\cite{qu2022jetclass}, and fine-tune it for jet tagging.

The true test of a self-supervised foundation model lies in its performance when labeled data is scarce. Demonstrating exceptional data-efficiency, our JP-JEPA Mini model pre-trained and fine-tuned on just 5\% of the JetClass data achieves an accuracy of 0.730.
Under the same input setting (kinematic variables only), our predictive approach outperforms the equivalent low-data performance reported by generative state-of-the-art MPMv2~\cite{leigh2025tokenization} by 0.8\%, despite relying on highly compressed latent representations rather than reconstructing full particle-level variables. 

To understand the evolution of this data-efficiency, we scale the architecture and compare the JP-JEPA Small variant (27.1M parameters) with the Mini variant (2.14M parameters). As shown in Figure~\ref{fig:jetclass-vs-dataset-size}, scaling the model significantly improves representational capacity and provides greater stability. Both models clearly outperform the supervised ParT baseline in the low-data regimes (up to 10\% of available labels), demonstrating that the self-supervised pre-training successfully captured reusable physical priors.

Finally, we evaluate the model's asymptotic performance when the entirety of the labeled datasets is available (100\% fine-tuning). On the 10-class JetClass benchmark (Table \ref{tab:jetclass}), JP-JEPA matches the fully supervised ParT baseline (0.8611 vs 0.8605 accuracy, and 0.9876 vs 0.9877 AUROC), while substantially outperforming other SSL approaches like HEP-JEPA (0.698)~\cite{bardhan2025hep} and MPMv2 (0.853). Regarding background rejection metrics (Figures \ref{fig:rejection:mini} and \ref{fig:rejection:small}), we remain competitive with ParticleNet~\cite{qu2020jet} and extremely close to the fully supervised ParT model.

\begin{table*}[ht!]
\resizebox{\linewidth}{!}{
\begin{tabular}{lcccccccccccc}
\toprule
\multirow{2}{*}{Methods} & \multirow{2}{*}{Cat.} & \multicolumn{2}{c}{All classes} & $H \rightarrow b\bar{b}$ & $H \rightarrow c\bar{c}$ & $H \rightarrow gg$ & $H \rightarrow 4q$ & $H \rightarrow l\nu qq'$ & $t \rightarrow bqq'$ & $t \rightarrow bl\nu$ & $W \rightarrow qq'$ & $Z \rightarrow q\bar{q}$ \\
& & Acc & AUC & $\text{Rej}_{\text{50\%}}$ & $\text{Rej}_{\text{50\%}}$ & $\text{Rej}_{\text{50\%}}$ & $\text{Rej}_{\text{50\%}}$ & $\text{Rej}_{\text{99\%}}$ & $\text{Rej}_{\text{50\%}}$ & $\text{Rej}_{\text{99.5\%}}$ & $\text{Rej}_{\text{50\%}}$ & $\text{Rej}_{\text{50\%}}$ \\
\midrule
\multicolumn{13}{l}{\textit{Supervised learning}} \\
PFN~\cite{komiske2019energy} & full & 0.772 & 0.9714 & 2924 & 841 & 75 & 198 & 265 & 797 & 721 & 189 & 159 \\
P-CNN~\cite{qu2020jet} & full & 0.809 & 0.9789 & 4890 & 1276 & 88 & 474 & 947 & 2907 & 2304 & 241 & 204 \\
ParticleNet~\cite{qu2020jet} & full & 0.844 & 0.9849 & 7634 & 2475 & 104 & 954 & 3339 & 10526 & 11173 & 347 & 283 \\
ParT~\cite{qu2022particle} & full & 0.861 & 0.9877 & 10638 & 4149 & 123 & 1864 & 5479 & 32787 & 15873 & 543 & 402 \\
MIParT~\cite{wu2025jet} & full & 0.861 & 0.9878 & 10753 & 4202 & 123 & 1927 & 5450 & 31250 & 16807 & 542 & 402 \\
L-GATr~\cite{spinner2024lorentz} & full & 0.865 & \textbf{\underline{0.9884}} & \textbf{\underline{12195}} & \textbf{\underline{4819}} & \textbf{\underline{128}} & \textbf{\underline{2304}} & 5764 & 37736 & \textbf{\underline{19231}} & \textbf{\underline{580}} & \textbf{\underline{427}} \\
LLoCa-ParticleNet~\cite{spinner2025lorentz} & full & 0.845 & 0.9852 & 7463 & 2833 & 105 & 1072 & 3155 & 10753 & 9302 & 403 & 306 \\
LLoCa-Transformer~\cite{spinner2025lorentz} & full & \textbf{\underline{0.864}} & 0.9882 & 11628 & 4651 & 125 & 2037 & 5618 & 39216 & 17241 & 548 & 410 \\
LLoCa-ParT~\cite{spinner2025lorentz} & full & \textbf{\underline{0.864}} & 0.9882 & 11561 & 4640 & 125 & 2037 & \textbf{\underline{5900}} & \textbf{\underline{41667}} & \textbf{\underline{19231}} & 552 & 419 \\
\midrule
\multicolumn{13}{l}{\textit{Self-Supervised learning}} \\
HEP-JEPA~\cite{bardhan2025hep} & kin & 0.698 & --- & --- & --- & --- & --- & --- & --- & --- & --- & --- \\
MPMv2~\cite{leigh2025tokenization} & full & 0.853 & --- & --- & --- & --- & --- & --- & --- & --- & --- & --- \\
JP-JEPA (Mini) & full & 0.858 & 0.9872 & 8368 & 3155 & 117 & 1566 & 4065 & 22989 & 11299 & 508 & 359 \\
JP-JEPA (Small) & full & \underline{0.861} & \underline{0.9876} & \underline{8511} & \underline{3184} & \underline{119} & \underline{1688} & \underline{4808} & \underline{32258} & \underline{13245} & \underline{528} & \underline{374} \\
\bottomrule
\end{tabular}
}
\caption{
\textbf{Comparison of performances of different methods on JetClass dataset.} For all metrics, the higher the better. 'kin', and 'full' categories respectively correspond to using only kinematics information, kinematics and particle identification, and all available categories. \textbf{Bolded} results denote the best results over all learning paradigms, and \underline{underlined} the best for its paradigm.
}
\label{tab:jetclass}
\end{table*}

When transferred to external, smaller-scale datasets (Table~\ref{tab:downstream})—Top Quark Tagging~\cite{kasieczka2019top} and Quark-Gluon Tagging~\cite{komiske_2019_3164691}—JP-JEPA yields highly competitive results (e.g., 0.939 accuracy on Top Quark and 0.848 on Quark-Gluon), though it slightly trails the fully supervised ParT (0.944 and 0.852, respectively).

\begin{table*}[ht!]
\resizebox{\linewidth}{!}{
\begin{tabular}{lcccccccccc}
\toprule
& \multicolumn{5}{c}{Top quark} & \multicolumn{5}{c}{Quark-gluon} \\
\cmidrule(lr){2-6}\cmidrule(lr){7-11}
Methods & Pretraining Cat. & Acc & AUC & $\text{Rej}_{\text{50\%}}$ & $\text{Rej}_{\text{30\%}}$ & Pretraining Cat. & Acc & AUC & $\text{Rej}_{\text{50\%}}$ & $\text{Rej}_{\text{30\%}}$ \\
\midrule
\multicolumn{6}{l}{\textit{Supervised learning}} \\
ParticleNet-f.t.~\cite{qu2020jet} & kin & 0.942 & 0.9866 & $487\pm9$ & $1771\pm80$ & --- & --- & --- & --- & --- \\
ParT-f.t.~\cite{qu2022particle} & kin & 0.944 & 0.9877 & $\boldsymbol{\underline{691\pm15}}$ & $2766\pm130$ & kinpid & 0.852 & 0.9230 & $50.6\pm0.2$ & $138.7\pm1.3$ \\
MIParT-f.t.~\cite{wu2025jet} & kin & 0.944 & 0.9878 & $640\pm10$ & $2789\pm133$ & kinpid &\textbf{ \underline{0.853}} & \textbf{\underline{0.9237}} & $\boldsymbol{\underline{51.9\pm0.5}}$ & $\boldsymbol{\underline{141.4\pm1.5}}$ \\
L-GATr-f.t.~\cite{spinner2024lorentz} & kin & \textbf{\underline{0.945}} & 0.9879 & $651\pm11$ & $2894\pm84$ & --- & --- & --- & --- & --- \\
OmniLearn~\cite{mikuni2025method,mikuni2025solving} & full & 0.942 & 0.9872 & $568\pm9$ & $2647\pm192$ & --- & --- & --- & --- & --- \\
OmniLearned-S~\cite{bhimji2025omnilearned} & full & 0.944 & \textbf{\underline{0.9880}} & $565\pm12$ & $2637\pm128$  & --- & --- & --- & --- & --- \\
OmniLearned-M~\cite{bhimji2025omnilearned} & full & 0.944 & \textbf{\underline{0.9880}} & $656\pm12$ & $3208\pm176$  & --- & --- & --- & --- & --- \\
OmniLearned-L~\cite{bhimji2025omnilearned} & full & 0.944 & \textbf{\underline{0.9880}} & $688\pm9$ & $\boldsymbol{\underline{3486\pm157}}$  & --- & --- & --- & --- & --- \\
\midrule
\multicolumn{6}{l}{\textit{Self-Supervised learning}} \\
J-JEPA~\cite{katel2024learning} & kin & 0.900 & --- & --- & --- & --- & --- & --- & --- & --- \\
HEP-JEPA~\cite{bardhan2025hep} (frozen) & kin & 0.928 & --- & --- & --- & kin & 0.821 & --- & --- & --- \\
HEP-JEPA~\cite{bardhan2025hep} & kin & 0.929 & --- & --- & --- & --- & --- & --- & --- & --- \\
JP-JEPA (Mini) & full & \underline{0.939} & \underline{0.9854} & $363\pm7$ & \underline{$1341\pm33$} & full & \underline{0.848} & \underline{0.9190} & $46.7\pm0.4$ & \underline{$127.3\pm2.4$} \\
JP-JEPA (Small) & full & \underline{0.939} & 0.9852 & \underline{$369\pm7$} & $1294\pm45$ & full & \underline{0.848} & 0.9185 & \underline{$46.8\pm1.8$} & $122.7\pm1.2$ \\
\bottomrule
\end{tabular}
}
\caption{
\textbf{Comparison of performances of different methods on top quark tagging and quark gluon tagging datasets.} For all metrics, the higher the better. The uncertainty quoted corresponds to the standard deviation of five runs with different random weight initialization for top quark and nine runs for quark-gluon. ParticleNet-f.t., ParT-f.t., MIParT-f.t., and L-GATr-f.t. correspond to pre-trained models on the JetClass dataset and fine-tuned on this dataset. For SSL methods, encoders are fine-tuned (default) or frozen (specified). J-JEPA is pre-trained on 1\% of JetClass (500k top jets and 500k QCD jets). \textbf{Bolded} results denote the best results over all learning paradigms, and \underline{underlined} the best for its paradigm.
}
\label{tab:downstream}
\end{table*}

Rather than a limitation, we hypothesize that this saturation in the 100\% data regime reflects JP-JEPA's reluctance to overfit to dataset-specific variables. In contrast, ParT is fully retrained for each benchmark on the corresponding set of dataset-specific input variables starting from JetClass, whereas we deliberately avoid any such retraining or adaptation from JetClass in order to preserve the generality of the learned representation. Unlike a supervised model that memorizes particular trajectory displacement or particle identification (PID) distributions to maximize strict benchmark accuracy, JP-JEPA relies on a generalized, holistic representation. As we demonstrate in the following section, this exact property is what grants JP-JEPA its unprecedented robustness when deployed in realistic, degraded detector environments.

\subsection{Bridging the reality gap: Robustness to detector anomalies}
\label{sec:results:jepa-robustness}

While supervised accuracy on idealized simulations is a standard benchmark, deploying models on real collider data requires extreme robustness to missing variables and sensor malfunctions—the core of the reality gap. We have shown that JP-JEPA pre-training structures the representation space according to underlying physical variables, but an important question is how this translates to resilience during inference. To investigate these aspects, we conducted a series of experiments and analyses using the model pre-trained on JetClass.

\begin{figure*}[ht!]
    \centering
    \begin{subfigure}{\textwidth}
        \centering
        \includegraphics[width=\textwidth]{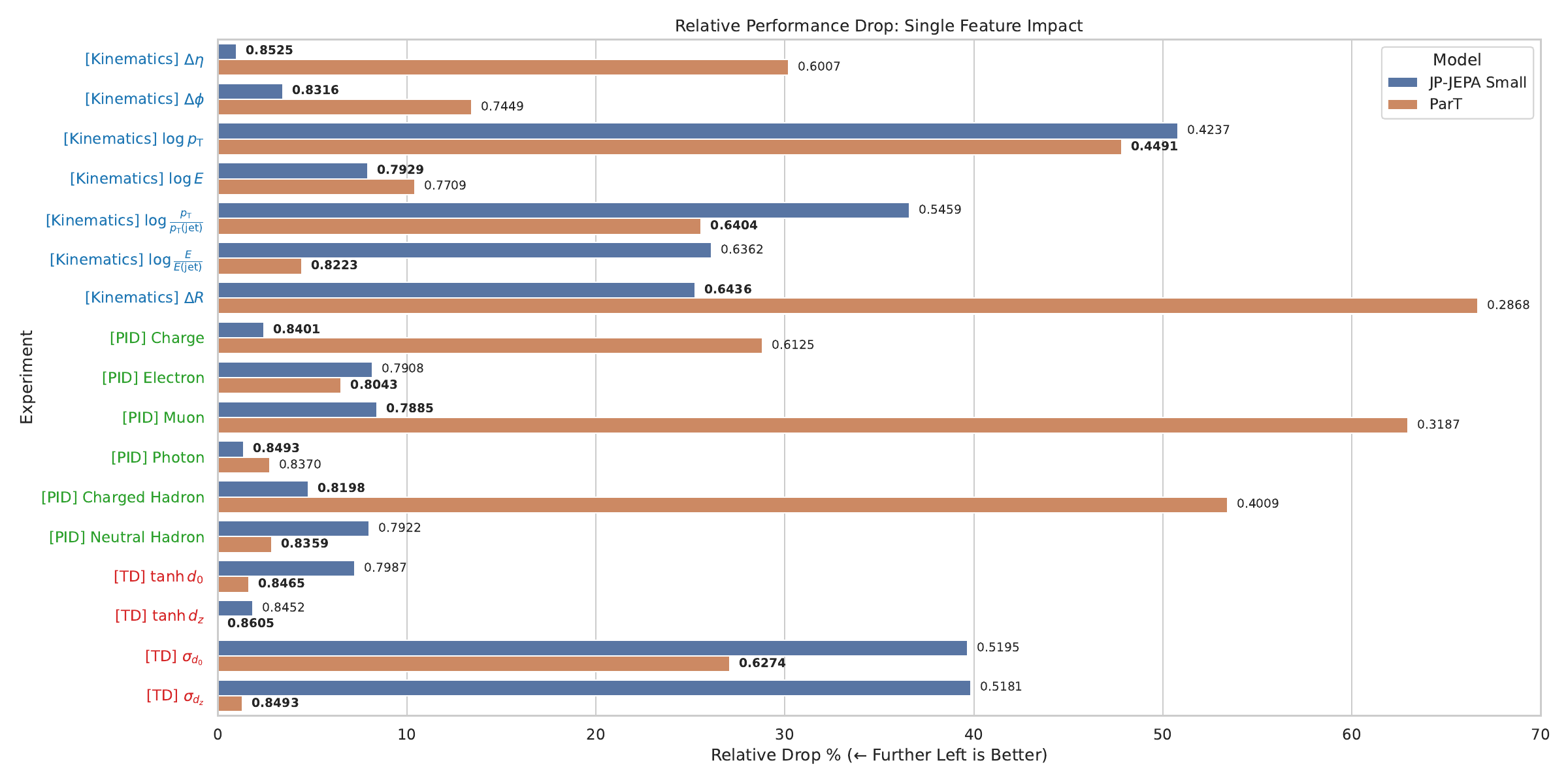}
        \caption{\textbf{Single variable drop.}}
        \label{fig:drop-single-variable}
    \end{subfigure}
    \begin{subfigure}{0.31\textwidth}
        \centering
        \includegraphics[width=\textwidth]{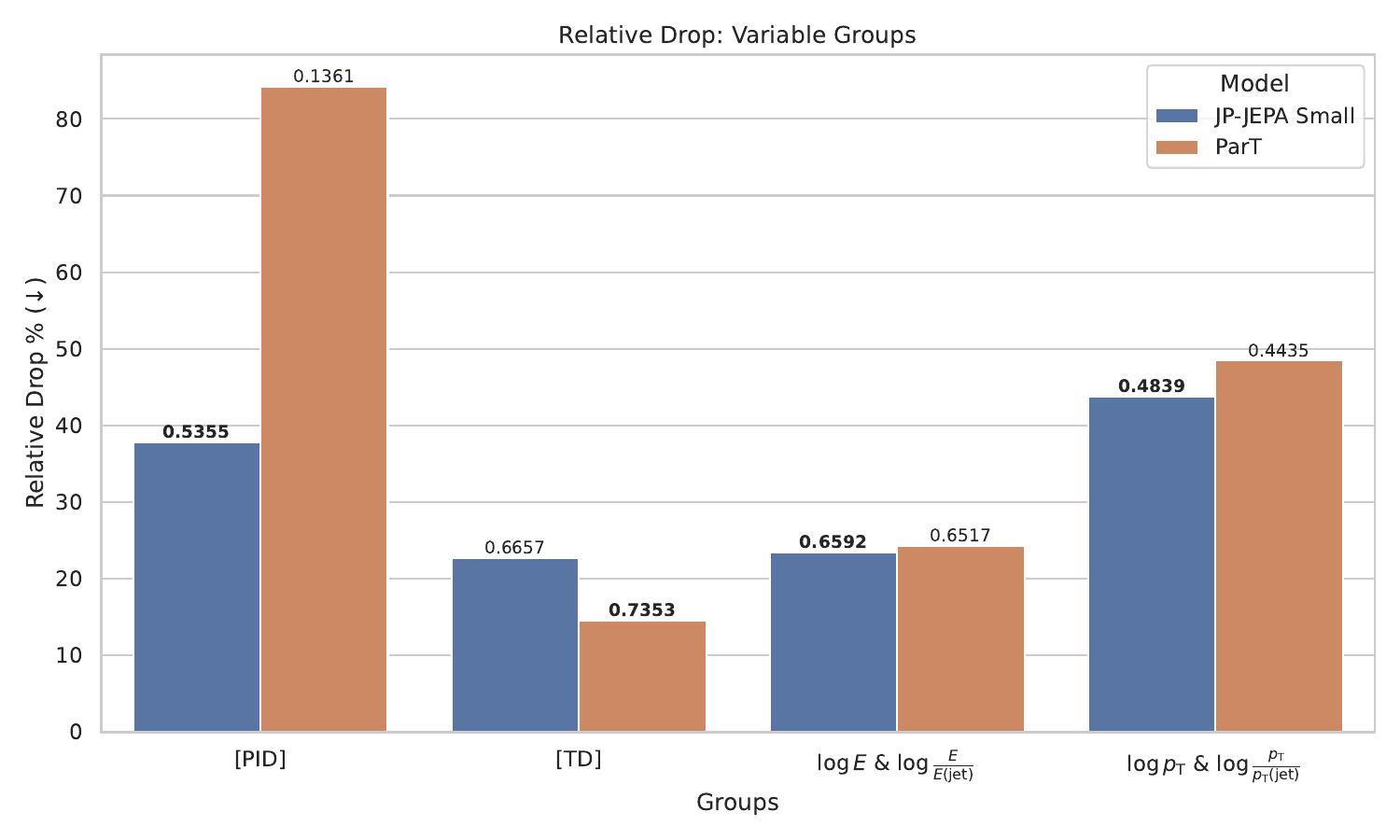}
        \caption{\textbf{Group variables drop.}}
        \label{fig:drop-group-variable}
    \end{subfigure}
    \begin{subfigure}{0.31\textwidth}
        \centering
        \includegraphics[width=\textwidth]{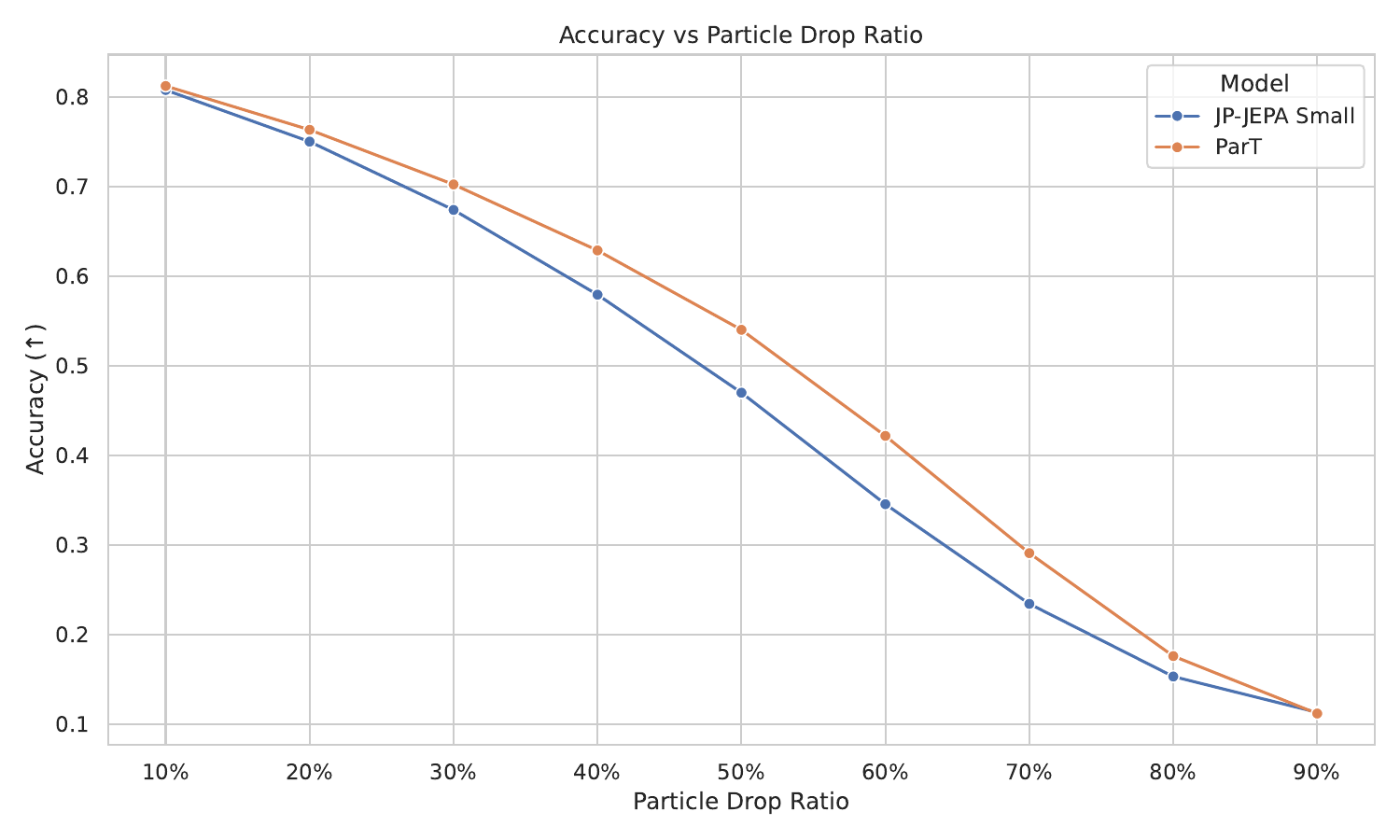}
        \caption{\textbf{Random particle drop.}}
        \label{fig:drop-random-particle}
    \end{subfigure}
    \begin{subfigure}{0.31\textwidth}
        \centering
        \includegraphics[width=\textwidth]{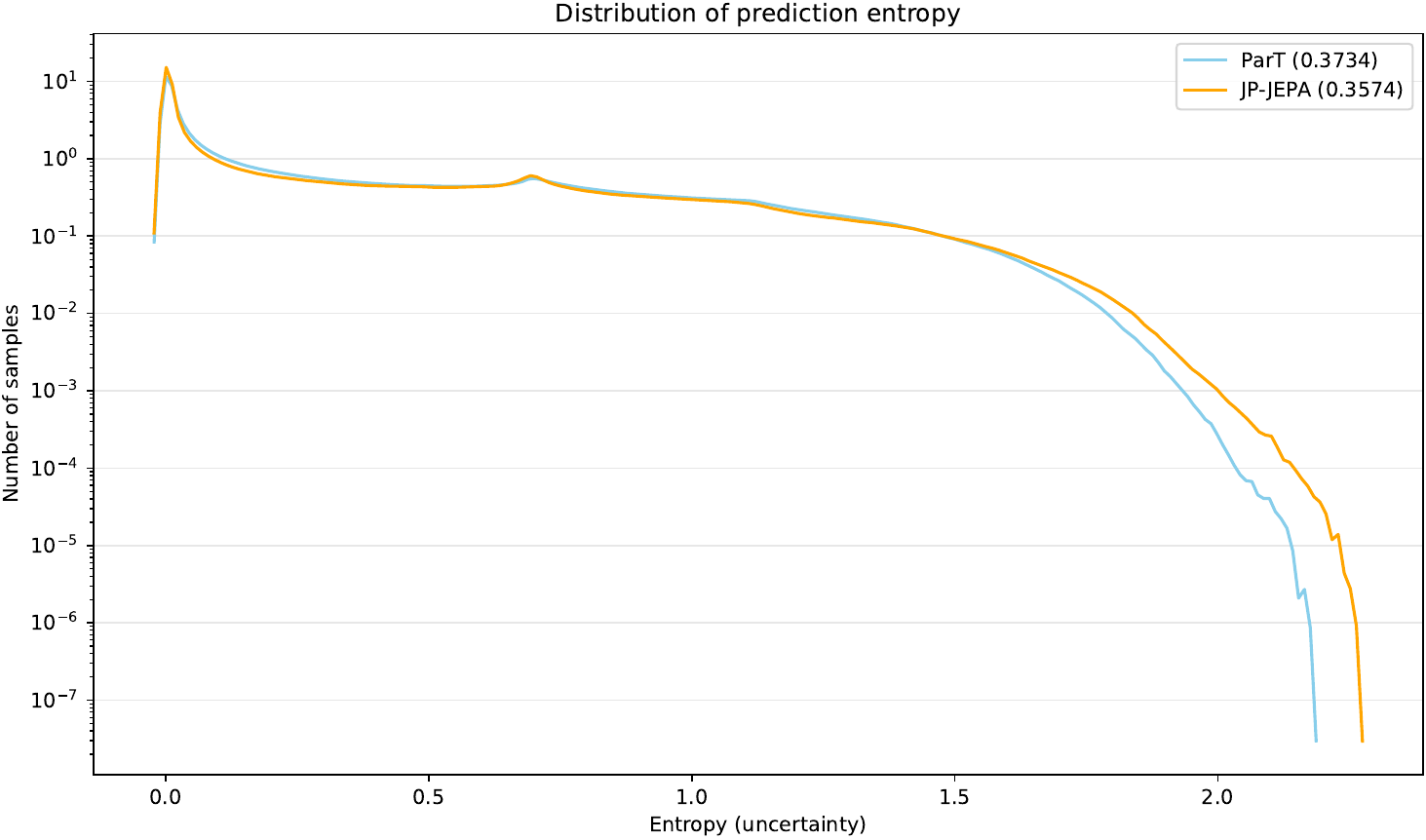}
        \caption{\textbf{Entropy distribution.}}
        \label{fig:entropy_distribution}
    \end{subfigure}
    \caption{\textbf{Quantitative results of our model compared to ParT.} (a) Quantitative inference results under the removal of an individual input variable. (b) Quantitative inference results under the removal of groups of correlated input variables. (c) Quantitative inference results on dropping random particles. (d) Distribution of the entropy of the predicted class probability distributions on the JetClass test set.}
\end{figure*}

First, Figure \ref{fig:drop-single-variable} shows the classification performance degradation as a function of the complete removal of individual input variables during inference. Under the extreme scenario where each of the 17 particle-level input variables available in JetClass is individually removed, JP-JEPA retains an average accuracy of 0.7229 compared to its 0.8611 baseline. Under the exact same conditions, the supervised ParT model collapses from 0.8605 to 0.6652. This much stronger degradation for the supervised baseline confirms that our predictive self-supervised method is significantly more robust to missing information.

Looking more closely at individual variables, we observe a striking divergence in how the models prioritize information. ParT is highly sensitive to particle identification (PID) annotations, especially for muons and charged hadrons, where its accuracy plummets to 0.3187 and 0.4009 upon their removal. In stark contrast, JP-JEPA is highly resilient to the absence of PID variables. This is a critical advantage for real-world deployment: PID is not a raw physical observable, but rather a fragile, algorithmically derived label generated by the detector software, which is prone to miscalibration in real data. By not memorizing these artificial tags, JP-JEPA avoids a major source of simulation bias.

Instead, JP-JEPA exhibits a strong dependence on fundamental kinematic variables (such as the fraction of the jet $p_T$ and energy carried by each particle) and, crucially, on trajectory displacement (TD) and its associated error rates. Unlike PID, TD is a pure geometric observable directly linked to the physical lifetime of decaying particles (such as B-hadrons). Relying heavily on spatial geometry rather than detector-assigned categories demonstrates that JP-JEPA has learned the underlying physics of the decay processes. This deep reliance on genuine physical geometry also elegantly explains why our model performs slightly worse on the external Top Quark and Quark-Gluon datasets (Section \ref{sec:results:main-results}):  these specific benchmarks do not provide TD information as input. While JP-JEPA is robust to randomly missing variables during inference on JetClass, transferring the pre-trained model to an entirely different, structurally impoverished dataset creates an extreme Out-Of-Distribution (OOD) shift, effectively starving JP-JEPA of the fundamental geometric correlations it successfully learned to exploit during pre-training.

Finally, we observe that our model is more invariant to particle positions on the detector surface ($\Delta\eta$ and $\Delta\phi$) compared to ParT. This suggests a better robustness of our model to variations in the collision position within the detector. Similar resilience is observed when removing an entire group of correlated variables together in Figure \ref{fig:drop-group-variable}, such as all particle identifiers, where our model still achieves 0.5355 compared to 0.1361 for ParT (corresponding to an almost random prediction for 10 classes).  Finally, we randomly removed particles to simulate defective detector cells (Figure \ref{fig:drop-random-particle}). Here, we observe that JP-JEPA experiences a steeper degradation in performance compared to ParT when more than 30\% of the particles are missing. Rather than a flaw, this behavior highlights a fundamental difference in information processing. Supervised models like ParT often rely on ``shortcuts''—such as disproportionately attending to a few high-$p_T$ leading particles—allowing them to maintain accuracy as long as these specific particles are not dropped. In contrast, JP-JEPA constructs a holistic, collective representation of the entire particle cloud geometry. Consequently, randomly destroying a large fraction of the cloud fundamentally alters the global kinematic distributions (e.g., mass and energy flow) that JP-JEPA relies upon, reflecting a more physics-aligned, albeit structurally sensitive, learned representation.

\subsection{Uncertainty profiling and discriminative confidence}
\label{sec:results:uncertainty}

Beyond robustness to missing inputs, the search for rare new physics demands models that provide clear, interpretable confidence profiles when faced with ambiguous signatures. When considering uncertainty-related metrics, JP-JEPA appears to exhibit a fundamentally different and more decisive discriminative profile compared to ParT. 

While the supervised ParT baseline achieves a slightly lower Expected Calibration Error (ECE) globally (0.0058 versus 0.0092 for JP-JEPA) and a marginally better log loss (0.3919 versus 0.3934), this is primarily due to JP-JEPA's tendency toward raw overconfidence on specific subsets. However, examining the entropy of the predicted probability distributions—which show comparable average values (0.3574 for JP-JEPA and 0.3734 for ParT)—reveals that JP-JEPA's underlying behavior is much less noisy. 

As shown in Figure~\ref{fig:entropy_distribution}, JP-JEPA exhibits a more pronounced peak at zero entropy, indicating highly confident and decisive predictions. Both models display similar densities in the low-entropy region (between 0 and 0.5). However, at the secondary peak around 0.7, which reflects increased uncertainty, ParT shows a significantly higher density. This indicates that ParT more frequently assigns moderate, hesitant probability scores distributed across several classes, leading to noisier predictions. In contrast, JP-JEPA tends to avoid getting trapped in these intermediate uncertainty states (as illustrated in the first and last examples of Figure~\ref{fig:qualitative-results} in the supplementary material).

Furthermore, JP-JEPA demonstrates greater stability at the extremes of the entropy spectrum (up to 1.5). This suggests that it better acknowledges its fundamental limitations: when it fails to extract decisive features from an input, it more readily expresses high uncertainty, whereas ParT tends to artificially reduce entropy even in highly ambiguous cases. Overall, JP-JEPA acts as a highly discriminant feature extractor.  Its strong capacity to separate highly ambiguous from highly clear topologies suggests that its confidence scores, while uncalibrated in their raw form, contain high-quality ordinal rankings. This makes JP-JEPA highly amenable to standard post-hoc recalibration techniques.

On individual JetClass samples (Figure~\ref{fig:qualitative-results} in the supplementary material), model errors or hesitations are generally consistent with the visible morphology of the jets. Misclassifications occur primarily between classes with similar topologies: two-prong bosonic jets ($Z/W/H$) or diffuse gluonic signatures ($H \to gg$ vs. $q/g$). Top jets, which possess a richer substructure, are well-identified when the three-prong structure remains discernible, but become more ambiguous when the prongs are partially merged. On the selected samples, ParT is more prone to confusing similar classes with hesitant scores; on the other hand, our model provides highly confident predictions on clear examples and legitimately high uncertainty on others. This phenomenon highlights the qualitative gap between ParT and JP-JEPA. ParticleNet, on the other hand, seems very uncertain about which class to predict, even though at least one typically stands out.

\section{Discussion}
\label{sec:discussion}
We have shown that JP-JEPA enables the training of jet representations whose downstream tagging performance is comparable to the supervised state of the art, while significantly outperforming existing SSL approaches. Beyond classification accuracy alone, our results indicate that the model learns physically meaningful and transferable representations of jets rather than optimizing exclusively for a fixed discrimination task.

A key aspect of this behavior is robustness. In contrast to supervised objectives based on cross-entropy losses, which tend to suppress ambiguity and focus narrowly on task-specific discriminative features, JP-JEPA preserves broader structural relationships present in the data. As a consequence, the learned representations remain stable under information degradation, in particular under reduced particle identification information, while still retaining competitive tagging performance. This property is especially relevant for realistic collider conditions, where detector effects, calibration mismodeling, pileup, or domain shifts between simulation and data can significantly impact supervised taggers.

As the development of JEPA methods originated in computer vision, it is instructive to compare our observations with trends previously identified in that domain. Interestingly, the performance ceiling that we observe is reminiscent of the behavior reported for SSL methods such as SimCLR~\cite[Figure 7]{chen2020simple} on ImageNet~\cite{russakovsky2015imagenet}. In particular, the authors show that reaching supervised-level performance in a self-supervised setting requires significantly larger model capacities. In our case, the observed scaling factor ($\times 12.5$) is consistent with the trends reported in computer vision. This behavior is physically well motivated: self-supervised learning aims to retain the full structure of the underlying data manifold instead of compressing the representation toward a single classification objective. While not all retained information is immediately useful for jet tagging, it becomes valuable for transferability, robustness, and uncertainty-aware inference.

DINO~\cite{caron2021emerging} reaches a similar conclusion, namely that surpassing supervised approaches with comparable model sizes remains difficult. If the parallel between computer vision and jet physics continues to hold, improving beyond current supervised methods may primarily require scaling both model capacity and pre-training datasets, similarly to recent developments such as DINOv3~\cite{simeoni2025dinov3} and V-JEPA 2~\cite{assran2025v}. In the context of particle physics, this scaling perspective is particularly relevant because one of the main promises of SSL lies in data efficiency. Our fine-tuned models exhibit improved behavior in low-label regimes, suggesting that a large fraction of physically relevant information can be extracted during pre-training without relying on large quantities of labeled data.

Another important direction concerns uncertainty estimation and ambiguity preservation. Since JEPA-based objectives do not force representations toward artificially sharp decision boundaries, they may provide a more suitable basis for downstream methods designed to quantify epistemic uncertainties or identify out-of-distribution jets. Such capabilities are becoming increasingly important in modern jet tagging applications, especially for searches involving poorly modeled signatures or rare new-physics processes.

Looking forward, the most immediate developments should focus on improving robustness and generalization rather than solely maximizing benchmark performance. This includes scaling the diversity and realism of pre-training datasets and studying transfer across detector conditions and energy regimes. Architectural improvements are also promising, particularly through recent advances in Particle Transformers such as Sophon~\cite{li2024accelerating}, as well as alternative architectures including PET~\cite{mikuni2025solving,mikuni2025method} and Lorentz-equivariant approaches such as L-GATr~\cite{spinner2024lorentz} or LLoCa~\cite{spinner2025lorentz}. More broadly, JP-JEPA should be viewed not as a task-specific tagger but as a pre-trained foundation backbone for particle physics, designed to support robust and data-efficient inference under realistic experimental conditions.

\section{Methods}
\label{sec:method}

\subsection{Pre-training}
\label{sec:method:pretraining}
The abundance of data from recent datasets~\cite{qu2022particle,li2024accelerating} allows using the Self-Supervised Learning (SSL) paradigm to pre-train potentially large encoders and build foundation models for high-energy physics. We pre-train a Particle Transformer (ParT)~\cite{qu2022particle} by using a JEPA-like framework with specific physics-based information and losses. Unlike HEP-JEPA~\cite{bardhan2025hep}, we use a standard ParT without grouping input particles. This choice is motivated by the relatively small number of particles per event compared to the point-cloud settings considered in their inspiration domains ($<200$ particles versus several thousands or millions of points). In our setting, the quadratic complexity of Transformers does not constitute a computational bottleneck. However, the grouping strategy introduces two drawbacks: (1) they group particles within the input event, introducing a compression that reduces the available information, (2) they bias the attention matrix by summing particle four-vectors within each group and using the resulting representations to compute pairwise token interactions. Overall, grouping particles introduces a two-level compression mechanism: first through the aggregation of input particles, and second through the physically biased pairwise interactions computed from the grouped representations. Using a standard ParT removes this additional compression.

\begin{figure*}[ht!]
    \centering
    \includegraphics[width=\linewidth]{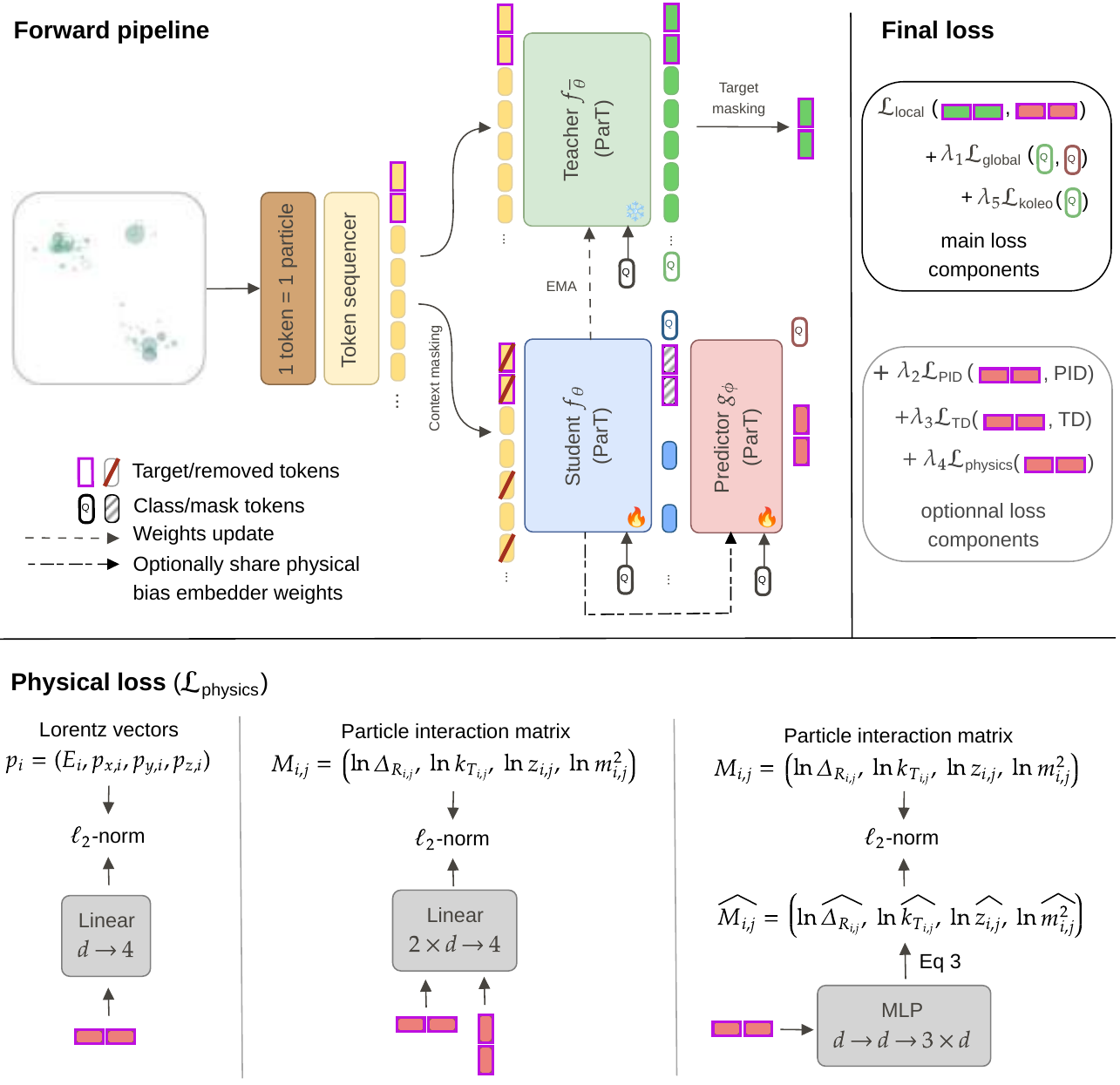}
    \caption{\textbf{Method overview.} The top left panel shows the forward pipeline from a particle cloud to the predictions produced by the teacher and student branches. The teacher branch has access to the full information, whereas the student branch must predict the missing part using the provided context. The top-right panel visually represents the final cost function across the different levels of granularity (particles and events). Finally, the bottom panel illustrates the different physics-informed losses implemented to recover interaction information between particles from the predictor outputs.}
    \label{fig:jpjepa:overview}
\end{figure*}

The top left part of Figure \ref{fig:jpjepa:overview} illustrates our framework pipeline. Firstly, we sequence the particles by applying the greedy algorithm introduced in Point-JEPA~\cite{saito2025point} and reused in HEP-JEPA~\cite{bardhan2025hep}. Briefly, we start from the point with the smallest global coordinates and iteratively append the nearest point to the last selected one. This sequencing allows us to apply contiguous masking to remove semantically coherent regions of the input. We have two branches: student and teacher, which have different roles. The teacher produces the target representations by using all the available input data of the event and by applying a mask afterward. The student tries to recover it by encoding a subset of particles called \textit{context} and, with the help of a predictor network, predicts the representation of the ground truth. Since the predictor is a ParT, we chose to add the physical bias to the attention to help with overall optimization. Since we also propagate the \texttt{[CLS]} token from the context encoder, we pad the bias matrix so that it does not add any physical bias to this token. We act on two levels of representation: event and particle. As in other JEPA methods, we do not backpropagate gradients through the teacher branch; therefore, a stop-gradient operation is applied to all losses that use teacher outputs.

For the event, we use a linear combination of two losses, $\mathcal{L}_{\text{global}}$, which matches the representation by using a smooth L1 loss and a KoLeo~\cite{DBLP:conf/iclr/SablayrollesDSJ19} regularizer. Given a set of $n$ normalized vectors $(x_1, \dots,x_n)$, it is defined as
\begin{equation}
    \mathcal{L}_{\text{koleo}} =-\frac{1}{n}\sum _{i=1}^{n}\log d_{n,i}
\end{equation}
where $d_{n,i} =\underset{j\neq i}{\min} ||x_{i} -x_{j} ||$ is the minimum distance between $x_i$ and any other point within the batch. It encourages the features to uniformly span the representation space within a batch. In our experiments, we also consider Sketched Isotropic Gaussian Regularization (SIGReg) as a plug-and-play replacement for KoLeo introduced in LeJEPA~\cite{DBLP:journals/corr/abs-2511-08544}. It is defined as follows:
\begin{equation}
    \mathcal{L}_{\text{sigreg}} = n \int_{-\infty}^{\infty} \left| \hat{\phi}_X(t) - \phi(t) \right|^2 \, \phi(t)\, dt
\end{equation}
where $\hat{\phi}_X(t)=\frac{1}{n} \sum_{j=1}^{n} e^{i t X_j}$ is the empirical characteristic function (ECF) of the samples, $\phi(t)=e^{-t^{2}/\sigma^{2}}$ the target CF with $\sigma$ commonly set to 1.
Similarly to KoLeo, it encourages a more uniform and isotropic organization of the feature space.

For particles, $\mathcal{L}_{\text{local}}$ is a smooth L1 loss for matching the representation of each predicted particle to those in the ground truth. To ensure that particle representations encode sufficient information, we can optionally use additional losses for recovering masked or unknown information derived solely from the inputs. $\mathcal{L}_{\text{PID}}$ encourages the recovery of particle identification (PID) by using a cross-entropy loss between predictor outputs passed into a projection layer and targets. We have five base particle types (photons, electrons, muons, charged hadrons, and neutral hadrons) with either positive, negative, or neutral charge. By combining these attributes, we obtain eight particle classes. Then, trajectory displacement (TD) corresponds to a quantity measuring the displacement of a particle trajectory with respect to the primary interaction point. It can notably help identify the origin of jets. $\mathcal{L}_{\text{TD}}$ regresses these values using an MSE on projected predictor outputs. Finally, we introduce the $\mathcal{L}_{\text{physics}}$ loss to evaluate whether the model captures physically meaningful particle interactions derived from the Standard Model. It regresses the quantities used to bias the attention matrix with particle interaction information. It can be performed in three ways, as shown in Figure \ref{fig:jpjepa:overview} (bottom). We investigate three variants for recovering particle interaction information solely from the predicted particle representations. The first illustrates a simple prediction setup where we regress the four-vector associated with each predicted particle. In the second, the head takes two particles and predicts the associated values from the particle information matrix. In the third, the head predicts each important component in another space with the same dimensionality as the representation space. The resulting outputs are then compared to the particle information matrix. We define the output of the module as:
\begin{align}
\widehat{M_{i,j}} &=\left(\ln\widehat{\Delta _{R_{i,j}}} ,\ln\widehat{k_{T_{i,j}}} ,\ln\widehat{z_{i,j}} ,\ln\widehat{m_{i,j}^{2}}\right)\\\label{eq:physical-loss:c}
&\text{with}\nonumber\\
\widehat{\Delta _{R_{i,j}}} &=||\text{pos}_{i} ||_{2}\\
\widehat{k_{T_{i,j}}} &=\min\left( ||\widehat{p_{T_{i}}} ||,\ ||\widehat{p_{T_{j}}} ||_{2}\right)\widehat{\Delta _{R_{i,j}}}\\
\widehat{z_{i,j}} &=\min\left( ||\widehat{p_{T_{i}}} ||,\ ||\widehat{p_{T_{j}}} ||_{2}\right) /\left( ||\widehat{p_{T_{i}}} ||+||\widehat{p_{T_{j}}} ||_{2}\right)\\
\widehat{m_{i,j}^{2}} &=||\text{lorentz}_{i} +\text{lorentz}_{j} ||_{2}
\end{align}
where $\text{pos}_{i}$ is the position of the $i$-th particle, $\widehat{p_{T_{i}}}$ its predicted $p_T$, and $\text{lorentz}_{i}$ its predicted four-vector. These quantities are represented in the latent space. The difference between (b) and (c) lies in the use of an MLP instead of a linear layer, allowing the model to learn a more flexible latent structure and to exploit a higher-dimensional representation space than the one used in (b), although the latter remains more computationally efficient. Let $V$ denote the number of target blocks (views) to recover in the latent space representation, the final optimization objective is defined as:
\begin{align}
    \mathcal{L} &= \mathcal{L}_{\text{local}} + \lambda_1\mathcal{L}_{\text{global}} + \lambda_2\mathcal{L}_{\text{PID}} + \lambda_3\mathcal{L}_{\text{TD}} \nonumber\\
    &+ \lambda_4\mathcal{L}_{\text{physics}} + \frac{\lambda_5}{V} \sum_V\mathcal{L}_{\text{koleo/sigreg}}.
\end{align}

During pre-training, we discard events containing fewer than two particles to ensure that masking leaves at least one particle to predict and at least one particle in the context. No additional data transformations are applied. Using this framework, we also investigate training with restricted subsets of the input features (kinematics, particle identification, and trajectory displacement). We can pre-train a single model that can handle inputs with only kinematics (\textbf{kin}), kinematics and PID (\textbf{kinpid}), or \textbf{full} data without any additional training process, which can later be applied to downstream tasks or datasets by simply padding missing variables. This feature dropout is applied only to the context provided by the student. This encourages the student to exploit the available information as efficiently as possible in order to match the richer teacher representations (due both to the fact that it has access to all particles and to all the associated information).

\subsection{Fine-tunings}
\label{sec:method:finetunings}
For jet tagging tasks, we only use the \texttt{[CLS]} token of the ParT, which encodes the event representation. This representation is fed into a 3-layer Multi-Layer Perceptron (MLP), composed of successive batch normalization layers~\cite{ioffe2015batch}, ReLU activations~\cite{DBLP:journals/corr/abs-1803-08375}, and dropout~\cite{DBLP:journals/jmlr/SrivastavaHKSS14}, with bias terms disabled in all linear layers except the final classification layer, which predicts the classes. When specified, we replace the MLP with a single linear classification head that predicts the output classes. We also specify whether the encoder is frozen during fine-tuning (by default, it is unfrozen from the first optimization step). Regarding input features, we use all variables available in the considered dataset (full for JetClass, kin for top quark tagging, and kin+pid for quark-gluon tagging). For the top quark tagging dataset, a pre-trained encoder with full categories can be used for fine-tuning by setting unused variables to zero, as done during pre-training. For jet tagging tasks, $\mathcal{L}_{\text{ft}}$ corresponds to either a cross-entropy or binary cross-entropy loss (resp. for JetClass and top quark/quark-gluon tagging datasets).

\subsection{Datasets}
\label{sec:method:datasets}

\noindent\textbf{JetClass.} This benchmark, introduced by Qu \textit{et al.}~\cite{qu2022particle}, is a large dataset containing 100M simulated jet events for the training set. The validation and test sets contain 5M and 20M jets, respectively. The events are divided into 10 jet categories with balanced class distributions. Two main categories can be distinguished: background jets initiated by light quarks or gluons ($q/g$), and signal jets originating from top quarks ($t$) or bosons ($W$, $Z$, or $H$). Several decay channels are considered for top quarks and Higgs bosons. This dataset is particularly valuable as a large-scale resource for pre-training foundation models in high-energy physics and demonstrates the effectiveness of deep learning methods for jet decay classification. As this is beyond the scope of this work, we do not detail the simulation procedure used to generate the events and refer interested readers to the original paper \citep[Section~2]{qu2022particle}. This dataset provides information about kinematics, particle identification, and trajectory displacement.

\noindent\textbf{Top quark tagging.} This dataset~\cite{kasieczka2019top} provides a dataset of 2M jets split into two classes ($t\rightarrow bqq'$ and $q/g$), with a clear distinction between training (1.2M), validation (0.4M), and test (0.4M) sets. As before, $q/g$ jets are treated as background. Only kinematic information is included as input features.

\noindent\textbf{Quark-gluon tagging.} We benchmark our method on the quark-gluon tagging dataset proposed by Komiske \textit{et al.}~\cite{komiske2019energy}. Similar to the top quark tagging dataset, it is a binary classification dataset distinguishing jets initiated by quarks (signal) from those initiated by gluons (background). We follow a 1.6M/0.2M/0.2M train/validation/test split. Input features include kinematic variables together with particle identification information\footnote{we use the \textit{full} scenario of Qu \textit{et al.}~\cite{qu2022particle} for a charged hadron (\texttt{(|pid|==211)+(|pid|==321)*0.5+(|pid|==2212*0.2)}) and neutral hadron (\texttt{(|pid|==130)+(|pid|==2212*0.2)}).}.

\subsection{Experimental setups}
\label{sec:method:setups}

\noindent\textbf{Protocol.} We conduct experiments on several JetClass subset sizes to study scaling behavior. When referring to X\% of the data, we use the first X\% of files for each class. For 5\% experiments, we use the first 5 files of each class, and the same files are used across all runs. Analyses of this JetClass subsampling strategy are provided in Section~\ref{sec:supp:datasets:jetclass} of the supplementary material. All experiments were conducted on NVIDIA A100 80GB GPUs, with the number of GPUs depending on the dataset size, and using \texttt{float32} precision. The AdamW optimizer~\cite{DBLP:conf/iclr/LoshchilovH19} is used, and weight decay is removed for tokens (\texttt{[MASK]}, \texttt{[CLS]}), biases, and normalization layer weights.

Concerning pre-training, we run experiments for approximately 10 epochs with a batch size of 2048 (other batch sizes are used depending on the subset size and are detailed in Table~\ref{tab:appendix:pretraining-hp} in the supplementary material). The peak learning rate is $10^{-3}$, with a linear warmup during the first 5\% of training steps from $10^{-5}$, followed by cosine annealing down to $10^{-6}$. The learning rate scheduler is updated at every optimization step. For ParT-Small, we updated the hyperparameters for stability by reducing the peak learning rate to $3\times10^{-4}$, increasing the predictor weight decay to favor minimal transformations through smaller predictor weights, adding gradient clipping, and making the prediction task harder by passing fewer context tokens. The teacher is updated via an exponential moving average (EMA) from student weights using the following updating rule:
\begin{equation}
    \bar{\theta}_t=\tau_{\text{ema}} \bar{\theta}_{t-1} + (1-\tau_{\text{ema}})\theta_t
\end{equation}
where $\theta_t$ is the student weight set and $\bar{\theta}_t$ the teacher weight set, both at the $t$-th step. We linearly increase $\tau_{\text{ema}}$ from 0.9995 to 0.99999 during the entire training process. Due to the small number of constituents per event, we use a single target to be predicted by the student. A detailed list of hyperparameters is available in Section~\ref{sec:supp:hp:arch} of the supplementary material. We set $\lambda_1=\lambda_2=\lambda_3=\lambda_4=1$ and $\lambda_5=0.1$ when using the KoLeo regularizer, $\lambda_5=0.05$ for SIGReg. For the physical loss, we perform online probing for all implementations, but only one variant is used when backpropagation is enabled through the predictor and encoder. An ablation study is available in Section~\ref{sec:supp:other-results:ablations} of the supplementary material, including an exploration of reconstruction loss coefficients and associated modules.

For fine-tuning, we train for 10 epochs on Top-Quark and Quark-Gluon tagging, and for 500k steps on JetClass; no data augmentation is performed. MLP heads have 3 layers with an intermediate hidden size of 128 and a dropout probability set to 0.5. The learning rate is the same as in ParT-f.t.~\cite{qu2022particle} for JetClass (peak at $10^{-3}$), and peaks at $10^{-4}$ for the other datasets, with the difference that we employ a cosine annealing scheduler. During fine-tuning, the learning rate warmup starts at $10^{-6}$ and increases during the first 500 steps for the quark-gluon tagging and top quark tagging datasets. For JetClass, warmup is applied during the first 5\% of training steps, similarly to pre-training. The batch size is set to 512 for all datasets, following Qu \textit{et al.}~\cite{qu2022particle}.

\noindent\textbf{Metrics.} For evaluating classification performances, we use mean accuracy (abbreviated as mAcc or simply Acc in our tables) and the Area Under the Curve (AUC) computed from the Receiver Operating Characteristic (ROC) curve. These metrics evaluate classification performance. Accuracy measures the proportion of correct predictions, while AUROC evaluates the trade-off between true-positive and false-positive rates across decision thresholds. Specifically for jet tagging evaluation, we use the background rejection at a given signal efficiency $X\%$, defined as:
\begin{equation}
    \text{Rej}_{\text{X\%}}=\left.\frac{1}{\mathrm{FPR}}\right|_{\mathrm{TPR}=X\%}
    \label{eq:bkg-rej}
\end{equation}
As this metric is defined for binary classification, for JetClass we compute it for each signal class (excluding $q/g$, considered as background) by adapting the score used in Equation \ref{eq:bkg-rej} through:
\begin{equation}
    \mathrm{score}_{SvsB} = \frac{\mathrm{score}(S)}{\mathrm{score}(S) + \mathrm{score}(B)}
\end{equation}
where $\mathrm{score}(S)$ and $\mathrm{score}(B)$ are the softmax outputs for signal and background classes, respectively.

Also, to evaluate the reliability of the probabilistic outputs, we conduct a comparative analysis using three complementary metrics: Log Loss (Cross-Entropy), Expected Calibration Error (ECE)~\cite{guo2017calibration}, and Shannon Entropy~\cite{shannon1948mathematical}. The Log Loss measures the global quality of the predicted distributions by penalizing the distance to the ground truth:
\begin{equation}
    \mathcal{L} = -\frac{1}{N} \sum_{i=1}^{N} \sum_{j=1}^{M} y_{ij} \log(p_{ij})
\end{equation}
While Log Loss provides a proper probabilistic scoring rule, the ECE specifically quantifies model calibration by measuring the gap between predicted confidence and observed accuracy across $K$ bins:
\begin{equation}
    ECE = \sum_{k=1}^{K} \frac{|B_k|}{n} \left| \mathrm{acc}(B_k) - \mathrm{conf}(B_k) \right|
\end{equation}
where $\mathrm{acc}(B_k)=\frac{1}{\left | B_k\right |}\sum_{i \in B_k} \mathbf{1}(\hat{y}_i=y_i)$ and $\mathrm{conf}(B_k)= \frac{1}{\left | B_k\right |}\sum_{i \in B_k} \hat{p}_i$, with $B_k$ the set of sample indices whose prediction confidence falls into the interval $I_k=\left ( \frac{k-1}{K},\frac{k}{K} \right ]$.
Finally, we use the Shannon Entropy, $H(P) = -\sum p_j \log p_j$, to assess the intrinsic uncertainty of the predictions.

\section{Data availability}

This study uses publicly available datasets that have been previously released. Specifically, we use the JetClass \cite{qu2022jetclass}, the Top Tagging \cite{kasieczka2019top}, and the Quark–Gluon \cite{komiske_2019_3164691} datasets, which are available from their original publications and the corresponding public repositories.

\section{Code availability}
The code will be made available upon request during the submission phase and will be publicly released together with the model weights after publication.

\section{Acknowledgements}
This work was supported by the Agence Nationale pour la Recherche (ANR) under award number ANR-19-CHIA-0017. The authors acknowledge Normandy Region (chaire ALPHA) for the access to the HPC resources at the LPC Caen. We would also like to thank Alexandre Perier for the fruitful and insightful discussions that greatly contributed to this work.

\bibliographystyle{plain}
\bibliography{refs}

\clearpage

\appendix
\begin{center}
  \Large\textbf{JetParticle-JEPA: An Efficient Self-Supervised Representation Learning method for Jet Tagging in High-Energy Physics}\\
  \addcontentsline{toc}{part}{Appendices}
  \vspace{0.5em}Supplementary Material \\
  \vspace{1.0em}
\end{center}
\localtableofcontents

\section{Datasets}
\label{sec:supp:datasets}
In this section, we define each variable for completeness.

\subsection{Input variables}
\label{sec:supp:datasets:inputs}

\noindent\textbf{Kinematic input features.} Seven features are used for JetClass~\cite{qu2022jetclass}, top quark tagging~\cite{kasieczka2019top} and quark-gluon~\cite{komiske2019energy} datasets. This set of variables can be derived from the 4-vector $(E, p_x, p_y, p_z)$. Positions are provided with $\Delta \eta$ and $\Delta \phi$, which are the difference in pseudorapidity $\eta$ and azimuthal angle $\phi$ between the particle and the jet axis. $\log p_T$ and $\log E$ are the logarithm of the particle's transverse momentum $p_T=(p_x^2+p_y^2)^{1/2}$ and particle's energy $E$. The energy and $p_T$ unit is GeV. Additionally, like other authors, we provide the logarithm of the particle's $p_T$/energy relative to the jet $p_T$/energy. We respectively note them $\log p_T/p_T(jet)$ and $\log E/E(jet)$. Finally, $\Delta R$ represents the angular separation between particle and jet axis.

\noindent\textbf{Particle identification features.} Six features can be passed if they are available (usable with JetClass and quark-gluon tagging datasets). The first is the electric charge of the particle. The five other correspond if the particle is an electron, muon, photon, charged hadron, or neutral hadron using a one-hot encoding. These variables can be used to derive a classification of particle identifiers, in particular to determine whether the particle is a particle or an antiparticle. Thus, we can have eight classes: electron, positron, muon, antimuon, photon, charged hadron, charged antihadron, and neutral hadron.

\noindent\textbf{Trajectory displacement input features.} They are only available in the JetClass dataset. $\tanh d_0$ and $\tanh d_z$ are the hyperbolic tangent of the transverse and longitudinal impact parameters. Respectively, $\sigma_{d_0}$ and $\sigma_{d_z}$ are their associated error of measurement.

\subsection{JetClass statistical analysis}
\label{sec:supp:datasets:jetclass}
This section focuses on JetClass data analysis for presenting the extracted statistics for efficiently doing our ablation studies and the majority of our experiments.

\begin{figure}[ht!]
    \centering
    \includegraphics[width=\linewidth]{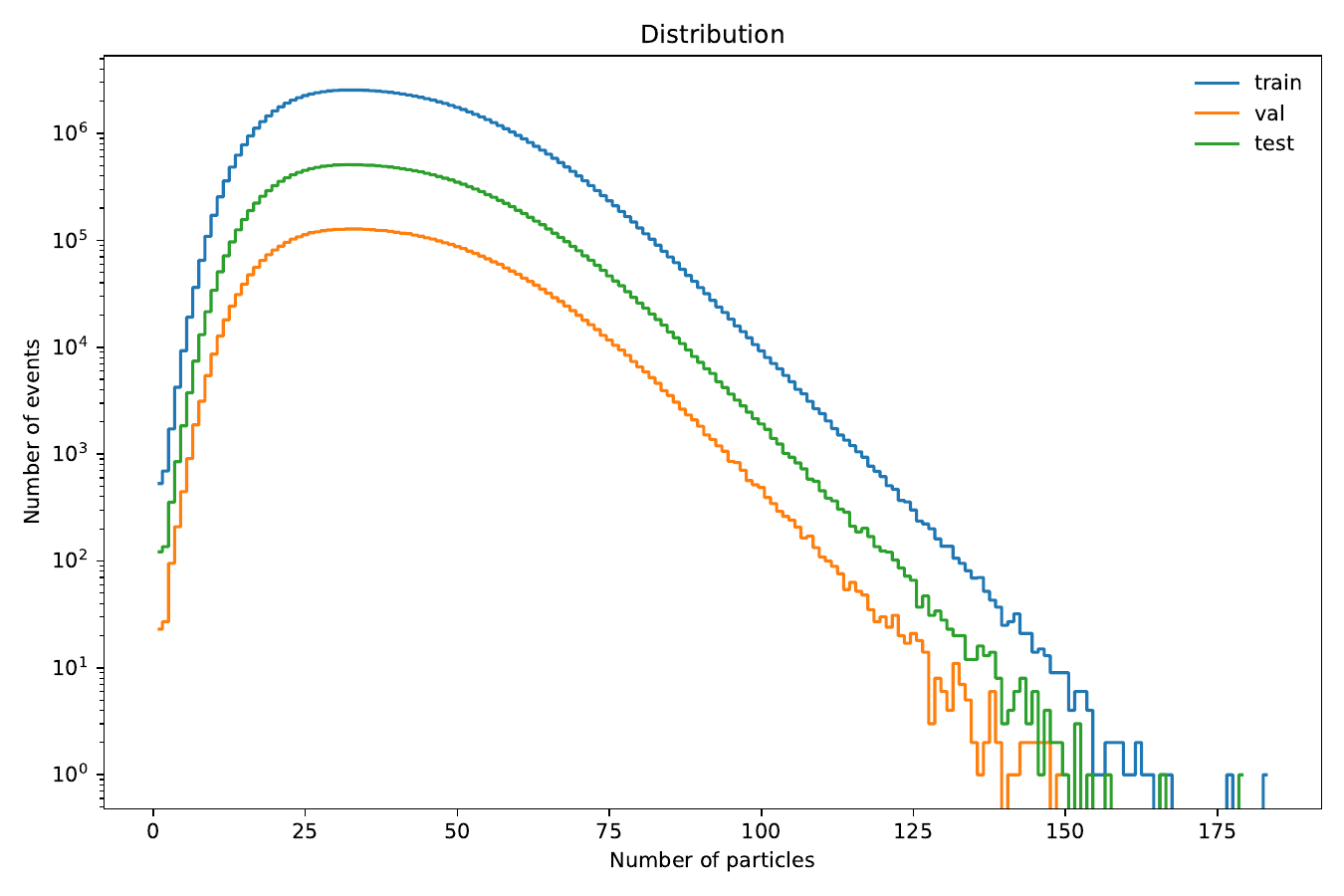}
    \caption{\textbf{Distribution of number of events w.r.t. number of constituents.} We compare the distribution of each set to provide a global comparison. We show that the three sets follow the same global distribution in terms of the number of particles.}
    \label{fig:supp:jetclass:global}
\end{figure}

\noindent\textbf{Set comparison.} The Figure \ref{fig:supp:jetclass:global} illustrates the distribution of the available sets by checking the number of particles. We observe that globally, the three sets follow the same global distribution. This dataset is correctly built.

\noindent\textbf{Set files comparison.} As the global distribution is similar for each of the three sets, we can now focus on those in the train set for comparing whether taking a subset of the available files is equivalent or not.

\captionof{figure}{\textbf{Distribution of number of events w.r.t. number of particles for train set.} We compare the distribution for each set of files from the train set. We clearly observe that each class follows the same distribution in terms of the number of particles. Differences in distribution tails are also observed, where variations may be more or less pronounced depending on the files analyzed.}
\label{fig:supp:jetclass:trainset}
\begin{center}
    \centering

    \begin{minipage}{0.45\textwidth}
        \centering
        \includegraphics[width=\textwidth,height=0.35\textheight,keepaspectratio]{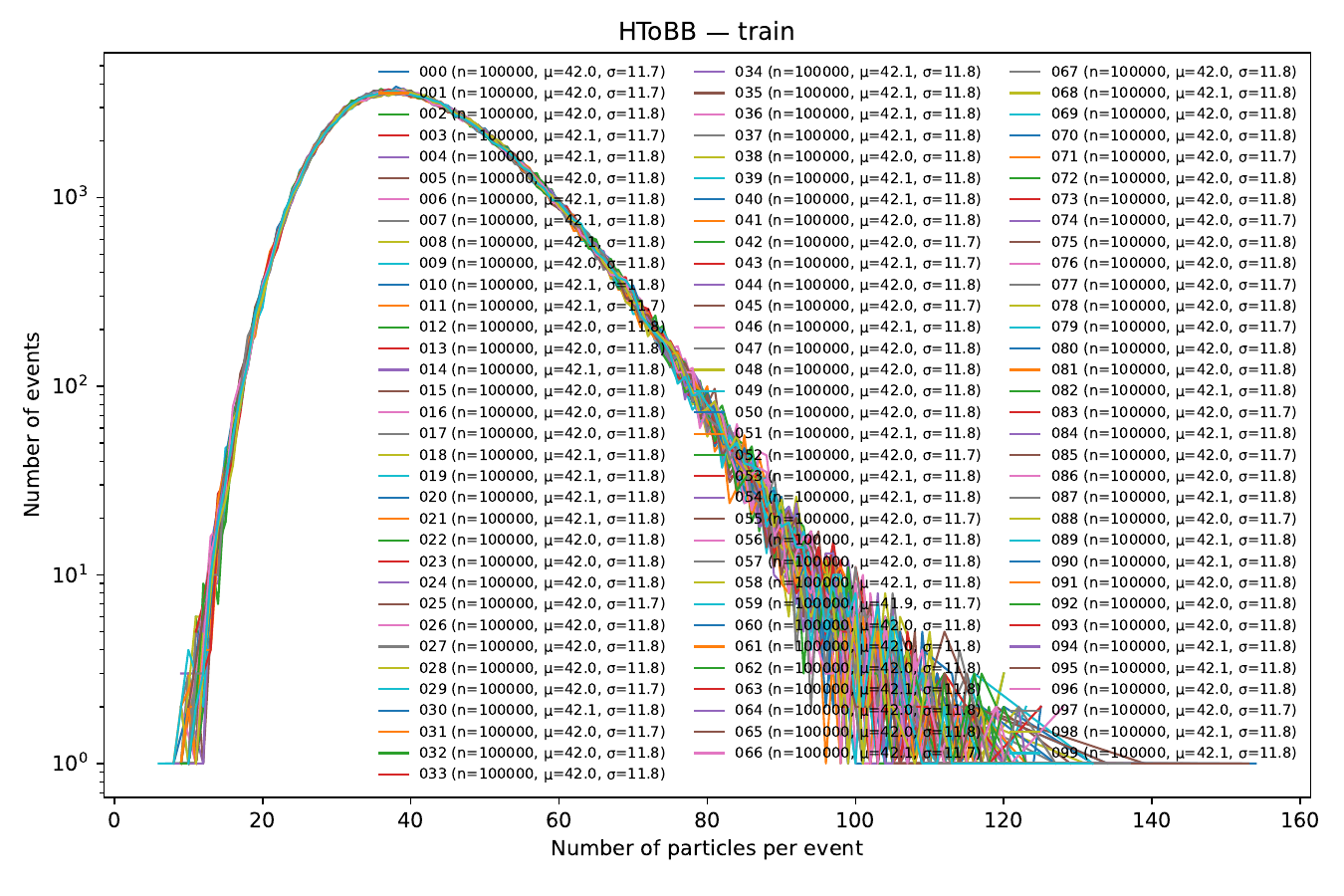}
    \end{minipage}
    \begin{minipage}{0.45\textwidth}
        \centering
        \includegraphics[width=\textwidth,height=0.35\textheight,keepaspectratio]{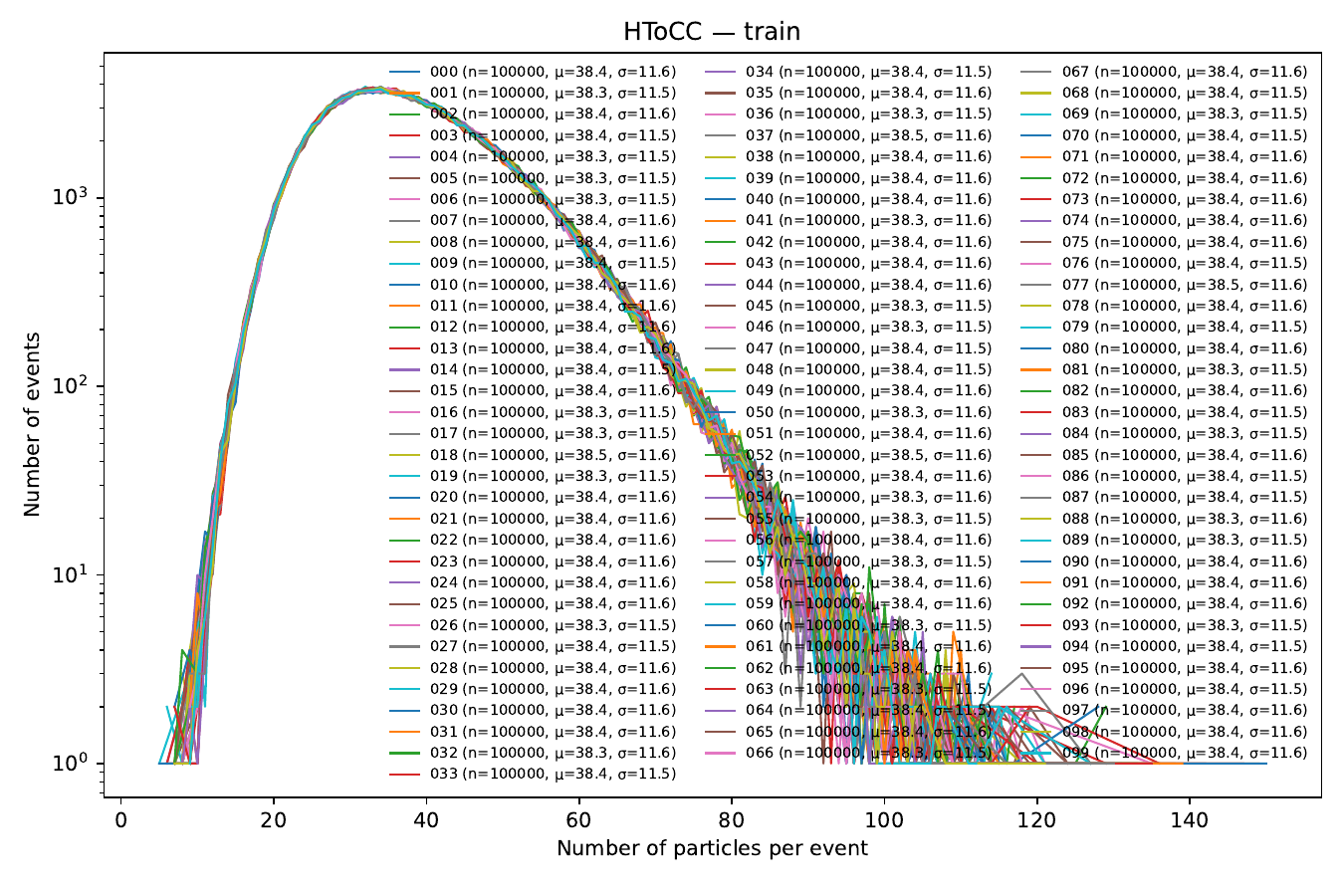}
    \end{minipage}
    \begin{minipage}{0.45\textwidth}
        \centering
        \includegraphics[width=\textwidth,height=0.35\textheight,keepaspectratio]{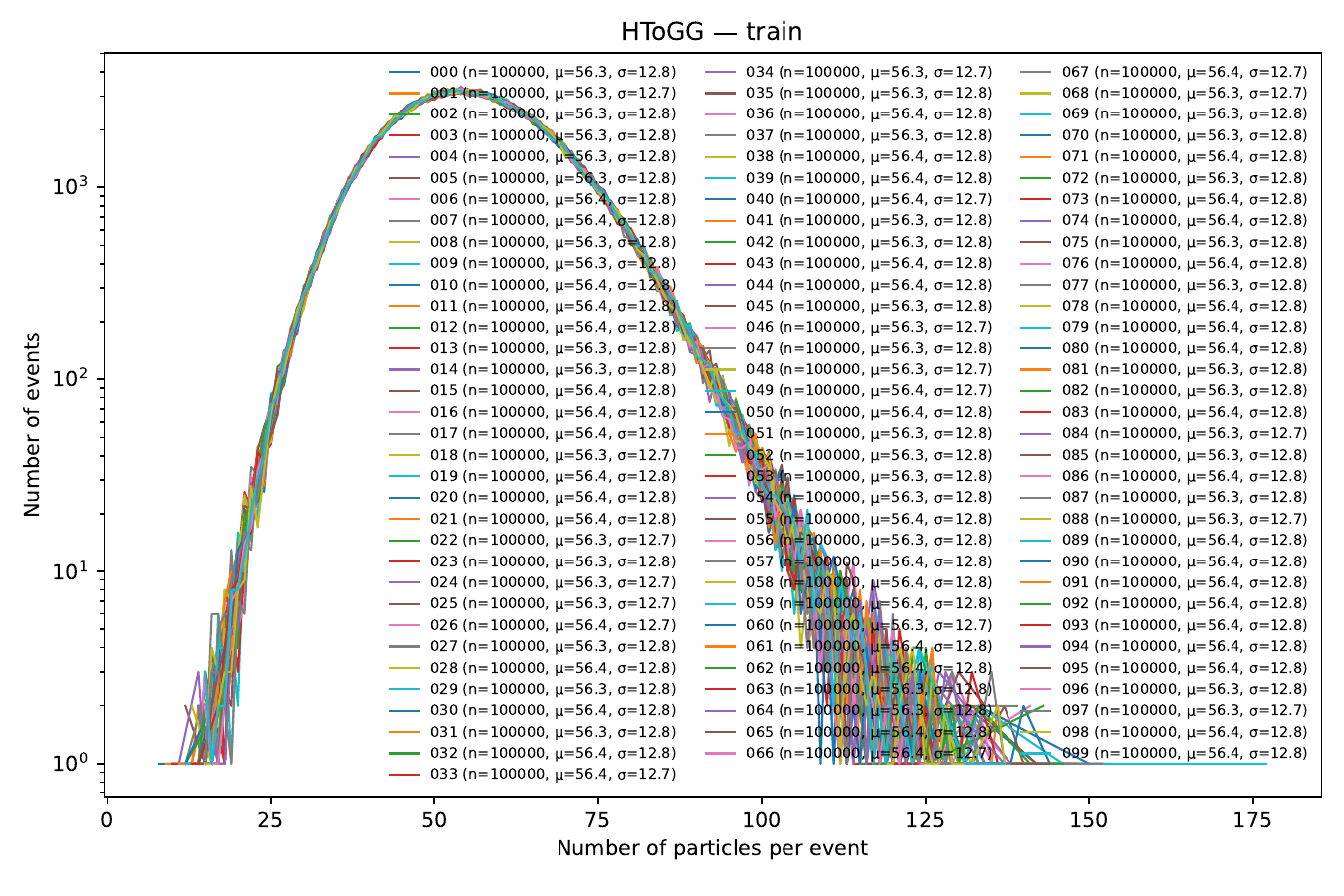}
    \end{minipage}
    \begin{minipage}{0.45\textwidth}
        \centering
        \includegraphics[width=\textwidth,height=0.35\textheight,keepaspectratio]{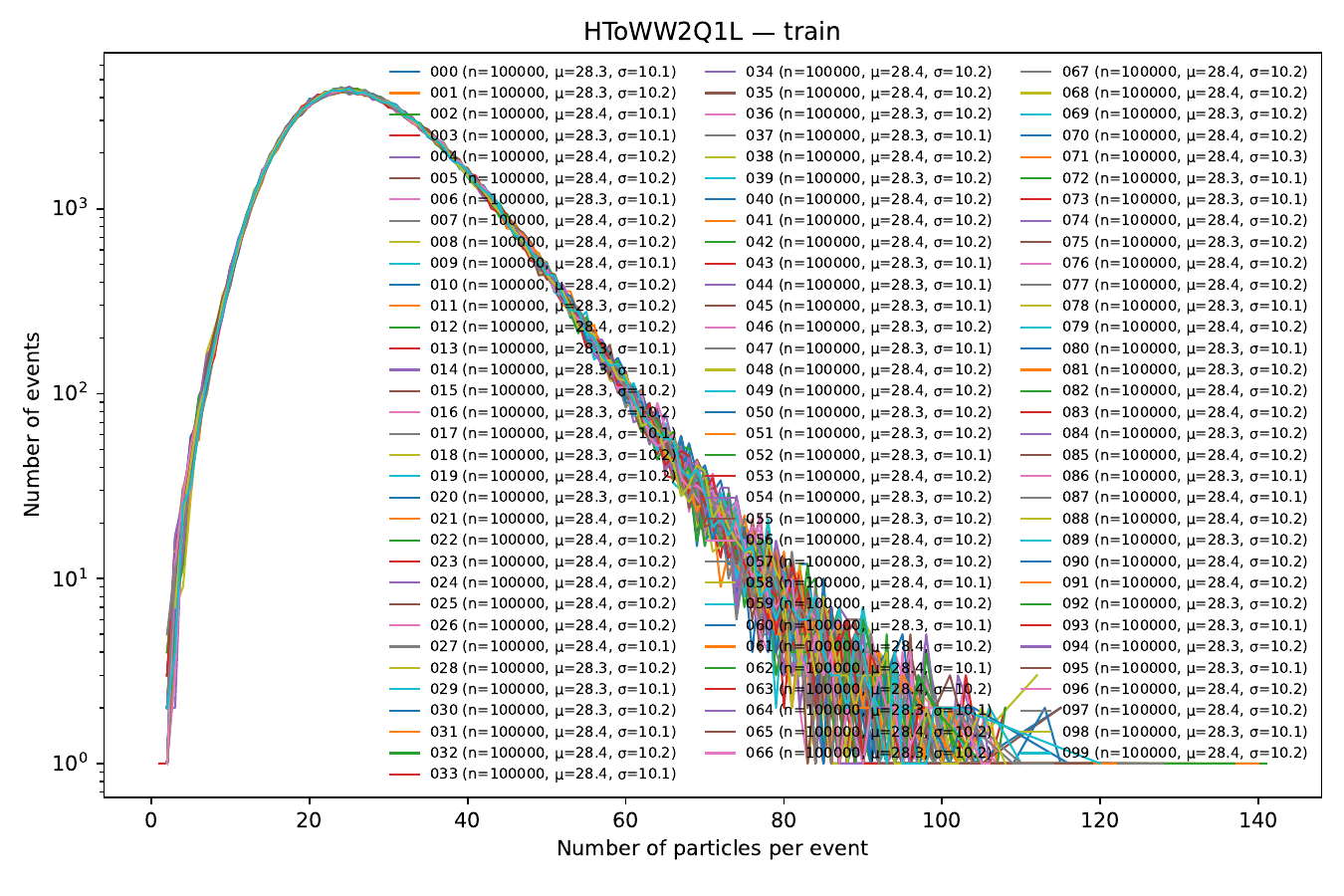}
    \end{minipage}
    \begin{minipage}{0.45\textwidth}
        \centering
        \includegraphics[width=\textwidth,height=0.35\textheight,keepaspectratio]{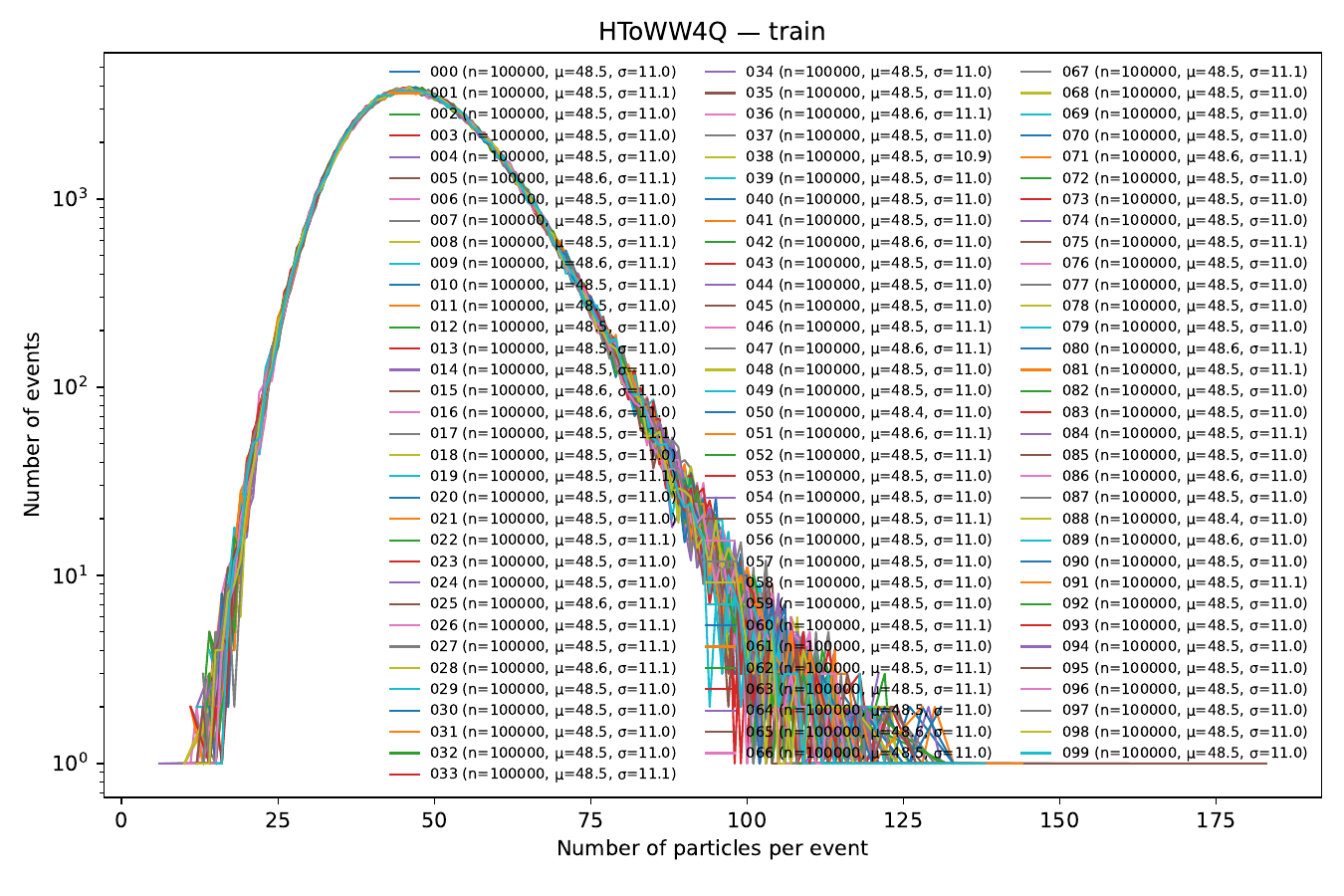}
    \end{minipage}
    \begin{minipage}{0.45\textwidth}
        \centering
        \includegraphics[width=\textwidth,height=0.35\textheight,keepaspectratio]{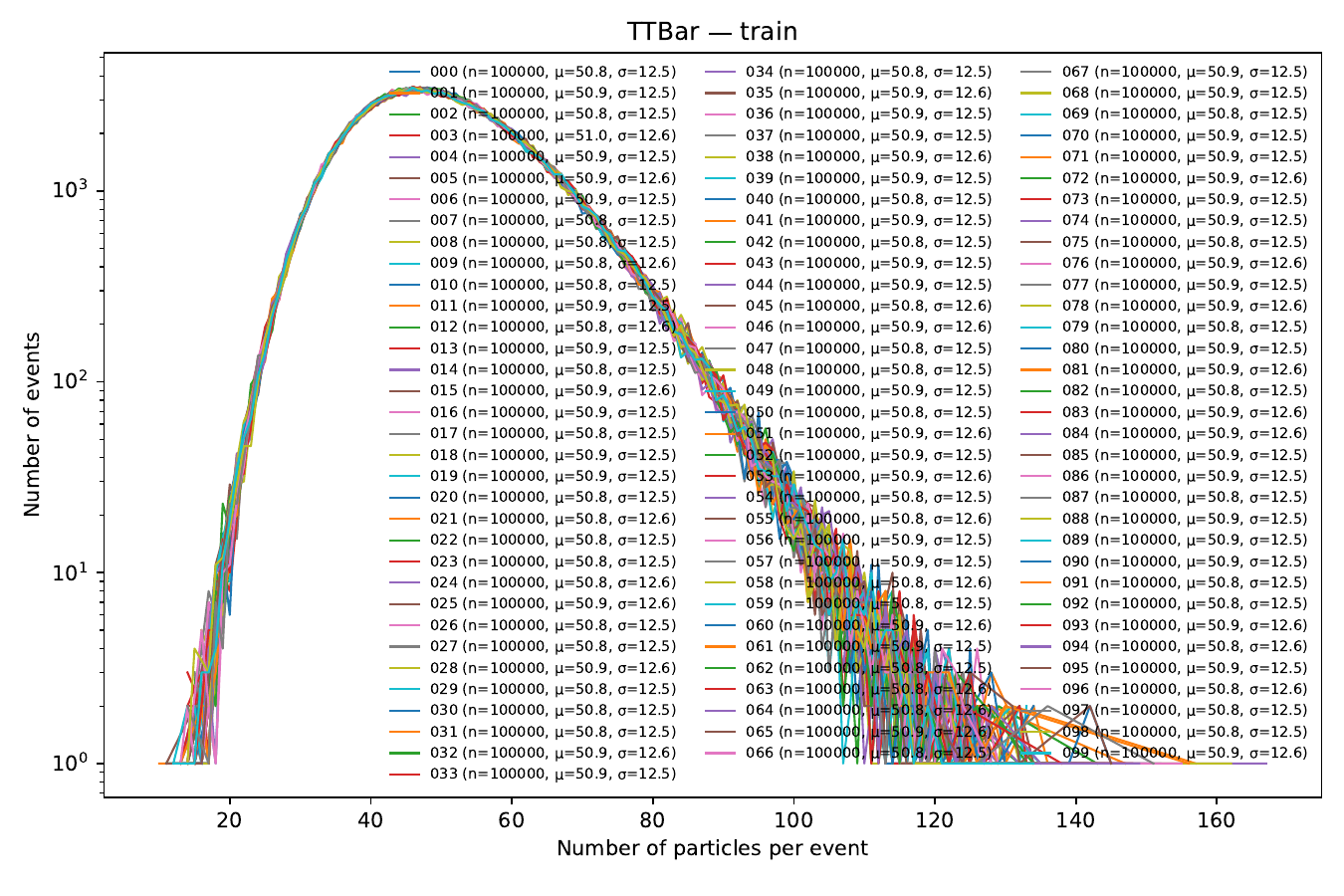}
    \end{minipage}
    \begin{minipage}{0.45\textwidth}
        \centering
        \includegraphics[width=\textwidth,height=0.35\textheight,keepaspectratio]{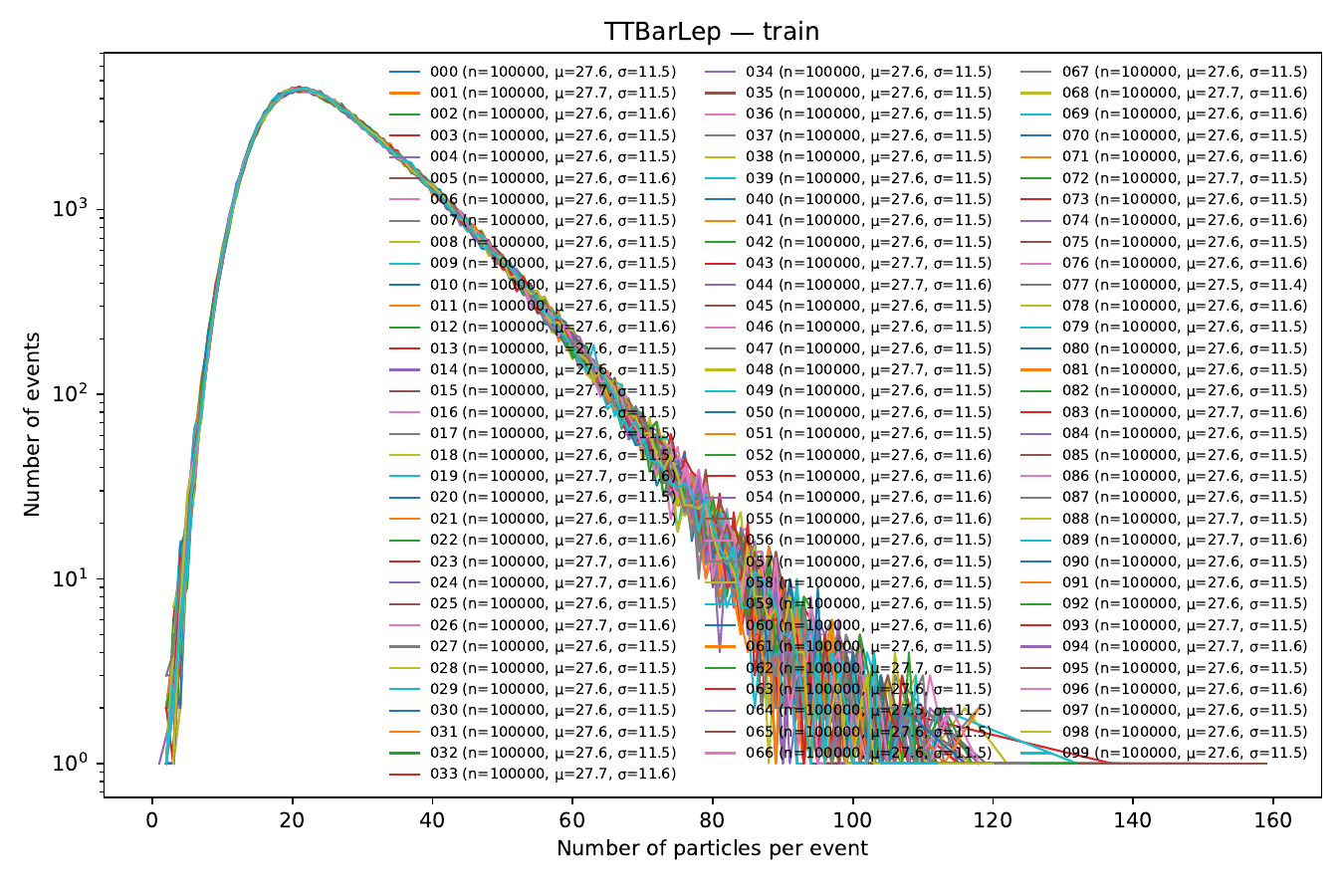}
    \end{minipage}
    \begin{minipage}{0.45\textwidth}
        \centering
        \includegraphics[width=\textwidth,height=0.35\textheight,keepaspectratio]{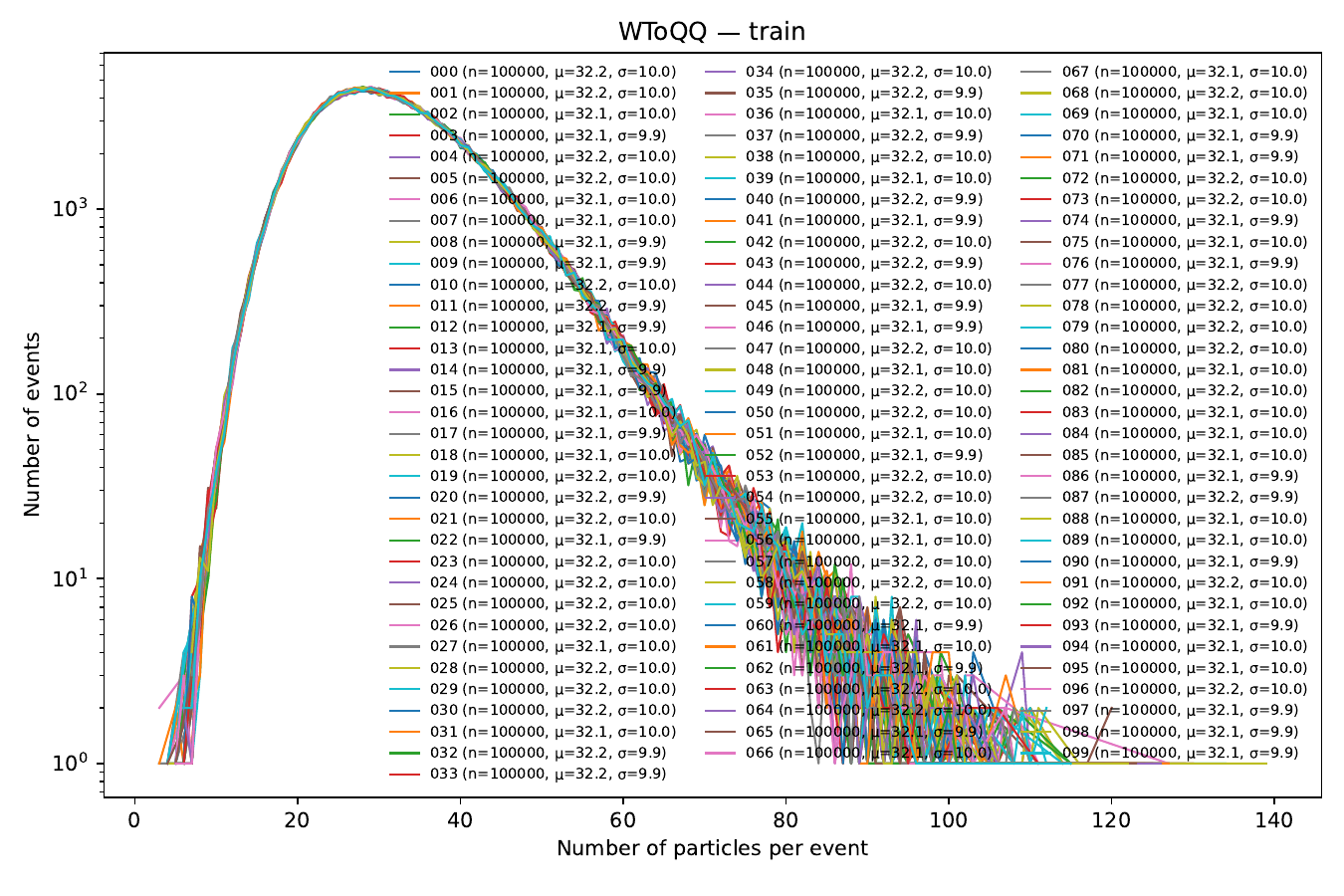}
    \end{minipage}
    \begin{minipage}{0.45\textwidth}
        \centering
        \includegraphics[width=\textwidth,height=0.35\textheight,keepaspectratio]{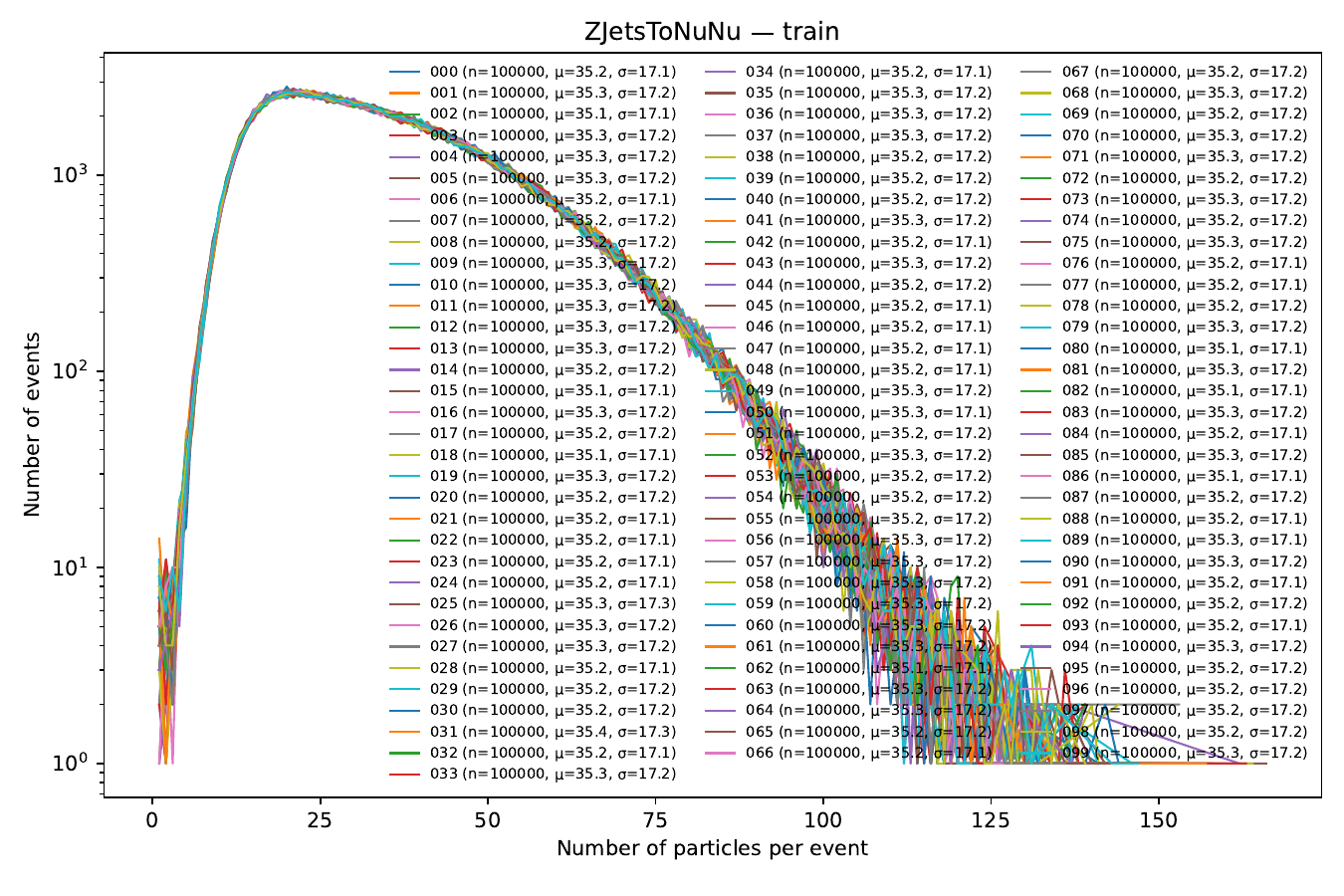}
    \end{minipage}
    \begin{minipage}{0.45\textwidth}
        \centering
        \includegraphics[width=\textwidth,height=0.35\textheight,keepaspectratio]{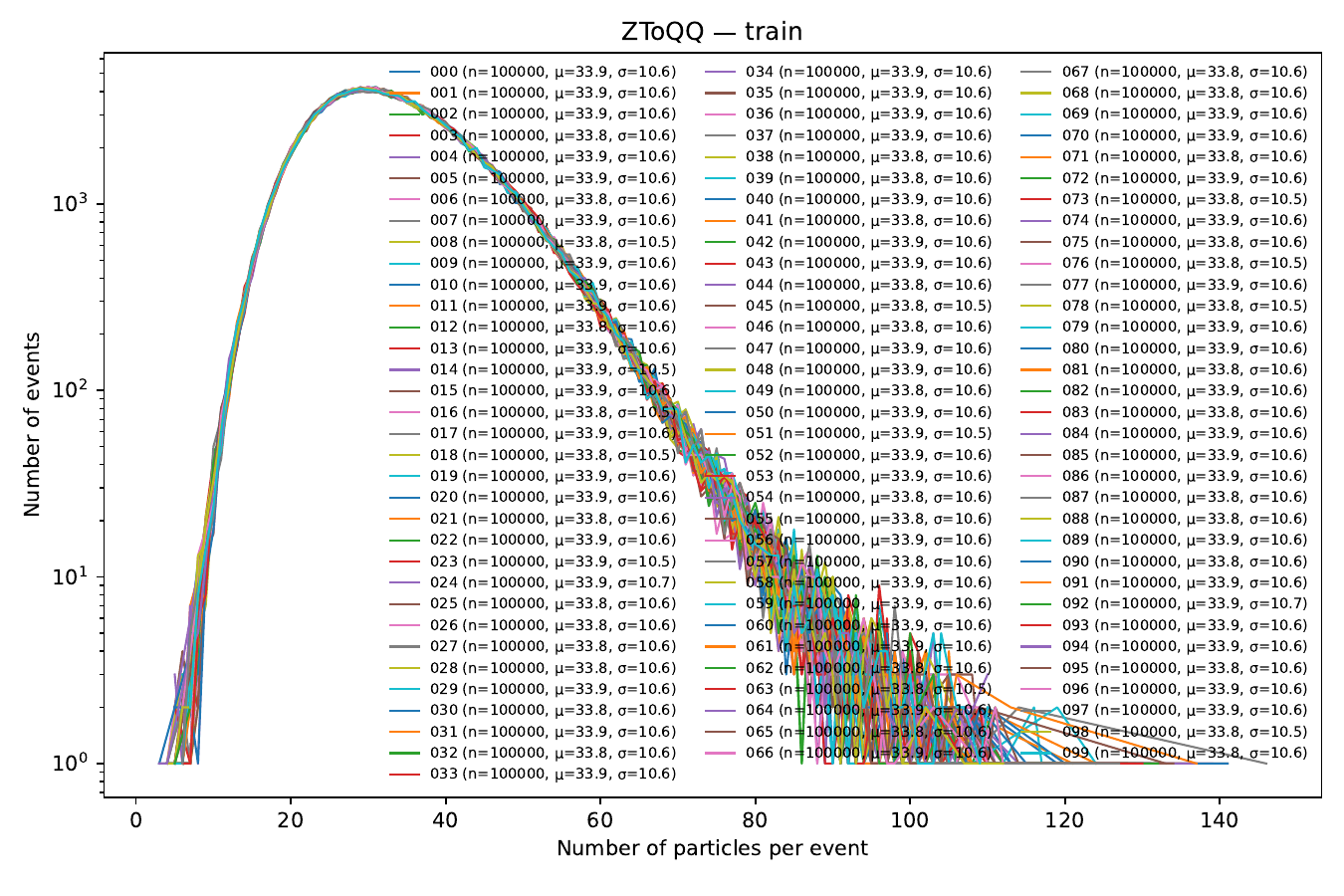}
    \end{minipage}
\end{center}
The Figure \ref{fig:supp:jetclass:trainset} shows the distributions. We observe that the distributions are very close to each other and that the differences occur at the tails of the distributions for a very small number of events. Looking at the details of the means and standard deviations (number of particles), we see that they are very similar between the files for each class. We observed the same phenomenon for the other sets (validation \ref{fig:supp:jetclass:valset} and test \ref{fig:supp:jetclass:testset}). What we can deduce from this is that the partitioning of file events has been carried out correctly. This shows that we can subsample files from different sets to reduce the duration and therefore the cost of training during method development (while ensuring that the same number of files per class is retained so that there is an equal number of events per class). Nevertheless, we can note that using more and more examples allows us to obtain better results and to take into account many more specific events, such as those found at the tails of these distributions (see \citep[Section~5.1]{qu2022particle}).

\captionof{figure}{\textbf{Distribution of number of events w.r.t. number of particles for val set.} We compare the distribution for each set of files from the val set. We clearly observe that each class follows the same distribution in terms of the number of particles. Differences in distribution tails are also observed, where variations may be more or less pronounced depending on the files analyzed.}
\label{fig:supp:jetclass:valset}
\begin{center}
    \centering

    \begin{minipage}{0.45\textwidth}
        \centering
        \includegraphics[width=\textwidth,height=0.35\textheight,keepaspectratio]{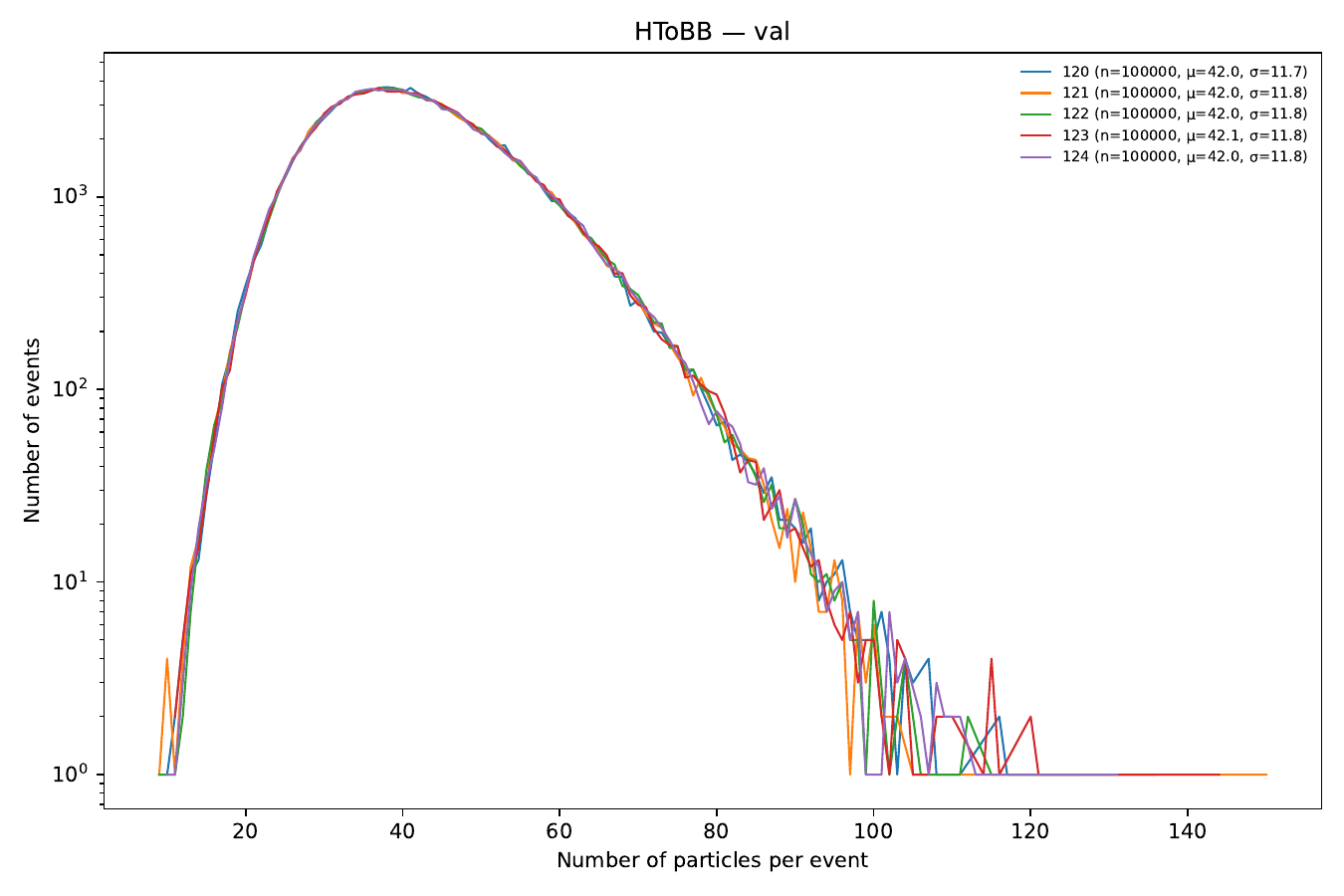}
    \end{minipage}
    \begin{minipage}{0.45\textwidth}
        \centering
        \includegraphics[width=\textwidth,height=0.35\textheight,keepaspectratio]{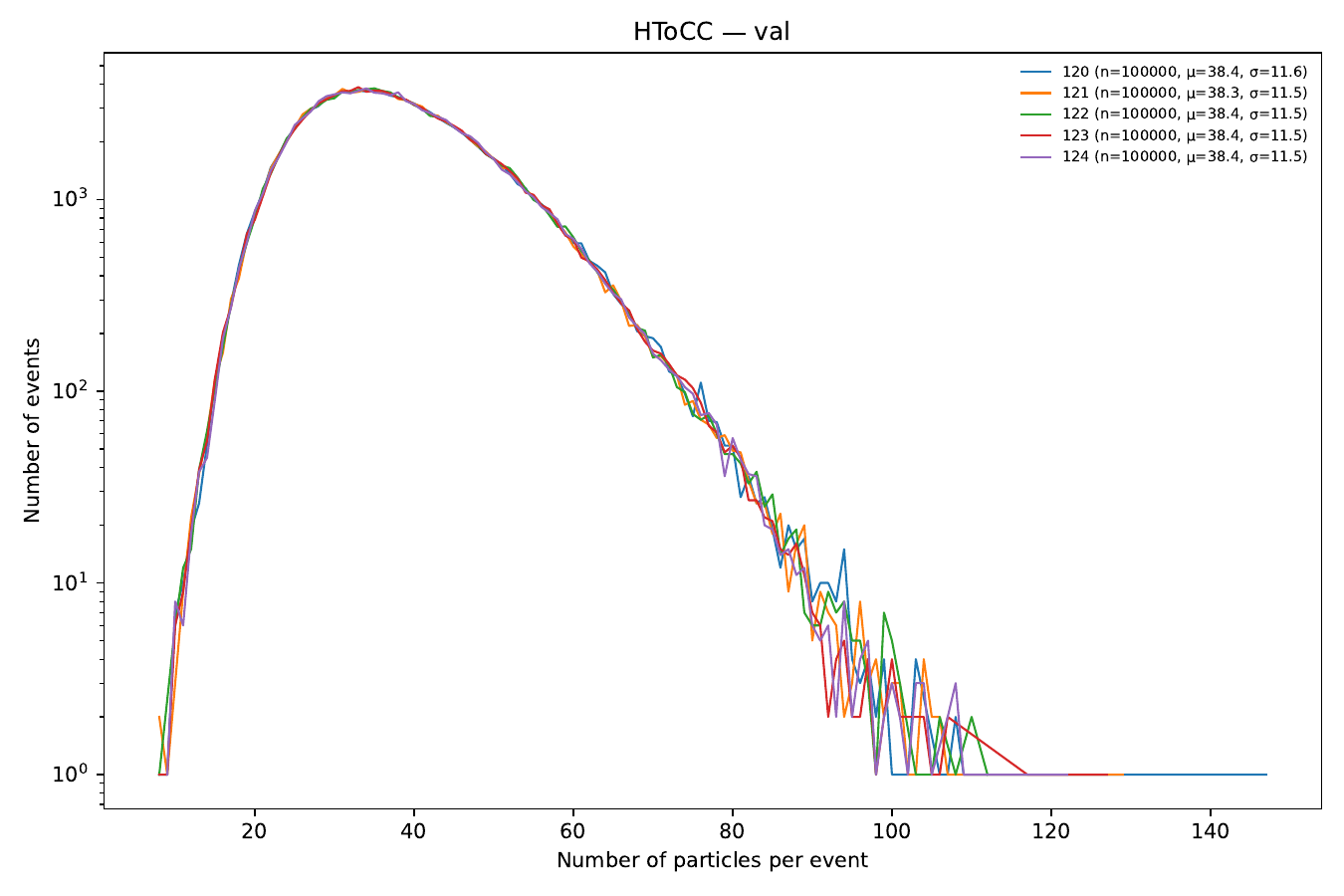}
    \end{minipage}
    \begin{minipage}{0.45\textwidth}
        \centering
        \includegraphics[width=\textwidth,height=0.35\textheight,keepaspectratio]{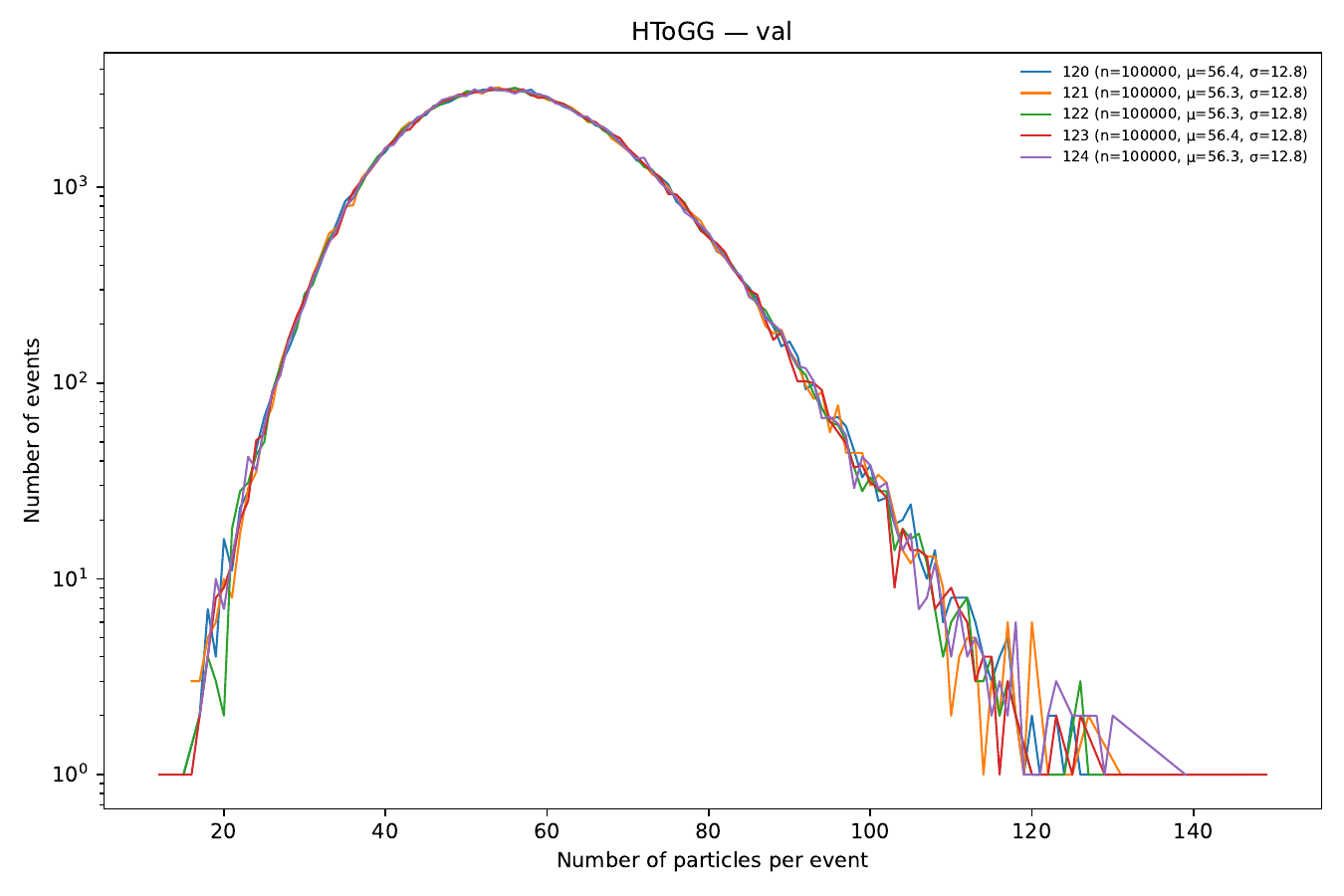}
    \end{minipage}
    \begin{minipage}{0.45\textwidth}
        \centering
        \includegraphics[width=\textwidth,height=0.35\textheight,keepaspectratio]{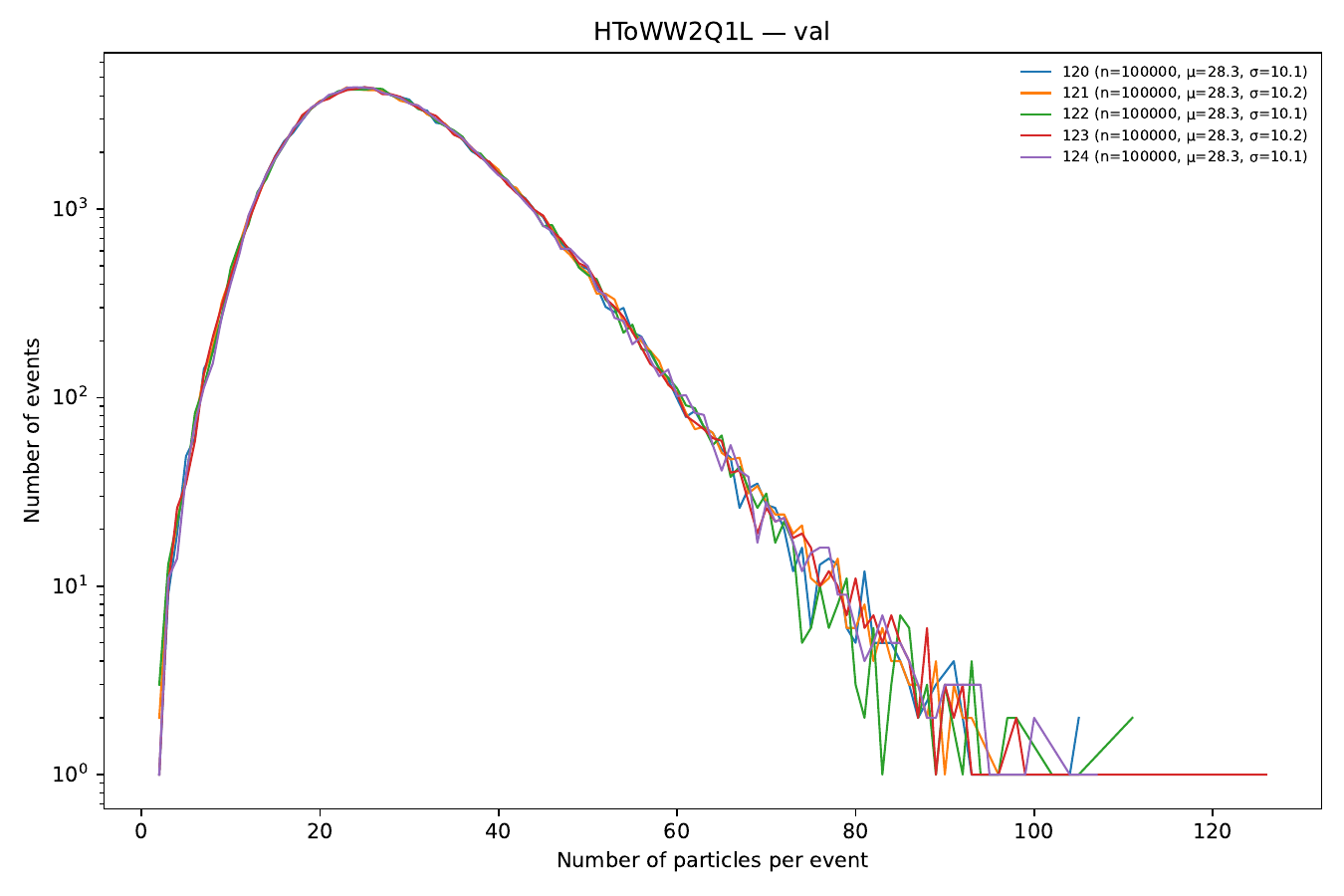}
    \end{minipage}
    \begin{minipage}{0.45\textwidth}
        \centering
        \includegraphics[width=\textwidth,height=0.35\textheight,keepaspectratio]{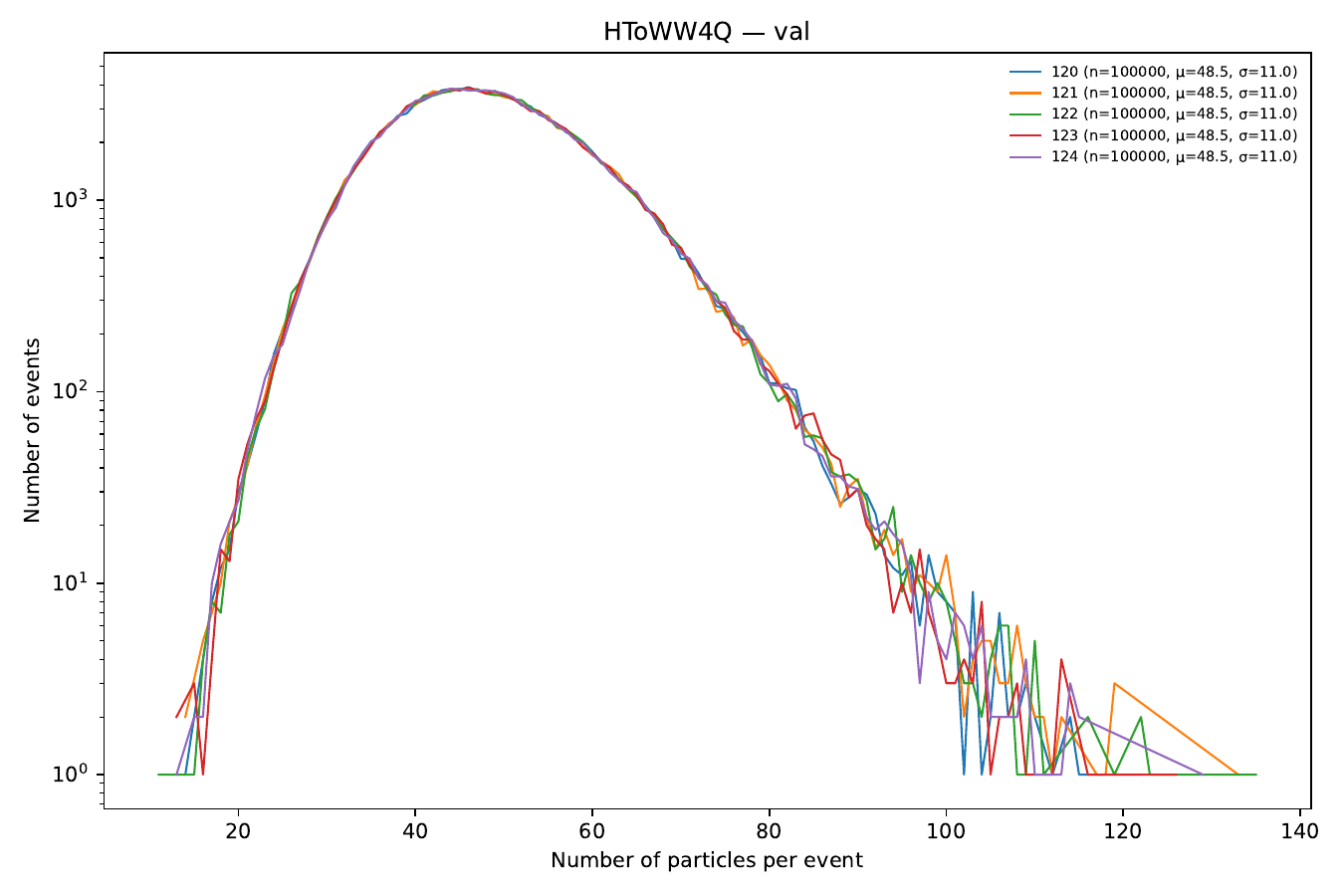}
    \end{minipage}
    \begin{minipage}{0.45\textwidth}
        \centering
        \includegraphics[width=\textwidth,height=0.35\textheight,keepaspectratio]{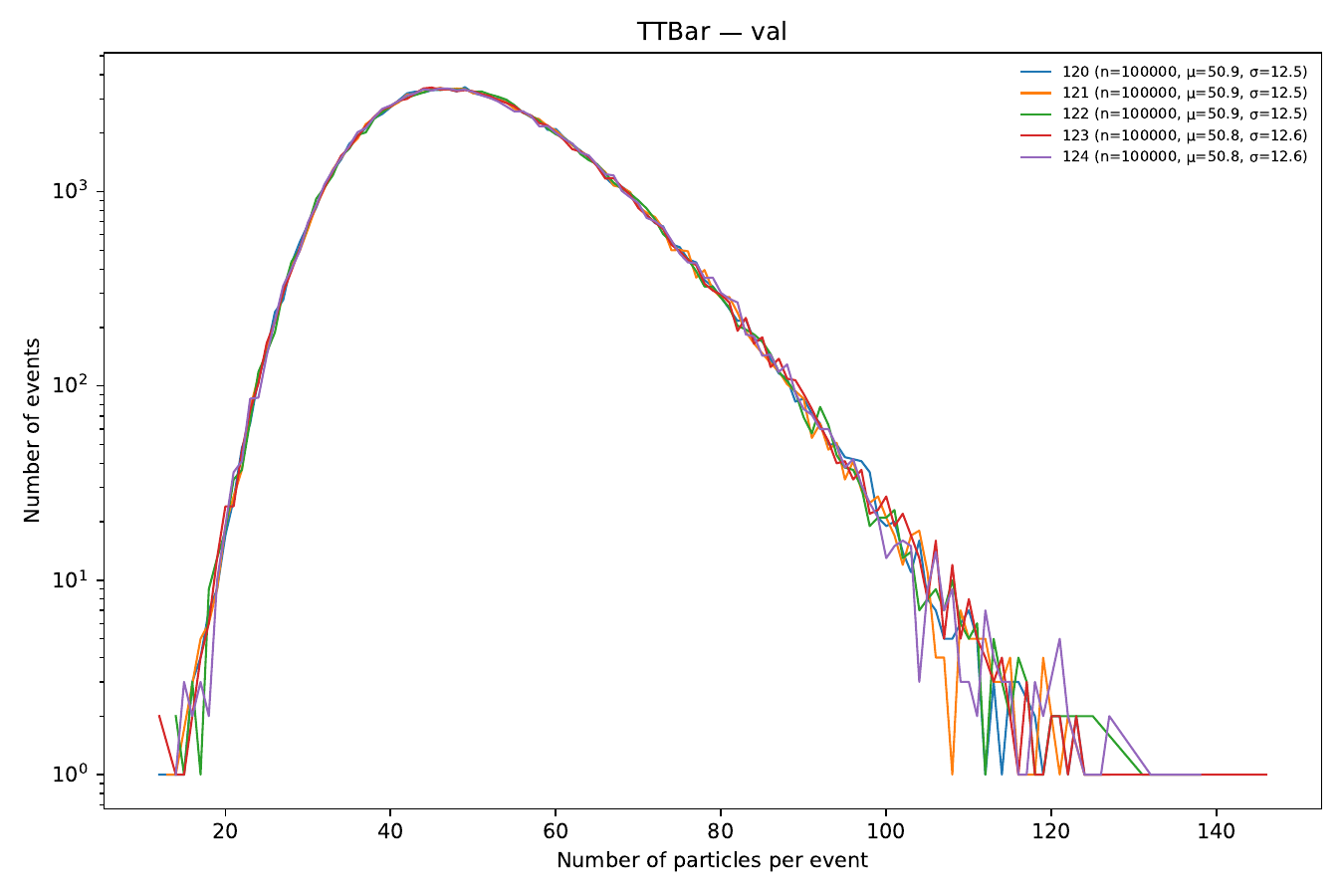}
    \end{minipage}
    \begin{minipage}{0.45\textwidth}
        \centering
        \includegraphics[width=\textwidth,height=0.35\textheight,keepaspectratio]{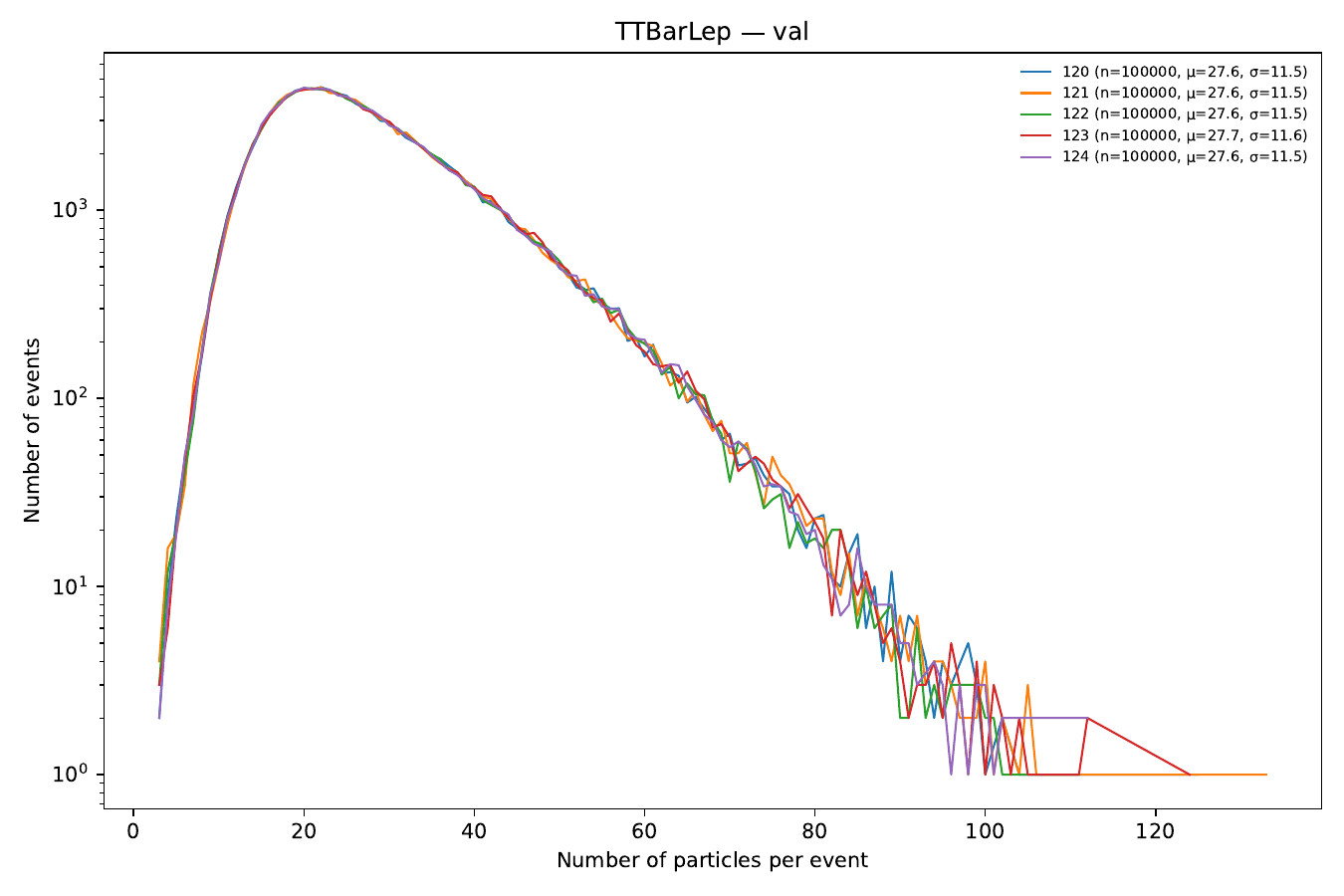}
    \end{minipage}
    \begin{minipage}{0.45\textwidth}
        \centering
        \includegraphics[width=\textwidth,height=0.35\textheight,keepaspectratio]{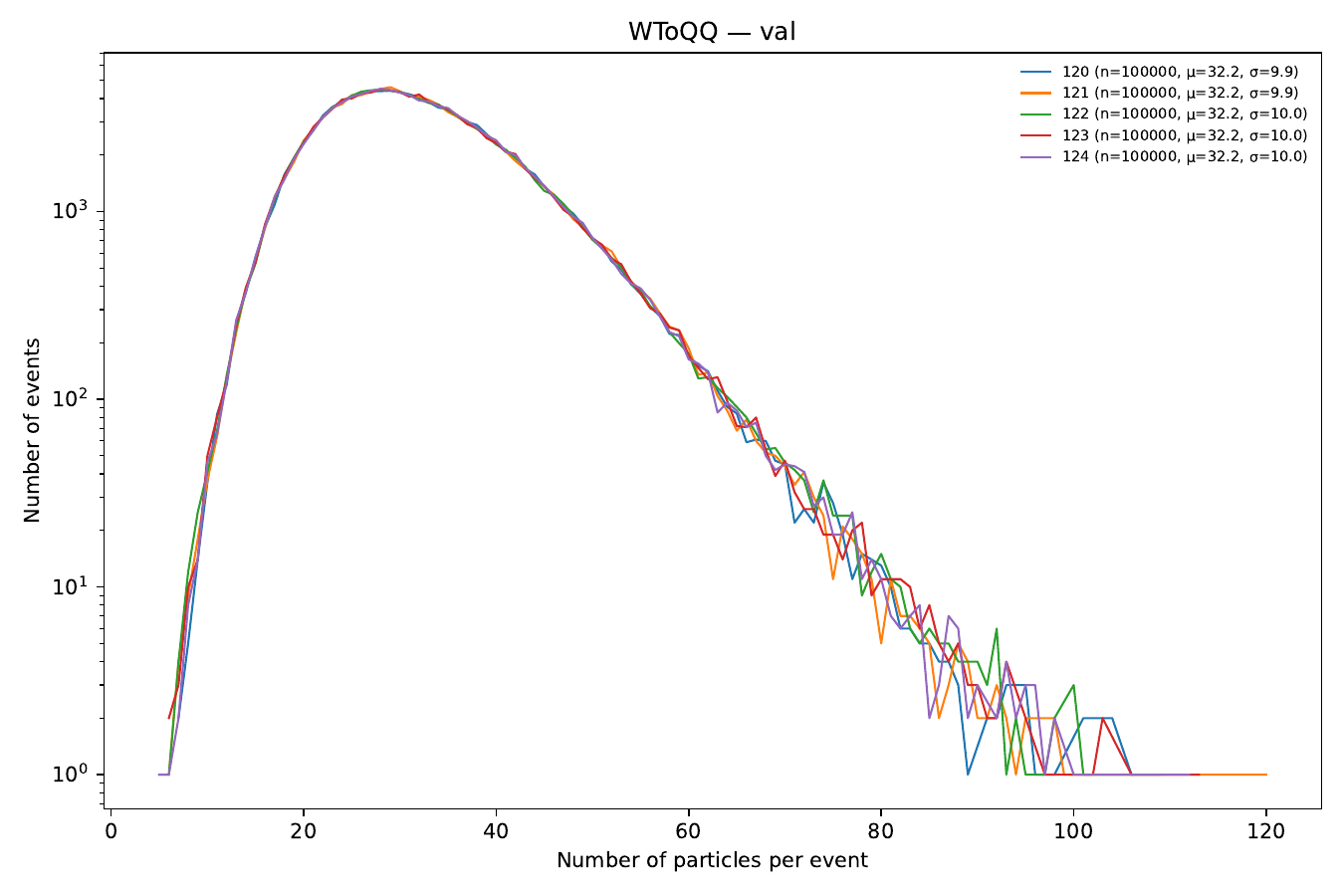}
    \end{minipage}
    \begin{minipage}{0.45\textwidth}
        \centering
        \includegraphics[width=\textwidth,height=0.35\textheight,keepaspectratio]{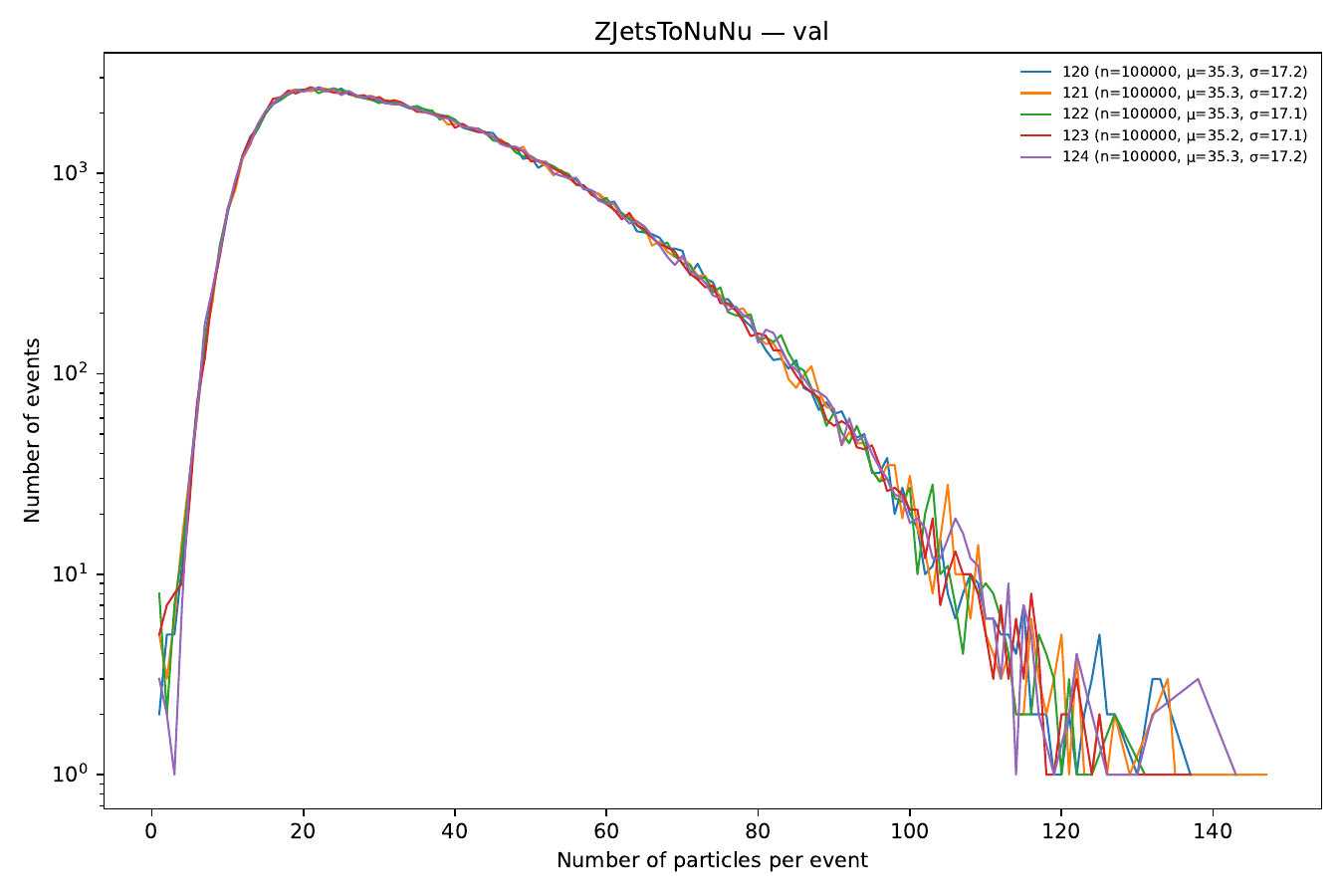}
    \end{minipage}
    \begin{minipage}{0.45\textwidth}
        \centering
        \includegraphics[width=\textwidth,height=0.35\textheight,keepaspectratio]{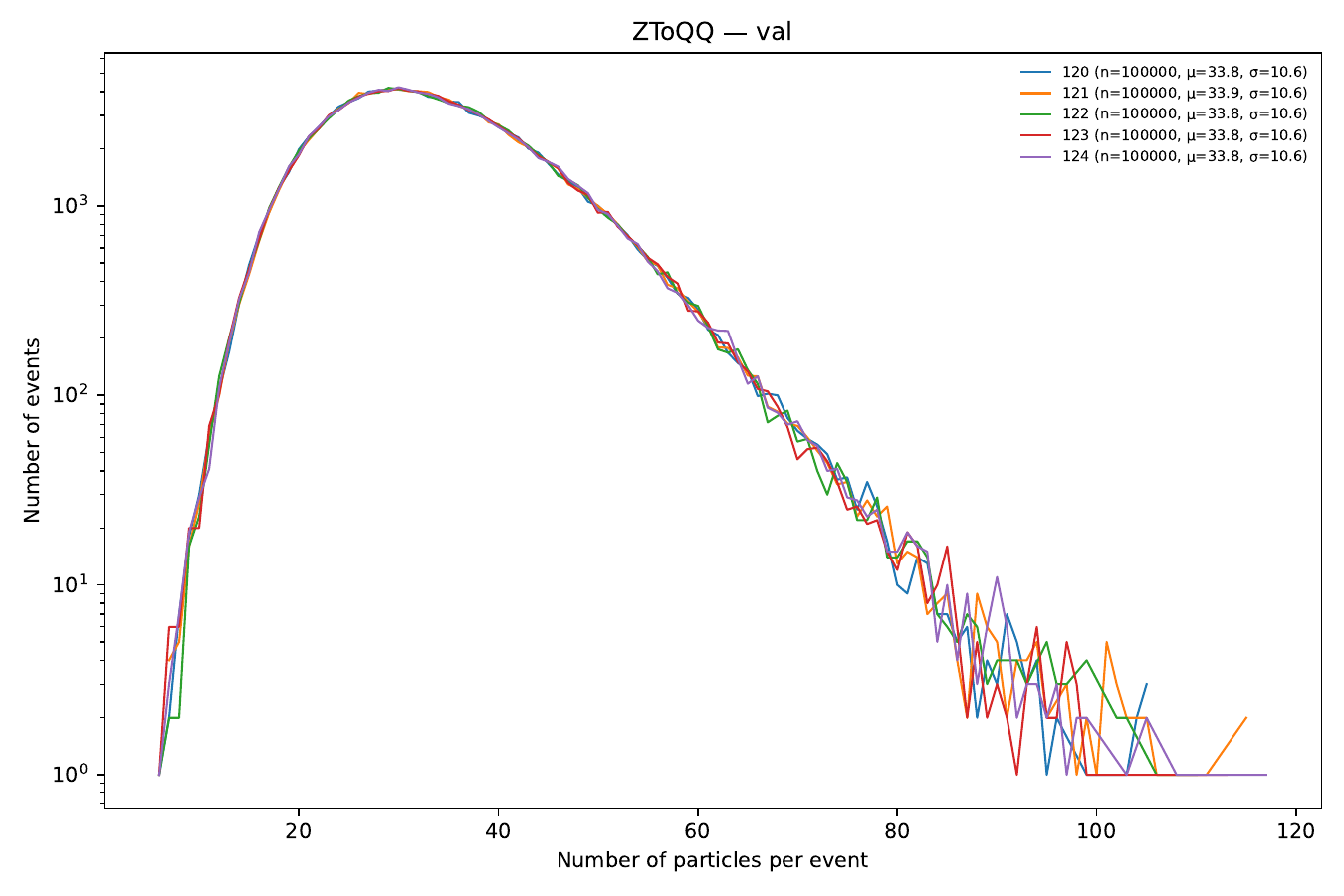}
    \end{minipage}
\end{center}
\captionof{figure}{\textbf{Distribution of number of events w.r.t. number of particles for test set.} We compare the distribution for each set of files from the test set. We clearly observe that each class follows the same distribution in terms of the number of particles. Differences in distribution tails are also observed, where variations may be more or less pronounced depending on the files analyzed.}
\label{fig:supp:jetclass:testset}
\begin{center}
    \centering

    \begin{minipage}{0.45\textwidth}
        \centering
        \includegraphics[width=\textwidth,height=0.35\textheight,keepaspectratio]{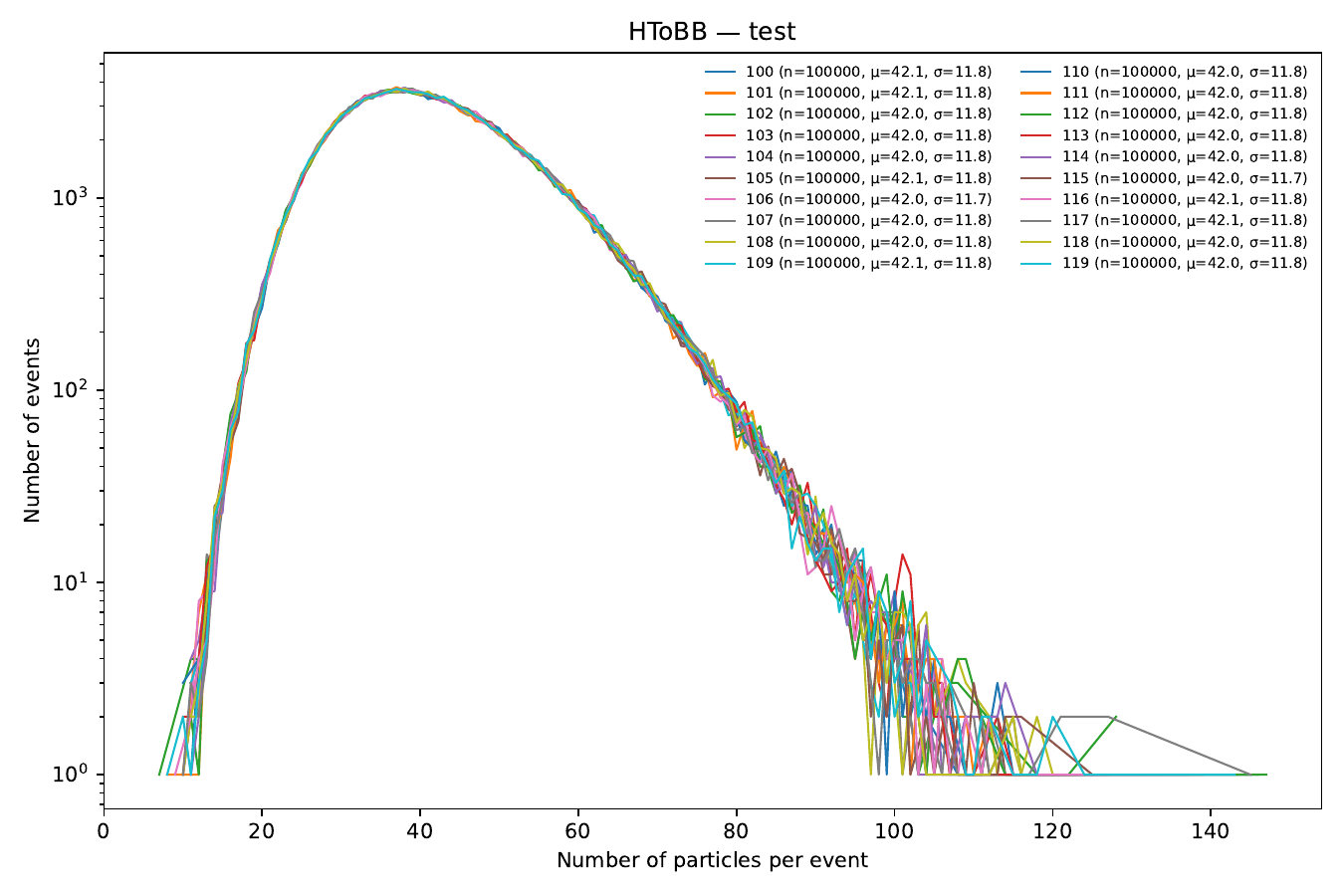}
    \end{minipage}
    \begin{minipage}{0.45\textwidth}
        \centering
        \includegraphics[width=\textwidth,height=0.35\textheight,keepaspectratio]{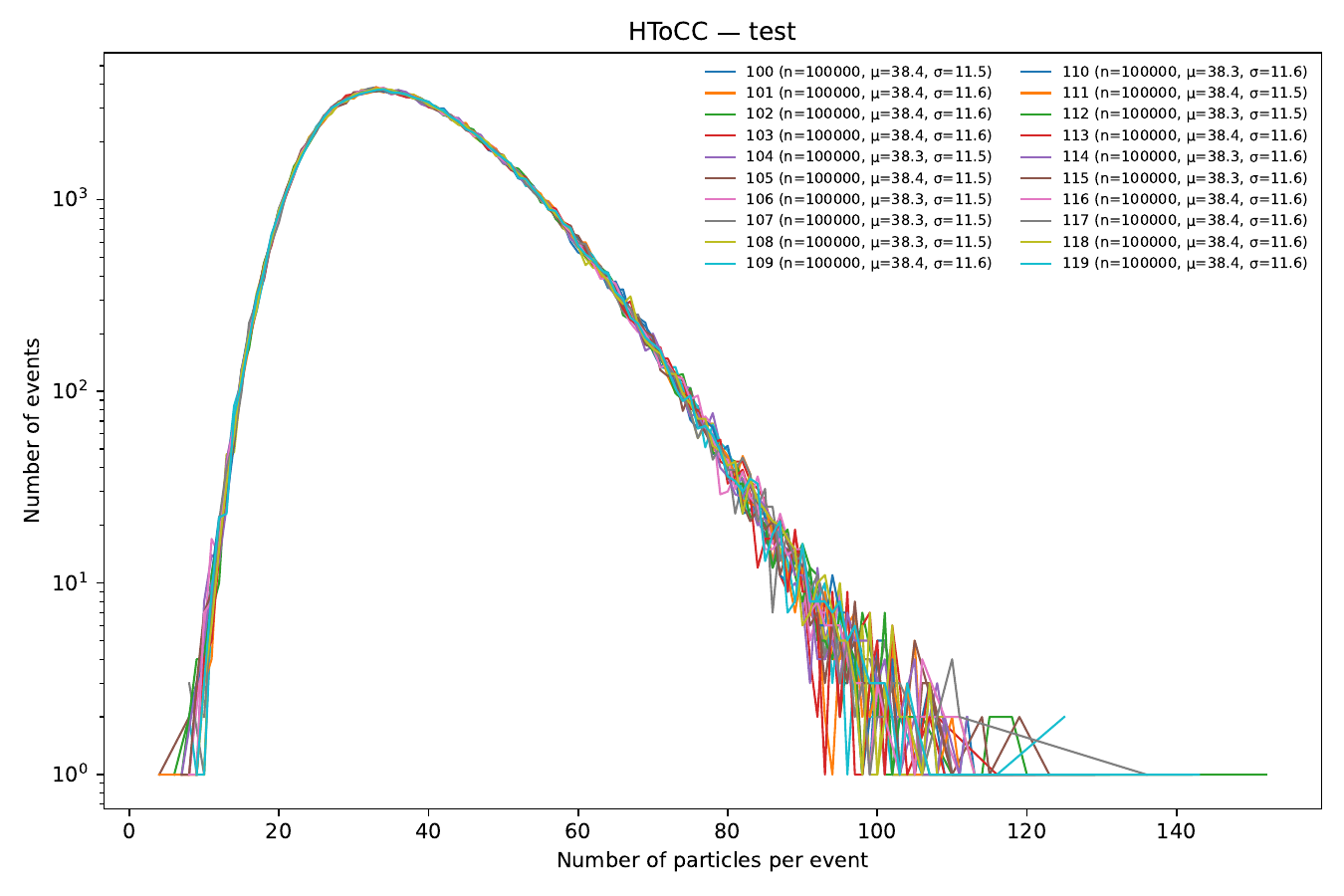}
    \end{minipage}
    \begin{minipage}{0.45\textwidth}
        \centering
        \includegraphics[width=\textwidth,height=0.35\textheight,keepaspectratio]{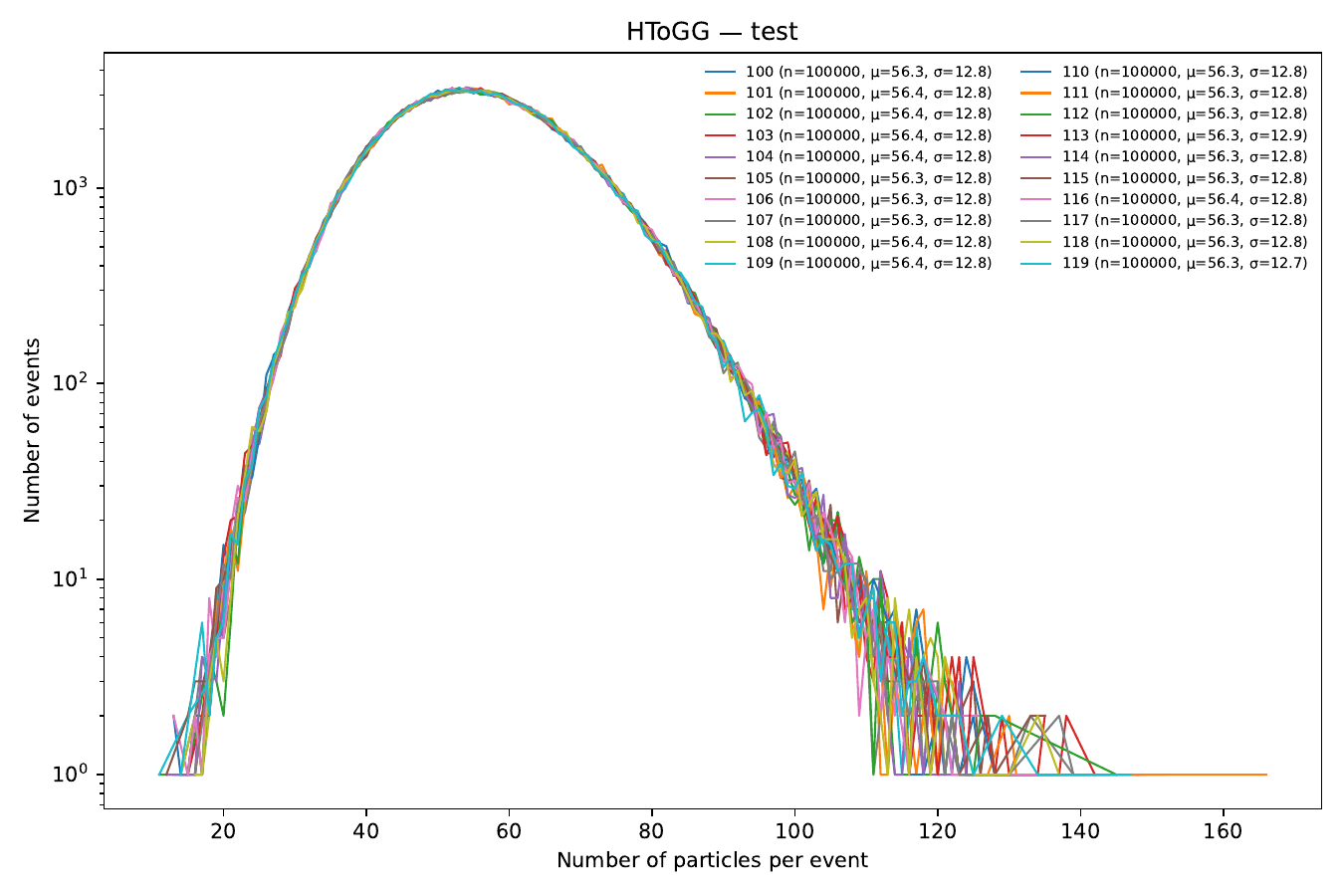}
    \end{minipage}
    \begin{minipage}{0.45\textwidth}
        \centering
        \includegraphics[width=\textwidth,height=0.35\textheight,keepaspectratio]{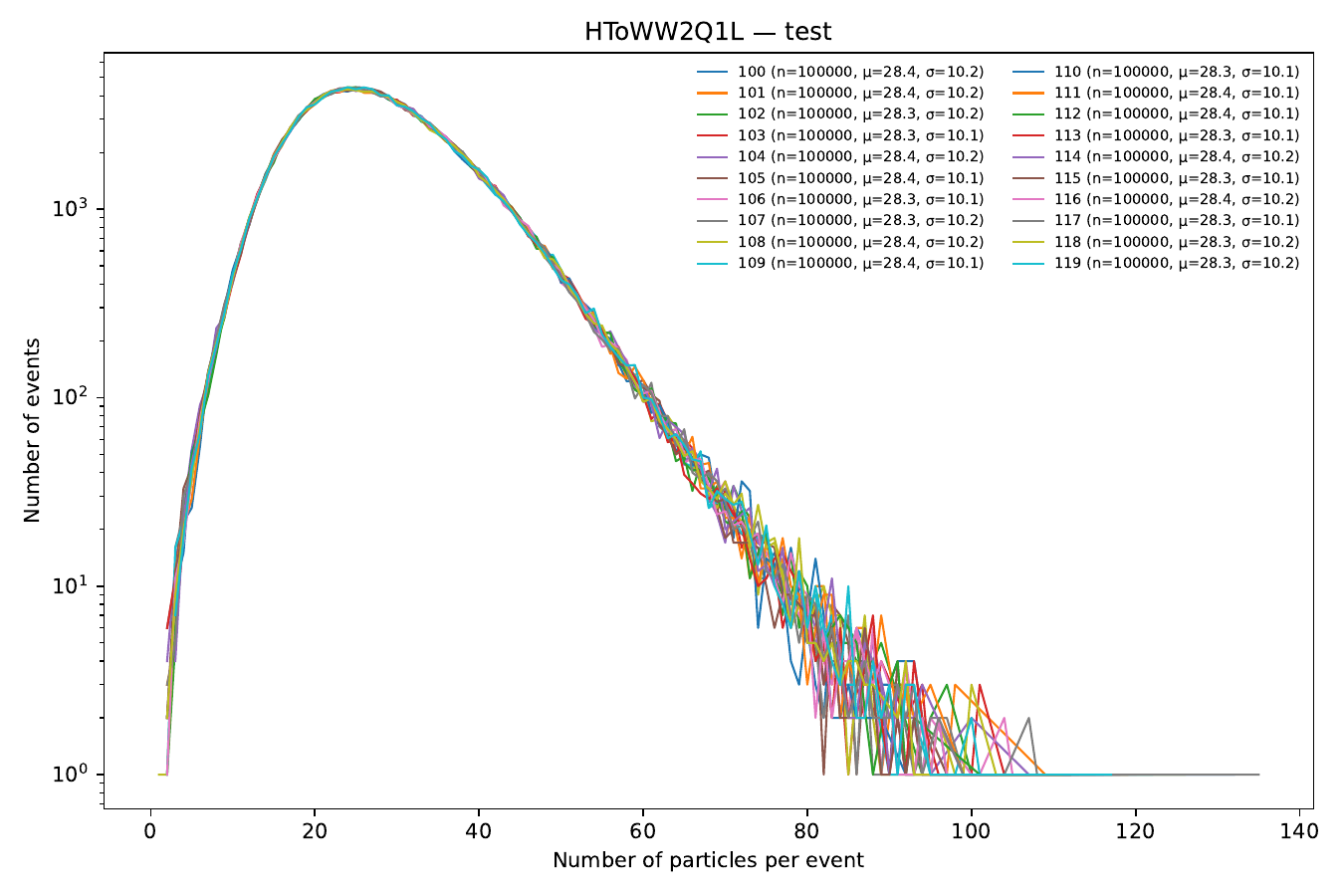}
    \end{minipage}
    \begin{minipage}{0.45\textwidth}
        \centering
        \includegraphics[width=\textwidth,height=0.35\textheight,keepaspectratio]{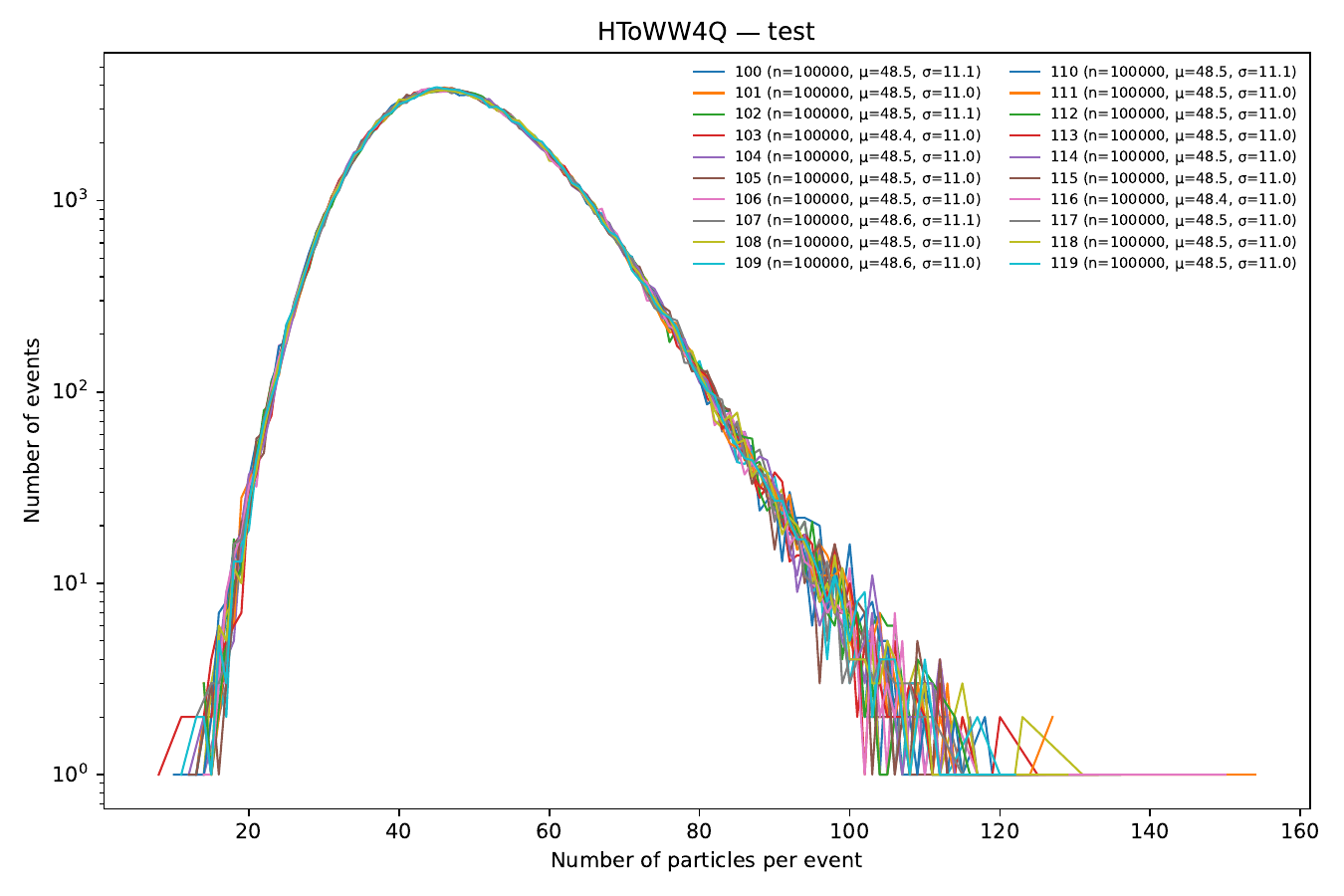}
    \end{minipage}
    \begin{minipage}{0.45\textwidth}
        \centering
        \includegraphics[width=\textwidth,height=0.35\textheight,keepaspectratio]{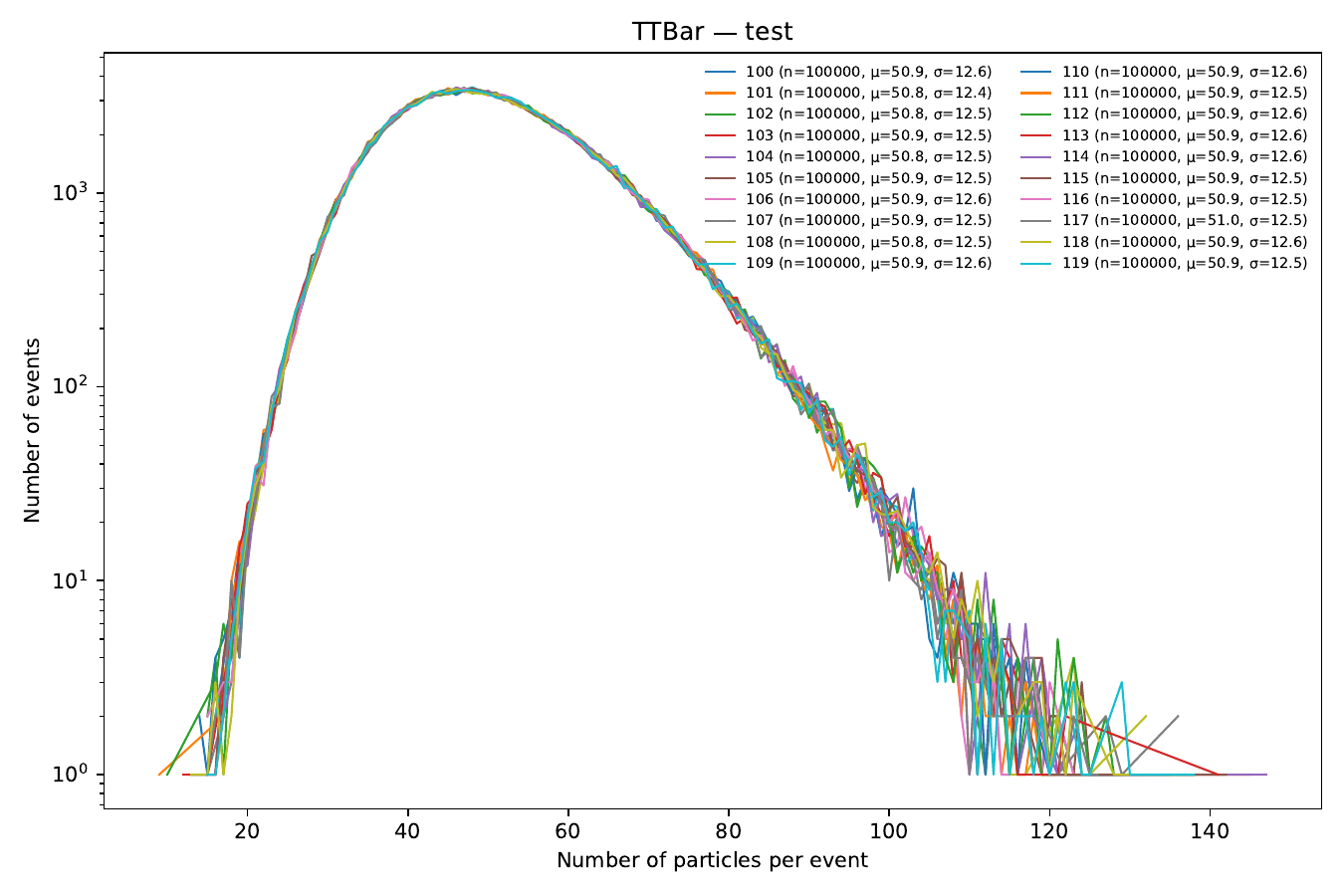}
    \end{minipage}
    \begin{minipage}{0.45\textwidth}
        \centering
        \includegraphics[width=\textwidth,height=0.35\textheight,keepaspectratio]{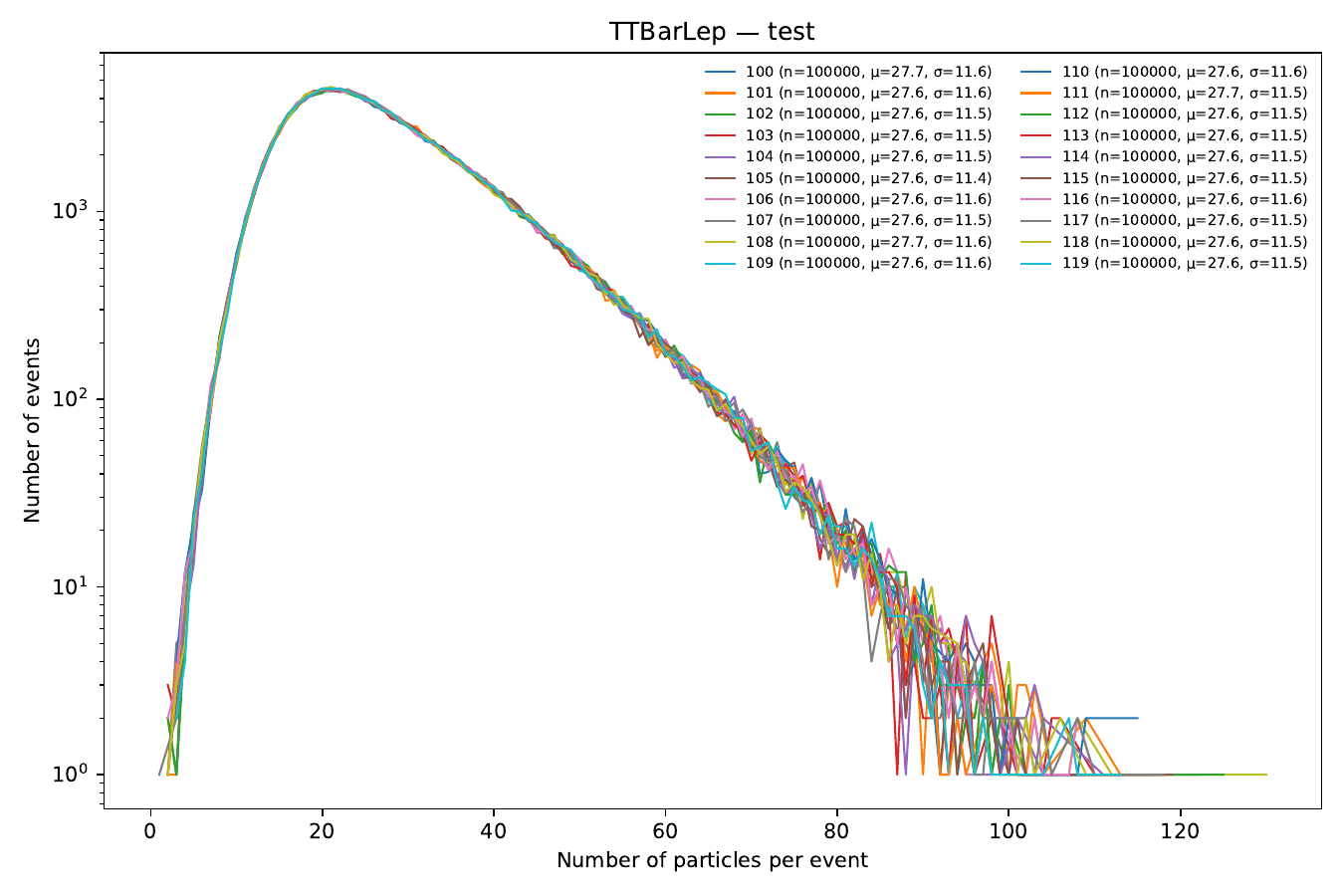}
    \end{minipage}
    \begin{minipage}{0.45\textwidth}
        \centering
        \includegraphics[width=\textwidth,height=0.35\textheight,keepaspectratio]{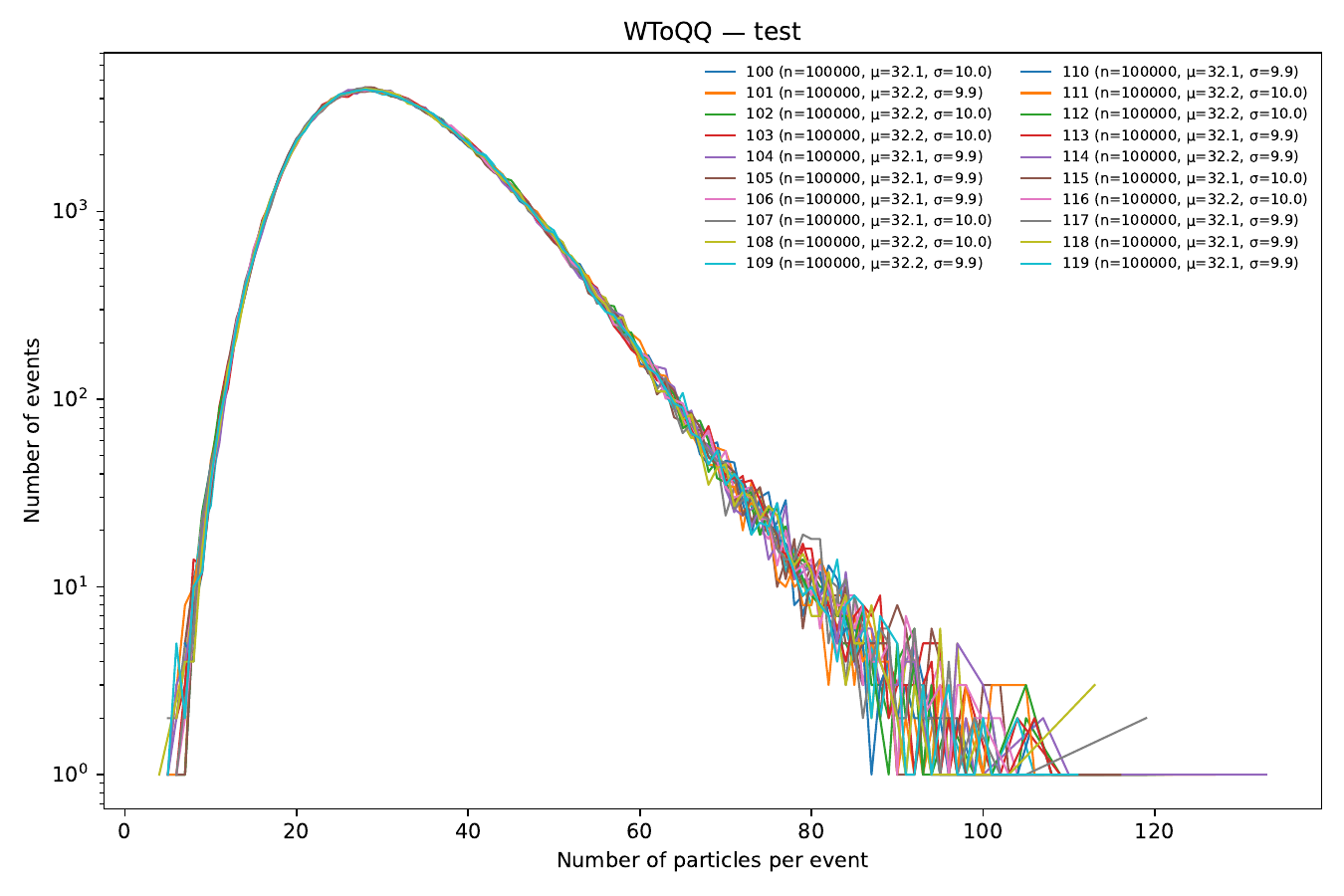}
    \end{minipage}
    \begin{minipage}{0.45\textwidth}
        \centering
        \includegraphics[width=\textwidth,height=0.35\textheight,keepaspectratio]{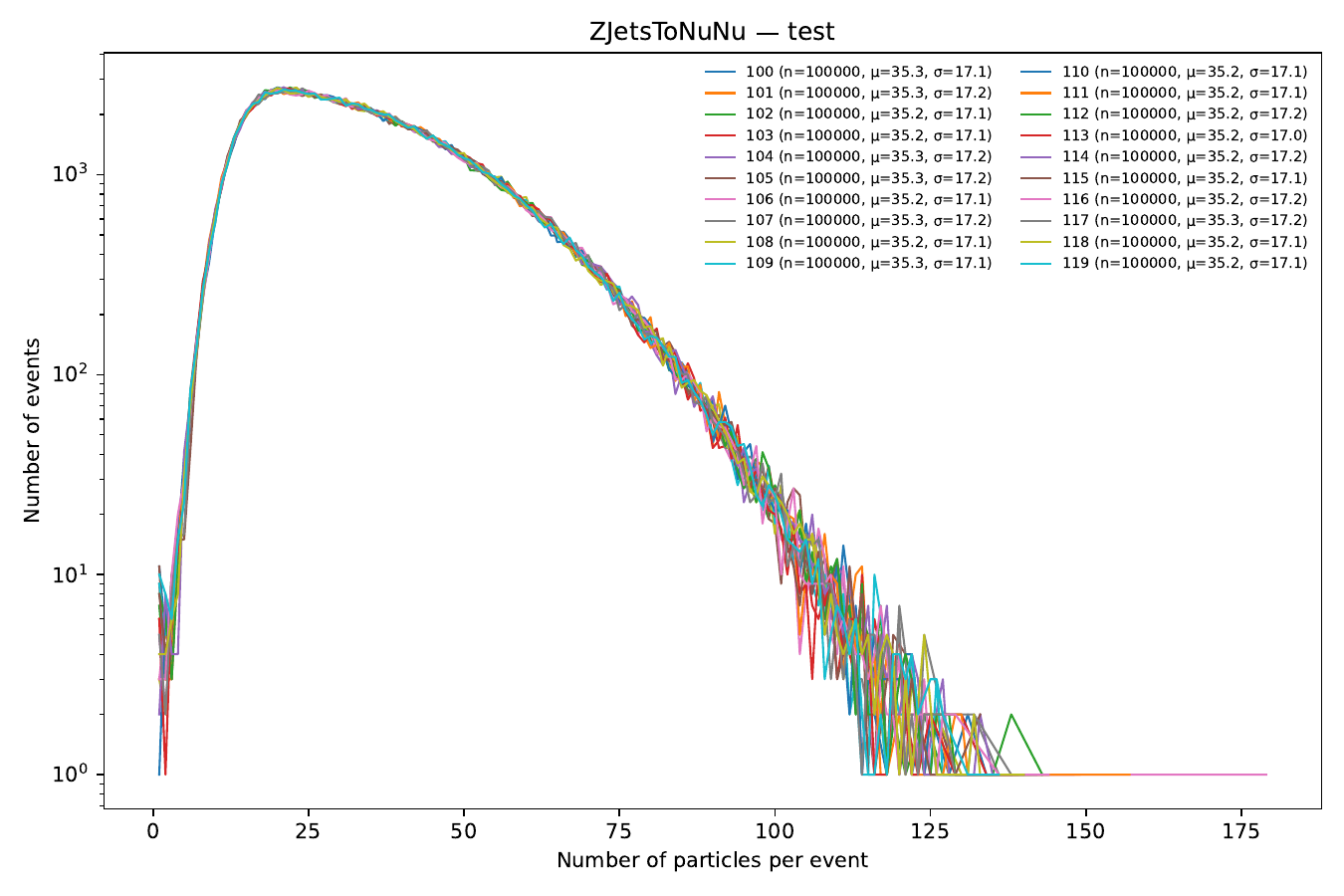}
    \end{minipage}
    \begin{minipage}{0.45\textwidth}
        \centering
        \includegraphics[width=\textwidth,height=0.35\textheight,keepaspectratio]{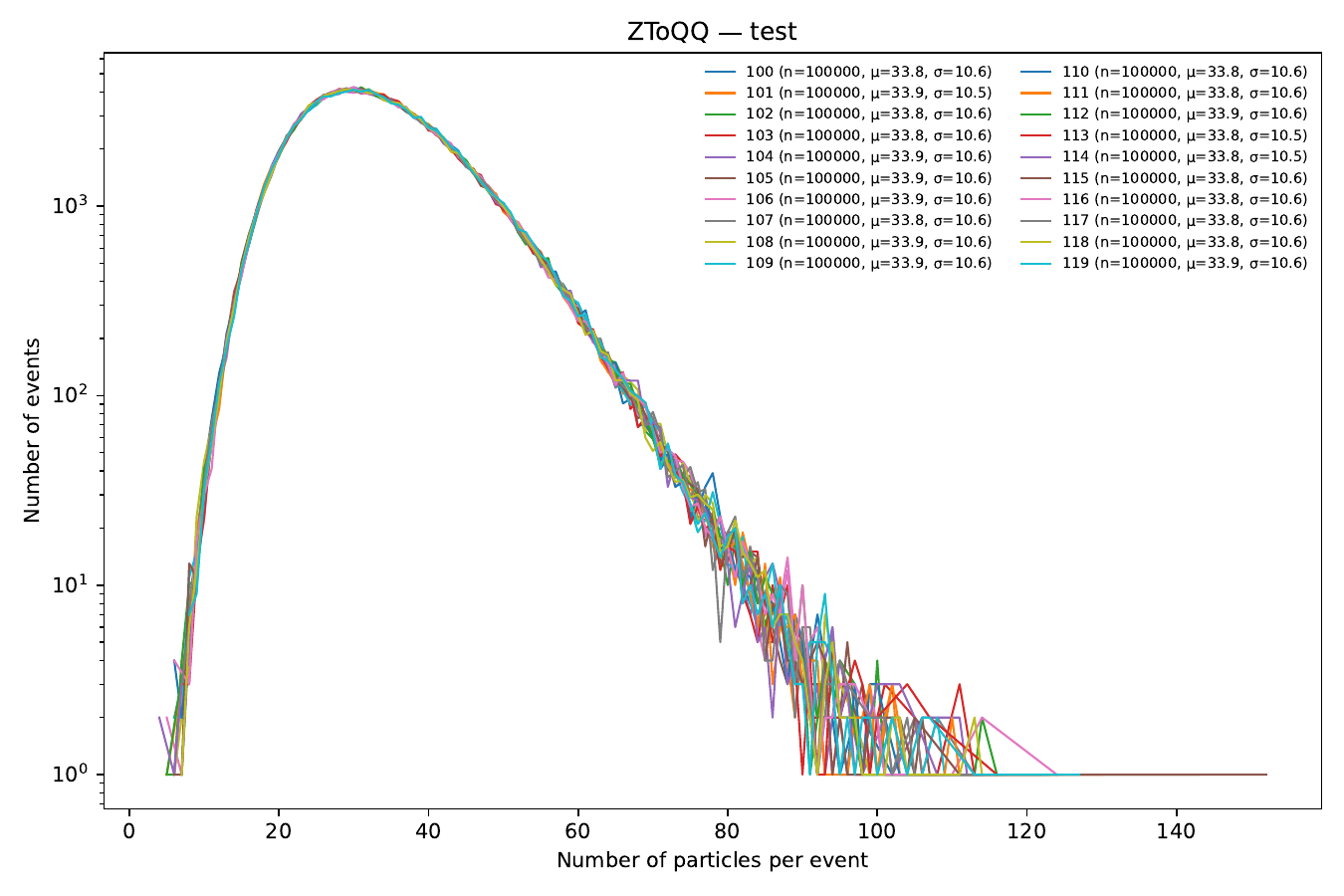}
    \end{minipage}
\end{center}

\begin{figure*}[ht!]
    \centering

    \begin{subfigure}{0.48\textwidth}
        \centering
        \includegraphics[width=\textwidth]{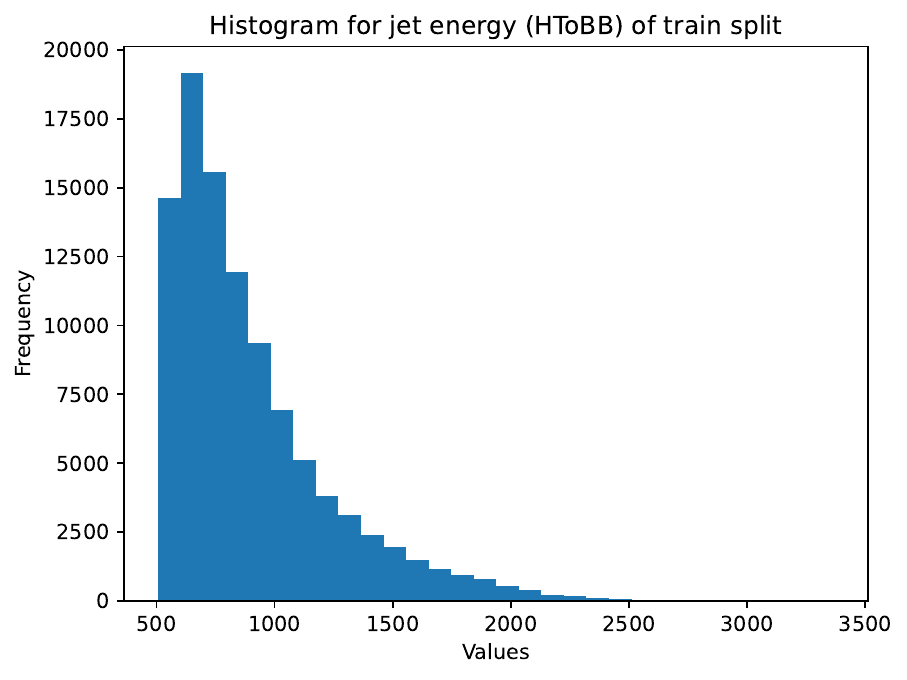}
    \end{subfigure}
    \hfill
    \begin{subfigure}{0.48\textwidth}
        \centering
        \includegraphics[width=\textwidth]{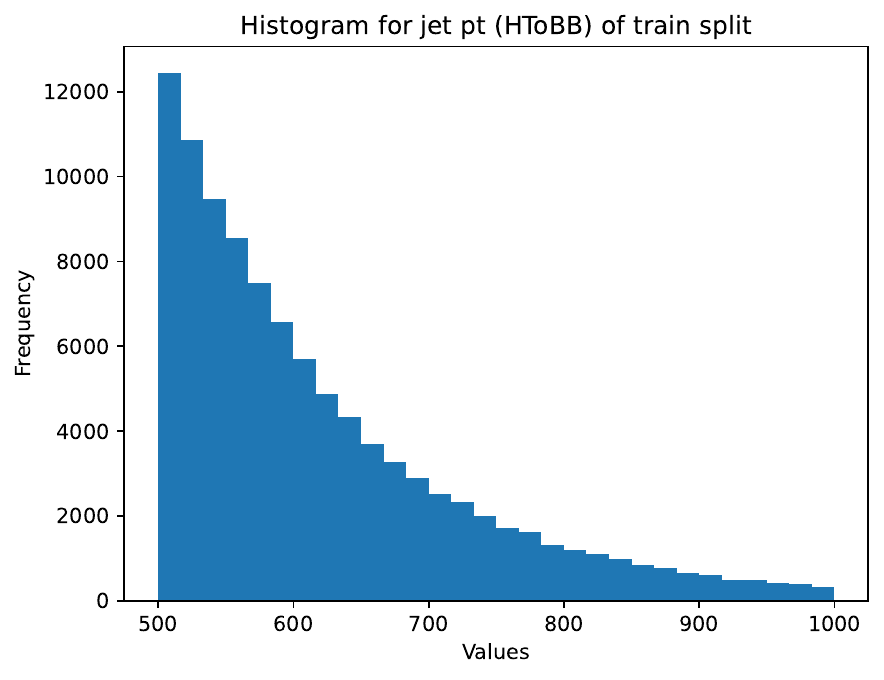}
    \end{subfigure}
    \caption{\textbf{Histograms of physics observables.} We plot the histograms of values of (left) jet energy and (right) jet $p_T$. We use 30 bins for both.}
    \label{fig:supp:jetclass:observables-distrib}
\end{figure*}

\noindent\textbf{Observables distribution.} We conduct a few other analyses, on jet $p_T$ and jet energy variables. We observe a Poisson distribution with a unique peak in the histograms at the minimum values, asymmetric, and a decreasing queue. We show this in Figure \ref{fig:supp:jetclass:observables-distrib}.

\section{Particle Transformer architecture}
\label{sec:supp:part-arch}
The ParT~\cite{qu2022particle} uses particle interaction information for biasing the attention matrices computed in the attention layers. We use the same biases as Qu \textit{et al.}~\cite{qu2022particle}, which is defined as:
\begin{align}
\Delta_{R_{ij}} &= \sqrt{(\eta_i\ - \eta_j)^2+(\phi_i - \phi_j)^2}\\
k_{T_{ij}} &= \min(p_{T_i}, p_{T_j}) \Delta_{R_{ij}}\\
z_{ij} &= \min(p_{T_i}, p_{T_j}) / (p_{T_i} + p_{T_j})\\
m^2_{ij} &= (E_i + E_j)^2 - \left\|p_i + p_j\right\|^2_2
\end{align}
$\Delta_{R_{ij}}$ is the angular distance in rapidity-azimuth space, $k_{T_{ij}}$ is the relative transverse momentum of the splitting that produced particles $i$ and $j$ (the larger $k_{T_{ij}}$, the better the splitting is), $z_{ij}$ measures the similarity of the momentum sharing between the two particles (0.5 is perfect splitting) and $m^2$ is the invariant mass squared of the two-particle system. As the particle interaction matrix is symmetric, it reduces the total amount of computation.

\section{Self-Supervised Learning metrics}
\label{sec:supp:ssl-metrics}
Unlike supervised tasks, it is not possible to directly evaluate the effectiveness of pre-training apart from implementing fine-tuning or probing (whether linear or kNN, online or offline). Nevertheless, metrics based on linear algebra and information theory have emerged that make it easier to estimate the success of pre-training and enable the detection of dimensional collapse.

\subsection{AUC of the Explained Variance Spectrum}
\label{sec:supp:ssl-metrics:auc}
To quantify the dimensional utilization and the degree of isotropy in the latent space, we employ the Area Under the Curve (AUC) of the normalized eigenvalue spectrum. This metric provides a scalar summary of how variance is distributed across the representation dimensions.

Let $\mathbf{Z} \in \mathbb{R}^{N \times d}$ denote a batch of $N$ embeddings in a $d$-dimensional latent space. Assuming $\mathbf{Z}$ is centered, the sample covariance matrix is defined as:
\begin{equation}
    \mathbf{C} = \frac{1}{N-1} \mathbf{Z}^\top \mathbf{Z}
\end{equation}

By performing an eigendecomposition on $\mathbf{C}$, or equivalently, computing the squared singular values of $\frac{\mathbf{Z}}{\sqrt{N-1}}$, we obtain the set of eigenvalues sorted in descending order:
\begin{equation}
    \lambda_1 \ge \lambda_2 \ge \dots \ge \lambda_d \ge 0
\end{equation}

To ensure scale invariance, we normalize the spectrum such that the total variance sums to unity. The relative explained variance $p_i$ for the $i$-th component is given by:
\begin{equation}
    p_i = \frac{\lambda_i}{\sum_{j=1}^d \lambda_j}
\end{equation}

We then define the cumulative explained variance $V_k$ for the first $k$ components as $V_k = \sum_{i=1}^k p_i$. The embedding AUC is calculated as the average of these cumulative values:
\begin{equation}
    \text{AUC}_{emb} = \frac{1}{d} \sum_{k=1}^d V_k
\end{equation}

The metric is bounded between $0.5$ and $1.0$. An $\text{AUC}_{emb} \approx 1.0$ indicates \textit{dimensional collapse}, where a single principal component captures nearly all the variance. Conversely, an $\text{AUC}_{emb} \approx 0.5$ signifies \textit{maximum isotropy}, where the variance is uniformly distributed across all $d$ dimensions, maximizing the representational capacity of the encoder.

\subsection{Effective rank}
\label{sec:supp:ssl-metrics:er}
The Effective Rank ($ER$)~\cite{roy2007effective} provides a continuous and robust measure of the intrinsic dimensionality of a linear operator, bypassing the numerical instability of the classical algebraic rank. In the study of Deep Neural Networks, the $ER$ is used to quantify the compression of information within weight matrices or latent representations. 

Let $\{\sigma_1, \sigma_2, \dots, \sigma_d\}$ be the singular values of $\mathbf{Z}$ obtained via Singular Value Decomposition (SVD). To assess the distribution of these values, we define a normalized distribution $p$ such that:
\begin{equation}
    p_i = \frac{\sigma_i}{\sum_{j=1}^{k} \sigma_j}
\end{equation}
The spectral entropy $H(p)$ of the matrix is then calculated using the Shannon entropy formula:
\begin{equation}
    H(p) = -\sum_{i=1}^{k} p_i \ln(p_i)
\end{equation}
The Effective Rank is finally defined as the exponential of this entropy:
\begin{equation}
    ER(W) = \exp\left( H(p) \right)
\end{equation}
In practice, a low $ER$ relative to the matrix dimensions indicates a high degree of redundancy, suggesting that the model operates in a low-dimensional manifold. This property is foundational for understanding generalization bounds.

\subsection{Pretraining Percent Error}
\label{sec:supp:ssl-metrics:ppe}
The Pretraining Percent Error (PPE)~\cite{schaeffer2024towards} is a diagnostic metric introduced to evaluate the convergence efficiency of the Maximum Manifold Capacity Representations (MMCR)~\cite{yerxa2023learning} framework but we use it for having comparable metrics between model sizes. The PPE quantifies the relative gap between the empirical pretraining loss—defined by the negative nuclear norm of the centroid matrix—and its theoretical lower bound. 

Given a batch of $N$ samples embedded in a space of dimension $d$, let $\mathbf{C}$ be the matrix of centroids (mean of embeddings of multiple views of the same sample). The pretraining objective in MMCR aims to maximize the nuclear norm $\|\mathbf{C}\|_*$, which is upper-bounded by $\sqrt{N \min(N, d)}$ under ideal conditions of unit-norm embeddings. The PPE is formally defined as:
\begin{equation}
    PPE = \frac{\sqrt{N \min(N, d)} - \|\mathbf{C}\|_*}{\sqrt{N \min(N, d)}}
\end{equation}
A PPE of $0$ indicates that the model has reached the absolute minimum of the loss function, achieving perfect alignment and uniformity of representations. This metric is particularly insightful for identifying "double descent" phenomena during pretraining, where the error non-monotonically peaks when the number of data points $N$ equals the embedding dimension $d$, providing a precise tool for hyperparameter tuning and compute scaling law analysis.

\subsection{Online probings}
\label{sec:supp:ssl-metrics:probs}
In JP-JEPA, we introduced losses for different elements (on PIDs, TDs, and Lorentz vectors). The heads used to make predictions can send gradients back to the predictor and ultimately to the encoder, but by applying stop-gradients, these heads perform online probing, i.e., they serve as metrics that we try to optimize without back-propagating them into the encoder. They can provide information on the model's ability to correctly reconstruct certain features. We chose not to use them because they do not adequately reflect the various factors tested during our ablation study.

\section{Other results}
\label{sec:supp:other-results}
We provide in this section more detailed results. For complete attention analysis, we refer readers to this excellent work by Legge \textit{et al.}~\cite{Legge:2025cnm} where they show the importance of the physics-inspired bias detailed in Section \ref{sec:supp:part-arch}.

\subsection{Ablation study}
\label{sec:supp:other-results:ablations}
We perform an ablation study on the core components of our framework. The results are listed in Table \ref{tab:ablations}. We change the learning rate to $10^{-4}$ during fine-tuning to analyze simply the differences between runs, because with the standard LR, the model quickly converged.

\begin{table*}
\resizebox{\linewidth}{!}{
\begin{tabular}{lccccccccc}
\toprule
Method                                                                    & Acc & $\text{AUC}_{emb}(\texttt{[CLS]})$ & $\text{AUC}_{emb}()$ & $\text{ER}(\texttt{[CLS]})$ & $\text{ER}()$ & $\text{PPE}(\texttt{[CLS]})$ & PL LMV & PL PIM & PL PFPIM \\
\midrule
Ours                                                                      & 0.8411     & 0.8966        & 0.9442         & 77.5074      & 64.5702       & 10.3118       & 270.6281   & 310.2128   & 206.8411     \\
- $\mathcal{L}_{\text{koleo}}$                                            & 0.8317     & 0.9579        & 0.9789         & 46.3963      & 41.2053       &  7.4615       & 259.6835   & 311.9423   & 208.0623     \\
- Context \texttt{[CLS]} in predictor                                     & 0.8381     & 0.9268        & 0.8512         & 68.6933      & 92.0879       &  6.9702       & 278.6076   & 312.4401   & 207.3481     \\
- Predictor w/o physical bias                                             & 0.8380     & 0.8877        & 0.9549         & 78.7243      & 57.8850       & 10.3326       & 276.1759   & 311.5761   & 206.7073     \\
- Pairwise interaction embedder shared                                & 0.8377     & 0.8905        & 0.9490         & 79.7707      & 62.5729       & 10.4003       & 272.3101   & 313.3375   & 206.6475     \\
+ Context \texttt{[CLS]} in $\mathcal{L}_{\text{global}}$                 & 0.8386     & 0.8833        & 0.8656         & 80.8311      & 92.0340       &  6.4053       & 265.6791   & 312.9879   & 208.7628     \\
+ Gradients from $\mathcal{L}_{\text{pid}}$ \\
\hspace{0.5cm} $\lambda_2=0.5$                                            & 0.8327     & 0.9094        & 0.8810         & 73.6109      & 86.5873       &  9.3430       & 264.9333   & 310.8937   & 206.6558     \\
\hspace{0.5cm} $\lambda_2=1$                                              & 0.8326     & 0.9260        & 0.8913         & 65.7085      & 81.9309       &  9.0409       & 266.4431   & 312.9175   & 207.9209     \\
\hspace{0.5cm} $\lambda_2=2$                                              & 0.8299     & 0.9550        & 0.9410         & 46.0997      & 60.5676       &  6.5255       & 266.7616   & 310.1133   & 208.6196     \\
+ Gradients from $\mathcal{L}_{\text{td}}$ \\
\hspace{0.5cm} $\lambda_3=0.5$                                            & 0.8405     & 0.8859        & 0.9434         & 80.5395      & 64.4144       & 10.6295       & 270.3683   & 310.6652   & 207.9948     \\
\hspace{0.5cm} $\lambda_3=1$                                              & 0.8380     & 0.8839        & 0.9423         & 81.7103      & 65.4002       & 10.7722       & 267.1892   & 312.8435   & 207.2185     \\
\hspace{0.5cm} $\lambda_3=2$                                              & 0.8387     & 0.8784        & 0.9573         & 82.6600      & 58.0945       & 10.5777       & 269.0010   & 313.2440   & 207.1877     \\
+ Gradients from $\mathcal{L}_{\text{physics}}$ \\
\hspace{0.5cm} lorentz-momentum-vector $\lambda_4=0.5$                    & 0.8261     & 0.9965        & 0.9949         &  5.8236      & 11.2178       &  3.2267       & 198.7960   & 318.8679   & 208.3633     \\
\hspace{0.5cm} lorentz-momentum-vector $\lambda_4=1, 2$                   & 0.8267     & 0.9965        & 0.9945         &  5.3833      & 11.7291       &  3.2216       & 203.3362   & 306.3163   & 205.5786     \\
\hspace{0.5cm} lorentz-momentum-vector $\lambda_4=2$                      & 0.8276     & 0.9972        & 0.9949         &  6.2145      & 11.7538       &  3.2590       & 203.1693   & 313.0284   & 208.8251     \\
\hspace{0.5cm} particle-interaction-matrix $\lambda_4=0.5$                & 0.8362     & 0.9916        & 0.9720         & 10.3231      & 39.8391       &  3.7761       & 351.0015   &  18.1017   & 210.3316     \\
\hspace{0.5cm} particle-interaction-matrix $\lambda_4=1$                  & 0.8342     & 0.9941        & 0.9874         &  8.7861      & 22.6538       &  3.5165       & 350.4854   &  17.9647   & 210.7742     \\
\hspace{0.5cm} particle-interaction-matrix $\lambda_4=2$                  & 0.8327     & 0.9939        & 0.9923         &  7.7651      & 16.5035       &  3.4265       & 352.9333   &  17.9819   & 208.5878     \\
\hspace{0.5cm} physics-forced-particle-interaction-matrix $\lambda_4=0.5$ & 0.8365     & 0.9963        & 0.9990         &  3.9484      &  3.9964       &  2.7862       & 352.3491   & 309.2774   &  75.9619     \\
\hspace{0.5cm} physics-forced-particle-interaction-matrix $\lambda_4=1$   & 0.8332     & 0.9995        & 0.9976         &  2.0456      &  1.5192       &  2.5706       & 352.3267   & 313.0298   &  77.1115     \\
\hspace{0.5cm} physics-forced-particle-interaction-matrix $\lambda_4=2$   & 0.8340     & 0.9992        & 1.0000         &  2.6291      &  1.9298       &  2.5561       & 352.3111   & 312.0340   &  86.6737     \\
+ $V=2$                                                                   & 0.8400     & 0.8857        & 0.9579         & 81.7491      & 57.8792       & 10.0781       & 265.3033   & 313.1923   & 207.9337     \\
+ $V=4$                                                                   & 0.8345     & 0.8855        & 0.9718         & 76.9152      & 46.9237       &  8.7861       & 248.1192   & 312.3864   & 207.5507     \\
+ SIGReg~\cite{DBLP:journals/corr/abs-2511-08544} $V=1$                                                            & 0.8314     & 0.9992        & 0.9865         &  2.3875      & 18.9717       &  1.9877       & 283.1487   & 311.4198   & 207.5980     \\
+ SIGReg~\cite{DBLP:journals/corr/abs-2511-08544} $V=2$                                                            & 0.8314     & 0.9992        & 0.9865         &  2.3875      & 18.9717       &  1.9877       & 283.1487   & 307.7398   & 208.5530     \\
+ SIGReg~\cite{DBLP:journals/corr/abs-2511-08544} $V=4$                                                            & 0.8293     & 0.9994        & 0.9901         &  4.2436      & 17.9901       &  2.4944       & 291.6442   & 312.3633   & 208.2003     \\

\bottomrule
\end{tabular}
}
\caption{
\textbf{Ablation of our method.} We pre-train the ParT-Mini size encoder using the specified hyperparameters and do a fine-tuning stage for having potential future performances. The pre-training and fine-tuning stages were performed on 5\% of JetClass (5M events out of 100M) on \textit{full} categories, during $\sim$10 epochs, and not tested on any other datasets or tasks. We only use accuracy and SSL metrics for choosing the best hyperparameter set on the validation set (first 1M events only out of 5M) and using the student encoder. No test samples have been used for performing the selection. $\text{AUC}_{emb}(\cdot)$, $\text{ER}(\cdot)$, and $\text{PPE}(\cdot)$ are the AUC of the explained variance spectrum, effective rank, and Pretraining Percent Error for the $\cdot$ matrix (which can be stacked \texttt{[CLS]} or mean of the encoded constituents if not specified). 'PL' denotes Physical Loss values with specific formulation ('LMV' stands for 'lorentz-momentum-vector', 'PIM' for 'particle-interaction-matrix', and 'PFPIM' for 'physics-forced-particle-interaction-matrix').
}
\label{tab:ablations}
\end{table*}

\noindent\textbf{KoLeo regularizer~\cite{DBLP:conf/iclr/SablayrollesDSJ19}.}
We observe the importance of this regularization, particularly in terms of the quality of the representation space at the final layer, both for representations at the constituent level and for those at the jet level. More specifically, we measure a lower $\text{AUC}_{\text{emb}}$ (0.8966 vs. 0.9579 for the global representation and 0.9442 vs. 0.9789 for the constituents) and an ER (77.51 vs. 46.40 for the global representation and 64.57 vs. 41.21 for the particles), indicating a more efficient use of latent space and a better distribution of information between its dimensions for both levels.

\noindent\textbf{Context \texttt{[CLS]} in predictor.}
Since the representation spaces of particles and jets may differ, we conducted an ablation study on whether to include the encoded jet token \texttt{[CLS]} at the input of the predictor, to provide additional conditioning to the contextual constituent representations. We observe that SSL metrics analyzing particle levels are much better (higher ER and lower $\text{AUC}_{\text{emb}}$), but there is a significant deterioration in overall representation. Including this token improves the final performance, although it is not a critical component of the framework.

\noindent\textbf{Physical bias in predictor attention layers.}
We also investigated the integration of physical biases based on particle interactions, implemented in the same way as in ParT's student and teacher encoders. Our results indicate that this component provides information for better optimization, even if the differences in SSL metrics are small. We decided to share the weights between embedders in order to limit the FLOPs costs. Since this information is available in all datasets and comes from the four-impulse vector $(E_i, p_{x,i}, p_{y,i}, p_{z,i})$, we do not consider it as supervised information injected a priori or a posteriori. We therefore emphasize its relevance for future SSL studies in high-energy physics.

\noindent\textbf{Context \texttt{[CLS]} in $\mathcal{L}_{\text{global}}$.}
The framework allows both local optimization at the particle level and global optimization at the jet level through the ParT architecture. However, we find that it is preferable not to include the context \texttt{[CLS]} in $\mathcal{L}_{\text{global}}$ and instead let it be optimized through the predictor and the KoLeo regularization. We attribute this to a competition effect between the global loss and the predictor, as the context \texttt{[CLS]} is less informative than the predictor \texttt{[CLS]}, which has access to both local and global encoder information, as well as particle positional information used for representation prediction.

\noindent\textbf{Enabling back-propagation from reconstruction losses.}
We introduced additional loss functions to reconstruct several input variables, such as particle type, track displacement, and a physics-based loss aimed at injecting or validating the physical information encoded by the model. Overall, we observe that removing the stop-gradient significantly degrades performance. The degradation is less pronounced for track displacement prediction compared to other tasks. Inspired by MPMv2~\cite{leigh2025tokenization}, we investigated whether these auxiliary losses could both verify and enhance representation learning. For PID, it helps to encode more information about the jet constituents, but not globally, as this loss only acts at this level. However, in most downstream tasks, only the jet representation is used, so we decided not to integrate it and to keep the stop-gradient. As for TD, fine-tuning performance is slightly degraded, as are SSL metrics on the overall representation, because it provides information about jets but does not seem to be essential for reconstruction. Locally, performance is poorer or, at best, equal. Finally, for physics-based losses, fine-tuning performance varies, but all give lower results, and in terms of SSL metrics, we observe that the space is hardly used at all on the two levels of data representation. We can therefore see that these reconstruction losses act on the space in different ways but force a certain modeling that is not optimal for downstream tasks. Consequently, the generalization capability of models pre-trained with reconstruction losses is reduced.

\begin{figure}[ht!]
\centering
\setlength{\tabcolsep}{1pt} 
\begin{tabular}{ccc}
\includegraphics[width=0.32\linewidth,page=1]{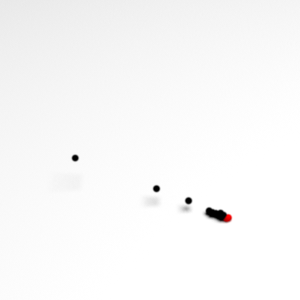} &
\includegraphics[width=0.32\linewidth,page=1]{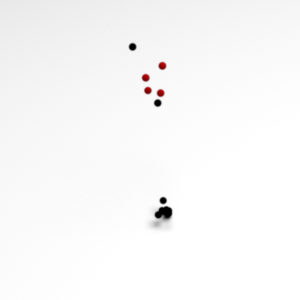} &
\includegraphics[width=0.32\linewidth,page=1]{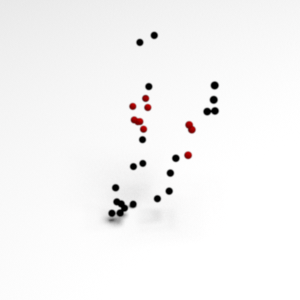} \\
\includegraphics[width=0.32\linewidth,page=1]{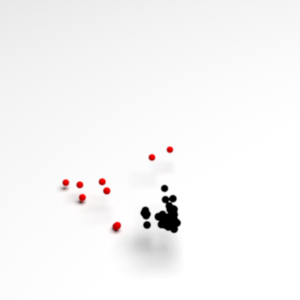} &
\includegraphics[width=0.32\linewidth,page=1]{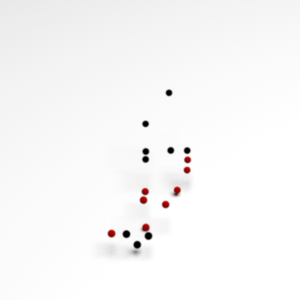} &
\includegraphics[width=0.32\linewidth,page=1]{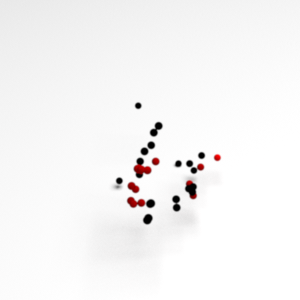} \\
\end{tabular}
\caption{\textbf{Visualization of the masking strategy.} Tokens highlighted in red correspond to the targets to be predicted from the black context.}
\label{fig:supp:masking}
\end{figure}

\noindent\textbf{Several targets.}
We employed a block masking strategy to remove semantic information from the student model input. Several examples are provided in Figure \ref{fig:supp:masking}, where masked tokens are shown in red and the context in black. We investigated the possibility of increasing the number of reconstruction targets in the latent space to two or four (with one target used by default, in accordance with HEP-JEPA~\cite{bardhan2025hep}). We observed that using multiple views degraded performance (from 84.11\% to 83.45\%). This is mainly due to the reduction in the number of particles available to form the context, which limits the encoder's learning capacity and leads to less informative gradients due to the small number of input constituents. This is confirmed by analyzing the effective rank metrics on particles (64.57 to 46.92) and the $\text{AUC}_{\text{emb}}$ on particles (0.9442 to 0.9718). They show that information is less encoded at the particle level but seems to be shifted towards the global representation ($\text{AUC}_{\text{emb}}$ from 0.8966 to 0.8855, ER from 77.51 to 81.75, and PPE from 10.31\% to 8.79\%). We kept the default parameter to achieve a balance between detailed and global information.

\noindent\textbf{Use of SIGReg instead of KoLeo.}
SIGReg and KoLeo are regularization losses designed to shape the representation space at a given level (here, only the final level). We observe that when KoLeo is applied to the \texttt{[CLS]} token, the regularization also propagates to the particle representations, whereas this effect is not observed with SIGReg. Both methods encourage the use of higher-dimensional representations, but KoLeo appears to be more effective in our configuration\footnote{We do not use data augmentations as in LeJEPA~\cite{DBLP:journals/corr/abs-2511-08544}, which itself is inspired by DinoV2~\cite{oquab2023dinov2}.}. Indeed, SSL-based metrics all confirm (except PPE) that the space remains largely underutilized, which also explains the poorer performance in fine-tuning. Furthermore, Balestriero \textit{et al.}~\cite{DBLP:journals/corr/abs-2511-08544} report that SIGReg benefits from a larger number of targets to achieve stronger regularization. By varying the number of targets $V$, we show that this does not lead to improvements in our framework.

\subsection{Additionnal t-SNE visualizations}
\label{sec:supp:other-results:tsne}
\begin{figure*}[ht!]
    \centering

    \begin{subfigure}{0.31\textwidth}
        \centering
        \includegraphics[width=\textwidth]{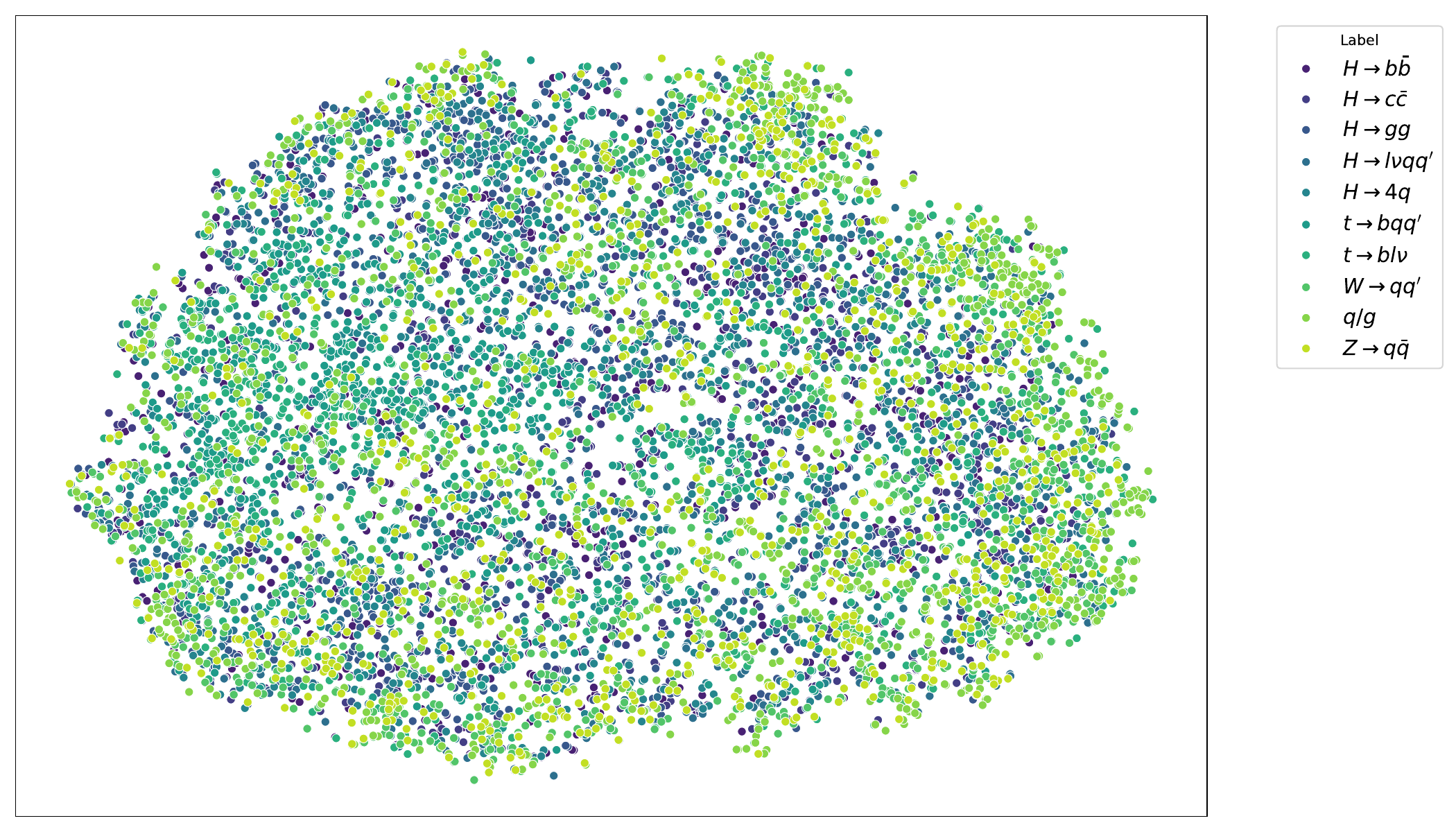}
        \caption{\textbf{Labels.}}
        \label{fig:embedding-viz-class-mini}
    \end{subfigure}
    \hfill
    \begin{subfigure}{0.31\textwidth}
        \centering
        \includegraphics[width=\textwidth]{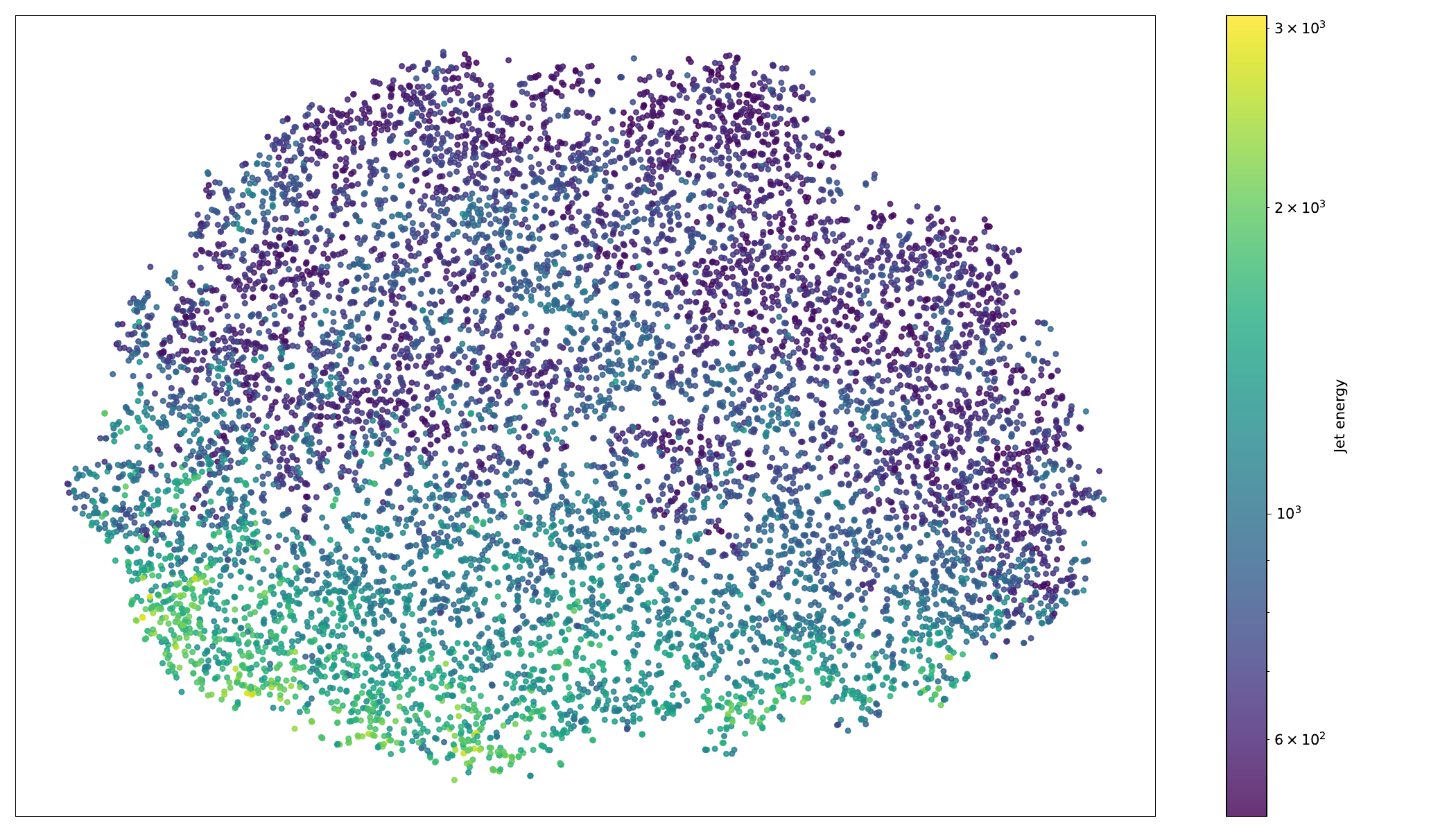}
        \caption{\textbf{Jet energy.}}
        \label{fig:embedding-viz-energy-mini}
    \end{subfigure}
    \hfill
    \begin{subfigure}{0.31\textwidth}
        \centering
        \includegraphics[width=\textwidth]{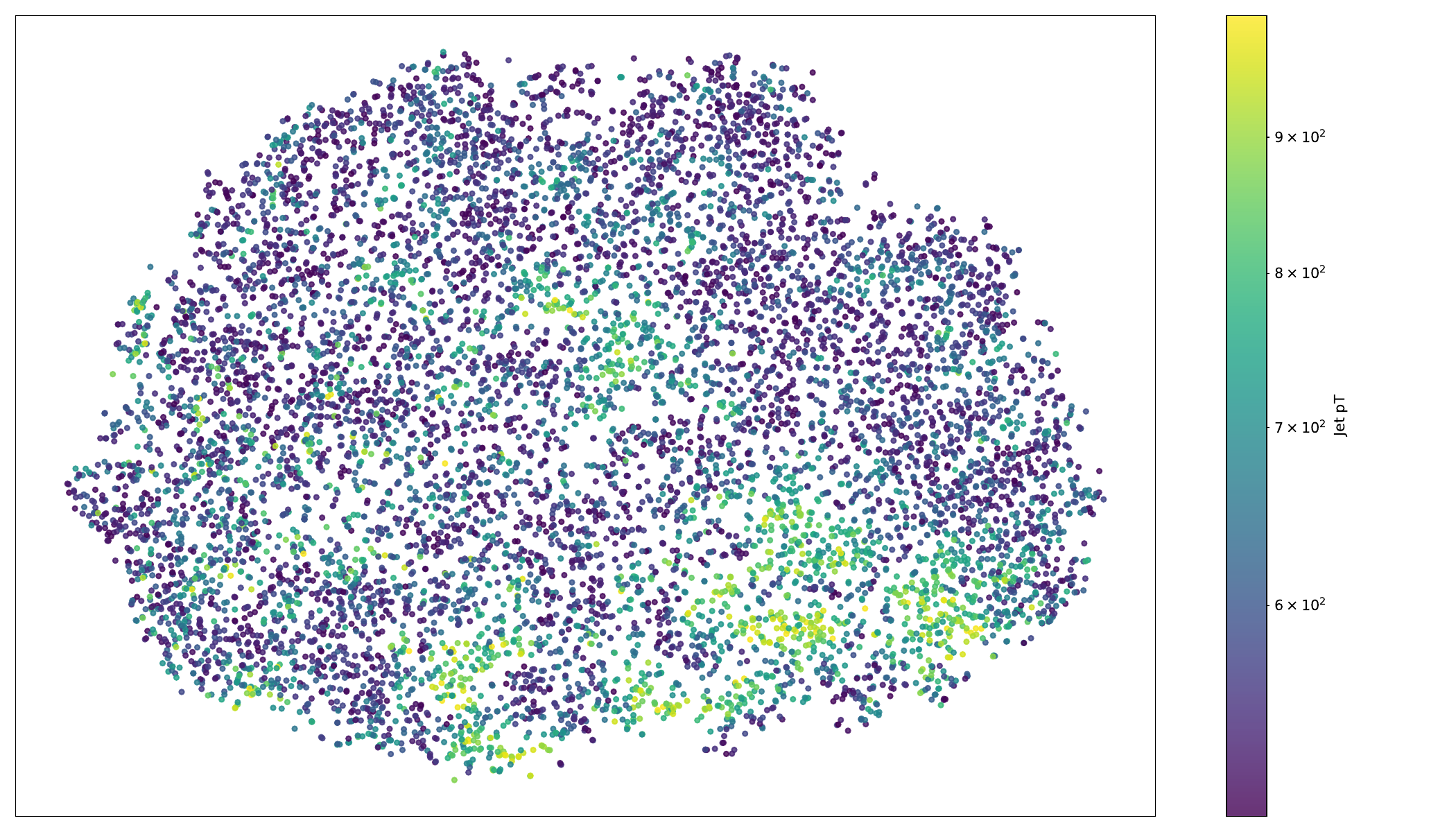}
        \caption{\textbf{Jet transverse momentum.}}
        \label{fig:embedding-viz-pt-mini}
    \end{subfigure}
    \caption{\textbf{Embedding projection via t-SNE of JP-JEPA Mini.}
    (a) Embedding projection via t-SNE of JP-JEPA Mini of 10{,}000 samples from JetClass, evenly drawn from the 10 classes, colored according to the ground-truth labels. (b) Embedding projection via t-SNE of JP-JEPA Mini on the same 10{,}000 samples colored by jet energy (in GeV). (c) Embedding projection via t-SNE of JP-JEPA Mini on the same 10{,}000 samples colored by jet transverse momentum $p_T$ (in GeV).}
\end{figure*}
Figures \ref{fig:embedding-viz-class-mini}, \ref{fig:embedding-viz-energy-mini}, and \ref{fig:embedding-viz-pt-mini} show the t-SNE projections of the embeddings onto a two-dimensional plane~\cite{van2008visualizing}. As observed for the Small variant in the main paper, the samples tend to cluster more strongly by physical observables than by semantic labels. However, compared to the Small model, the Mini variant exhibits a reduced ability to form well-separated clusters when colored by annotation labels. Similar grouping patterns are observed for energy and transverse momentum. These results therefore suggest that increasing the model size is more beneficial for learning higher-level semantic representations than for improving the physical structuring of the latent space.

\subsection{Confusion matrices}
\label{sec:supp:other-results:confusion-matrix}
\begin{figure*}[ht!]
    \centering
    \includegraphics[width=\textwidth]{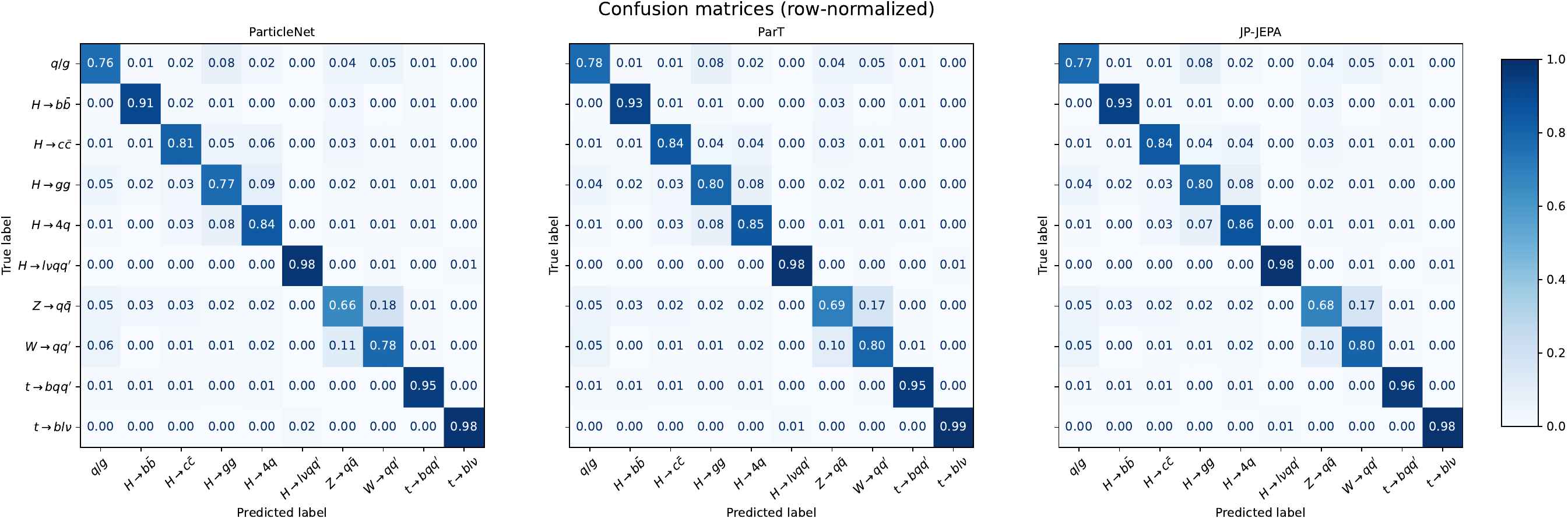}
    \caption{\textbf{Detailed comparison of accuracy per class via confusion matrices.} We compare the confusion matrices of trained models. (Left) ParticleNet, (middle) ParT, and (right) JP-JEPA Small. We observe that ParT improves results on all classes compared to ParticleNet in a similar way. JP-JEPA results are very close to ParT ones.}
    \label{fig:jetclass:confusion-matrix}
\end{figure*}
Across all models, including ParticleNet, ParT, and JP-JEPA, class separability issues appear to follow similar patterns at different scales, as illustrated by the confusion matrices in Figure \ref{fig:jetclass:confusion-matrix}. In particular, we clearly observe stronger separability for jets whose decay involves particles other than quarks or gluons. For example, the classes $H \to b\bar{b}$, $H \to \ell\nu qq'$, $t \to bqq'$ and $t \to b\ell \nu$ achieve classification rates above $90\%$ across all models. This is consistent with the nature of the final-state particles and the resulting jet structures, whose physical characteristics make them easier to distinguish. In contrast, the remaining JetClass categories ($q/g$, $H \to c\bar{c}$, $H \to gg$, $H \to 4q$, $Z \to q\bar{q}$, $W \to qq'$) are more challenging to separate, as they share similar local structures. In particular, Higgs boson ($H$) decay modes are harder to distinguish from one another, as are the $Z$ and $W$ bosons. In the former case, the Higgs boson can decay into several configurations involving only quarks/anti-quarks and gluons. In the latter, decays involve only quarks, with differences arising from quark flavors (for $W$) or the presence of an anti-quark (for $Z$). Finally, these decay modes, composed exclusively of quarks and gluons, further complicate separation from the $q/g$ background.

\subsection{Metrics vs. number of constituents}
\label{sec:supp:other-results:metrics-vs-constituents}
\begin{figure*}[ht!]
    \centering

    \begin{subfigure}{0.48\textwidth}
        \centering
        \includegraphics[width=\textwidth]{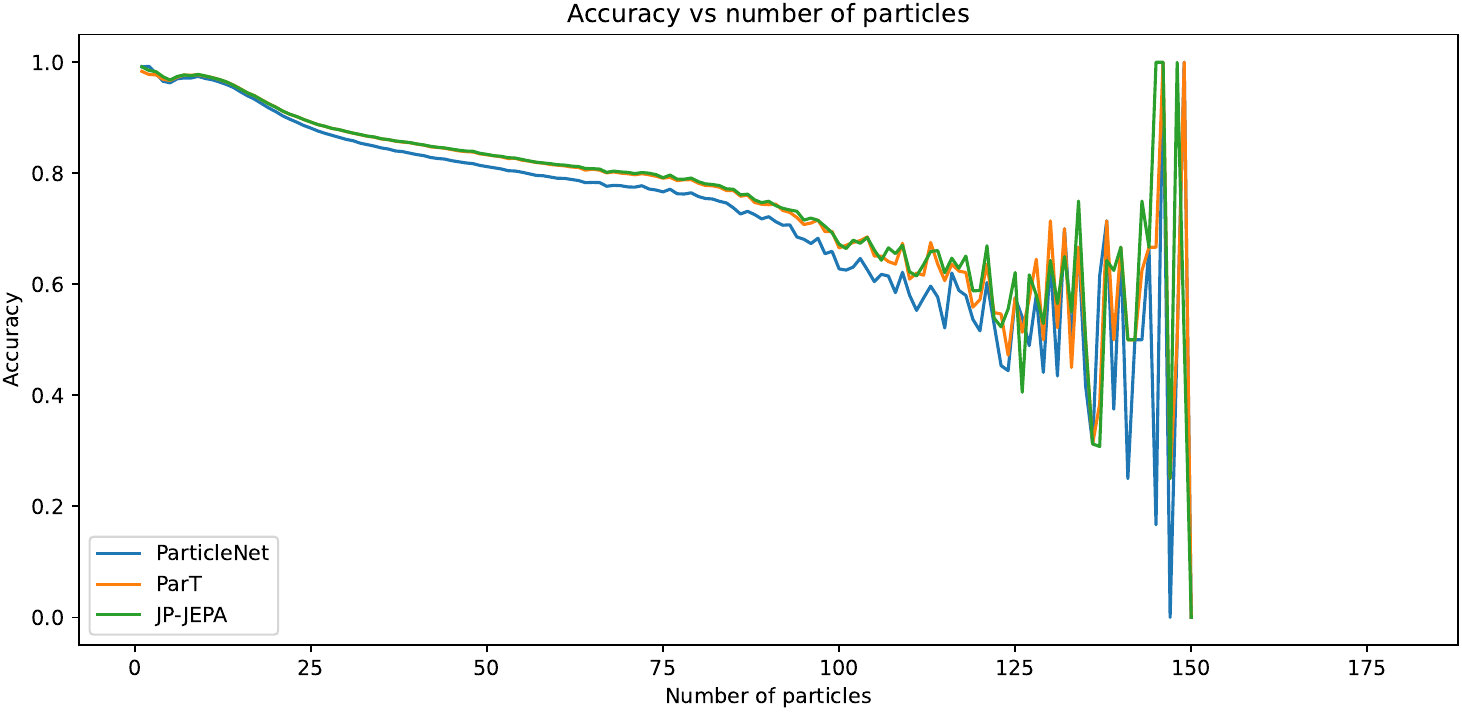}
    \end{subfigure}
    \hfill
    \begin{subfigure}{0.48\textwidth}
        \centering
        \includegraphics[width=\textwidth]{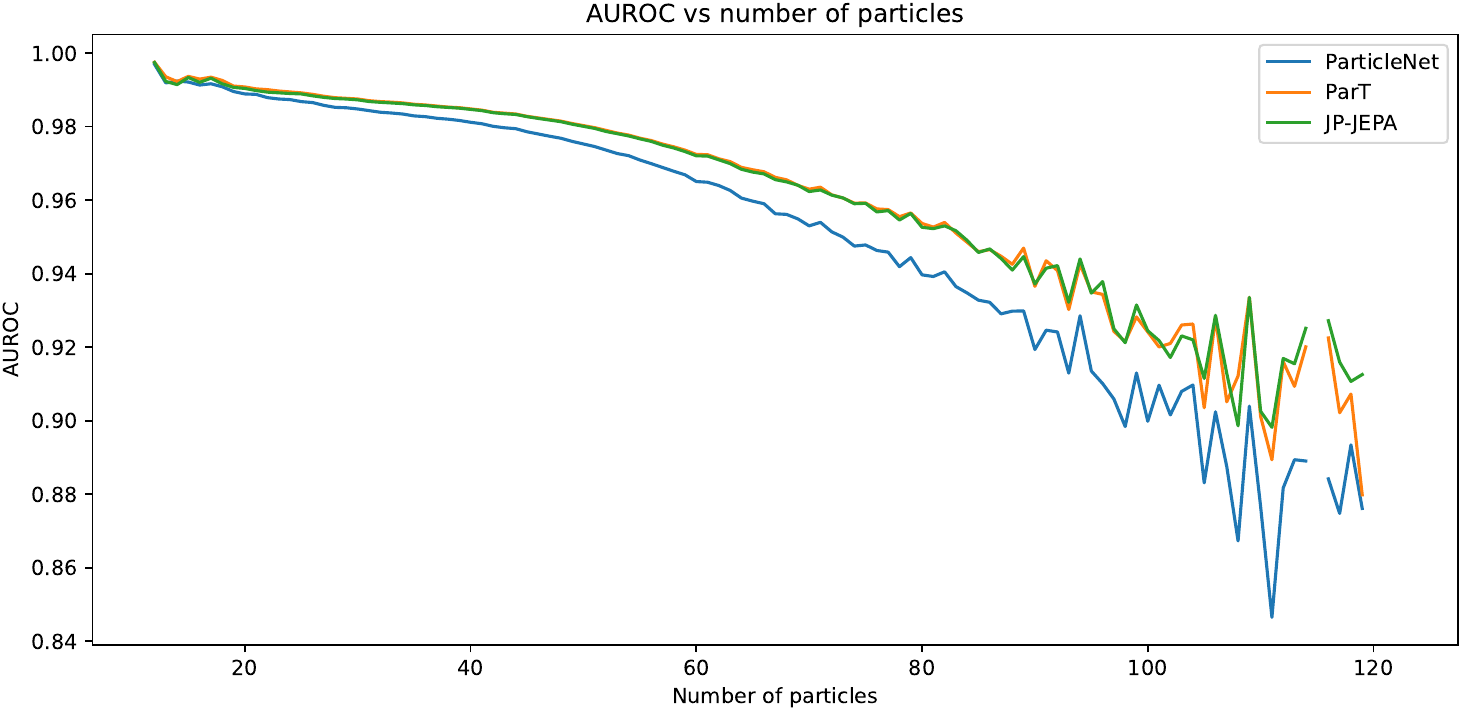}
    \end{subfigure}
    \caption{\textbf{Detailed comparison of accuracy and AUROC per number of constituents.} We compare the metrics per particle number of trained models. (Left) Accuracy (right) AUROC. We can easily remark that the more particles you have in your event, the more difficult it is to correctly classify it.}
    \label{fig:supp:jetclass:particle-number}
\end{figure*}

The Figure \ref{fig:supp:jetclass:particle-number} compares two metrics, accuracy and AUROC, based on the number of particles contained in the event. We clearly observe poorer metrics as the number of constituents increases. We believe this is mainly due to the low number of particles in the vast majority of examples (see Section \ref{sec:supp:datasets:jetclass}), which makes it more difficult to classify these jets correctly.

\subsection{Metrics vs. physics-based observables}
\label{sec:supp:other-results:metrics-vs-observables}
\begin{figure*}[ht!]
    \centering

    \begin{subfigure}{0.48\textwidth}
        \centering
        \includegraphics[width=\textwidth]{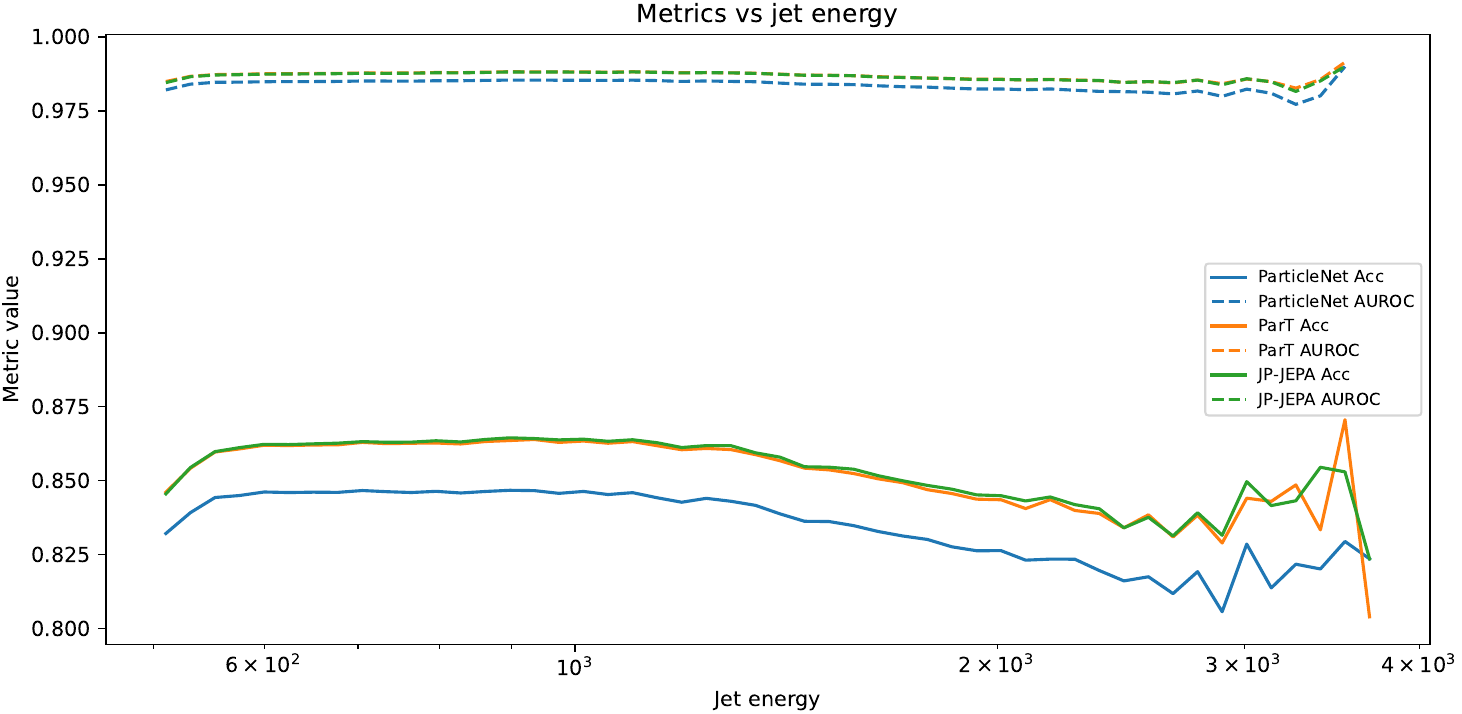}
    \end{subfigure}
    \hfill
    \begin{subfigure}{0.48\textwidth}
        \centering
        \includegraphics[width=\textwidth]{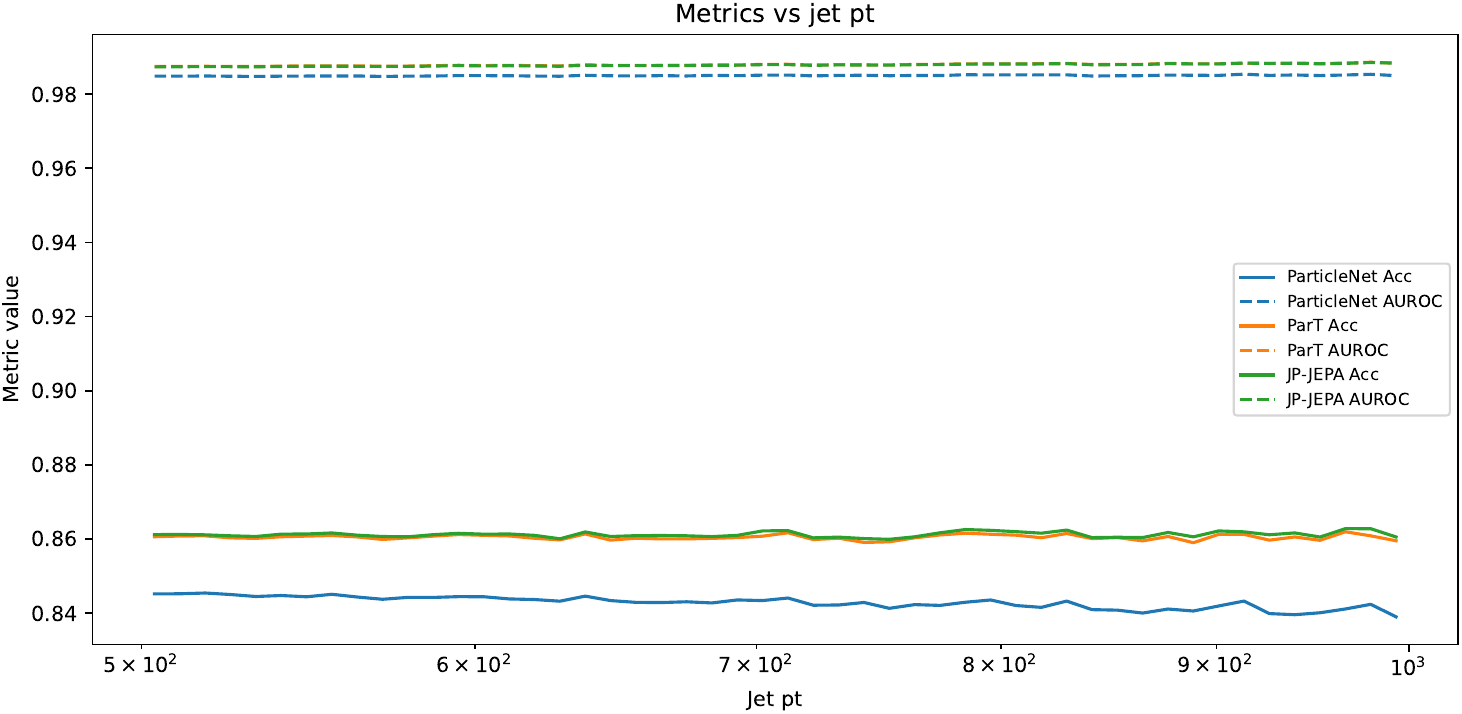}
    \end{subfigure}
    \caption{\textbf{Detailed comparison of accuracy and AUROC w.r.t. physics observables.} We compare the metrics of trained models compared to (left) energy and (right) $p_T$. The set has been split into 50 bins.}
    \label{fig:supp:jetclass:observables}
\end{figure*}

The Figure \ref{fig:supp:jetclass:observables} provides comparisons on classification metrics based on physical values such as jet energy and jet transverse momentum. We see that for $p_T$, the metrics are very stable with small variations observed. However, this contrasts with the details when looking at each class individually. Furthermore, we see that depending on the energy, decreases are observed at the extremes, when the energy is less than 600 GeV or greater than 2 TeV. Using the analysis provided in Section \ref{sec:supp:datasets:jetclass}, the decrease at 3 TeV and more for jet energy can be explained by a small number of jets with low energy, thereby introducing more variations. Between 2 and 3 TeV, this explains part of the problem, but the lack of data impacts performance at high energies because the model cannot correctly refine its separations. To improve this, precise losses must be used, such as the focal loss, or the number of high-energy events must be increased. For $p_T$, the small variations are explained by the previous analysis, where the amount of data is better distributed across the range, particularly because the generation of JetClass jets comes from this variable, as mentioned in \citep[Section~2]{qu2022particle} (only jets with $500 \le p_T \le 1000$ GeV are considered).

\subsection{Study on number of steps when fine-tuning}
\label{sec:supp:other-results:ft-steps-study}
\begin{figure*}[ht!]
    \centering
    \begin{subfigure}{0.49\linewidth}
        \centering
        \includegraphics[width=\linewidth]{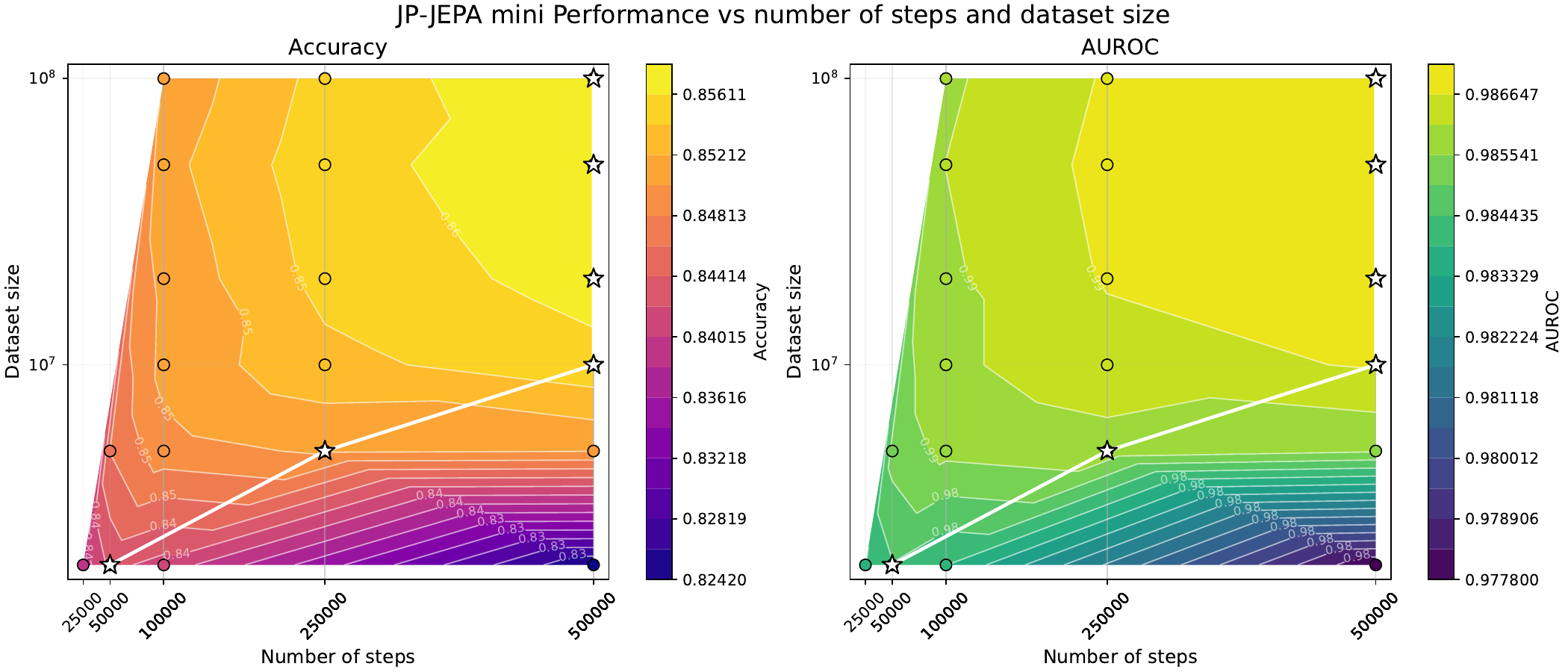}
        \caption{JP-JEPA Mini}
        \label{fig:scaling-ft:mini}
    \end{subfigure}
    \hfill
    \begin{subfigure}{0.49\linewidth}
        \centering
        \includegraphics[width=\linewidth]{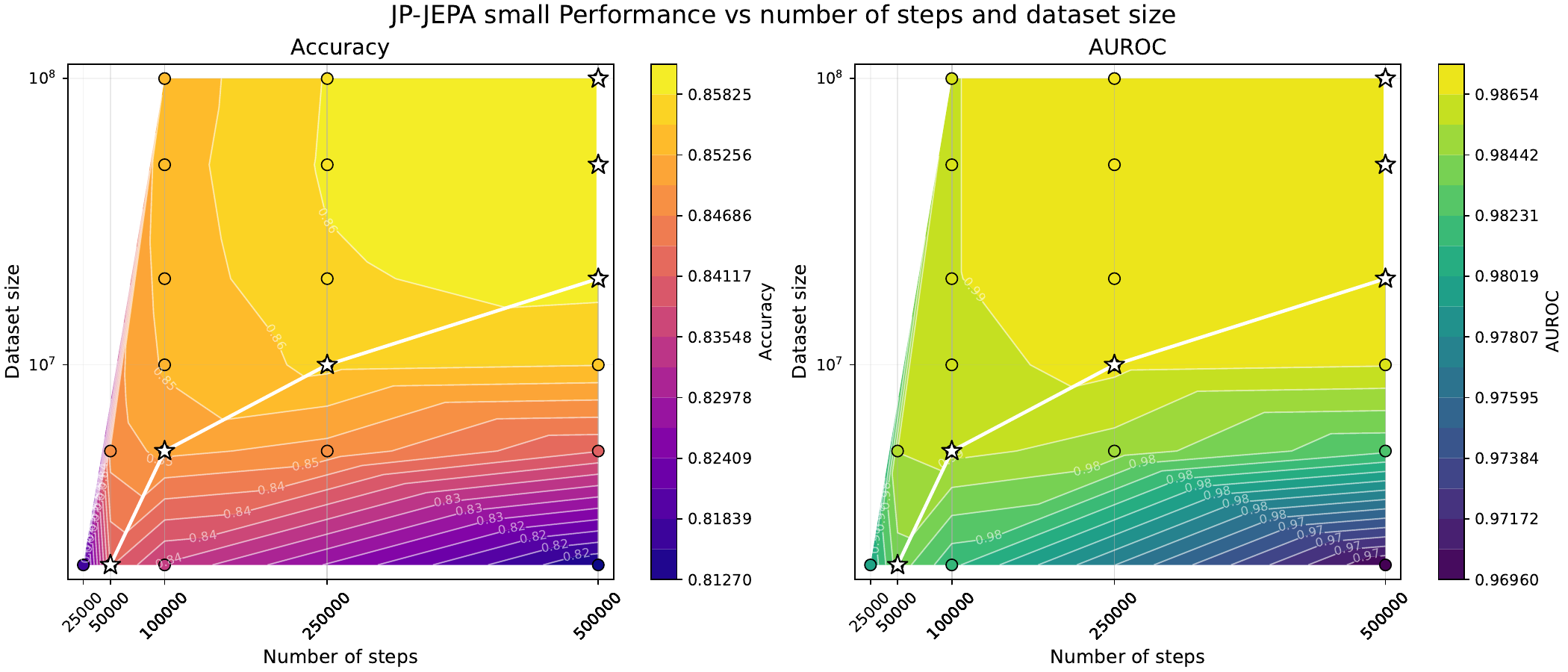}
        \caption{JP-JEPA Small}
        \label{fig:scaling-ft:small}
    \end{subfigure}
    \caption{\textbf{Detailed fine-tuning classification performances of JP-JEPA models on JetClass.} For each dataset size, the subsample set is the same. Only one single fine-tuning was done. Optimal points are drawn with a white star.}
    \label{fig:scaling-ft}
\end{figure*}

Figure \ref{fig:scaling-ft} (and detailed values in Table \ref{tab:jetclass-steps-study}) present the results for each number of fine-tuning steps tested on the JetClass dataset. We observe that for each point on the scaling curve based on dataset size, there is an optimal number of steps for both avoiding under- and over-fitting. It is evident that the larger the dataset, the more necessary it is to increase the number of fine-tuning iterations. Nevertheless, when compared to the subsample containing 2 million jets, it is clear that JP-JEPA requires fewer fine-tuning iterations than ParT, its direct competitor. Furthermore, the larger the model in terms of the number of parameters, the fewer iterations it requires to achieve equivalent or better results.
\begin{table*}[ht!]
\resizebox{\linewidth}{!}{
\begin{tabular}{lcccccccccccccc}
\toprule
\multirow{2}{*}{Methods} & \multirow{2}{*}{Cat.} & \multirow{2}{*}{FT size} & \multirow{2}{*}{\# steps} & \multicolumn{2}{c}{All classes} & $H \rightarrow b\bar{b}$ & $H \rightarrow c\bar{c}$ & $H \rightarrow gg$ & $H \rightarrow 4q$ & $H \rightarrow l\nu qq'$ & $t \rightarrow bqq'$ & $t \rightarrow bl\nu$ & $W \rightarrow qq'$ & $Z \rightarrow q\bar{q}$ \\
& & & & Acc & AUC & $\text{Rej}_{\text{50\%}}$ & $\text{Rej}_{\text{50\%}}$ & $\text{Rej}_{\text{50\%}}$ & $\text{Rej}_{\text{50\%}}$ & $\text{Rej}_{\text{99\%}}$ & $\text{Rej}_{\text{50\%}}$ & $\text{Rej}_{\text{99.5\%}}$ & $\text{Rej}_{\text{50\%}}$ & $\text{Rej}_{\text{50\%}}$ \\
\midrule
ParT~\cite{qu2022particle} & full & 2M   & 500k    & 0.8379 & 0.9833 & 4425 & 1706 &  95 & 709 & 1627 & 6826 & 3683 & 300 & 227 \\

JP-JEPA (Mini)             & full & 2M   & 25k     & 0.8390 & 0.9838 & 4717 & 1840 &  99 & 812 & 1959 & 8163 & 3968 & 325 & 247 \\
JP-JEPA (Mini)             & full & 2M   & 50k (s) & \textbf{\underline{0.8427}} & \textbf{\underline{0.9844}} & \textbf{\underline{4831}} & \textbf{\underline{1951}} & \textbf{\underline{101}} & \textbf{\underline{862}} & \textbf{\underline{1998}} & \textbf{\underline{9217}} & \underline{4202} & \textbf{\underline{357}} & \textbf{\underline{260}} \\
JP-JEPA (Mini)             & full & 2M   & 100k    & 0.8414 & 0.9839 & 4608 & 1791 &  97 & 800 & 1537 & 7752 & 3745 & 335 & 245 \\
JP-JEPA (Mini)             & full & 2M   & 500k    & 0.8242 & 0.9778 & 1608 &  688 &  62 & 372 &  642 & 2685 & 2829 & 161 & 123 \\

JP-JEPA (Small)            & full & 2M   & 25k     & 0.8173 & 0.9801 & 3663 & 1234 &  86 & 539 & 1085 & 3643 & 2882 & 232 & 192 \\
JP-JEPA (Small)            & full & 2M   & 50k (s) & \underline{0.8415} & \underline{0.9840} & \underline{4505} & \underline{1837} &  \underline{99} & \underline{788} & \underline{1866} & \underline{7168} & \textbf{\underline{4415}} & \underline{331} & \underline{247} \\
JP-JEPA (Small)            & full & 2M   & 100k    & 0.8351 & 0.9815 & 3110 & 1325 &  84 & 563 & 1022 & 4310 & 2714 & 235 & 187 \\
JP-JEPA (Small)            & full & 2M   & 500k    & 0.8127 & 0.9696 &  567 &  399 &  40 & 326 &  184 &  511 &  938 &  86 &  63 \\
\midrule
ParT~\cite{qu2022particle} & full & 5M   & 500k     & 0.8478 & 0.9854 & 5464 & 2356 & 105 & 1070 & 2751 & 13889 & \textbf{6803} & 394 & 290 \\

JP-JEPA (Mini)             & full & 5M   & 50k      & 0.8462 & 0.9852 & 5464 & 2237 & 105 &  993 & 2743 & 12903 & 6079 & 386 & 285 \\
JP-JEPA (Mini)             & full & 5M   & 100k     & 0.8494 & 0.9857 & 5917 & 2460 & \underline{\textbf{109}} & \underline{\textbf{1147}} & \underline{\textbf{2990}} & \underline{\textbf{15152}} & 6231 & 412 & 305 \\
JP-JEPA (Mini)             & full & 5M   & 250k (s) & \underline{0.8504} & \underline{\textbf{0.9859}} & 6494 & \underline{\textbf{2677}} & \underline{109} & 1142 & 2878 & 12579 & \underline{6452} & \underline{\textbf{419}} & \underline{\textbf{312}} \\
JP-JEPA (Mini)             & full & 5M   & 500k     & 0.8502 & 0.9856 & \textbf{\underline{6536}} & 2516 & 107 & 1106 & 2395 & 11494 & 5970 & 404 & 299 \\

JP-JEPA (Small)            & full & 5M   & 50k      & 0.8481 & 0.9855 & 5634 & 2301 & 105 & 1030 & 2577 & 12500 & 5831 & 394 & 286 \\
JP-JEPA (Small)            & full & 5M   & 100k (s) & \underline{\textbf{0.8505}} & \underline{0.9858} & \underline{5731} & \underline{2567} & \underline{108} & \underline{1110} & \underline{2699} & \underline{13986} & \underline{6098} & \underline{416} & \underline{305} \\
JP-JEPA (Small)            & full & 5M   & 250k     & 0.8486 & 0.9850 & 5063 & 2193 & 102 &  869 & 1957 &  9217 & 5348 & 368 & 277 \\
JP-JEPA (Small)            & full & 5M   & 500k     & 0.8413 & 0.9825 & 2460 & 1220 &  86 &  566 & 1209 &  3540 & 4219 & 252 & 204 \\
\midrule
ParT~\cite{qu2022particle} & full & 10M  & 500k     & 0.8485 & 0.9856 & 6061 & 2361 & 106 & 1110 & 3130 & 13986 & 7042 & 408 & 298 \\

JP-JEPA (Mini)             & full & 10M  & 100k     & 0.8509 & 0.9860 & 5797 & 2513 & 109 & 1185 & 3215 & 17391 & 7463 & 438 & 313 \\
JP-JEPA (Mini)             & full & 10M  & 250k     & 0.8535 & 0.9864 & 7463 & 2759 & 113 & 1285 & 3484 & 16000 & 8511 & 457 & 337 \\
JP-JEPA (Mini)             & full & 10M  & 500k (s) & \underline{0.8555} & \underline{0.9867} & \underline{\textbf{8097}} & \underline{2924} & \underline{\textbf{114}} & \underline{\textbf{1363}} & \underline{3676} & \underline{18349} & \underline{8850} & \underline{\textbf{470}} & \underline{\textbf{345}} \\

JP-JEPA (Small)            & full & 10M  & 100k     & 0.8530 & 0.9863 & 6623 & 2545 & 110 & 1237 & 3546 & 17699 & 7722 & 440 & 322 \\
JP-JEPA (Small)            & full & 10M  & 250k (s) & \underline{\textbf{0.8562}} & \underline{\textbf{0.9868}} & \underline{7246} & \underline{\textbf{3026}} & \underline{113} & \underline{1347} & \underline{\textbf{3697}} & \underline{\textbf{20000}} & \underline{\textbf{9302}} & \underline{467} & 343 \\
JP-JEPA (Small)            & full & 10M  & 500k     & 0.8555 & 0.9866 & 7117 & 2770 & 110 & 1270 & 3231 & 16393 & 7273 & 461 & \underline{344} \\
\midrule
ParT~\cite{qu2022particle} & full & 20M  & 1M       & 0.8576 & 0.9871 & 7968 & \textbf{3200} & \textbf{117} & 1493 & 4211 & 22989 & 10753 & 504 & 360 \\

JP-JEPA (Mini)             & full & 20M  & 100k     & 0.8507 & 0.9860 & 6024 & 2418 & 110 & 1229 & 3252 & 14815 &  8929 & 435 & 318 \\
JP-JEPA (Mini)             & full & 20M  & 250k     & 0.8548 & 0.9867 & 7605 & 2821 & 114 & 1372 & 3831 & 18182 &  \underline{9662} & 470 & 340 \\
JP-JEPA (Mini)             & full & 20M  & 500k     & \underline{0.8569} & \underline{0.9870} & \underline{8130} & \underline{3017} & \underline{116} & \underline{1508} & \underline{4082} & \underline{21505} &  9259 & \underline{491} & \underline{354} \\

JP-JEPA (Small)            & full & 20M  & 100k     & 0.8537 & 0.9865 & 6135 & 2729 & 112 & 1308 & 3373 & 18692 &  8696 & 446 & 322 \\
JP-JEPA (Small)            & full & 20M  & 250k     & 0.8579 & 0.9871 & 7782 & 2972 & 115 & 1437 & 4211 & 23529 & 10417 & 499 & 357 \\
JP-JEPA (Small)            & full & 20M  & 500k (s) & \underline{\textbf{0.8593}} & \underline{\textbf{0.9873}} & \underline{\textbf{8230}} & \underline{3096} & \underline{\textbf{117}} & \underline{\textbf{1541}} & \underline{\textbf{4301}} & \underline{\textbf{24390}} & \underline{\textbf{10989}} & \underline{\textbf{511}} & \underline{\textbf{363}} \\
\midrule
ParT~\cite{qu2022particle} & full & 50M  & 1M       & 0.8586 & 0.9873 & 8475 & 3284 & \textbf{118} & 1559 & 4396 & \textbf{28169} & 11494 & 510 & 368 \\

JP-JEPA (Mini)             & full & 50M  & 100k     & 0.8515 & 0.9861 & 6515 & 2564 & 110 & 1245 & 3378 & 16807 &  7782 & 442 & 317 \\
JP-JEPA (Mini)             & full & 50M  & 250k     & 0.8554 & 0.9868 & 7326 & 2825 & 114 & 1431 & 3899 & 19231 &  9709 & 479 & 347 \\
JP-JEPA (Mini)             & full & 50M  & 500k     & \underline{0.8576} & \underline{0.9871} & \underline{8163} & \underline{2981} & \underline{116} & \underline{1513} & \underline{4141} & \underline{21277} & \underline{10152} & \underline{500} & \underline{357} \\

JP-JEPA (Small)            & full & 50M  & 100k     & 0.8542 & 0.9865 & 6154 & 2625 & 111 & 1284 & 3565 & 18692 &  8772 & 457 & 329 \\
JP-JEPA (Small)            & full & 50M  & 250k     & 0.8586 & 0.9872 & 7407 & 3170 & 117 & 1486 & 4396 & 25974 & 10929 & 501 & 359 \\
JP-JEPA (Small)            & full & 50M  & 500k (s) & \underline{\textbf{0.8603}} & \underline{\textbf{0.9875}} & \underline{\textbf{8621}} & \underline{\textbf{3413}} & \underline{\textbf{118}} & \underline{\textbf{1617}} & \underline{\textbf{4545}} & \underline{26316} & \underline{\textbf{12121}} & \underline{\textbf{524}} & \underline{\textbf{380}} \\
\midrule
ParT~\cite{qu2022particle} & full & 100M & 1M       & 0.8605 & \textbf{0.9877} & \textbf{10638} & \textbf{4149} & \textbf{123} & \textbf{1864} & \textbf{5479} & \textbf{32787} & \textbf{15873} & \textbf{543} & \textbf{402} \\

JP-JEPA (Mini)             & full & 100M & 100k     & 0.8507 & 0.9860 &  6270 & 2454 & 110 & 1219 & 3466 & 18349 &  7519 & 443 & 319 \\
JP-JEPA (Mini)             & full & 100M & 250k     & 0.8548 & 0.9867 &  7299 & 2805 & 114 & 1390 & 3788 & 20833 &  9524 & 475 & 340 \\
JP-JEPA (Mini)             & full & 100M & 500k (s) & \underline{0.8581} & \underline{0.9872} &  \underline{8368} & \underline{3155} & \underline{117} & \underline{1566} & \underline{4065} & \underline{22989} & \underline{11299} & \underline{508} & \underline{359} \\

JP-JEPA (Small)            & full & 100M & 100k     & 0.8538 & 0.9865 &  6289 & 2625 & 111 & 1332 & 3752 & 19231 &  9615 & 452 & 330 \\
JP-JEPA (Small)            & full & 100M & 250k     & 0.8584 & 0.9872 &  7576 & 3135 & 116 & 1533 & 4141 & 22989 & 12270 & 498 & 360 \\
JP-JEPA (Small)            & full & 100M & 500k (s) & \underline{\textbf{0.8611}} & \underline{0.9876} &  \underline{8511} & \underline{3184} & \underline{119} & \underline{1688} & \underline{4808} & \underline{32258} & \underline{13245} & \underline{528} & \underline{374} \\
\bottomrule
\end{tabular}
}
\caption{
\textbf{Study of the number of iterations for fine-tuning on JetClass.} For all metrics, the higher the better. 'full' category corresponds to using all available information. \textbf{Bolded} results denote the best results over all methods in its subsampling, and \underline{underlined} the best for each method in its sampling (only for JP-JEPA).
}
\label{tab:jetclass-steps-study}
\end{table*}

Table \ref{tab:tq-qg-epochs-study} presents a small study examining performance variations across different maximum epoch values. Overall, better results are observed when the number of epochs is close to 10, a choice that differs from that of ParT, which uses 20.
\begin{table*}[ht!]
\resizebox{\linewidth}{!}{
\begin{tabular}{lccccccccc}
\toprule
\multirow{3}{*}{Methods} & \multirow{3}{*}{\# epochs} & \multicolumn{4}{c}{Top Quark} & \multicolumn{4}{c}{Quark-Gluon} \\
\cmidrule(lr){3-6}\cmidrule(lr){7-10}
& & Acc & AUC & $\text{Rej}_{\text{50\%}}$ & $\text{Rej}_{\text{30\%}}$ & Acc & AUC & $\text{Rej}_{\text{50\%}}$ & $\text{Rej}_{\text{30\%}}$ \\
\midrule
JP-JEPA (Mini) & 10 (s) & \textbf{0.938} & \textbf{0.9854} & \textbf{369} & \textbf{1364} & 0.847 & \textbf{0.9190} & 46.6 & \textbf{128.4} \\
JP-JEPA (Mini) & 20 & 0.938 & 0.9846 & 300 & 1030 & \textbf{0.848} & 0.9188 & \textbf{46.8} & 125.3 \\
\midrule
JP-JEPA (Small) & 10 (s) & \textbf{0.939} & \textbf{0.9851} & \textbf{357} & \textbf{1286} & \textbf{0.848} & \textbf{0.9184} & \textbf{47.1} & \textbf{123.6} \\
JP-JEPA (Small) & 20 & 0.929 & 0.9782 & 139 & 326 & 0.838 & 0.9034 & 34.5 & 77.8 \\
\bottomrule
\end{tabular}
}
\caption{
\textbf{Study of the number of epochs for fine-tuning on Top Quark Tagging and Quark-Gluon datasets.} For all metrics, the higher the better. \textbf{Bolded} results denote the best results over all methods in its number of epochs. One training has been performed.
}
\label{tab:tq-qg-epochs-study}
\end{table*}

\subsection{Additionnal qualitative visualizations}
\label{sec:supp:other-results:qualitative}
Electrons are represented by upward-pointing colored triangles outlined in black. Muons are depicted as downward-pointing colored triangles, also with a black outline. Photons are illustrated as transparent pentagons with a colored outline. Finally, hadrons are represented by circles: charged hadrons appear as solid colored disks without an outline, whereas neutral hadrons are shown as transparent circles with a colored outline. The left column shows the overall result for the considered example, where the blue color denotes the method with the highest probability associated with the predicted label. The boxes are hatched when the final prediction is incorrect. For all figures, the second, third, and fourth columns respectively provide the detailed output probability distributions of ParticleNet, ParT, and JP-JEPA.

\begin{figure*}[ht!]
\centering
\setlength{\tabcolsep}{1pt} 
\begin{tabular}{cccccccc}
\includegraphics[width=0.24\textwidth,page=1]{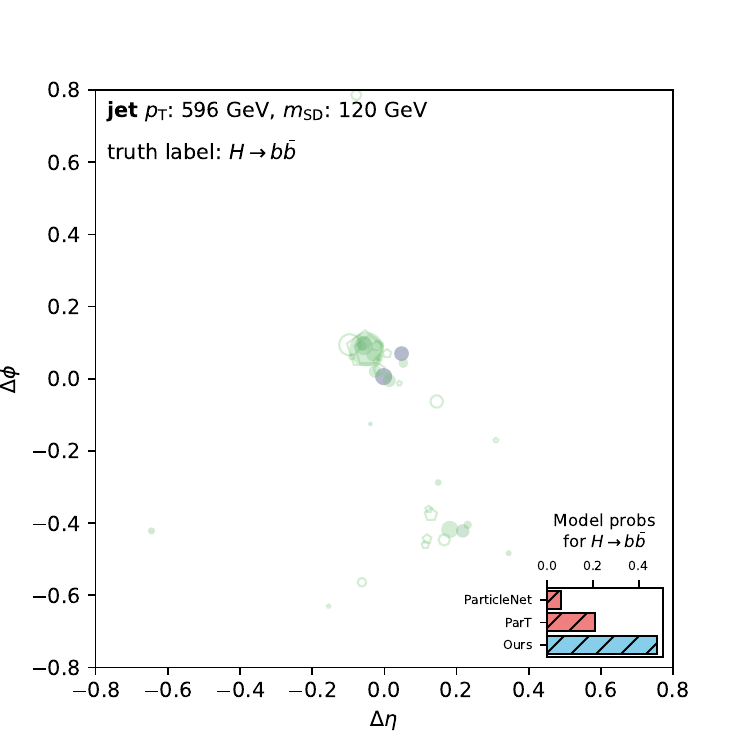} &
\includegraphics[width=0.24\textwidth,page=1]{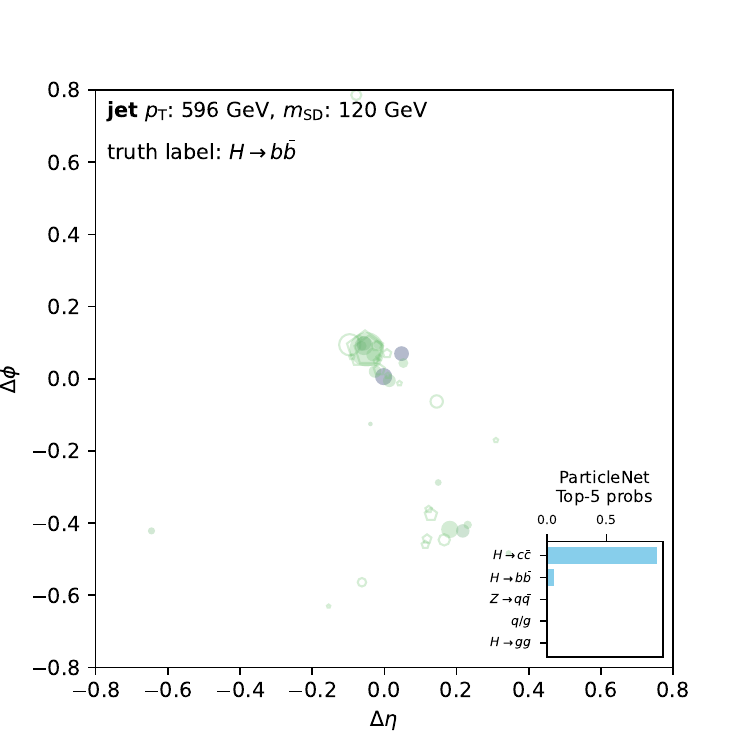} &
\includegraphics[width=0.24\textwidth,page=1]{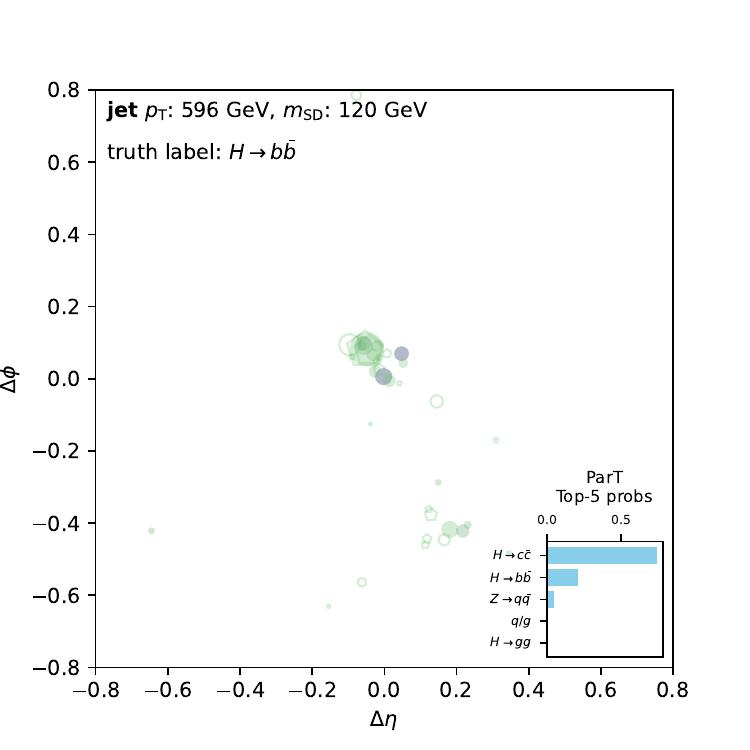} &
\includegraphics[width=0.24\textwidth,page=1]{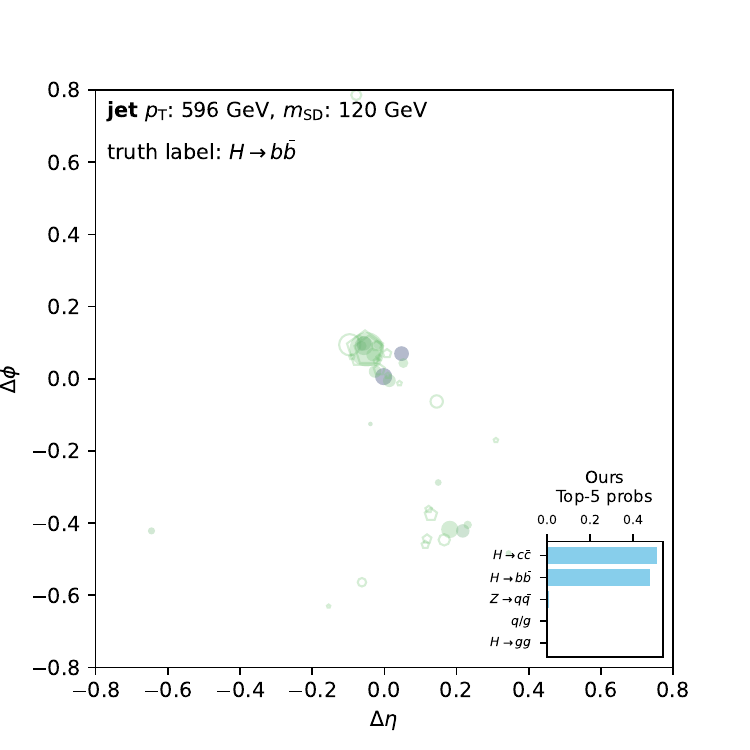} \\

\includegraphics[width=0.24\textwidth,page=1]{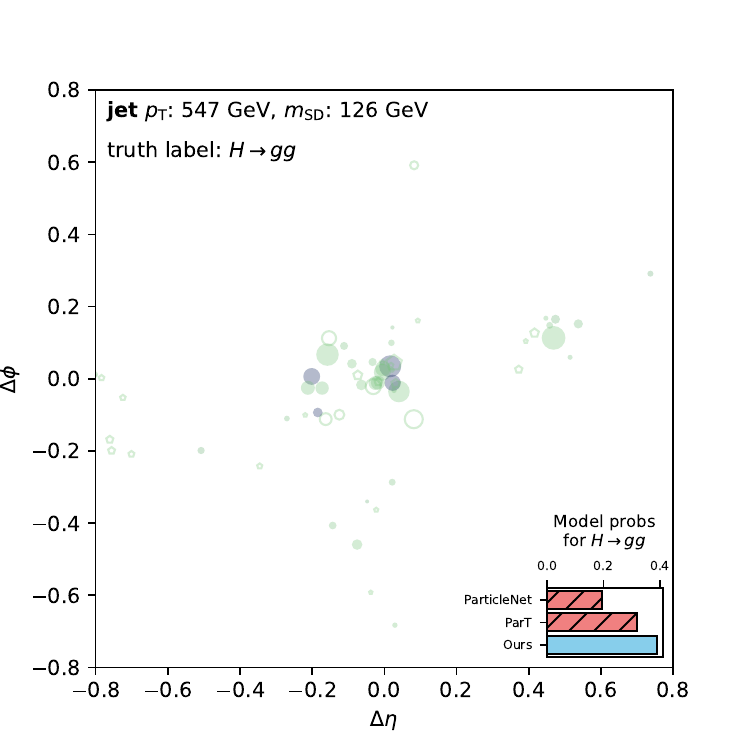} &
\includegraphics[width=0.24\textwidth,page=1]{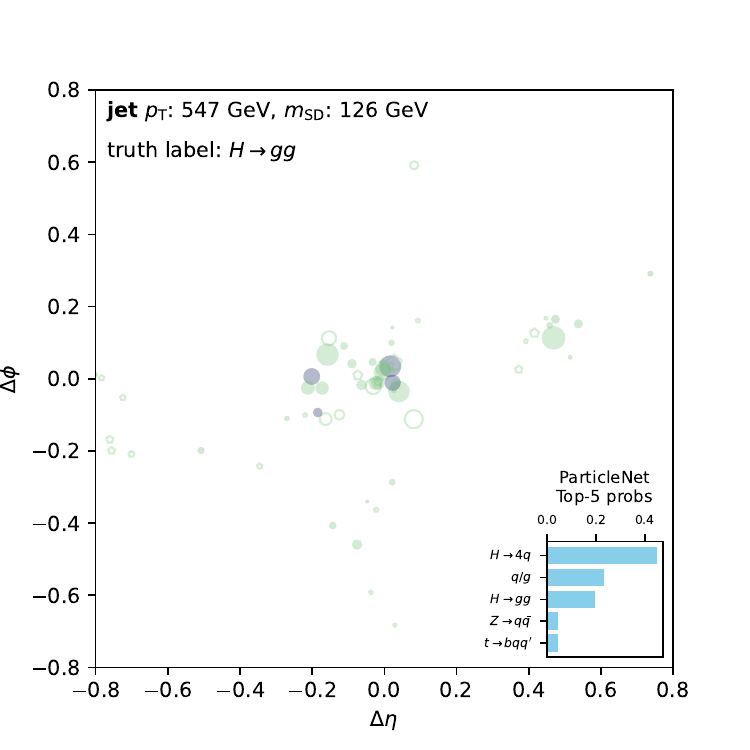} &
\includegraphics[width=0.24\textwidth,page=1]{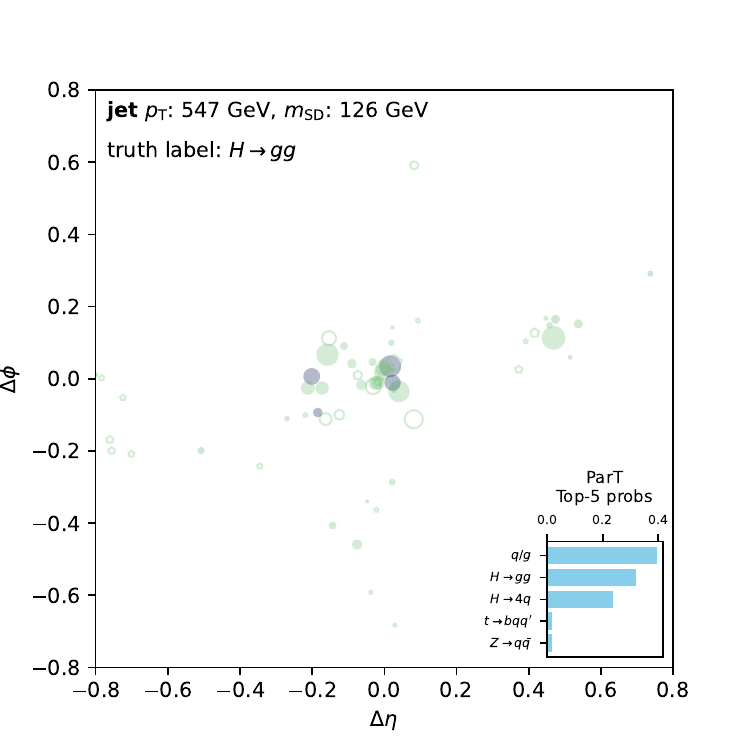} &
\includegraphics[width=0.24\textwidth,page=1]{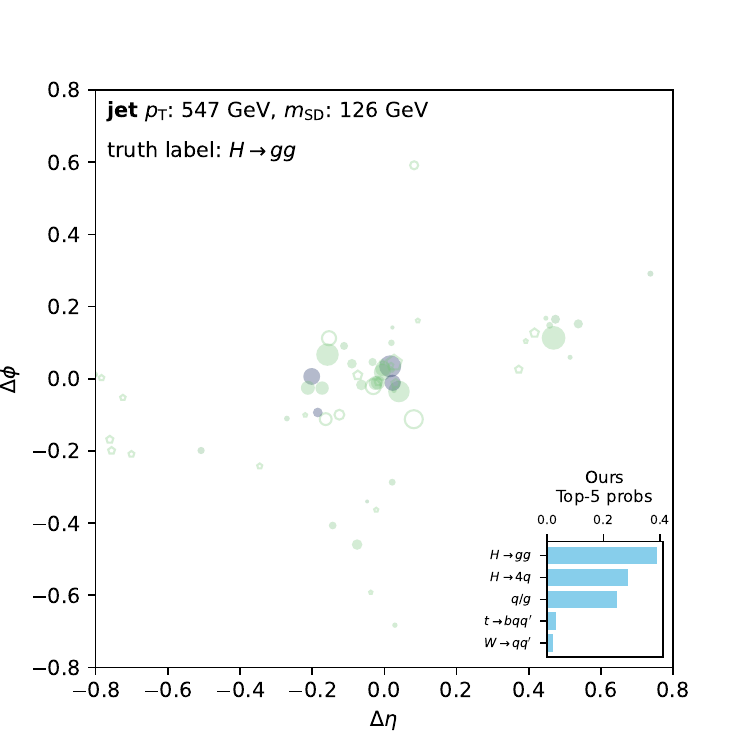} \\

\includegraphics[width=0.24\textwidth,page=1]{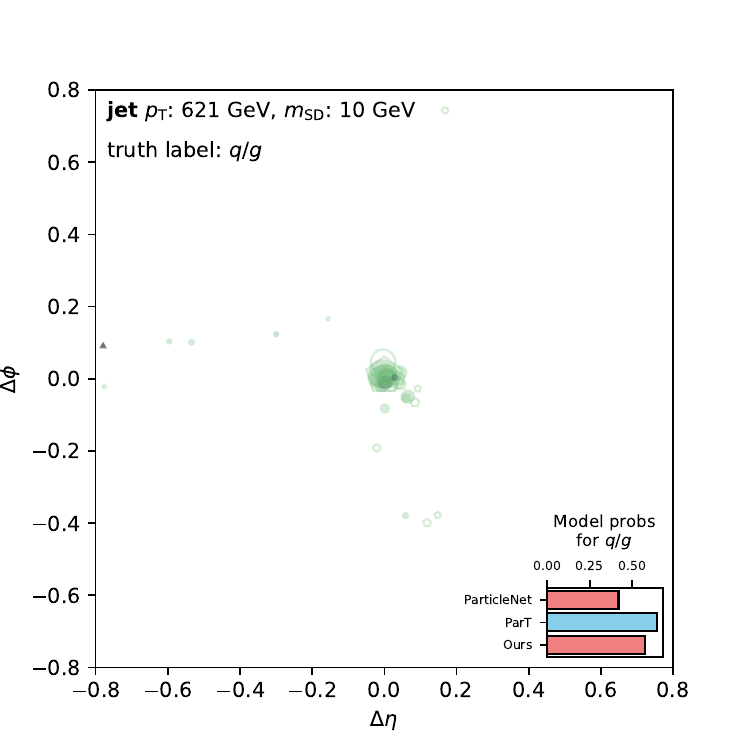} &
\includegraphics[width=0.24\textwidth,page=1]{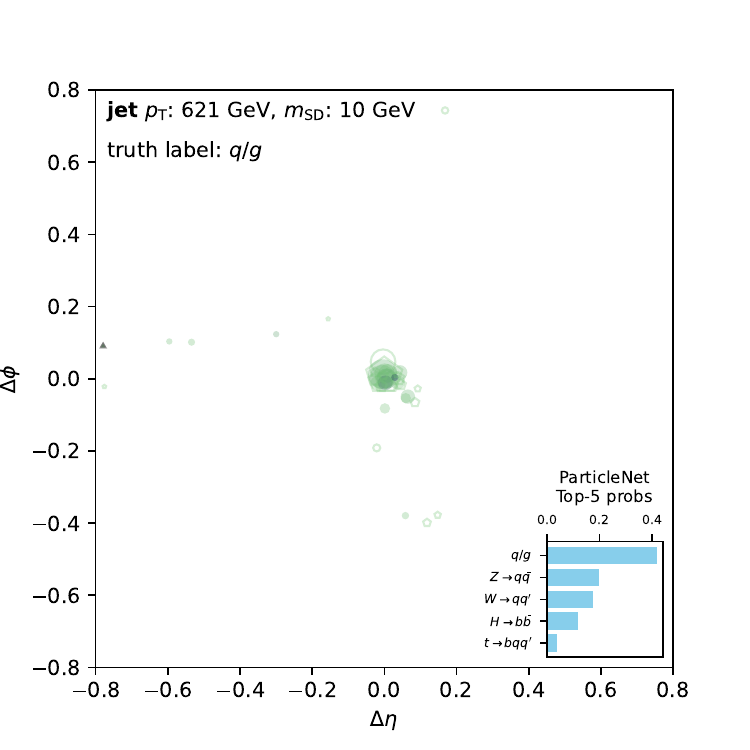} &
\includegraphics[width=0.24\textwidth,page=1]{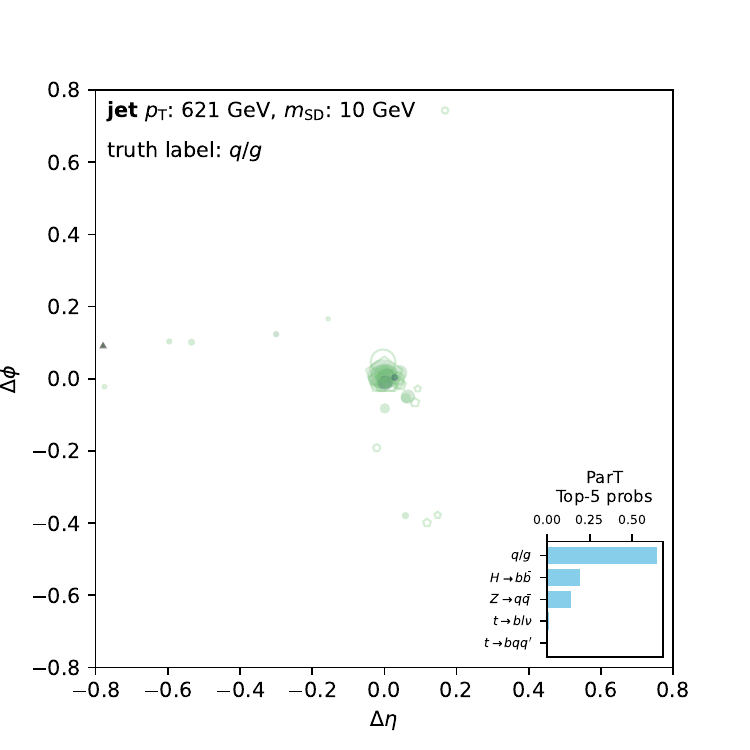} &
\includegraphics[width=0.24\textwidth,page=1]{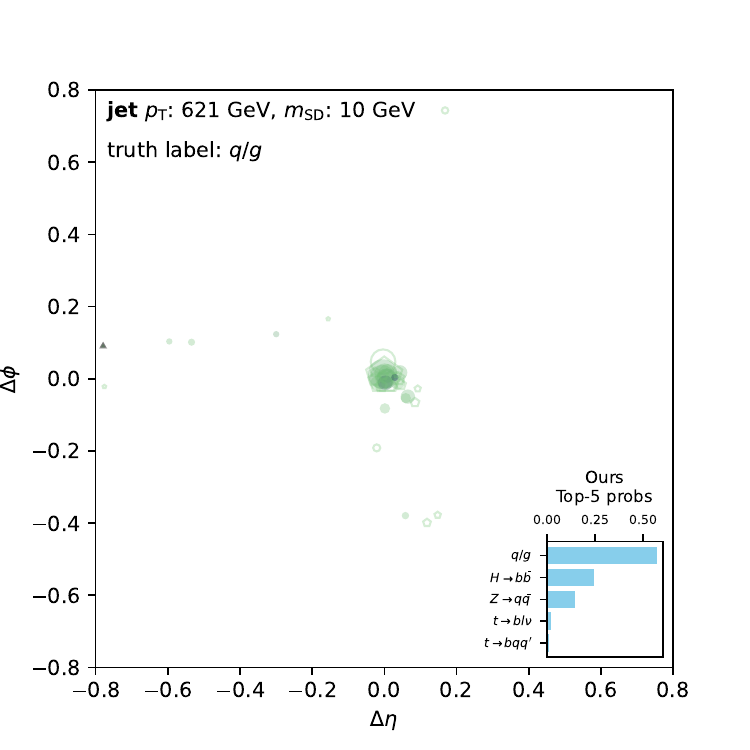} \\

\includegraphics[width=0.24\textwidth,page=1]{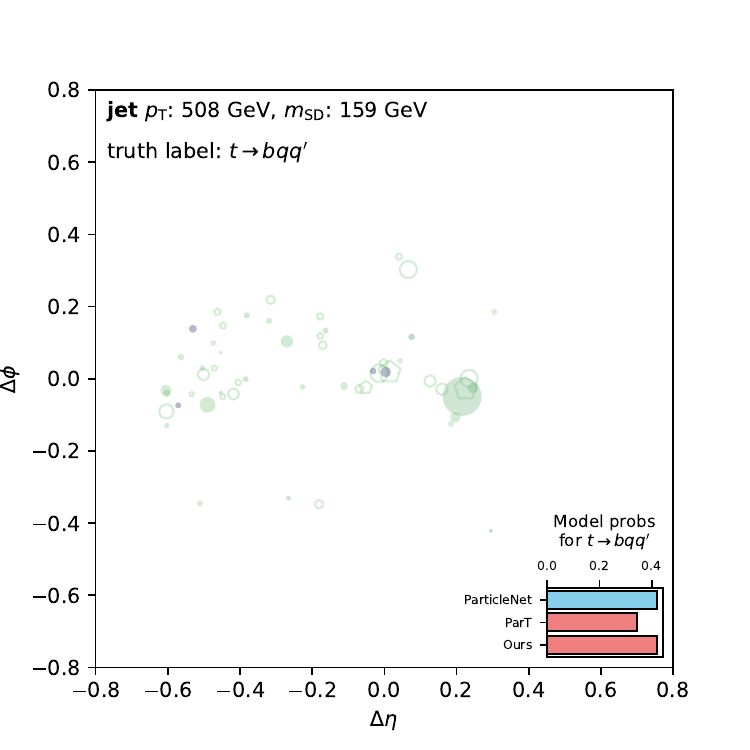} &
\includegraphics[width=0.24\textwidth,page=1]{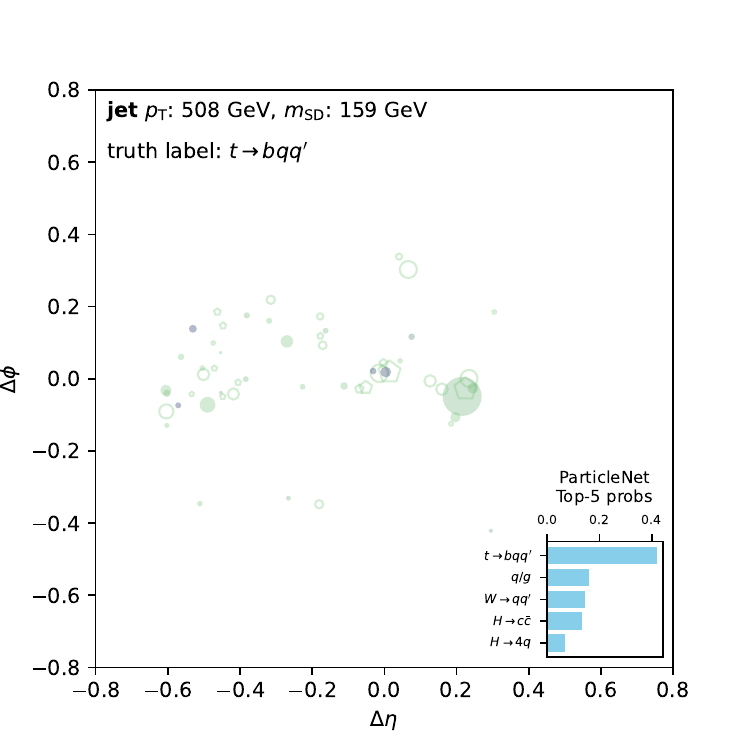} &
\includegraphics[width=0.24\textwidth,page=1]{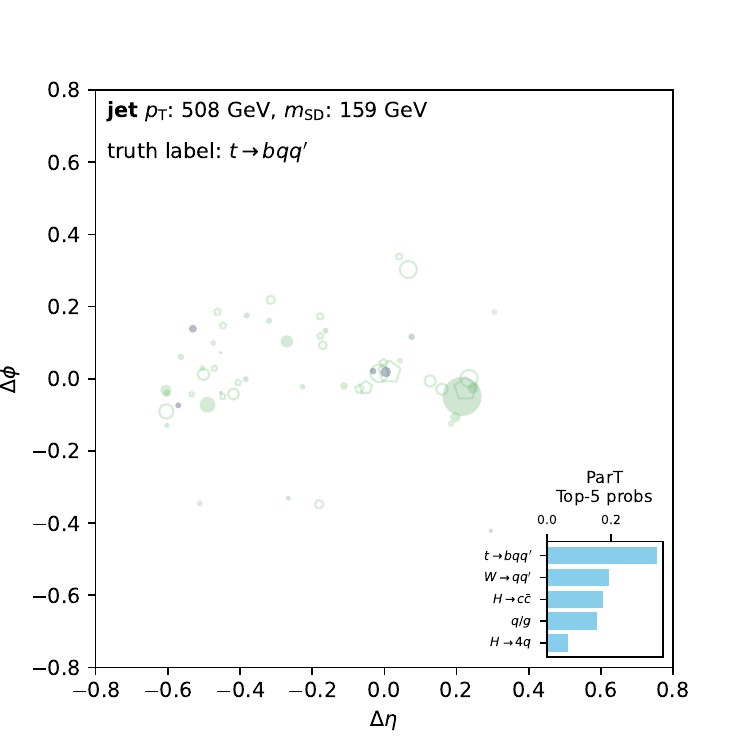} &
\includegraphics[width=0.24\textwidth,page=1]{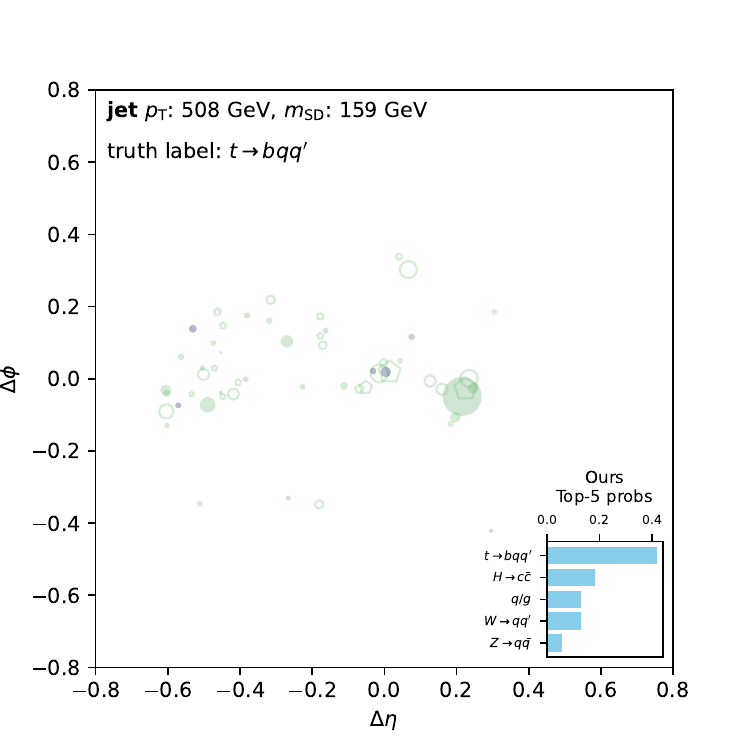} \\

\includegraphics[width=0.24\textwidth,page=1]{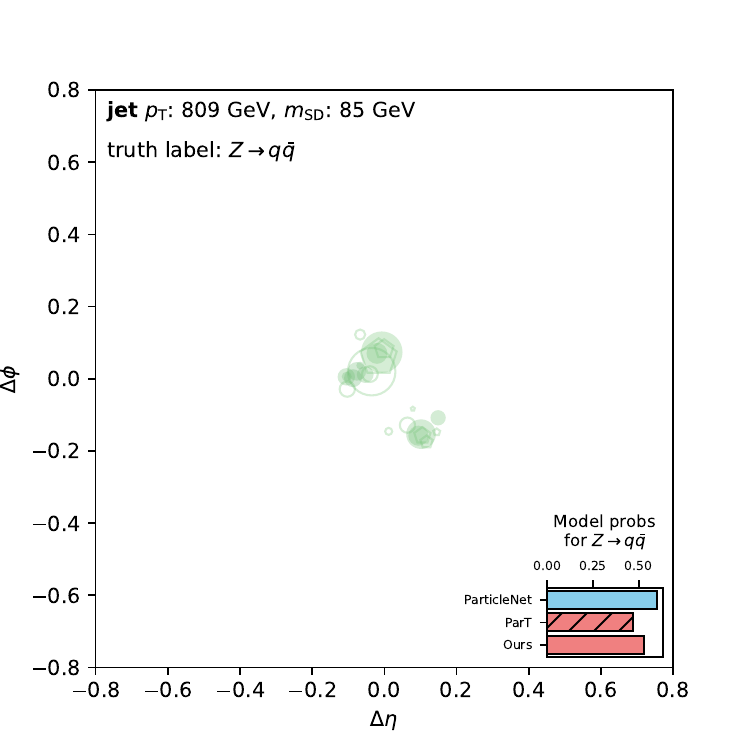} &
\includegraphics[width=0.24\textwidth,page=1]{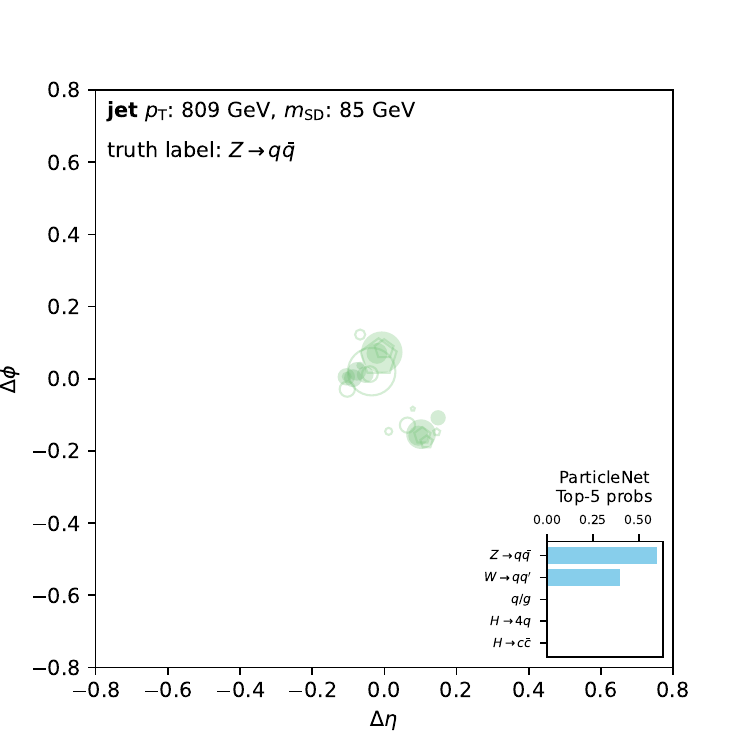} &
\includegraphics[width=0.24\textwidth,page=1]{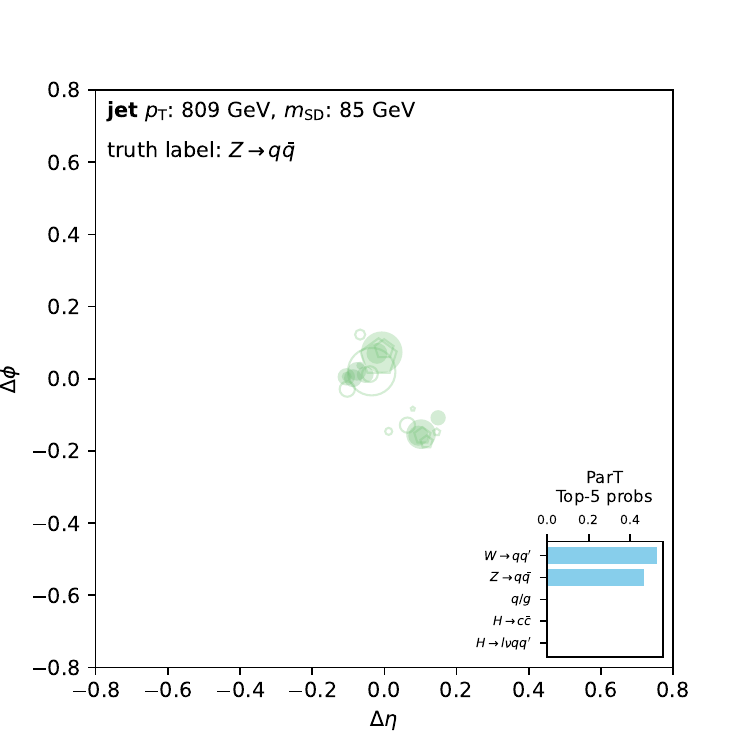} &
\includegraphics[width=0.24\textwidth,page=1]{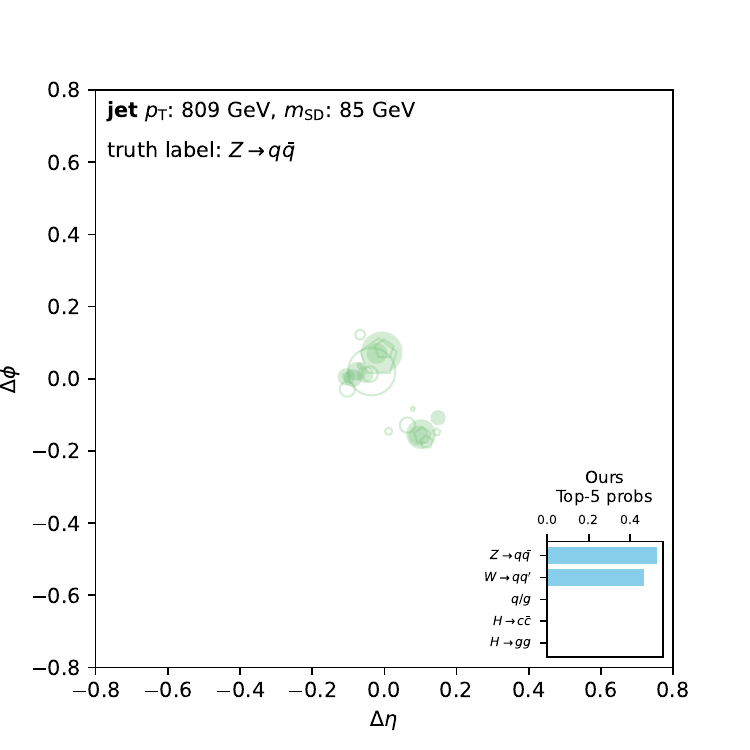} \\
\end{tabular}
\caption{\textbf{Qualitative results on JetClass.}}
    \label{fig:qualitative-results}
\end{figure*}

Figure \ref{fig:qualitative-results} shows qualitative results for illustrating the results part on uncertainty.

\begin{figure*}[ht!]
\centering
\setlength{\tabcolsep}{1pt} 
\begin{tabular}{cccccccc}
\includegraphics[width=0.24\textwidth,page=1]{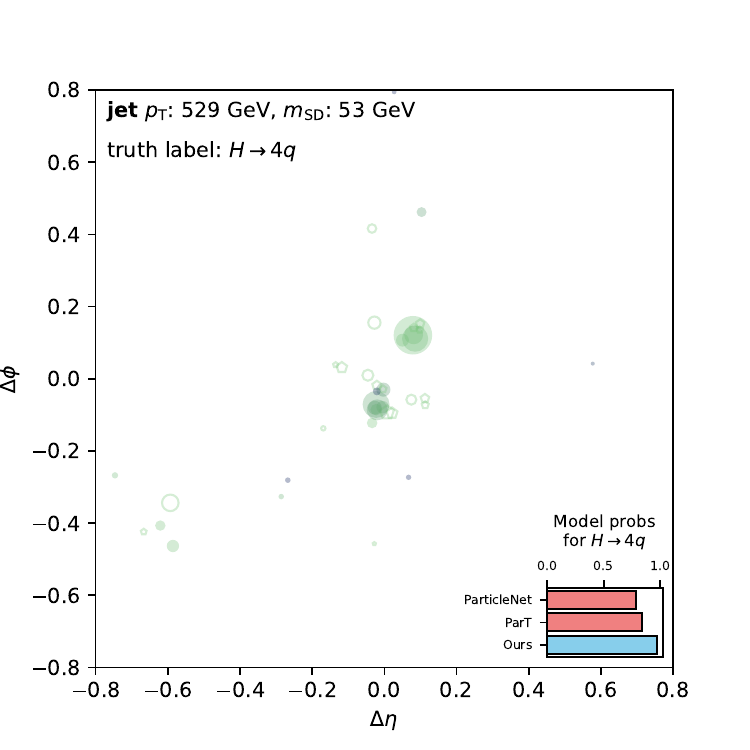} &
\includegraphics[width=0.24\textwidth,page=1]{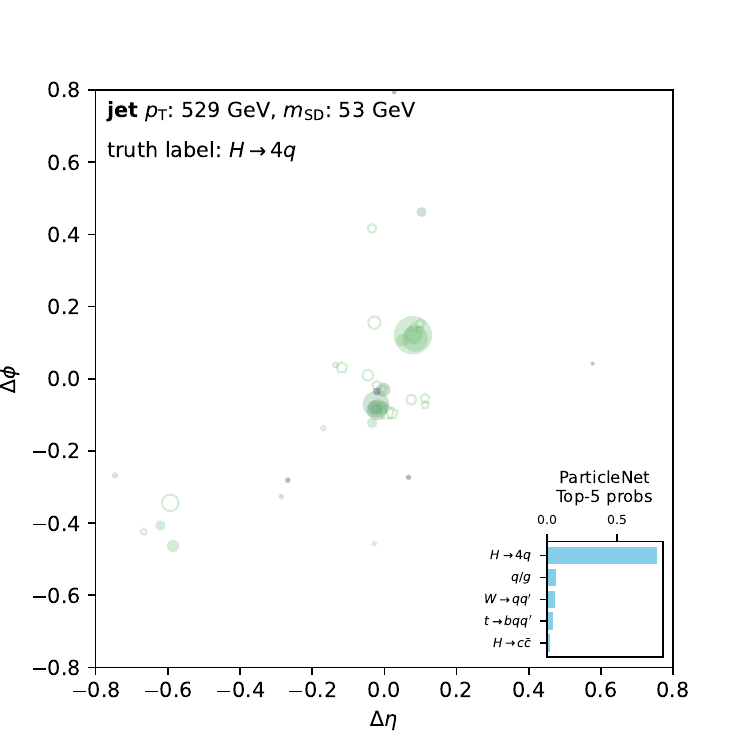} &
\includegraphics[width=0.24\textwidth,page=1]{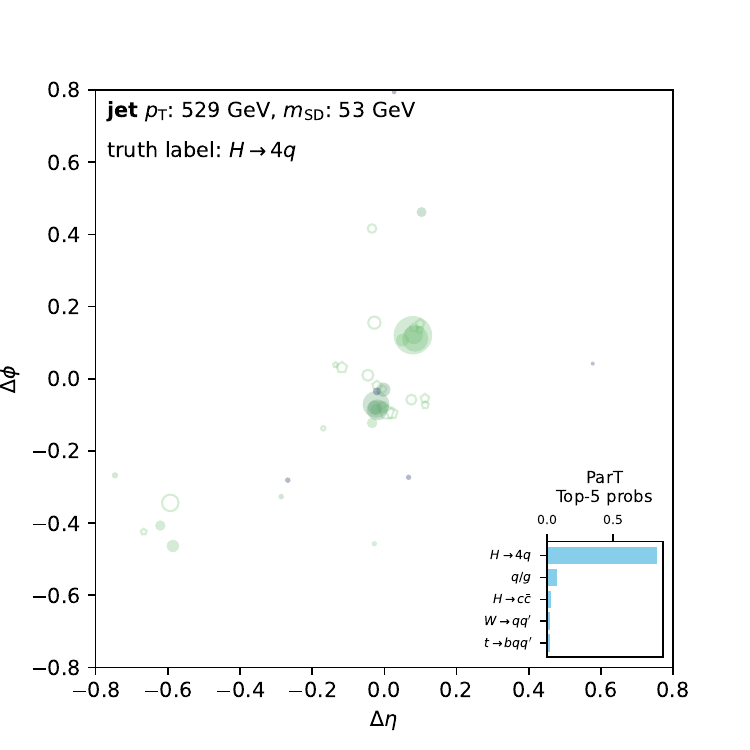} &
\includegraphics[width=0.24\textwidth,page=1]{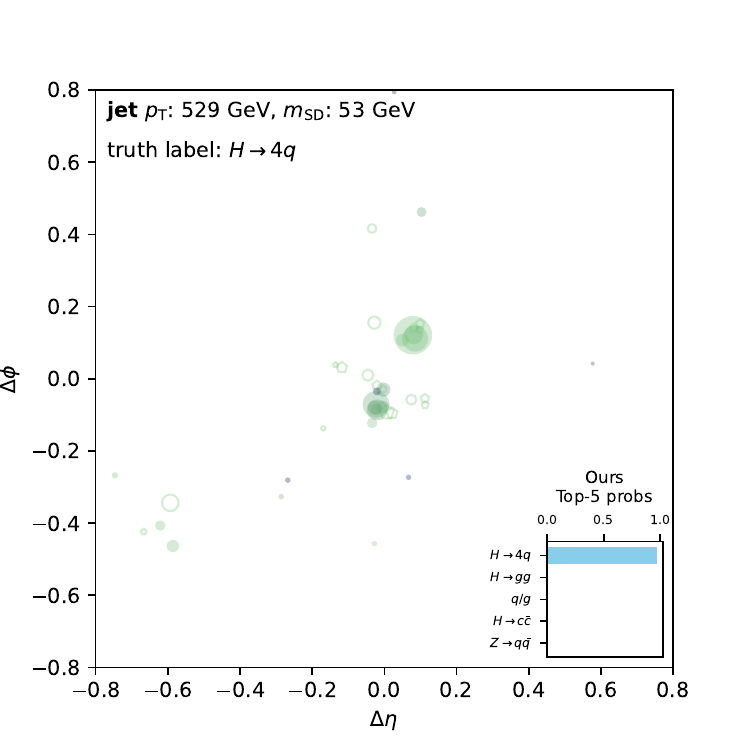} \\

\includegraphics[width=0.24\textwidth,page=1]{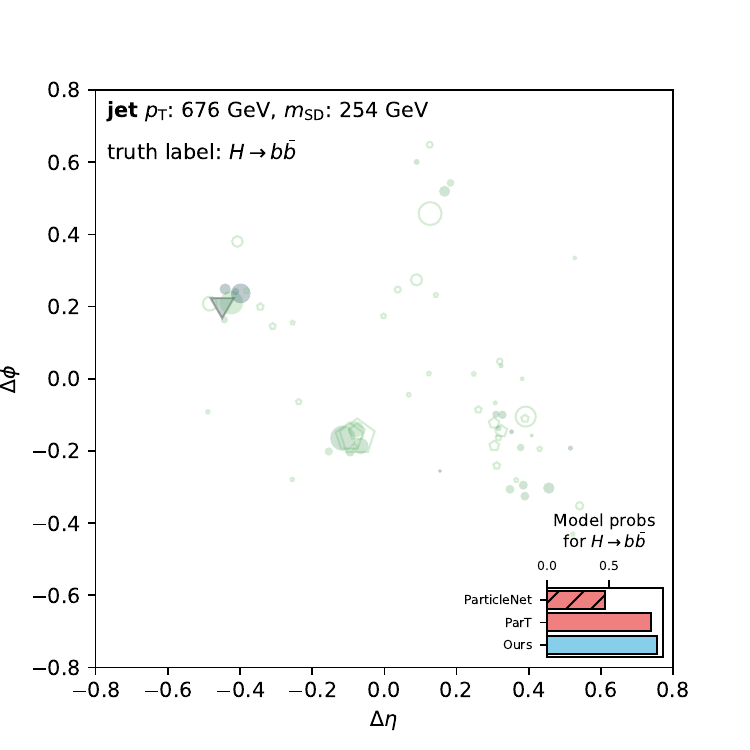} &
\includegraphics[width=0.24\textwidth,page=1]{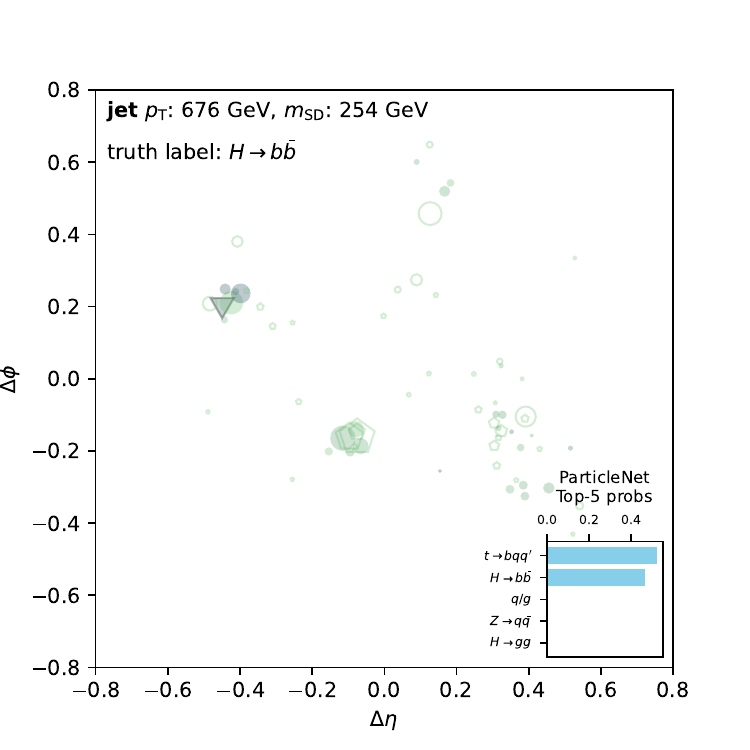} &
\includegraphics[width=0.24\textwidth,page=1]{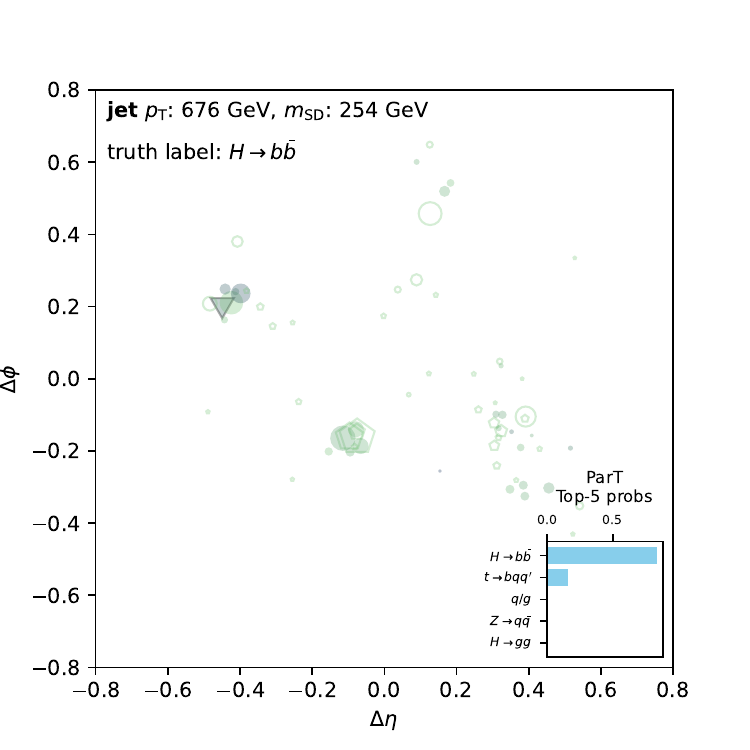} &
\includegraphics[width=0.24\textwidth,page=1]{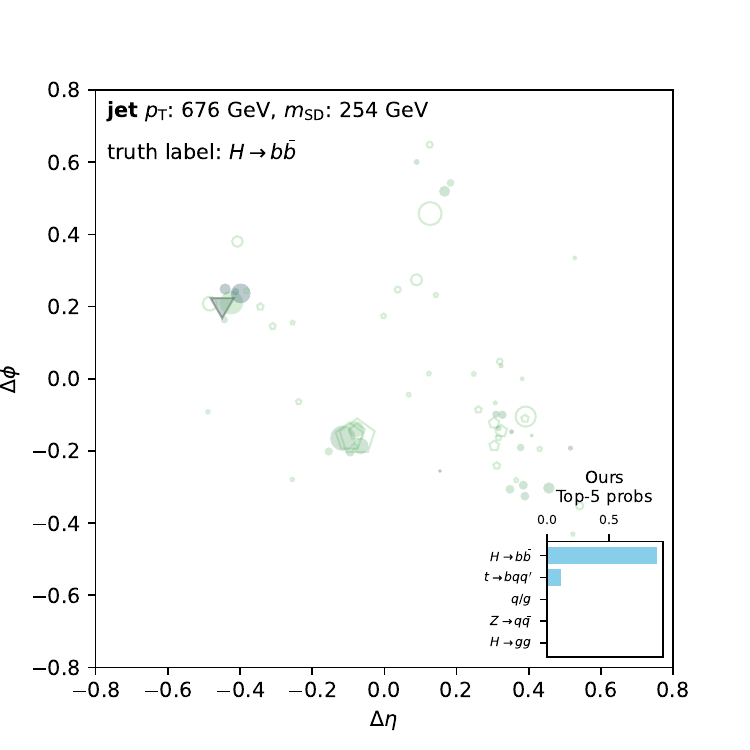} \\

\includegraphics[width=0.24\textwidth,page=1]{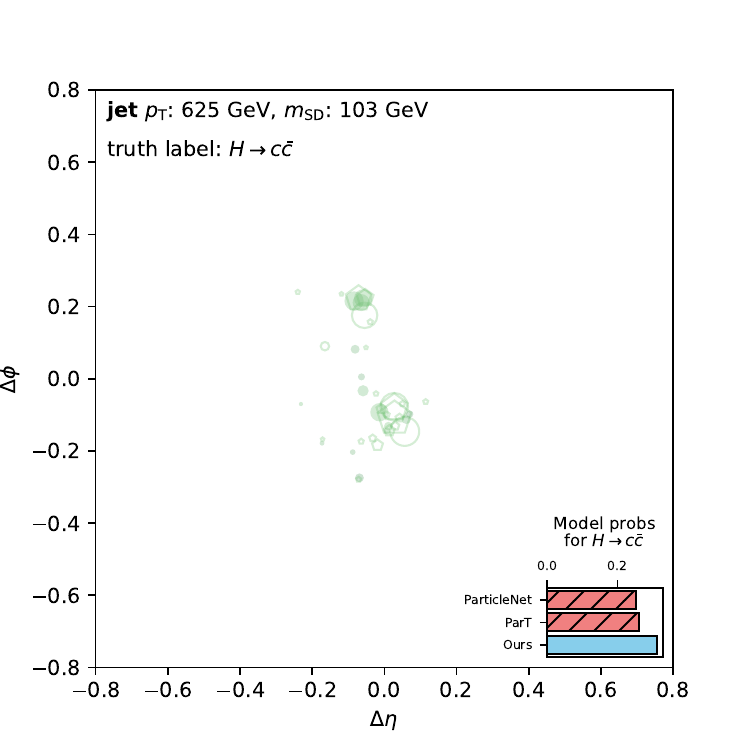} &
\includegraphics[width=0.24\textwidth,page=1]{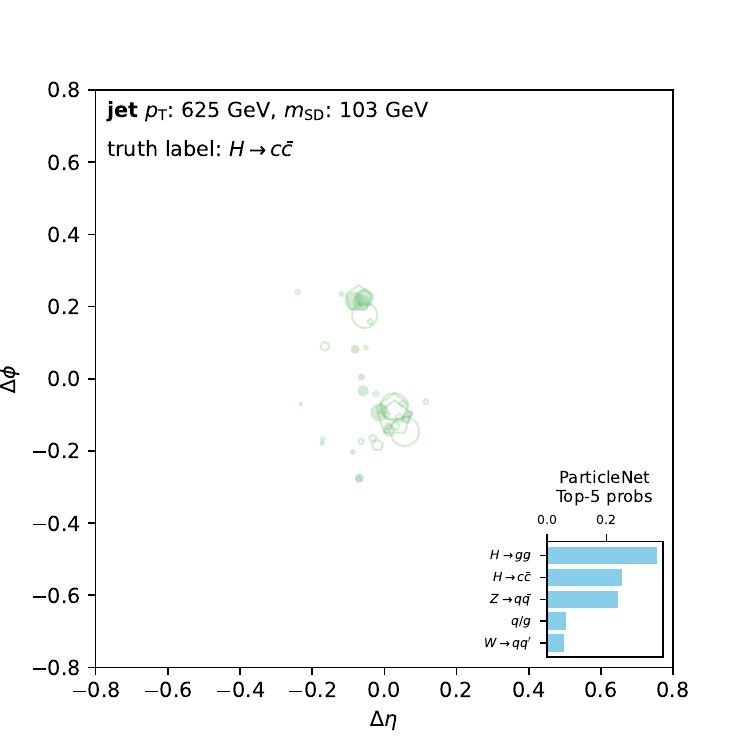} &
\includegraphics[width=0.24\textwidth,page=1]{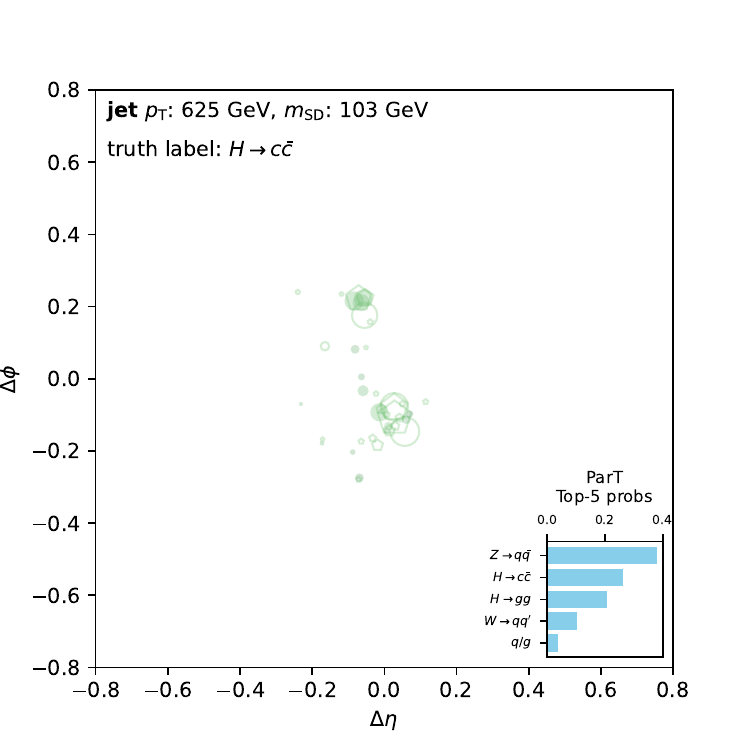} &
\includegraphics[width=0.24\textwidth,page=1]{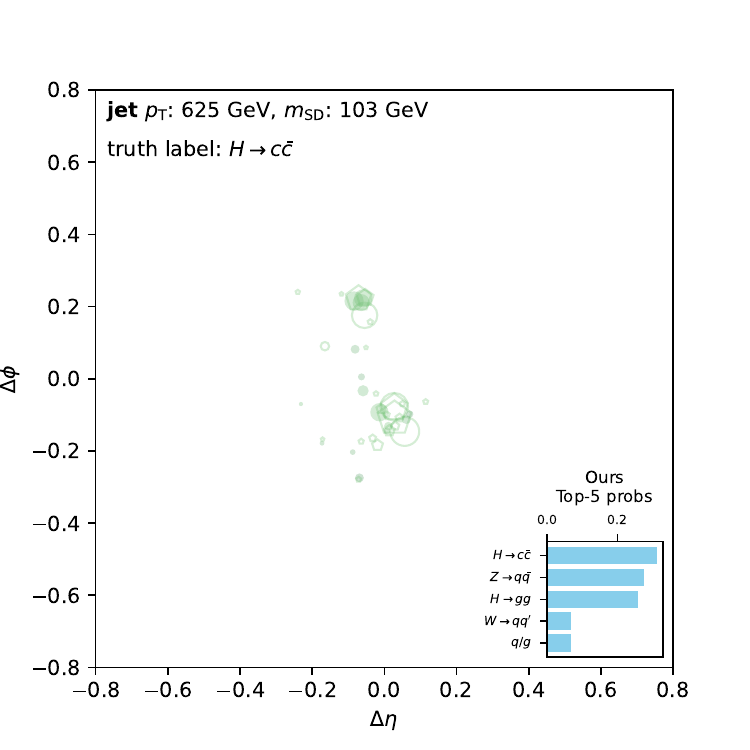} \\

\includegraphics[width=0.24\textwidth,page=1]{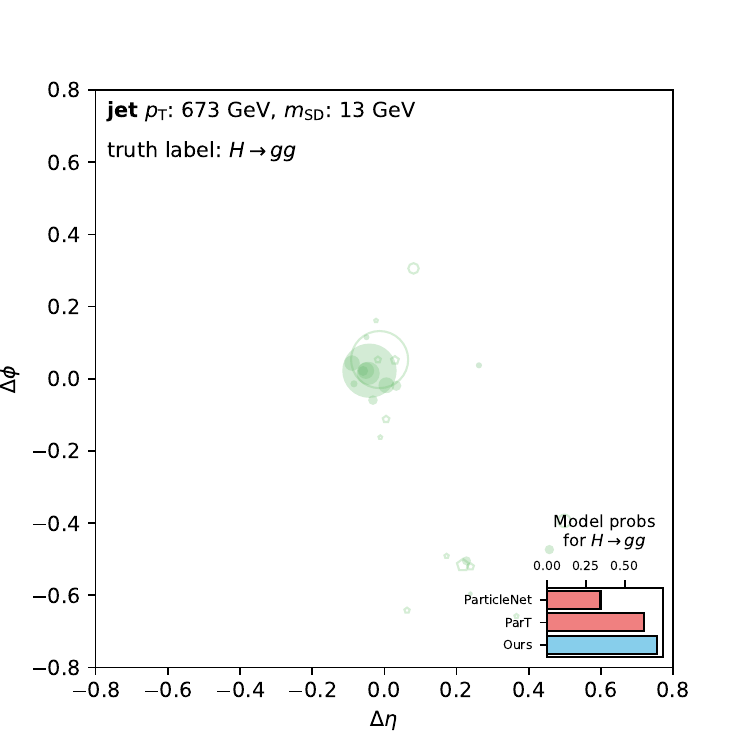} &
\includegraphics[width=0.24\textwidth,page=1]{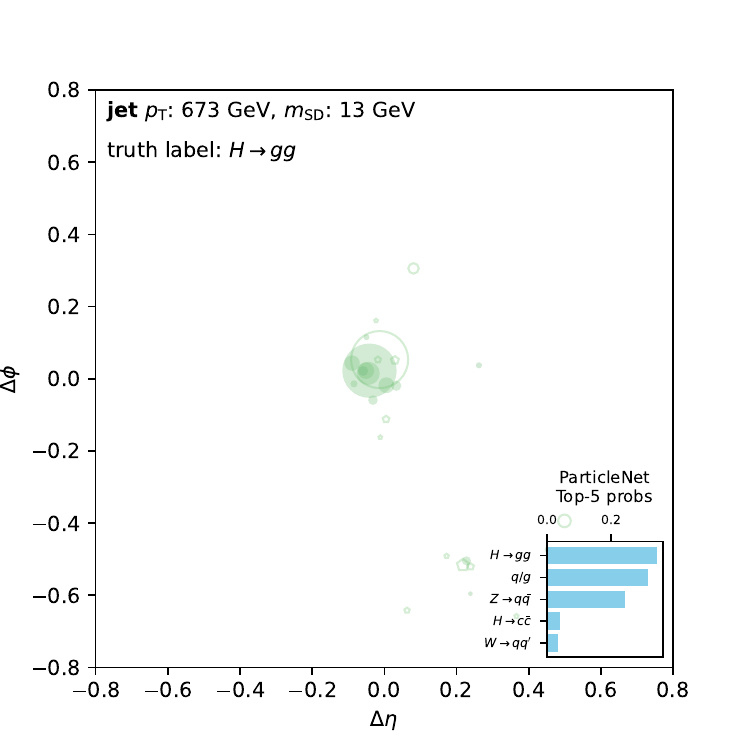} &
\includegraphics[width=0.24\textwidth,page=1]{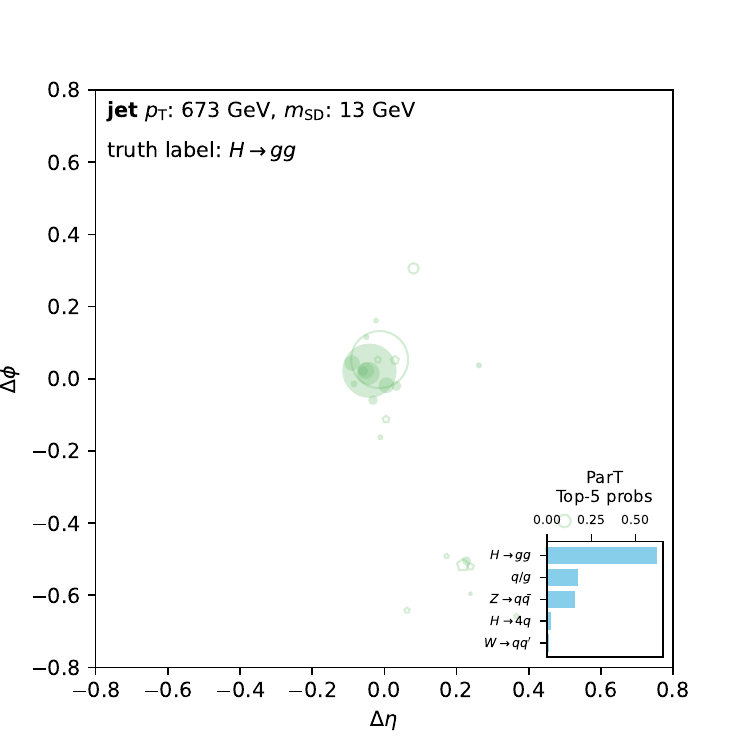} &
\includegraphics[width=0.24\textwidth,page=1]{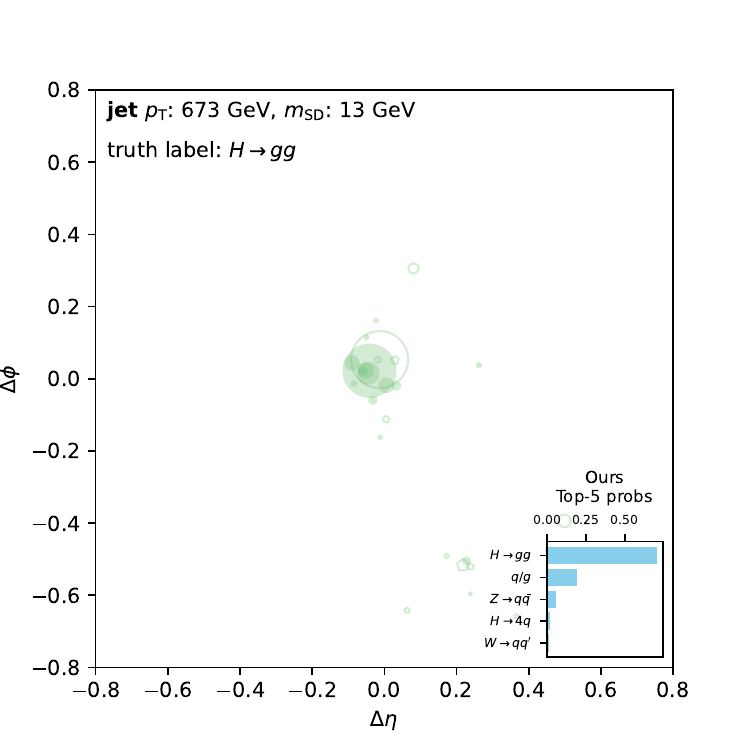} \\

\includegraphics[width=0.24\textwidth,page=1]{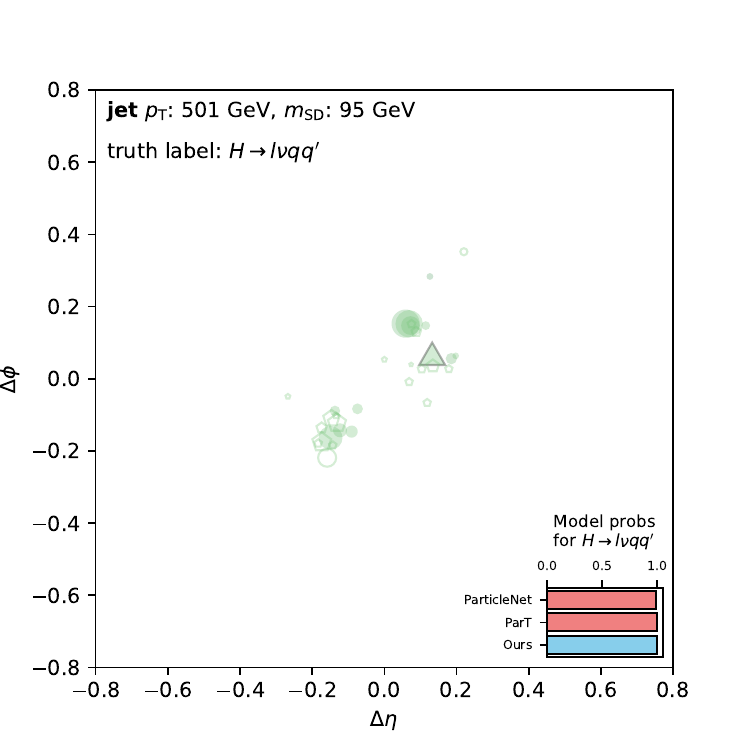} &
\includegraphics[width=0.24\textwidth,page=1]{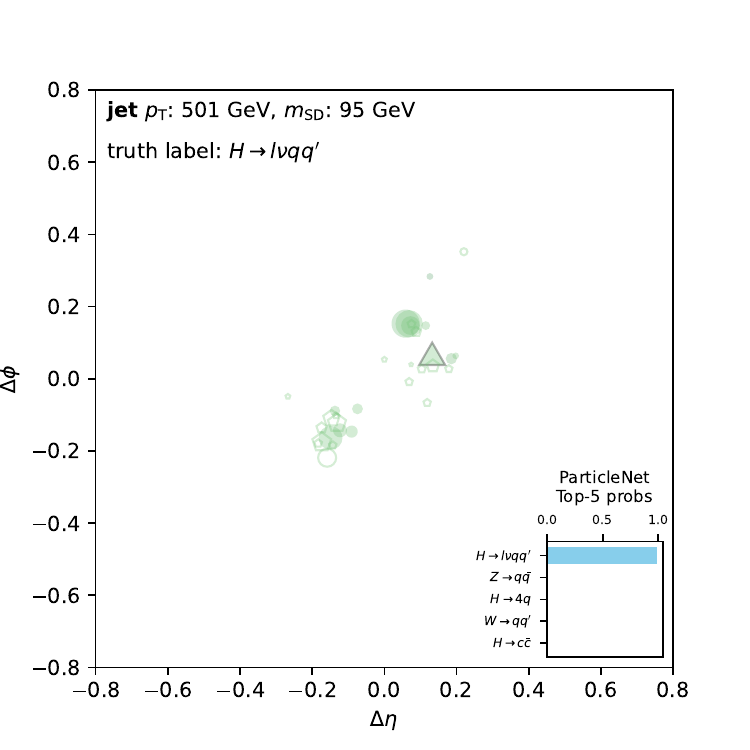} &
\includegraphics[width=0.24\textwidth,page=1]{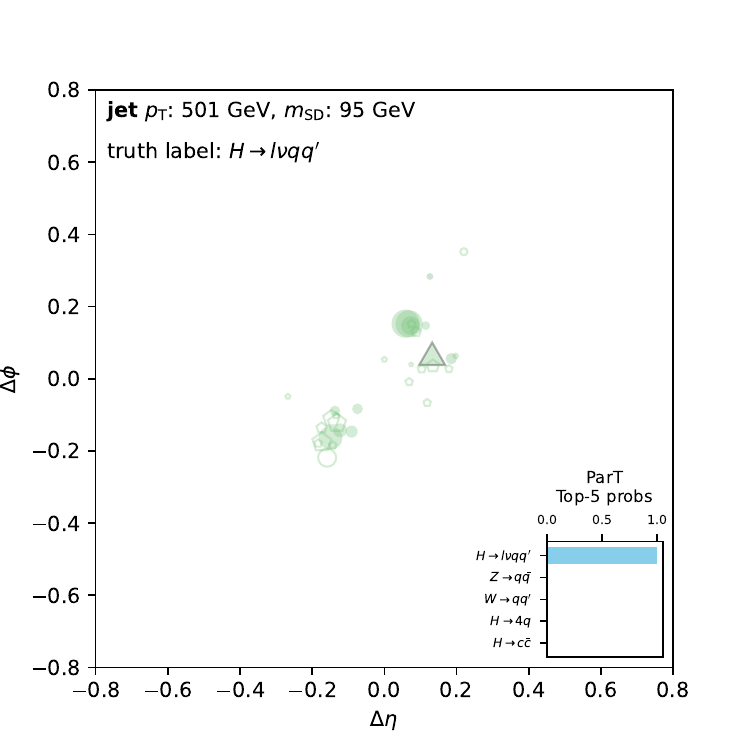} &
\includegraphics[width=0.24\textwidth,page=1]{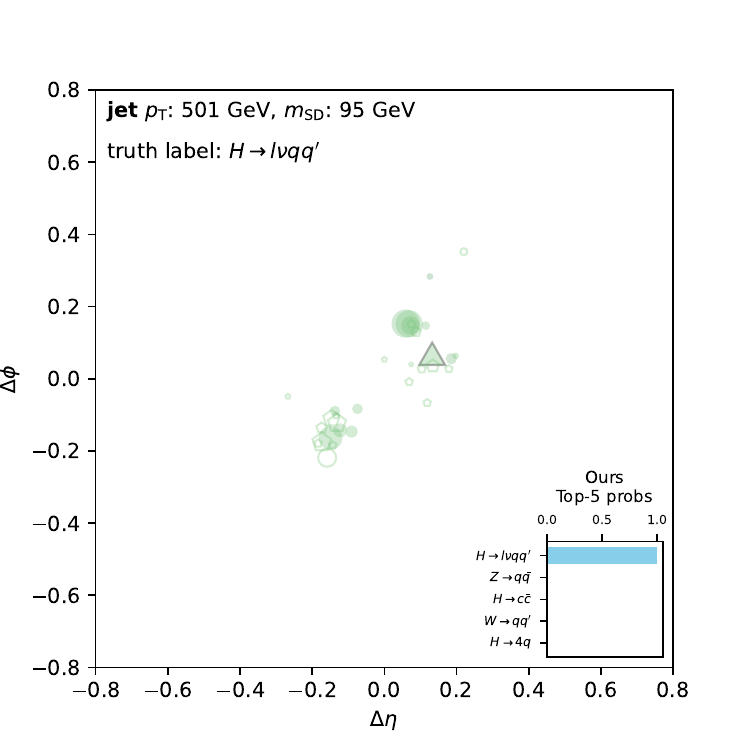} \\
\end{tabular}
\caption{\textbf{Qualitative results on JetClass highlighting successful JP-JEPA predictions (part 1).}}
\label{tab:qualitative:good1}
\end{figure*}

\begin{figure*}[ht!]
\centering
\setlength{\tabcolsep}{1pt} 
\begin{tabular}{cccccccc}
\includegraphics[width=0.24\textwidth,page=1]{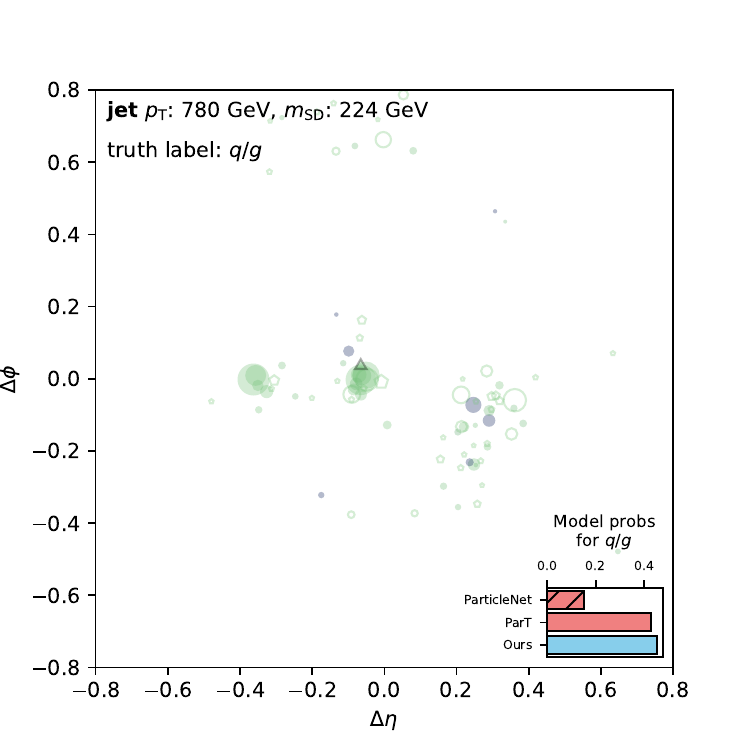} &
\includegraphics[width=0.24\textwidth,page=1]{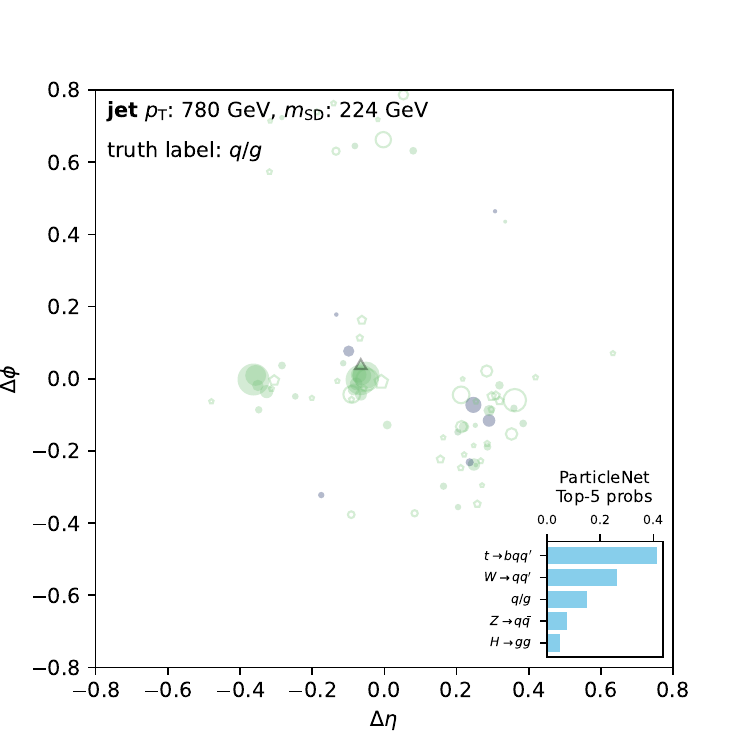} &
\includegraphics[width=0.24\textwidth,page=1]{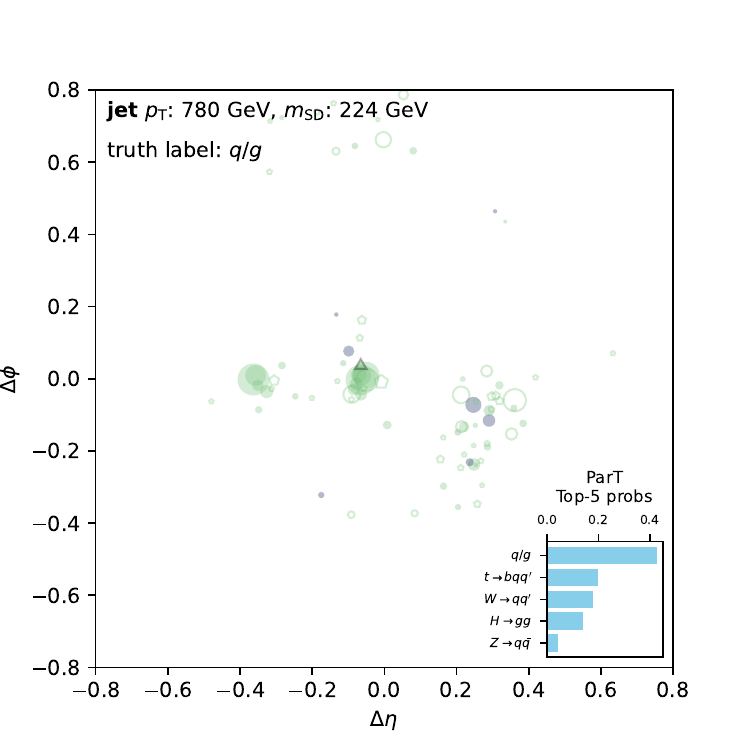} &
\includegraphics[width=0.24\textwidth,page=1]{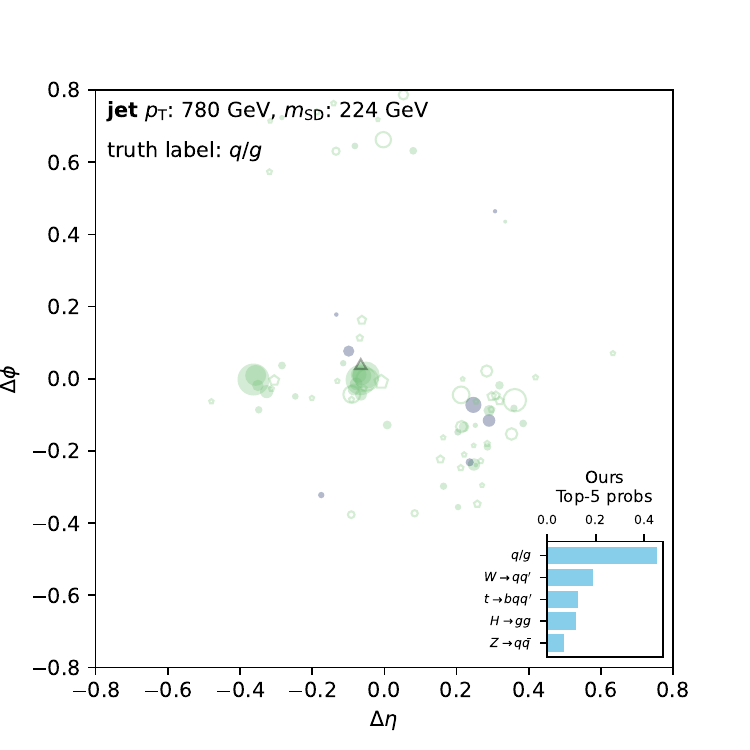} \\

\includegraphics[width=0.24\textwidth,page=1]{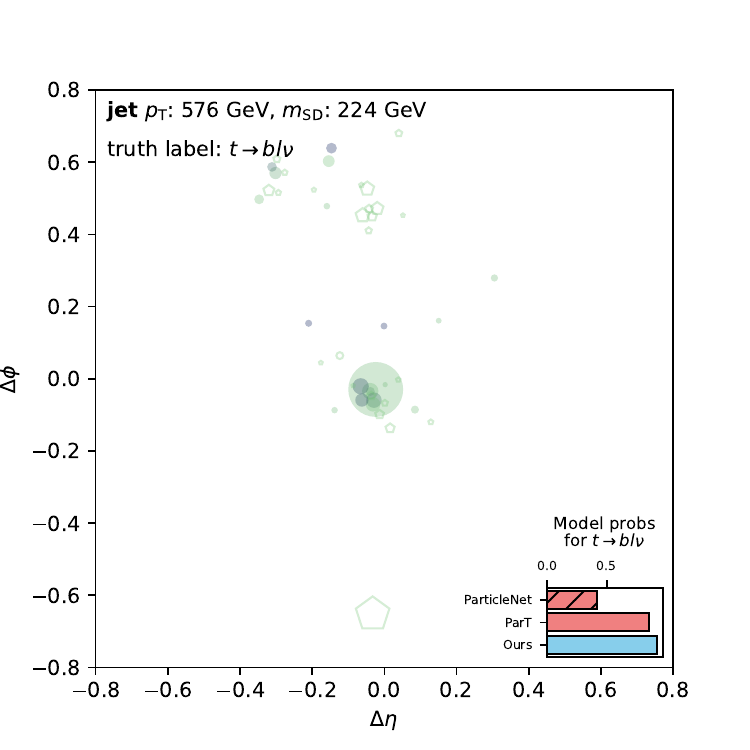} &
\includegraphics[width=0.24\textwidth,page=1]{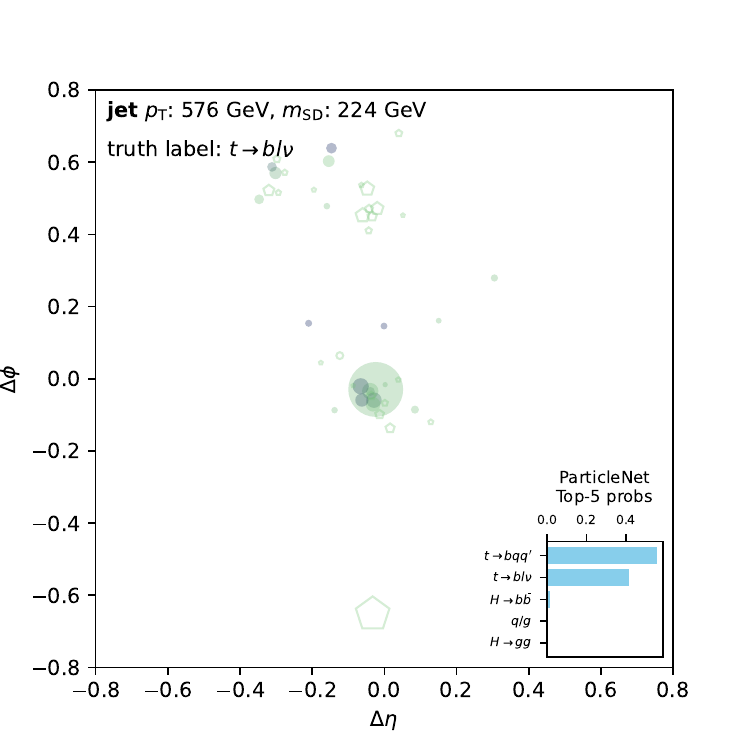} &
\includegraphics[width=0.24\textwidth,page=1]{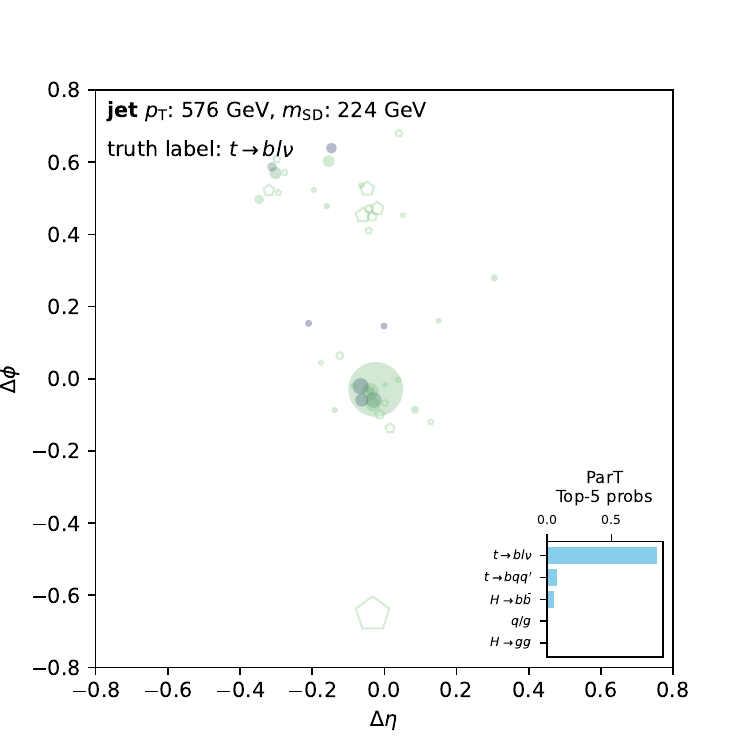} &
\includegraphics[width=0.24\textwidth,page=1]{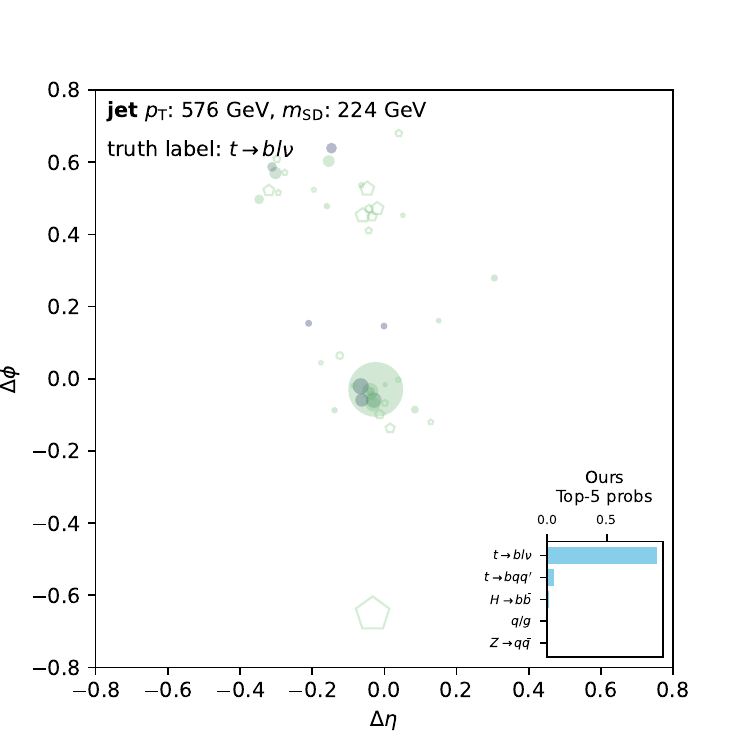} \\

\includegraphics[width=0.24\textwidth,page=1]{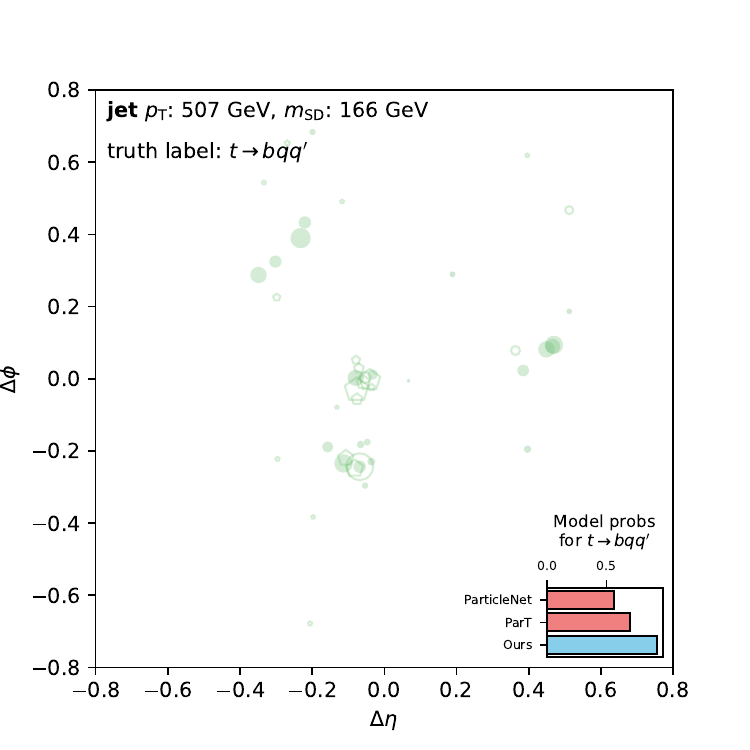} &
\includegraphics[width=0.24\textwidth,page=1]{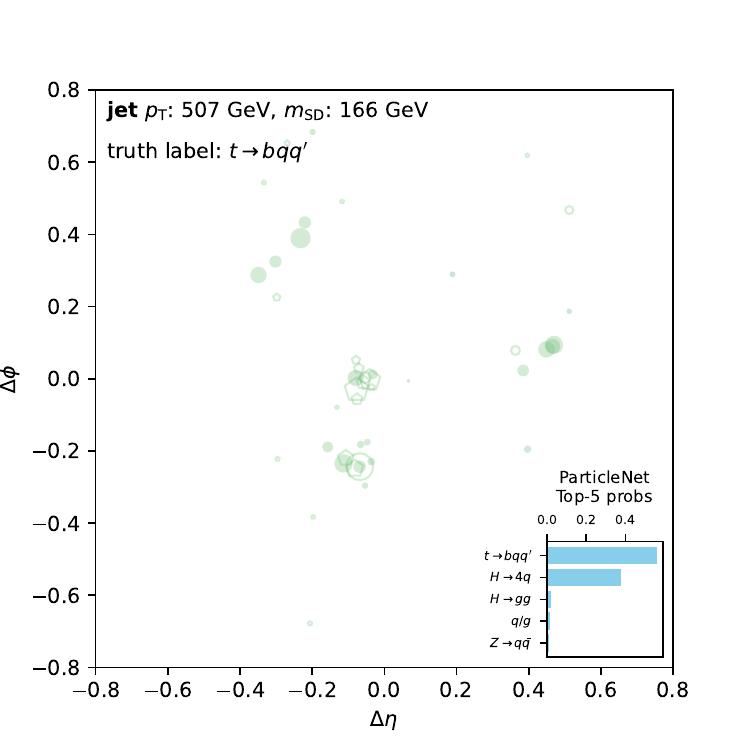} &
\includegraphics[width=0.24\textwidth,page=1]{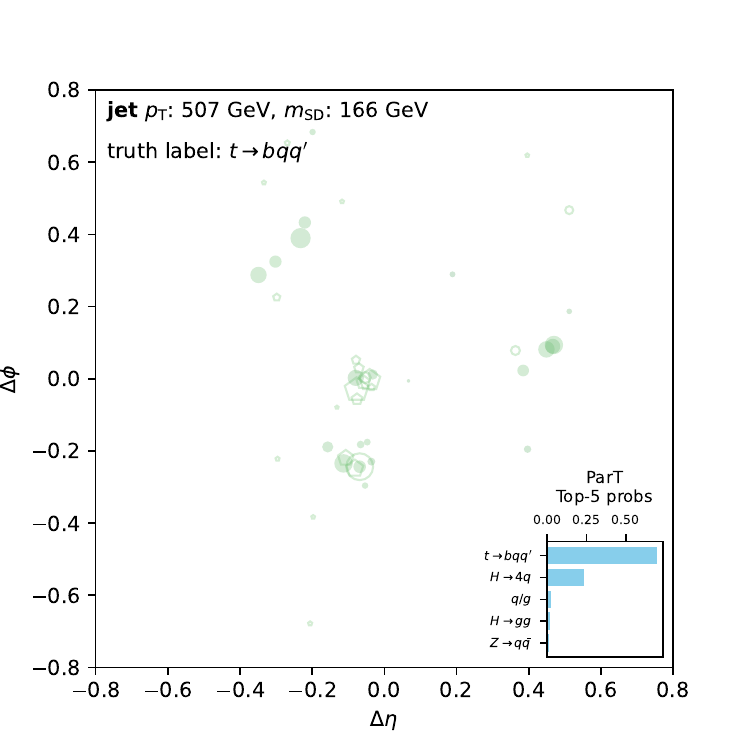} &
\includegraphics[width=0.24\textwidth,page=1]{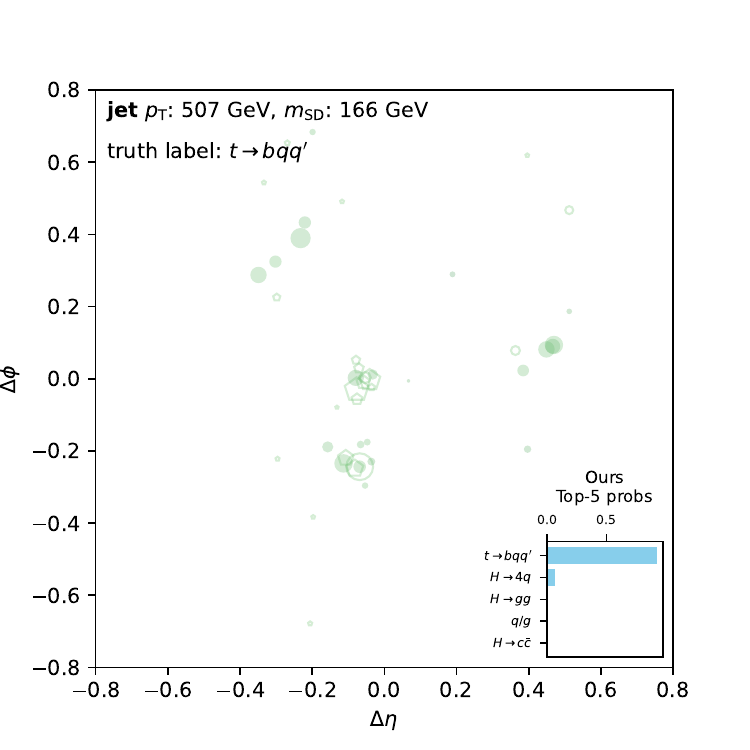} \\

\includegraphics[width=0.24\textwidth,page=1]{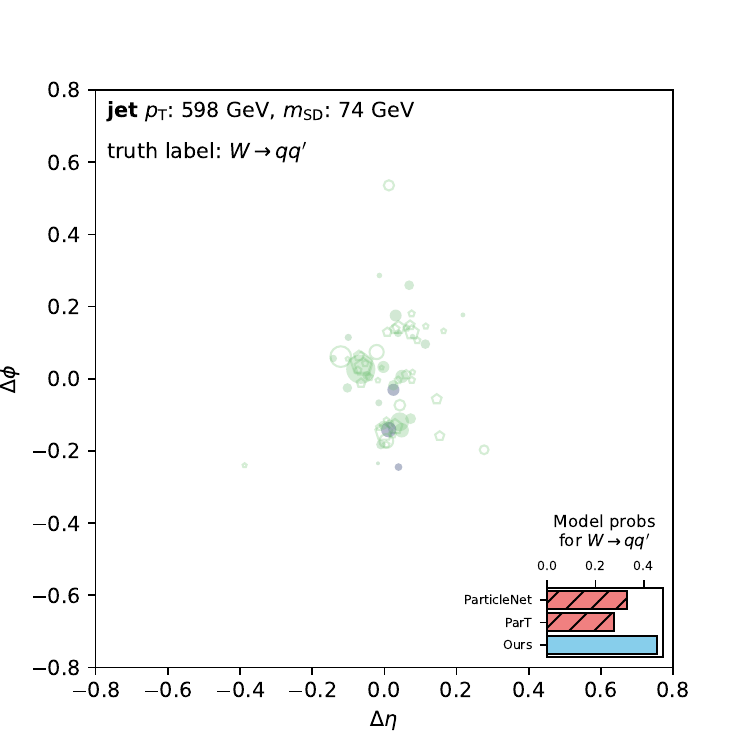} &
\includegraphics[width=0.24\textwidth,page=1]{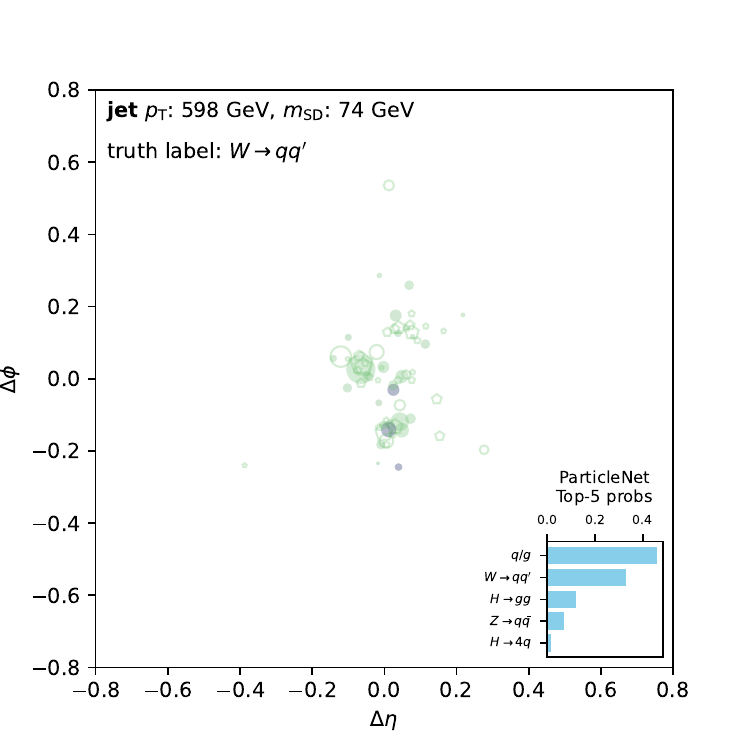} &
\includegraphics[width=0.24\textwidth,page=1]{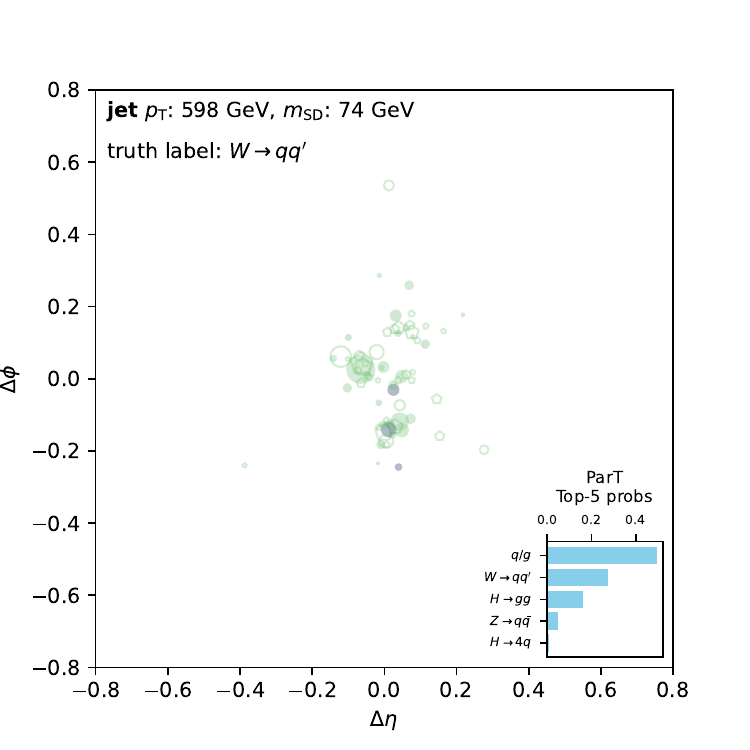} &
\includegraphics[width=0.24\textwidth,page=1]{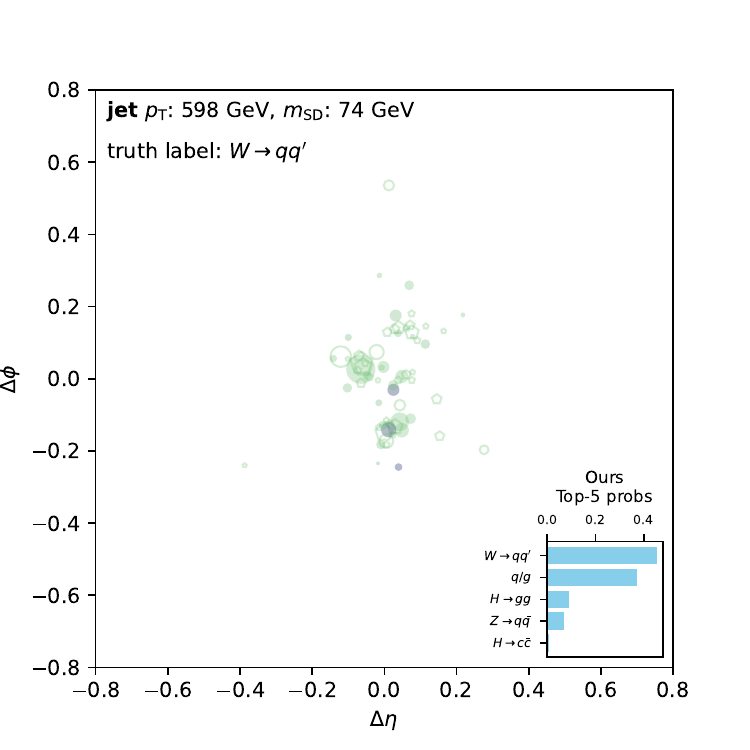} \\

\includegraphics[width=0.24\textwidth,page=1]{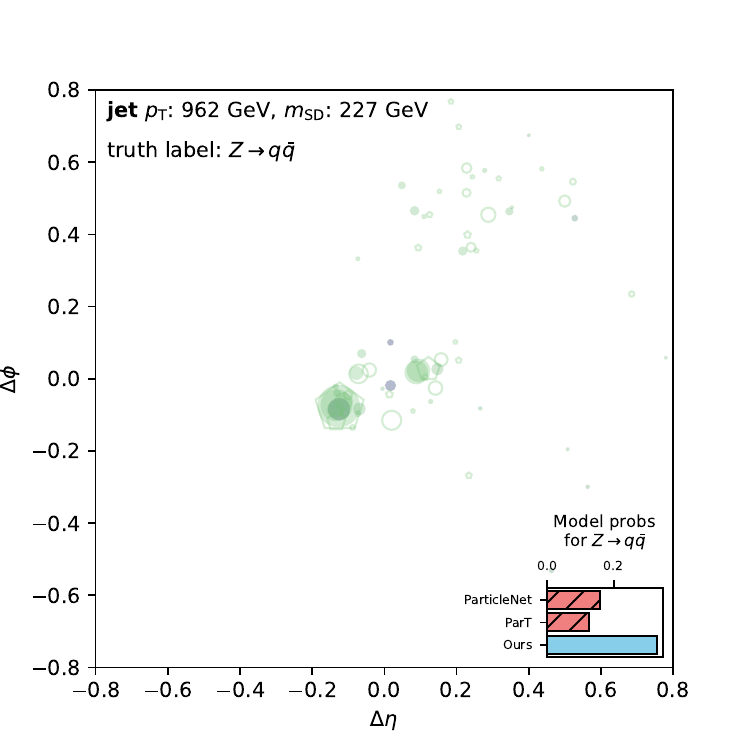} &
\includegraphics[width=0.24\textwidth,page=1]{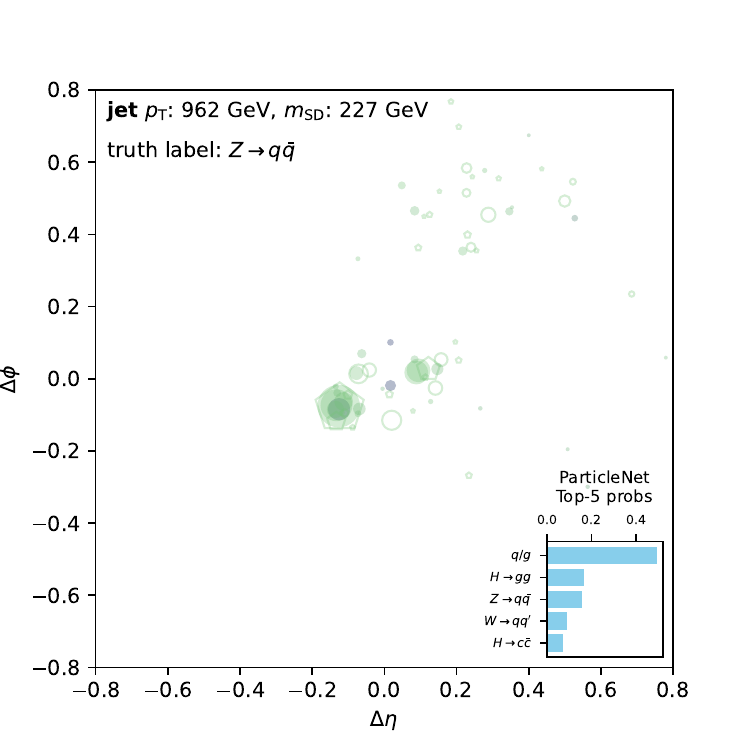} &
\includegraphics[width=0.24\textwidth,page=1]{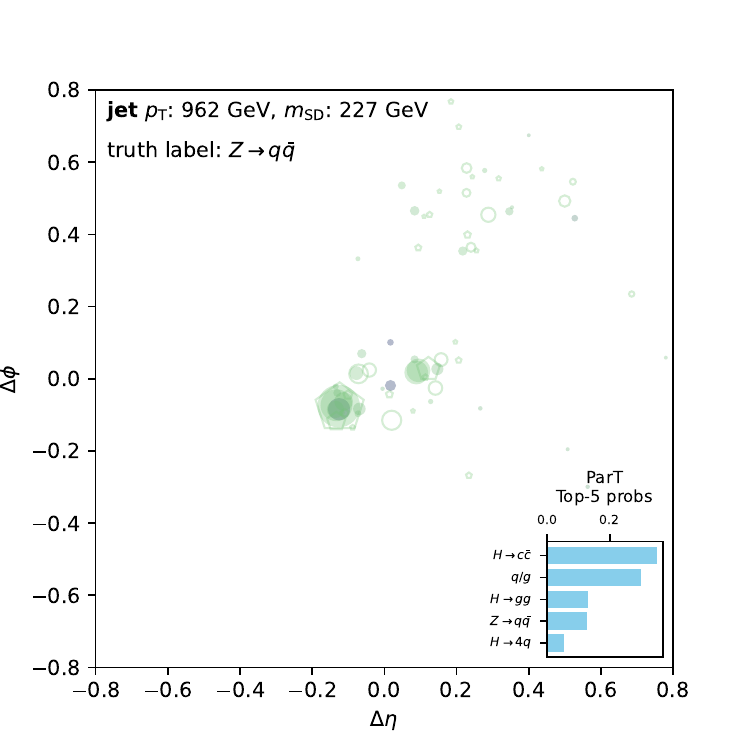} &
\includegraphics[width=0.24\textwidth,page=1]{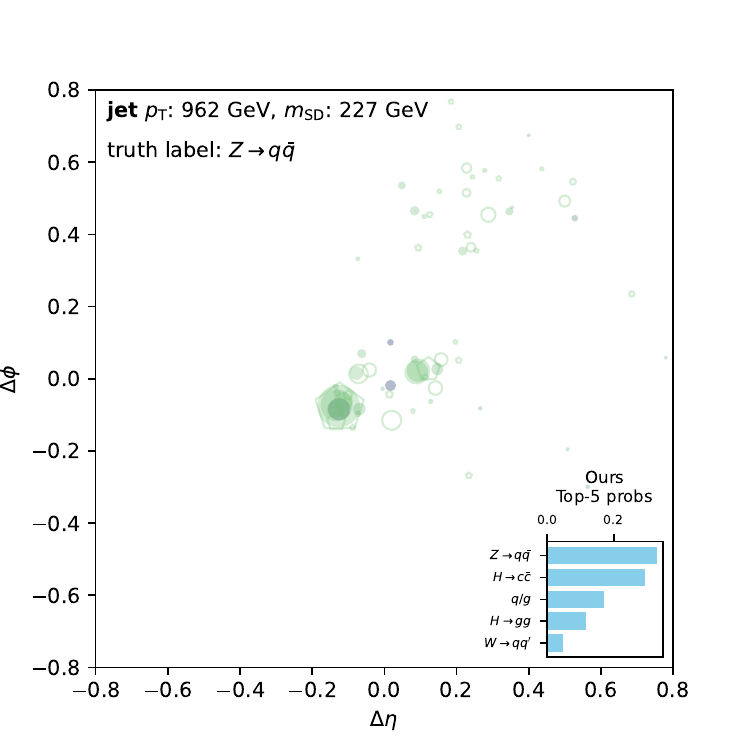} \\
\end{tabular}
\caption{\textbf{Qualitative results on JetClass highlighting successful JP-JEPA predictions (part 2).}}
\label{tab:qualitative:good2}
\end{figure*}

\begin{figure*}[ht!]
\centering
\setlength{\tabcolsep}{1pt} 
\begin{tabular}{cccccccc}
\includegraphics[width=0.24\textwidth,page=1]{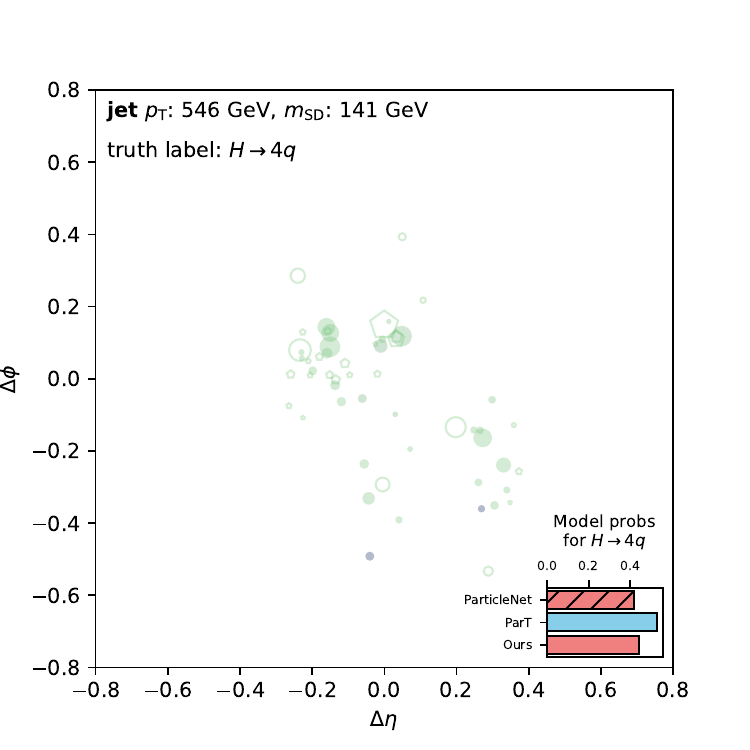} &
\includegraphics[width=0.24\textwidth,page=1]{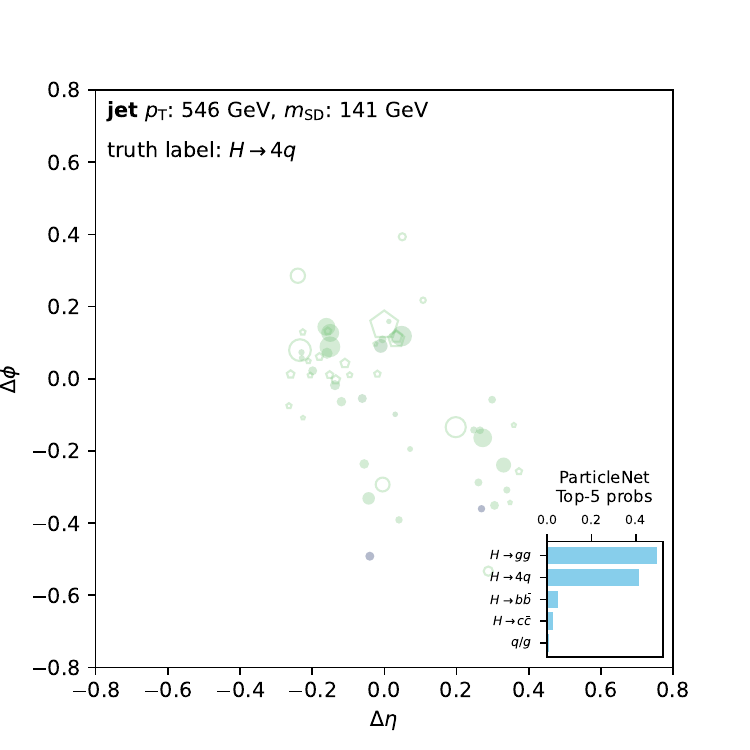} &
\includegraphics[width=0.24\textwidth,page=1]{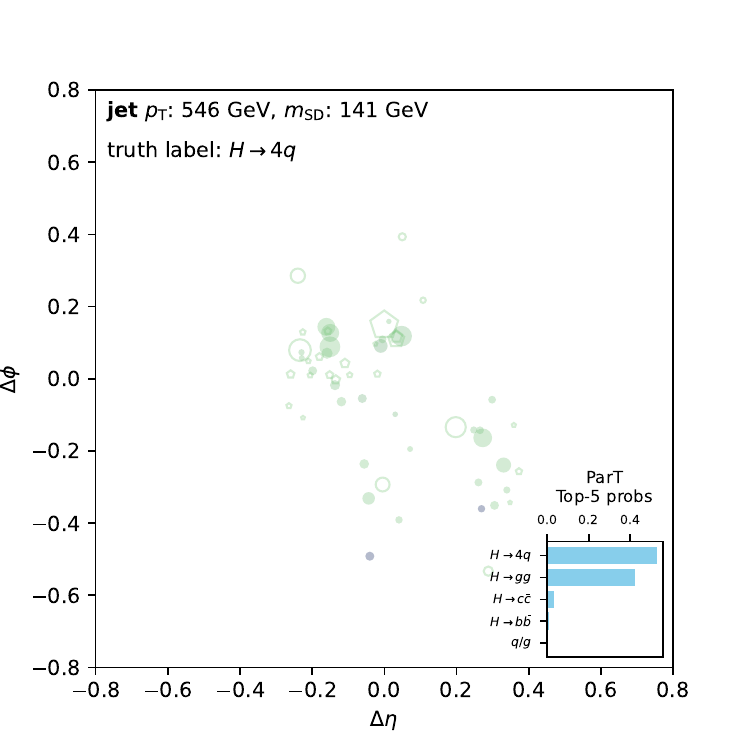} &
\includegraphics[width=0.24\textwidth,page=1]{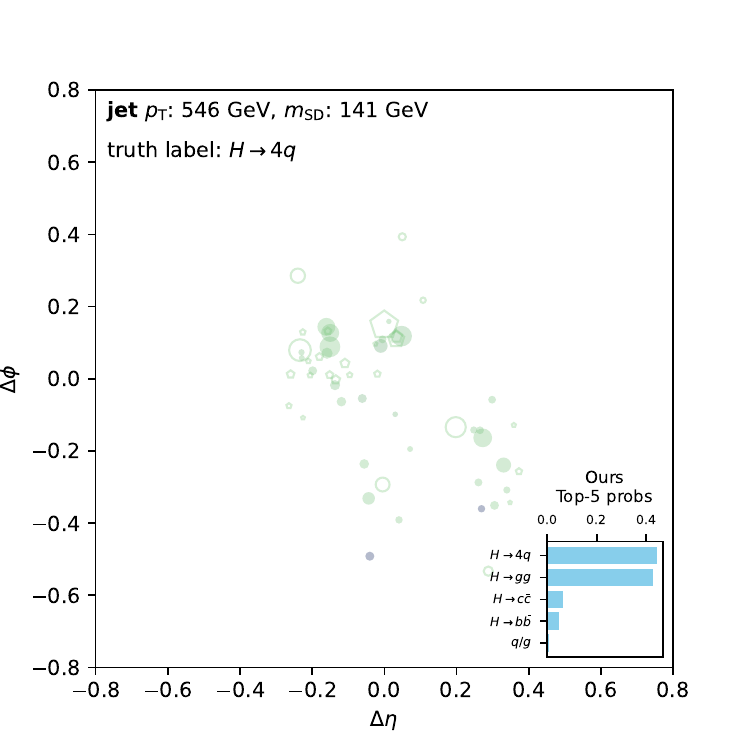} \\

\includegraphics[width=0.24\textwidth,page=1]{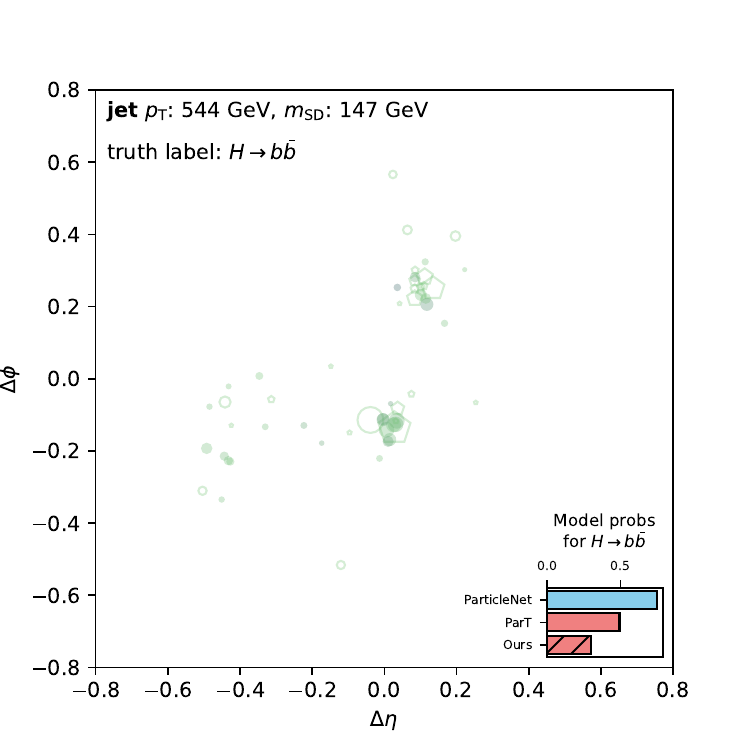} &
\includegraphics[width=0.24\textwidth,page=1]{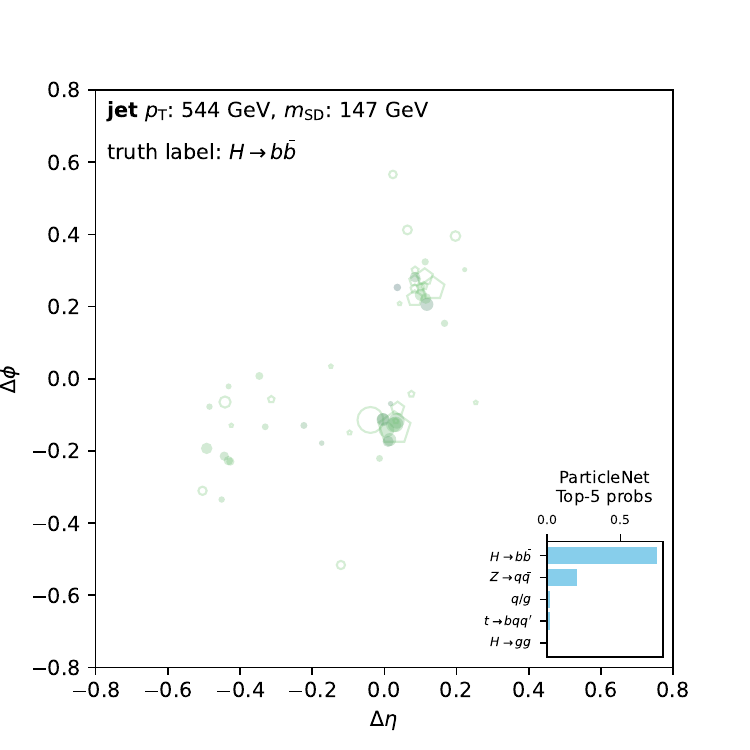} &
\includegraphics[width=0.24\textwidth,page=1]{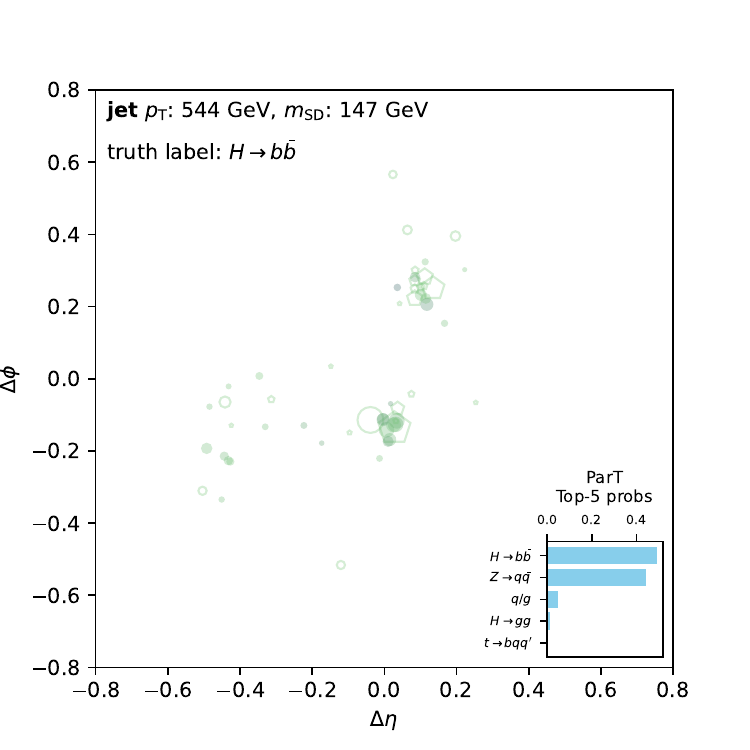} &
\includegraphics[width=0.24\textwidth,page=1]{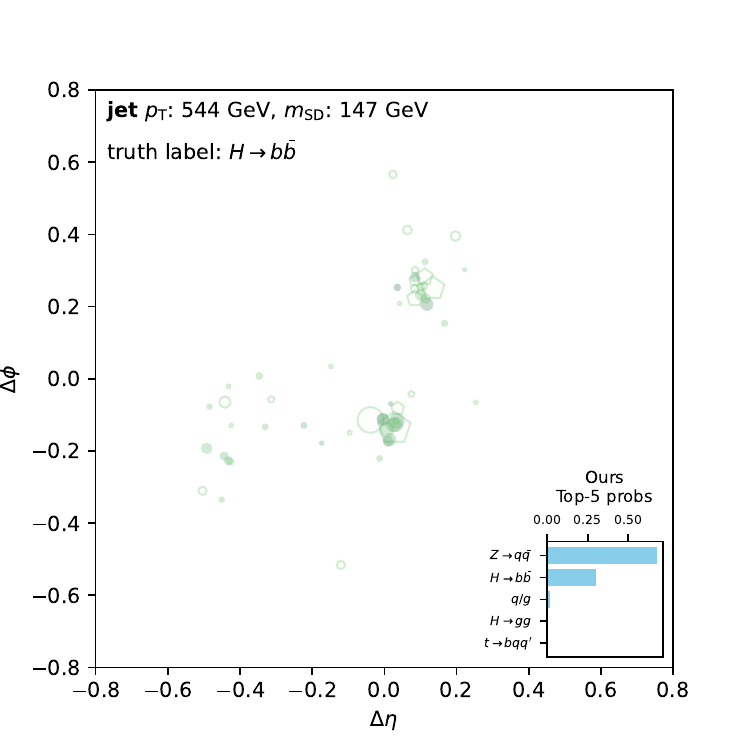} \\

\includegraphics[width=0.24\textwidth,page=1]{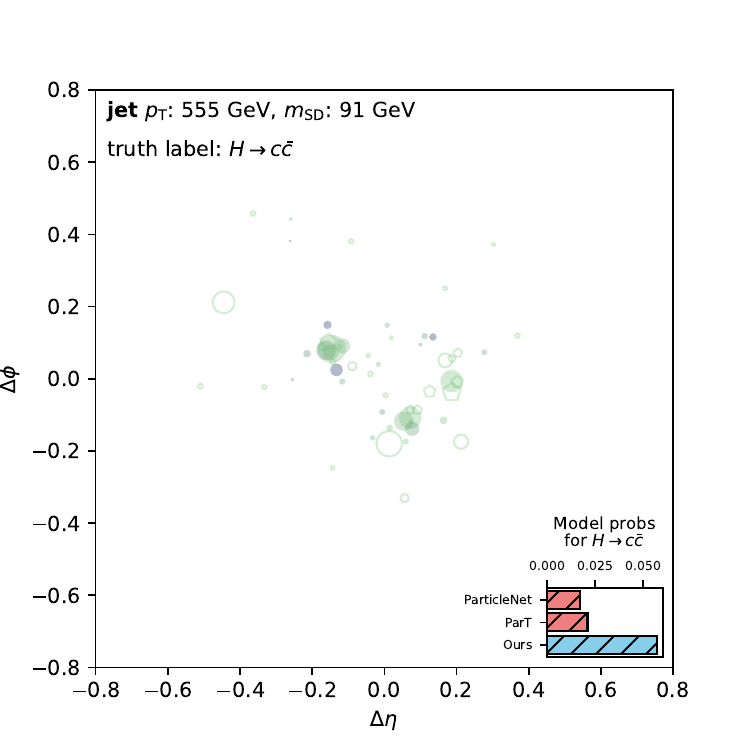} &
\includegraphics[width=0.24\textwidth,page=1]{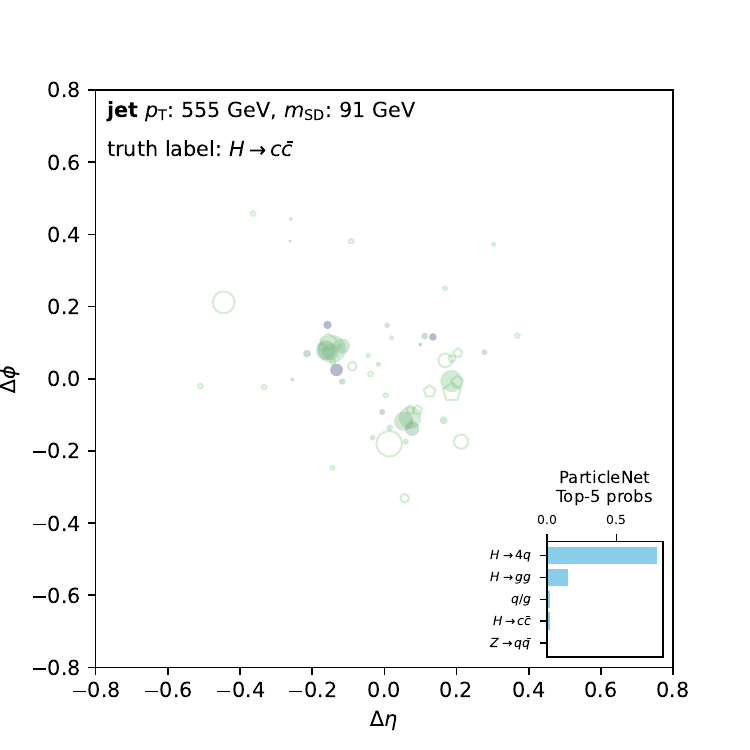} &
\includegraphics[width=0.24\textwidth,page=1]{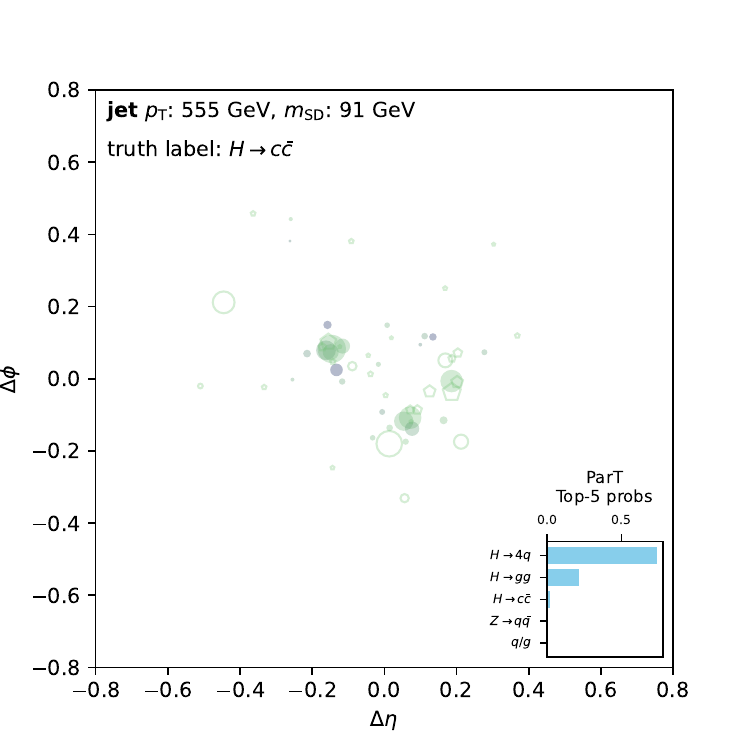} &
\includegraphics[width=0.24\textwidth,page=1]{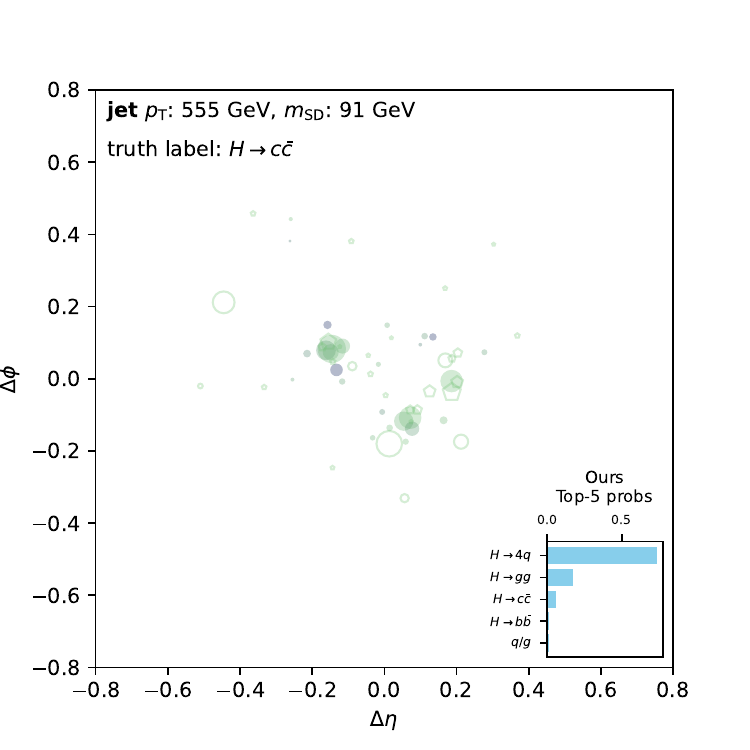} \\

\includegraphics[width=0.24\textwidth,page=1]{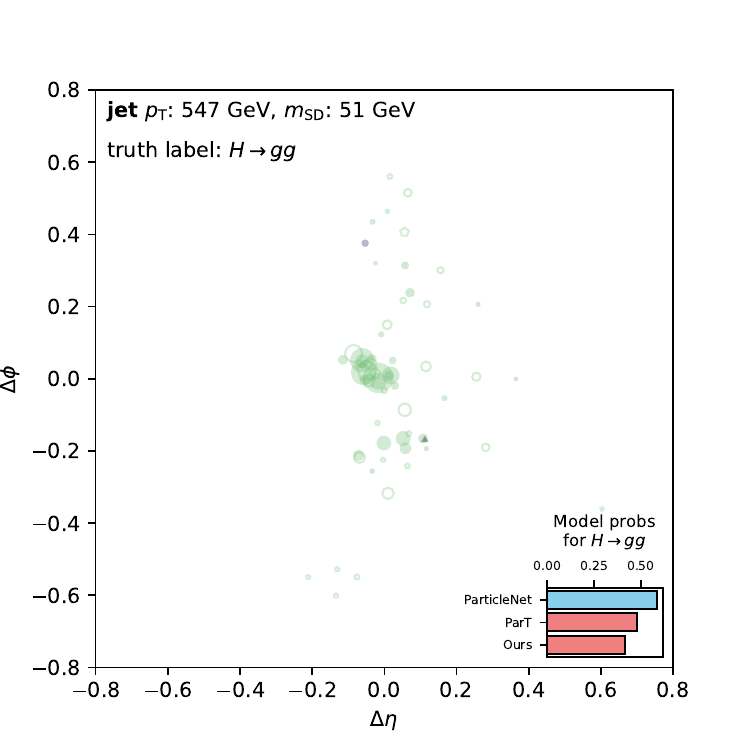} &
\includegraphics[width=0.24\textwidth,page=1]{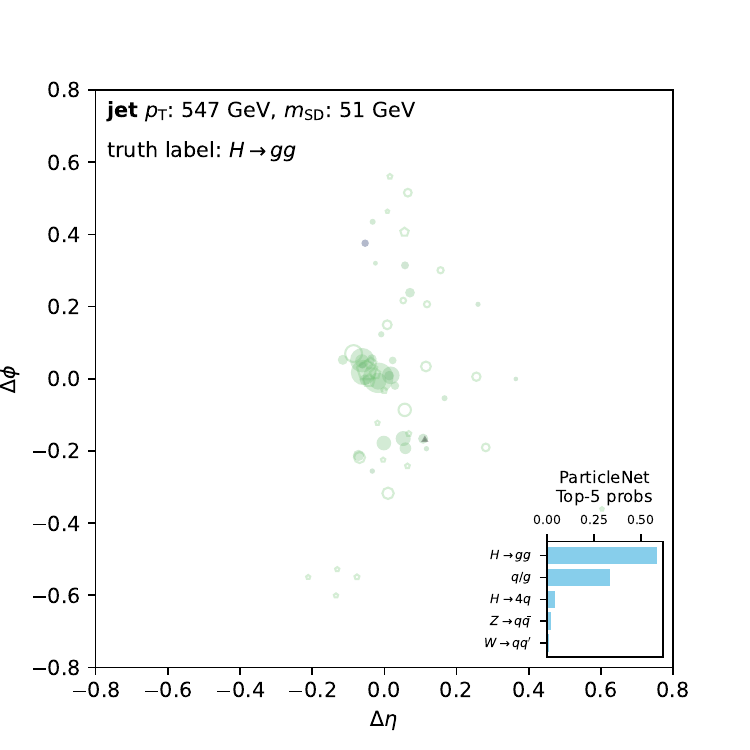} &
\includegraphics[width=0.24\textwidth,page=1]{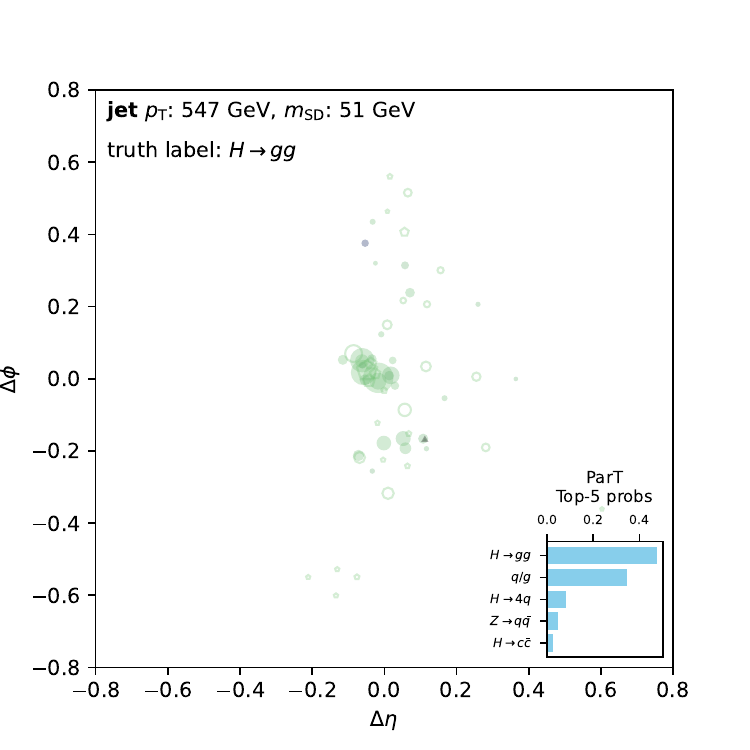} &
\includegraphics[width=0.24\textwidth,page=1]{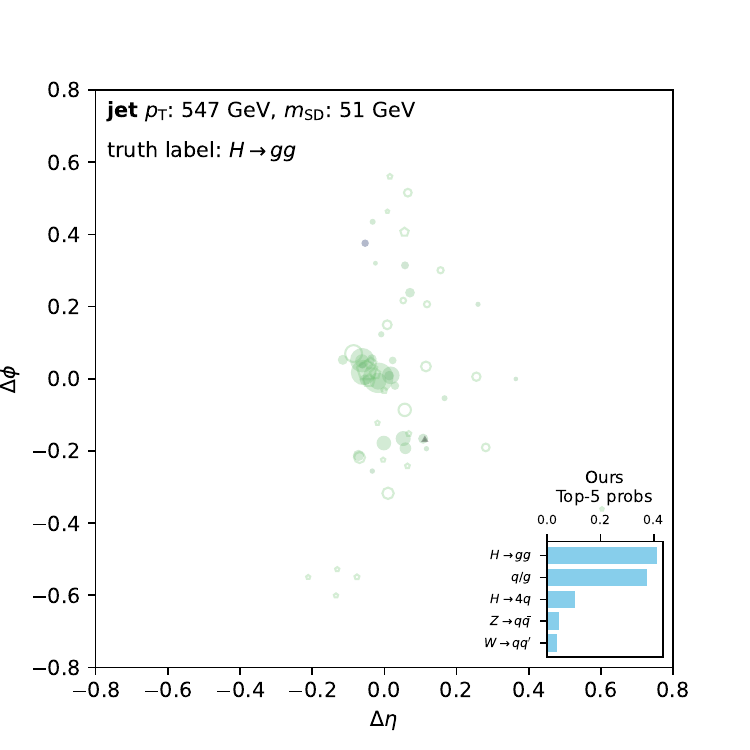} \\

\includegraphics[width=0.24\textwidth,page=1]{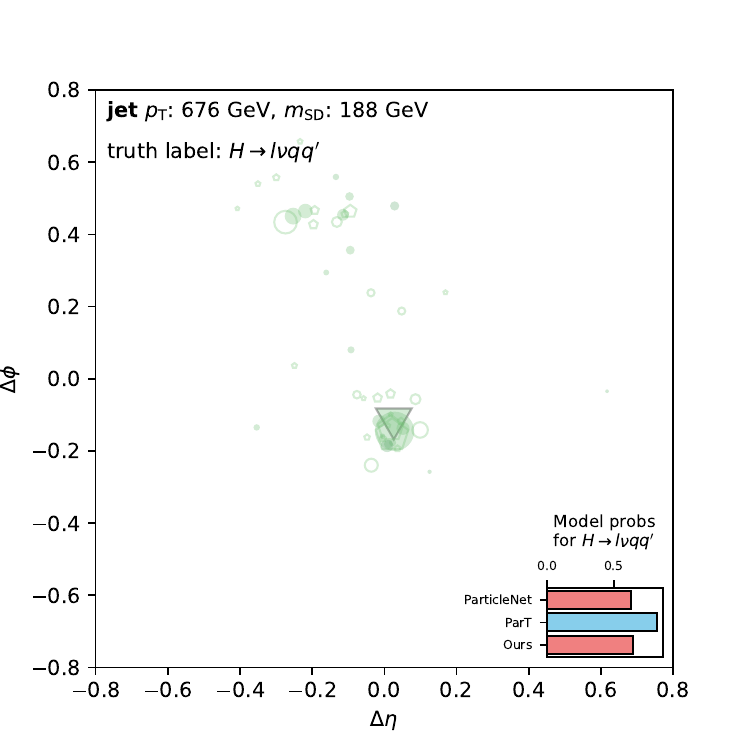} &
\includegraphics[width=0.24\textwidth,page=1]{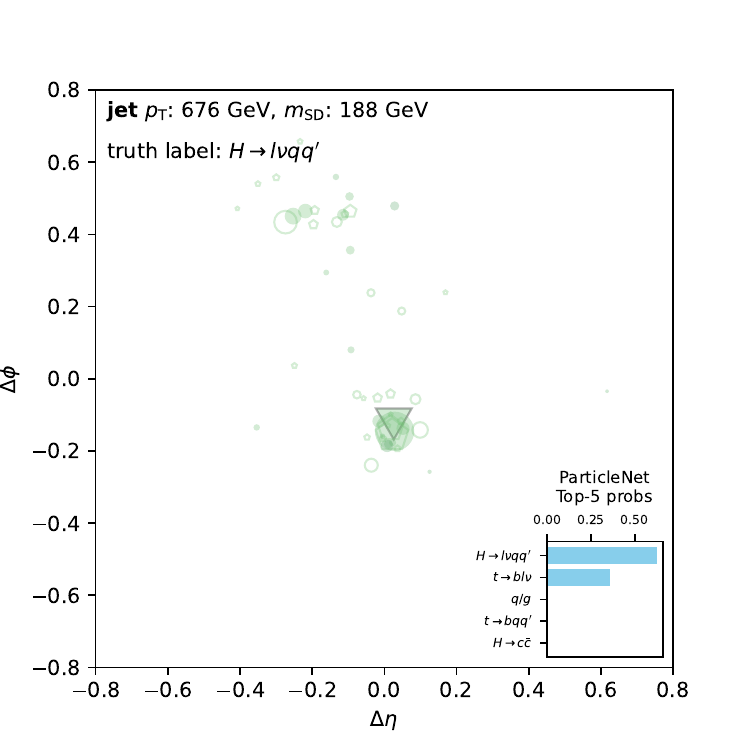} &
\includegraphics[width=0.24\textwidth,page=1]{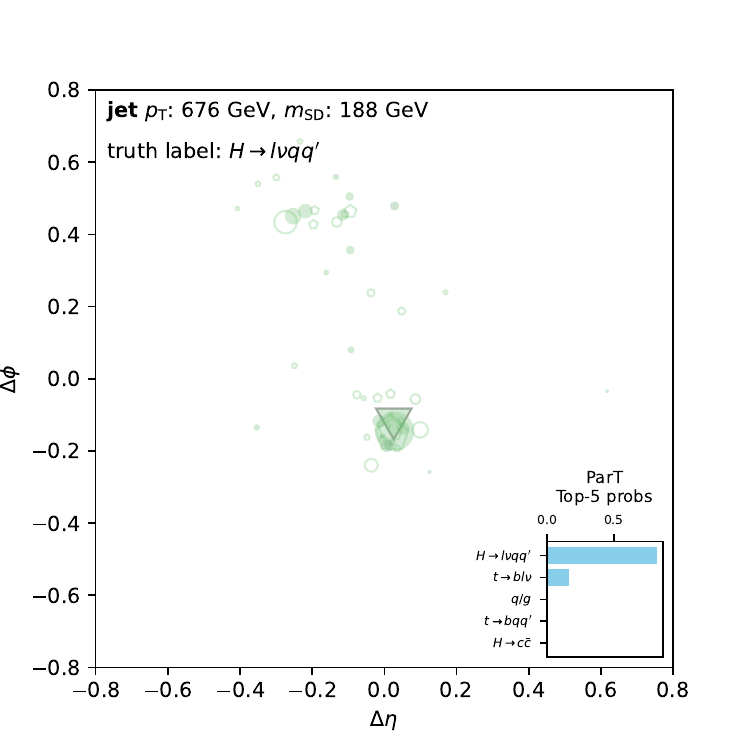} &
\includegraphics[width=0.24\textwidth,page=1]{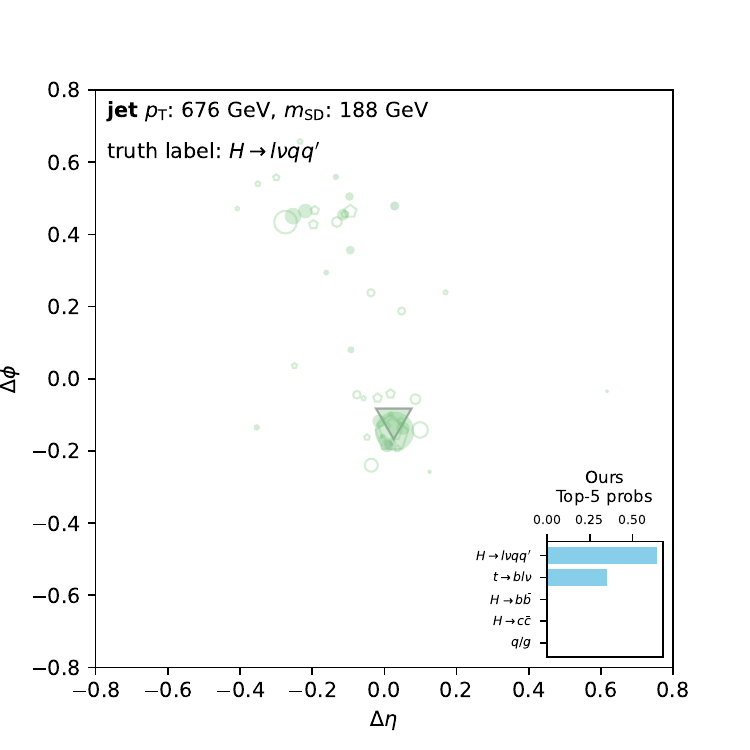} \\
\end{tabular}
\caption{\textbf{Qualitative results on JetClass highlighting JP-JEPA failure cases (part 1).}}
\label{tab:qualitative:bad1}
\end{figure*}

\begin{figure*}[ht!]
\centering
\setlength{\tabcolsep}{1pt} 
\begin{tabular}{cccccccc}
\includegraphics[width=0.24\textwidth,page=1]{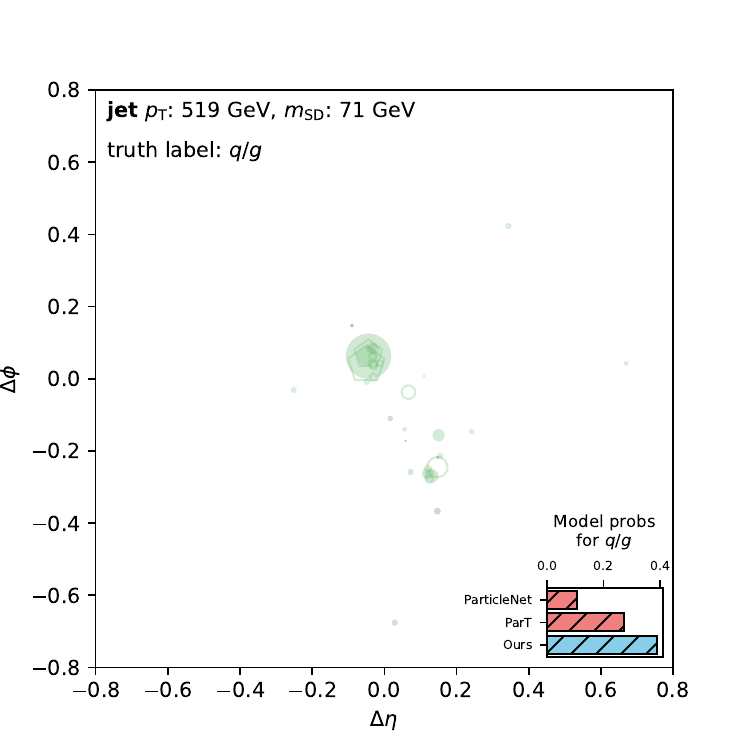} &
\includegraphics[width=0.24\textwidth,page=1]{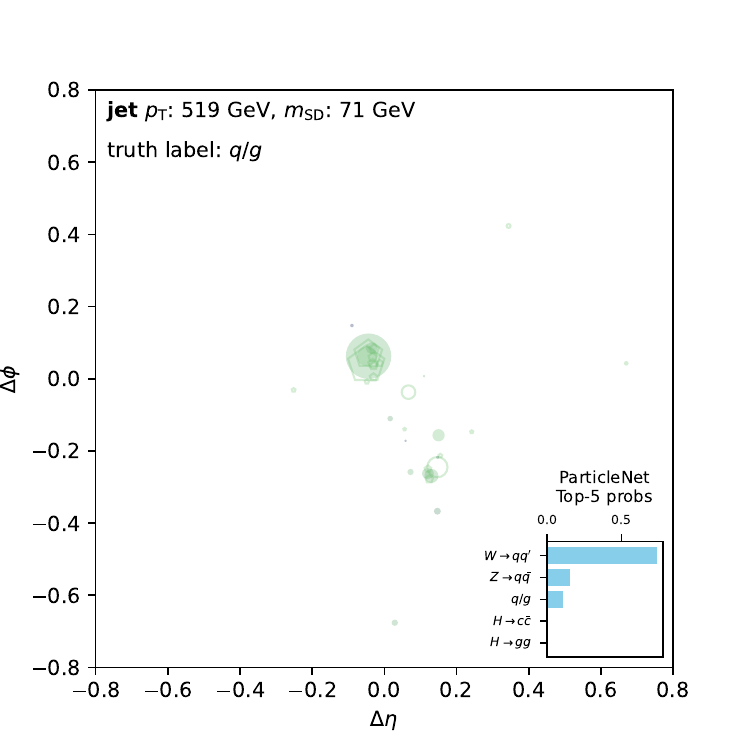} &
\includegraphics[width=0.24\textwidth,page=1]{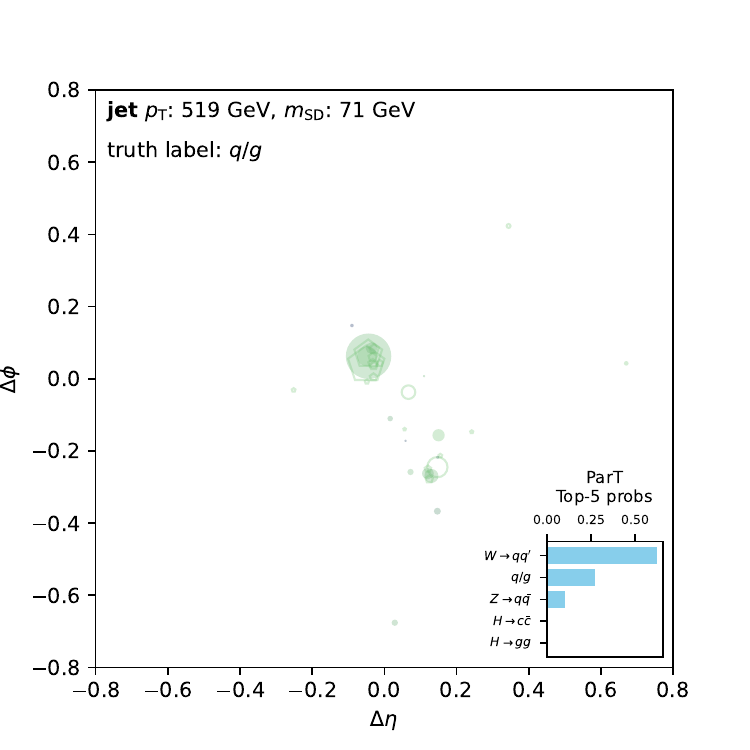} &
\includegraphics[width=0.24\textwidth,page=1]{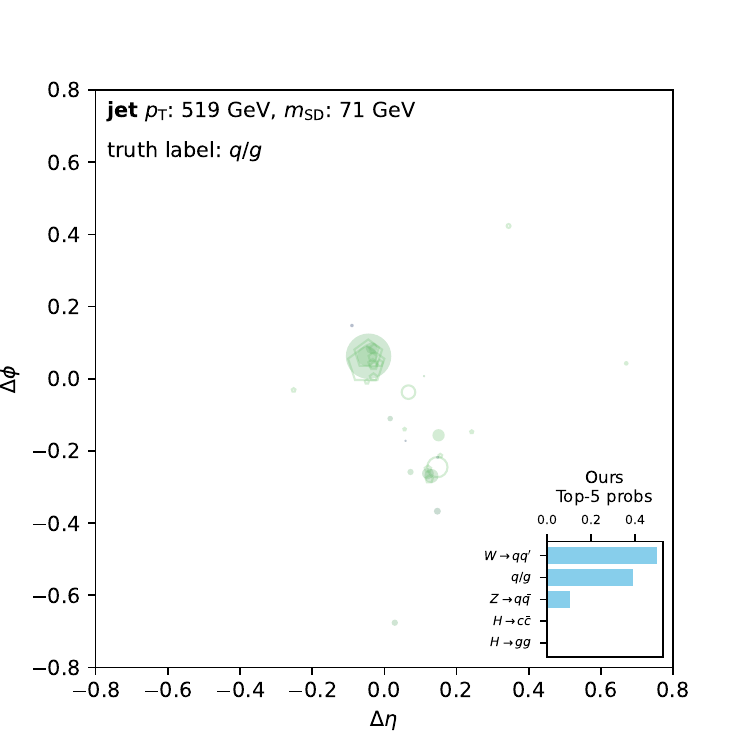} \\

\includegraphics[width=0.24\textwidth,page=1]{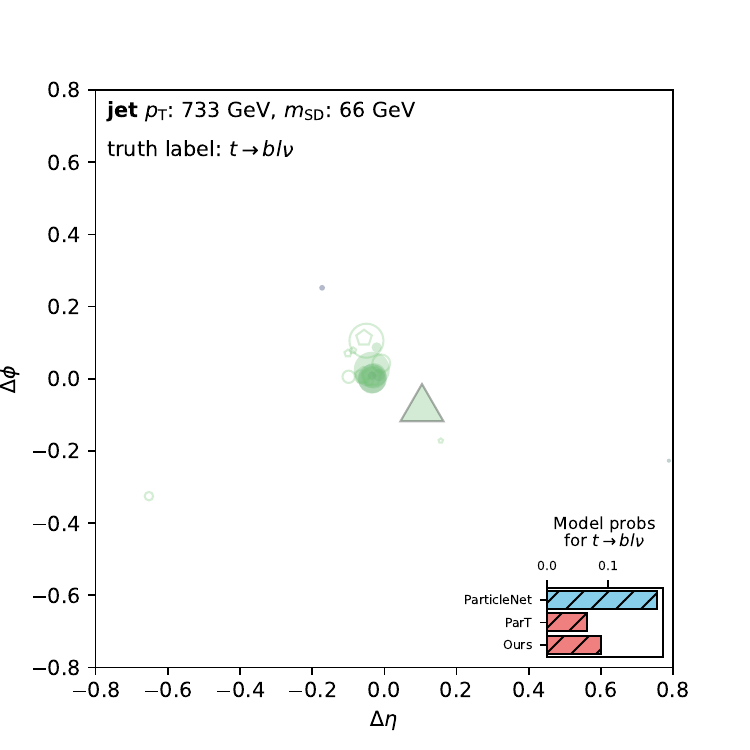} &
\includegraphics[width=0.24\textwidth,page=1]{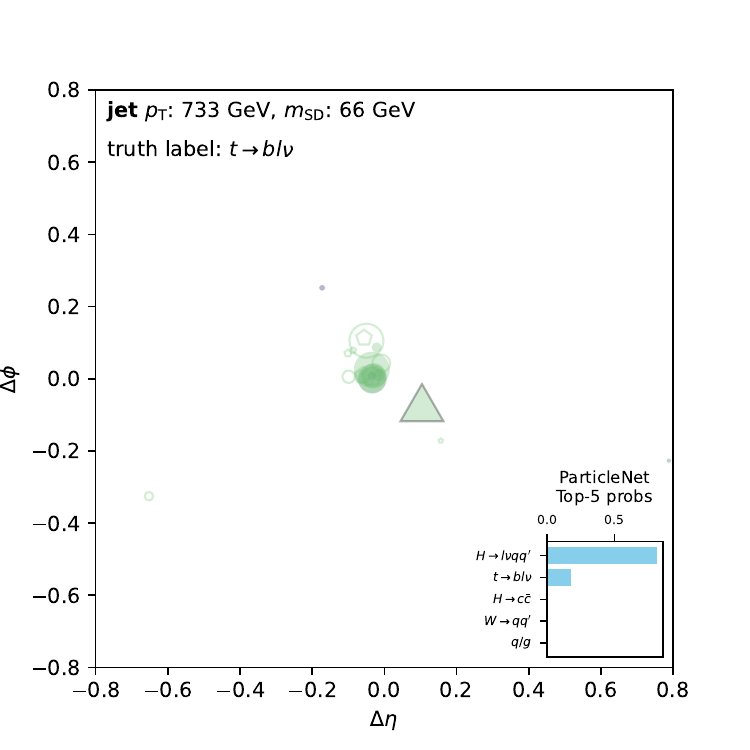} &
\includegraphics[width=0.24\textwidth,page=1]{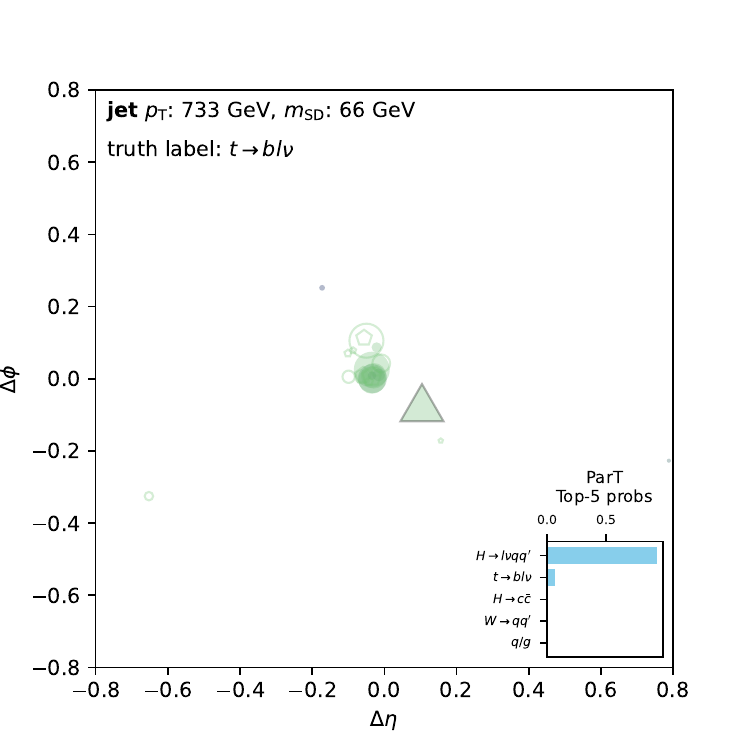} &
\includegraphics[width=0.24\textwidth,page=1]{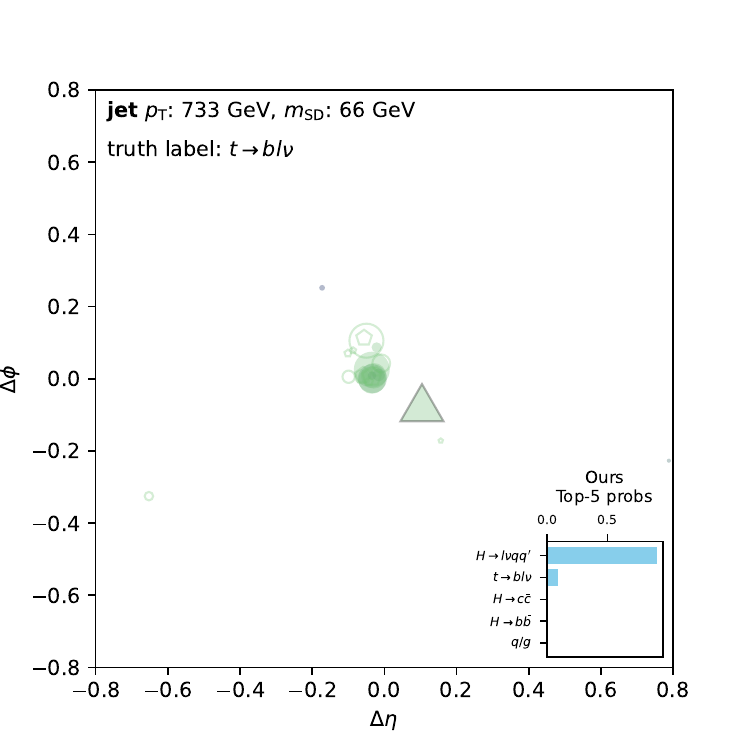} \\

\includegraphics[width=0.24\textwidth,page=1]{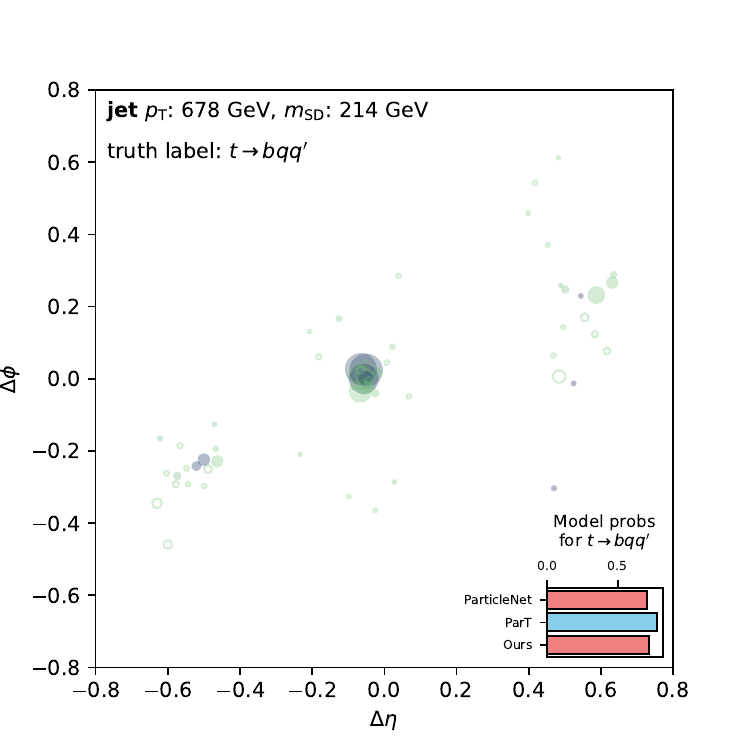} &
\includegraphics[width=0.24\textwidth,page=1]{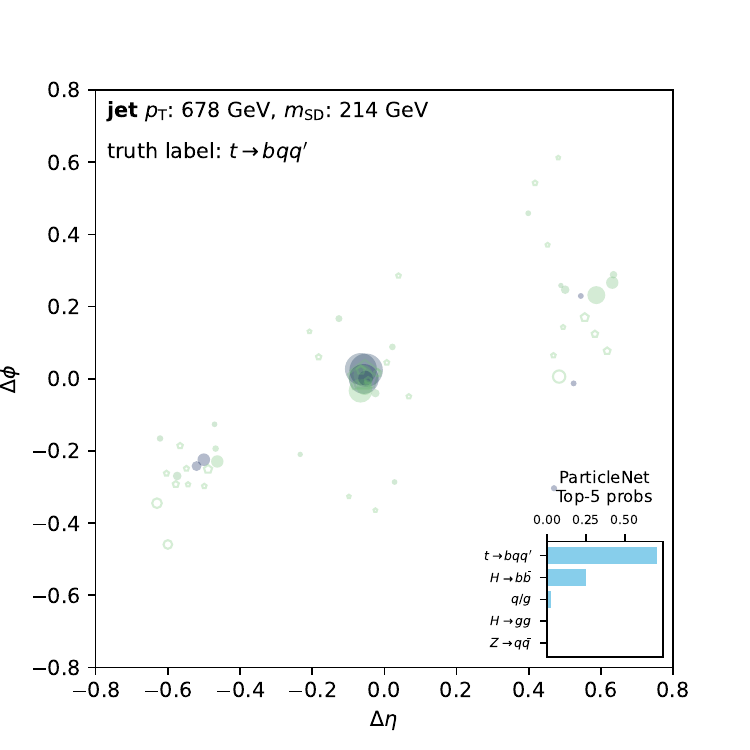} &
\includegraphics[width=0.24\textwidth,page=1]{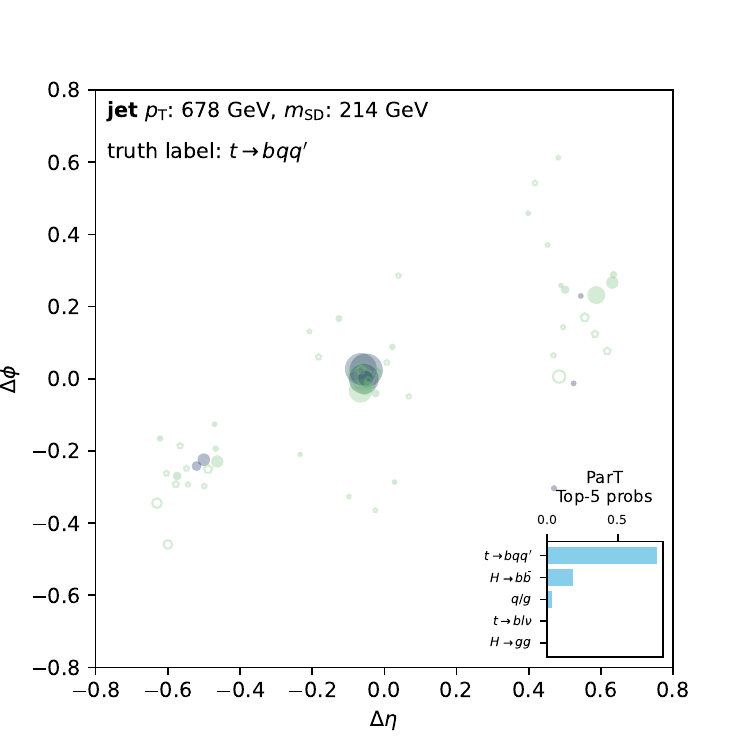} &
\includegraphics[width=0.24\textwidth,page=1]{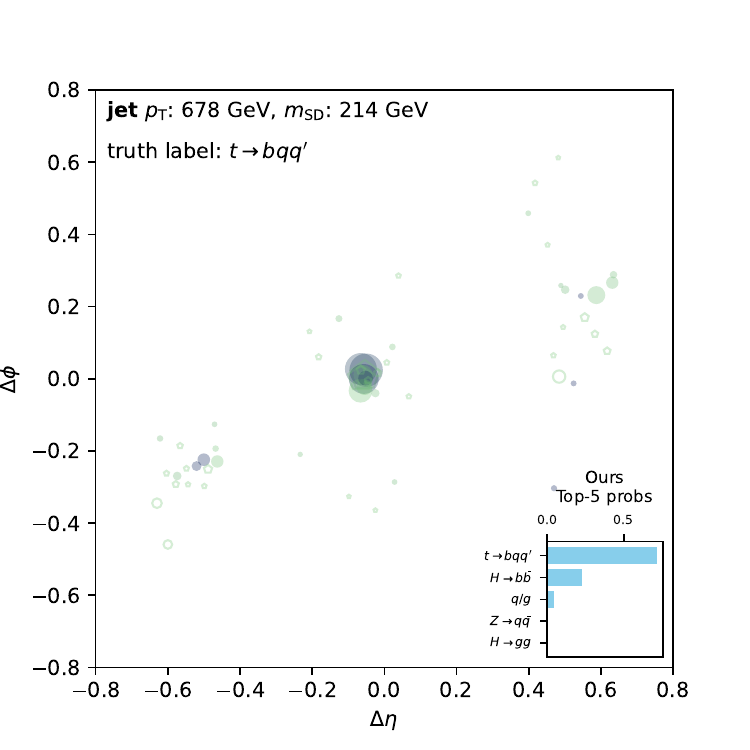} \\

\includegraphics[width=0.24\textwidth,page=1]{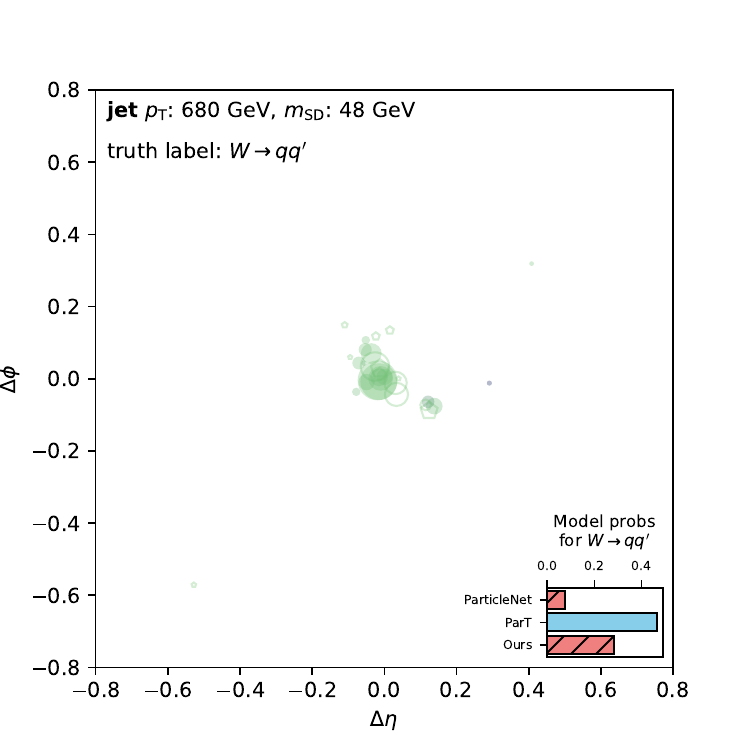} &
\includegraphics[width=0.24\textwidth,page=1]{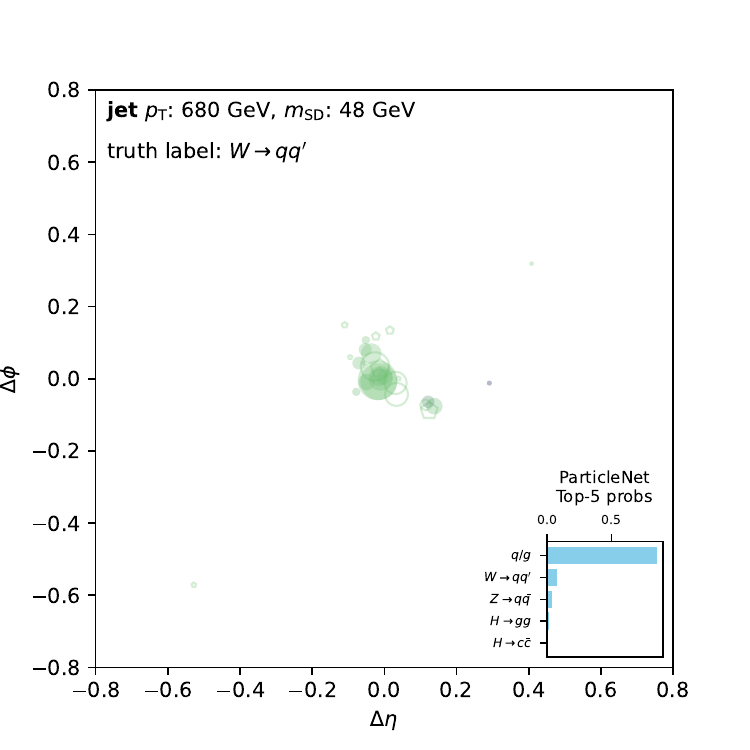} &
\includegraphics[width=0.24\textwidth,page=1]{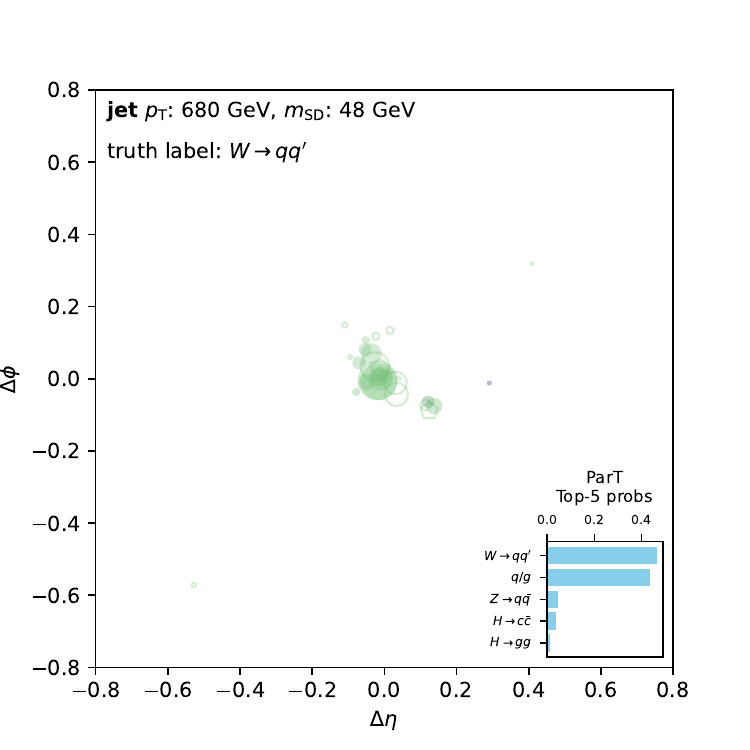} &
\includegraphics[width=0.24\textwidth,page=1]{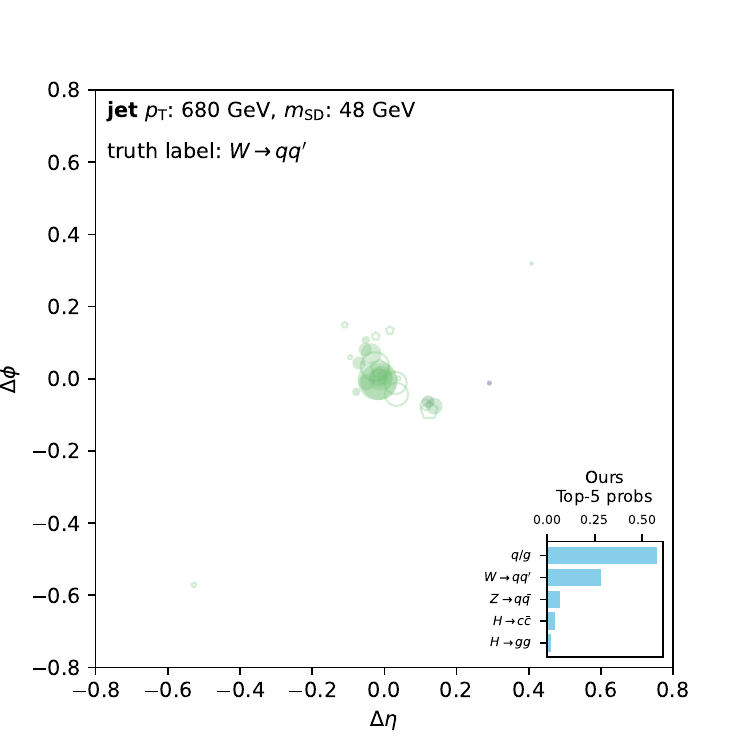} \\

\includegraphics[width=0.24\textwidth,page=1]{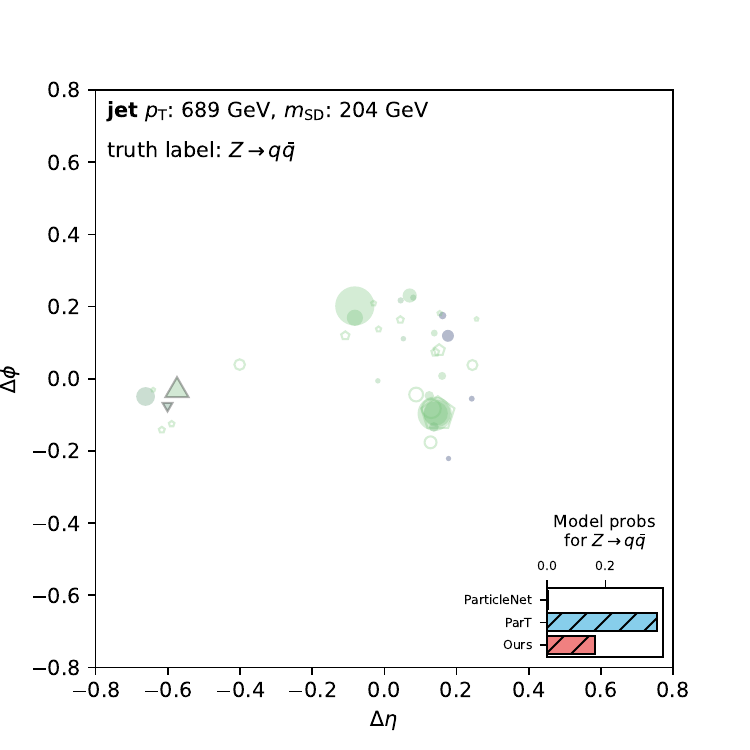} &
\includegraphics[width=0.24\textwidth,page=1]{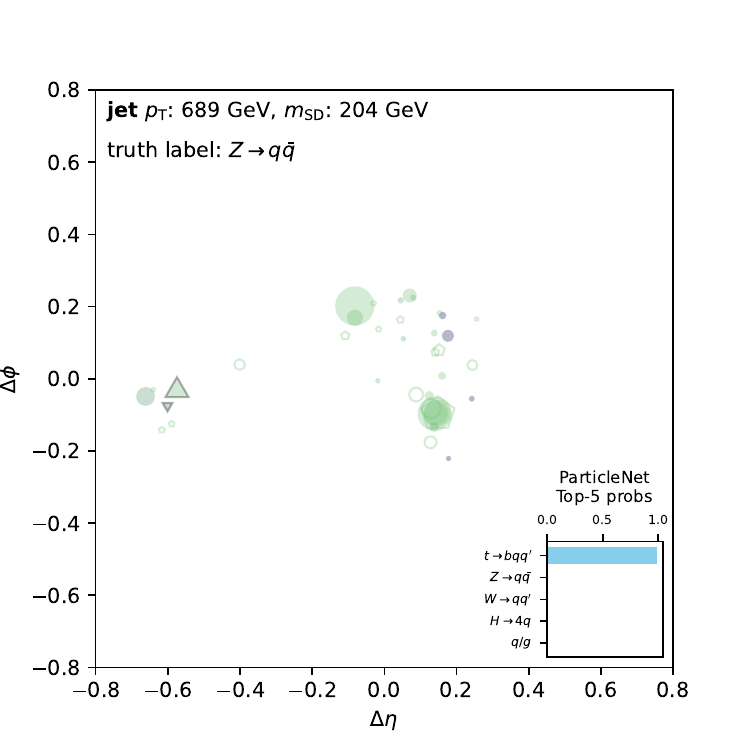} &
\includegraphics[width=0.24\textwidth,page=1]{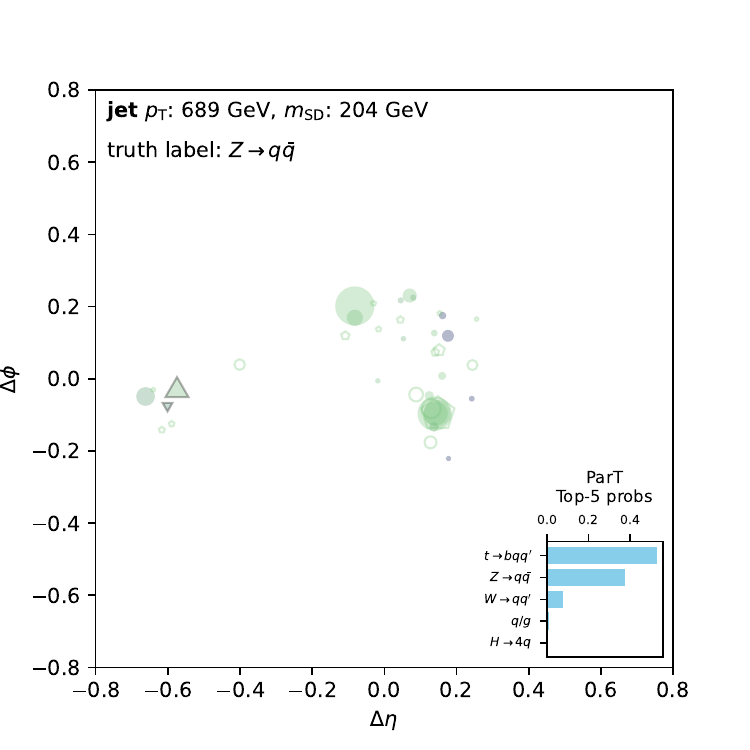} &
\includegraphics[width=0.24\textwidth,page=1]{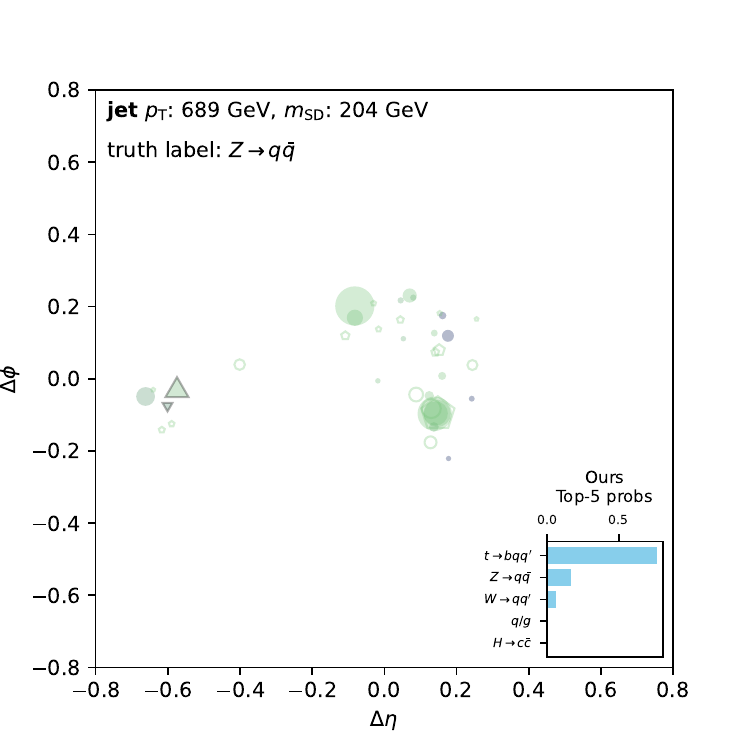} \\
\end{tabular}
\caption{\textbf{Qualitative results on JetClass highlighting JP-JEPA failure cases (part 2).}}
\label{tab:qualitative:bad2}
\end{figure*}

Figures \ref{tab:qualitative:good1} and \ref{tab:qualitative:good2} illustrate examples for which our method outperforms ParticleNet and ParT. Figures \ref{tab:qualitative:bad1} and \ref{tab:qualitative:bad2}, on the other hand, show examples of failure cases. One example from each class is provided.

\section{Hyper-parameters}
\label{sec:supp:hp}

\subsection{JP-JEPA architecture variants}
\label{sec:supp:hp:arch}
Table \ref{tab:appendix:part-sizes-hp} presents the hyperparameters for each size of ParT mentioned in the main paper.
\begin{table*}[ht!]
    \centering
    \resizebox{\textwidth}{!}{
    \begin{tabular}{lp{0.0001cm}cccccc}
        \toprule
        Hyper-parameter && ParT-Mini & ParT-Tiny & ParT-Small & ParT-Base & ParT-Large & ParT-Huge \\
        \midrule
        latent dimension && 128(64) & 192 & 384 & 768 & 1024 & 1280 \\
        embed dims && [128, 512, 128] & [192, 512, 192] & [384, 512, 384] & [768, 768, 768] & [1024, 1024, 1024] & [1280, 1280, 1280] \\
        num heads && 8(8) & 8(8) & 8(8) & 12(12) & 16(16) & 16(16) \\
        PAB depth && 8(4) & 12(6) & 12(6) & 12(6) & 24(12) & 32(16) \\
        CAB depth && 2(1) & 3(1) & 3(1) & 3(1) & 6(2) & 8(4) \\
        pair embedder dims && \multicolumn{6}{c}{[64, 64, 64]} \\
        \bottomrule
    \end{tabular}
    }
    \caption{\textbf{Hyper-parameters of different ParT sizes}. In \texttt{(-)} are values for associated predictor when pre-training.}
    \label{tab:appendix:part-sizes-hp}
\end{table*}

\subsection{Pre-training}
\label{sec:supp:hp:pretraining}
Table \ref{tab:appendix:pretraining-hp} shows hyperparameters for pre-training models w.r.t. the JetClass subset size for completeness.
\begin{table*}[ht!]
    \centering
    \resizebox{\textwidth}{!}{
    \begin{tabular}{lp{0.0001cm}ccc}
        \toprule
        Hyper-parameter && JP-JEPA Mini (ablation) & JP-JEPA Mini & JP-JEPA Small \\
        \midrule
        \textit{data} \\
        datasets && JetClass~\cite{qu2022jetclass} (5\%) & \multicolumn{2}{c}{JetClass~\cite{qu2022jetclass} (100\%)} \\
        filtering && \multicolumn{3}{c}{jets with $>V$ constituents} \\
        \midrule
        \textit{augmentation} \\
        PID dropout && \multicolumn{3}{c}{0.3 (when \textit{kinpid} or \textit{full})} \\
        TD dropout && \multicolumn{3}{c}{0.6 (when \textit{full})} \\
        \# targets ($V$) && \{1, 2, 4\} & \multicolumn{2}{c}{1} \\
        target ratio && \multicolumn{2}{c}{[0.15, 0.2]} & [0.25, 0.35] \\
        target sampling && \multicolumn{3}{c}{contiguous} \\
        context ratio && \multicolumn{2}{c}{[0.4, 0.75]} & [0.4, 0.7] \\
        context sampling && \multicolumn{3}{c}{contiguous} \\
        \midrule
        \textit{optimization} \\
        total batch size && 1024 & \multicolumn{2}{c}{2048} \\
        max steps && 48000 & \multicolumn{2}{c}{500000 ($\sim$ 10 epochs)} \\
        optimizer && \multicolumn{3}{c}{AdamW~\cite{DBLP:conf/iclr/LoshchilovH19}} \\
        encoder weight decay && \multicolumn{3}{c}{$5\times 10^{-2}$} \\
        predictor weight decay && \multicolumn{2}{c}{$5\times 10^{-2}$} & $1.5 \times 10^{-1}$ \\
        momentum && \multicolumn{3}{c}{$0.9$} \\
        lr && $5\times 10^{-4}$ & $10^{-3}$ & $3\times 10^{-4}$ \\
        start lr && \multicolumn{3}{c}{$10^{-5}$} \\
        final lr && \multicolumn{3}{c}{$10^{-6}$} \\
        lr scheduler && \multicolumn{3}{c}{Cosine annealing} \\
        warmup type && \multicolumn{3}{c}{linear} \\
        warmup steps && \multicolumn{3}{c}{5\% of total steps} \\
        $\tau_{\text{ema}}$ && \multicolumn{3}{c}{$0.9995 \rightarrow 0.99999$} \\
        $\tau_{\text{ema}}$ scheduler && \multicolumn{3}{c}{linear on total steps} \\
        gradient clipping value && \multicolumn{2}{c}{-} & 0.5 (norm) \\
        $\lambda_1$ && \multicolumn{3}{c}{1.0} \\
        $\mathcal{L}_{\text{global}}$ && \{predictor \texttt{[CLS]}, encoder \texttt{[CLS]} + predictor \texttt{[CLS]}\} & \multicolumn{2}{c}{predictor \texttt{[CLS]} only} \\
        $\lambda_2$ && \{0.5,1,2\} (\texttt{sg} enabled and disabled) & \multicolumn{2}{c}{1.0 (\texttt{sg} enabled)} \\
        $\lambda_3$ && \{0.5,1,2\} (\texttt{sg} enabled and disabled) & \multicolumn{2}{c}{1.0 (\texttt{sg} enabled)} \\
        $\lambda_4$ && \{0.5,1,2\} (\texttt{sg} enabled and disabled) & \multicolumn{2}{c}{1.0 (\texttt{sg} enabled)} \\
        $\mathcal{L}_{\text{physics}}$ && \{LMV,PIM,PFPIM\} & \multicolumn{2}{c}{lorentz-momentum-vector (LMV)} \\
        $\lambda_5$ && 0.1 for KoLeo \& \{0.01, 0.05, 0.1\} for SIGReg & \multicolumn{2}{c}{0.1 for KoLeo \& 0.05 for SIGReg} \\
        distributed KoLeo && \multicolumn{3}{c}{no} \\
        \midrule
        \textit{architecture} \\
        encoder && \multicolumn{2}{c}{ParT-Mini (Tab. \ref{tab:appendix:part-sizes-hp})} & ParT-Small (Tab. \ref{tab:appendix:part-sizes-hp}) \\
        predictor && \multicolumn{2}{c}{ParT-Mini-Predictor (Tab. \ref{tab:appendix:part-sizes-hp})} & ParT-Small-Predictor (Tab. \ref{tab:appendix:part-sizes-hp}) \\
        predictor pair embedder && \multicolumn{3}{c}{enabled} \\
        predictor use encoder pair embedder && \multicolumn{3}{c}{enabled} \\
        context \texttt{[CLS]} in predictor input && \multicolumn{3}{c}{enabled} \\
        \midrule
        \textit{hardware} \\
        dtype && \multicolumn{3}{c}{float32} \\
        accelerator && 1 A100 80G & 2 A100 80G & 4 A100 80G \\
        mutli-device strategy && - & \multicolumn{2}{c}{DDP} \\
        \bottomrule
    \end{tabular}
    }
    \caption{\textbf{Hyper-parameters for pre-training JP-JEPA on JetClass}.}
    \label{tab:appendix:pretraining-hp}
\end{table*}

\subsection{Fine-tunings}
\label{sec:supp:hp:finetuning}
Table \ref{tab:appendix:finetuning-hp} presents hyperparameters for fine-tuning models w.r.t. the JetClass subset size for completeness and for the other two jet tagging datasets. For retraining ParticleNet~\cite{qu2020jet} and ParT~\cite{qu2022particle}, we use the same HPs as presented here, but just the model is randomly initialized without pre-trained weights and biases.
\begin{table*}[t!]
    \centering
    \begin{tabular}{lp{0.0001cm}ccc}
        \toprule
        Hyper-parameter && \multicolumn{3}{c}{JP-JEPA} \\
        \midrule
        \textit{data} \\
        datasets && JetClass~\cite{qu2022jetclass} & Top quark~\cite{kasieczka2019top} & Quark-gluon~\cite{komiske2019energy} \\
        \midrule
        \textit{optimization} \\
        batch size && \multicolumn{3}{c}{512} \\
        max epochs or steps && 500k (More details in Tab. \ref{tab:jetclass-steps-study}) & \multicolumn{2}{c}{10 epochs} \\
        optimizer && \multicolumn{3}{c}{AdamW~\cite{DBLP:conf/iclr/LoshchilovH19}} \\
        weight decay && \multicolumn{3}{c}{$5\times 10^{-2}$} \\
        momentum && \multicolumn{3}{c}{$0.9$} \\
        lr encoder && $10^{-3}$ & \multicolumn{2}{c}{$10^{-4}$} \\
        lr head && $10^{-3}$ & \multicolumn{2}{c}{$10^{-4}$} \\
        start lr && \multicolumn{3}{c}{$10^{-6}$} \\
        final lr && \multicolumn{3}{c}{$10^{-6}$} \\
        lr scheduler && \multicolumn{3}{c}{Cosine annealing} \\
        warmup type && \multicolumn{3}{c}{linear} \\
        warmup steps && 5\% of total steps & \multicolumn{2}{c}{500} \\
        loss && \multicolumn{3}{c}{cross-entropy} \\
        encoder unfreeze && \multicolumn{3}{c}{no freeze} \\
        \midrule
        \textit{architecture} \\
        encoder && \multicolumn{3}{c}{Part-\{Mini,Small\}} \\
        mlp structure && \multicolumn{3}{c}{[linear, BN, ReLU, dropout]} \\
        mlp output dimensions && \multicolumn{3}{c}{[128, 128, \# classes]} \\
        mlp dropout && \multicolumn{3}{c}{0.5} \\
        \midrule
        \textit{hardware} \\
        dtype && \multicolumn{3}{c}{float32} \\
        accelerator && 1 A100 80G & \multicolumn{2}{c}{1 A100 80G} \\
        \bottomrule
    \end{tabular}
    \caption{\textbf{Hyper-parameters for fine-tuning JP-JEPA on downstream datasets on jet tagging}.}
    \label{tab:appendix:finetuning-hp}
\end{table*}

\end{document}